\documentclass[10pt,twocolumn]{article}
\usepackage{authblk}
\usepackage{url}
\usepackage{wrapfig}
\usepackage{color}
\usepackage{graphicx}
\usepackage{amsmath,amsfonts,amssymb,amsthm}
\usepackage{wrapfig,comment}
\usepackage{mathdefs}
\usepackage{proposal}
\usepackage{multirow}
\usepackage{tikz}
\usepackage{changepage}
\usepackage{my_macros}
\usepackage{float}
\graphicspath{{figures/}}
\usepackage[justification=justified,font=small]{caption}
\usepackage{subcaption}
\usepackage{graphicx}
\usepackage{units}
\usepackage{upgreek}
\usepackage{lipsum}
\usepackage[export]{adjustbox}
\usepackage{placeins}
\usepackage{cite}
\usepackage{booktabs}
\usepackage[switch]{lineno}	
\usepackage[a4paper, total={7in, 9in}]{geometry}
\usepackage{subfiles} 

\title{Non-invasive and noise-robust light focusing using confocal wavefront shaping}

\author[]{Dror Aizik}
\author[]{Anat Levin}
\affil[]{ Department of Electrical and Computer Engineering, Technion, Haifa, Israel}%

\begin{document}
	
	\twocolumn[
	\begin{@twocolumnfalse}
		\maketitle
		
		\begin{abstract}
			Wavefront-shaping is a promising approach for imaging fluorescent targets deep inside scattering tissue despite strong aberrations. It enables focusing an incoming illumination into a single spot inside tissue, as well as correcting the outgoing light scattered from the tissue, by modulating the incoming and/or outgoing wavefronts. Previously, wavefront shaping modulations have been successively estimated using feedback from strong fluorescent beads, which have been manually added to a sample. However, ideally, such feedback should be provided by the fluorescent components of the tissue itself, whose emission is orders of magnitude weaker than the one provided by beads. When a low number of photons is spread over multiple sensor pixels, the image is highly susceptible to noise, and the feedback signal required for
			previous algorithms cannot be detected. In this work, we suggest a wavefront shaping approach that works with a confocal modulation of both the illumination and imaging arms. Since the aberrations are corrected in the optics before the detector, the low photon budget can be directed into a single sensor spot and detected with high SNR. We derive a score function for modulation evaluation from mathematical principles, and successfully use it to image EGFP labeled neurons, despite scattering through thick tissue.
		\end{abstract}
		\vspace*{1cm}
	\end{@twocolumnfalse}
	]

	\section{Introduction}
\label{sec:intro}
Scattering forms one of the hardest barriers on 
 light-based approaches for tissue imaging.
A promising approach for overcoming  scattering  is a wavefront-shaping  correction.
By using a spatial light modulator (SLM) device, one can reshape the coherent wavefront illuminating the sample, such that its aberration  is conjugate to the aberration that will happen inside the tissue. When
such a wavefront propagates through the sample, all incoming light can be focused into a small spot. 
In the same way, one can modulate the outgoing wavefront so that light photons emerging from a single target point are brought into a single sensor point, despite the tissue aberration. The main advantage of this approach is that all light photons are used, unlike in ballistic-filtering approaches where scattered light is rejected.

Earlier adaptive optics approaches used such modulations to correct aberrations in the optical path~\cite{Booth2014,Ji2017review,HampsonBooth21review}.
More recently, wavefront shaping techniques~\cite{Horstmeyer15,YU2015632,Gigan22} have shown that it is possible to focus light through thick, highly-scattering layers~\cite{Vellekoop:07,Yaqoob2008,Vellekoop2010,Vellekoop2012}.

Despite the large potential of the idea, finding the desired shape of the modulation correction is rather challenging. 
The desired modulation varies between different  tissue samples and even varies spatially between different positions of the same tissue. 
For thick tissue, the modulation is a complex pattern containing a large number of free modes.
\begin{figure*}[t!]
	\begin{center}
		\includegraphics[width= 0.95\textwidth]{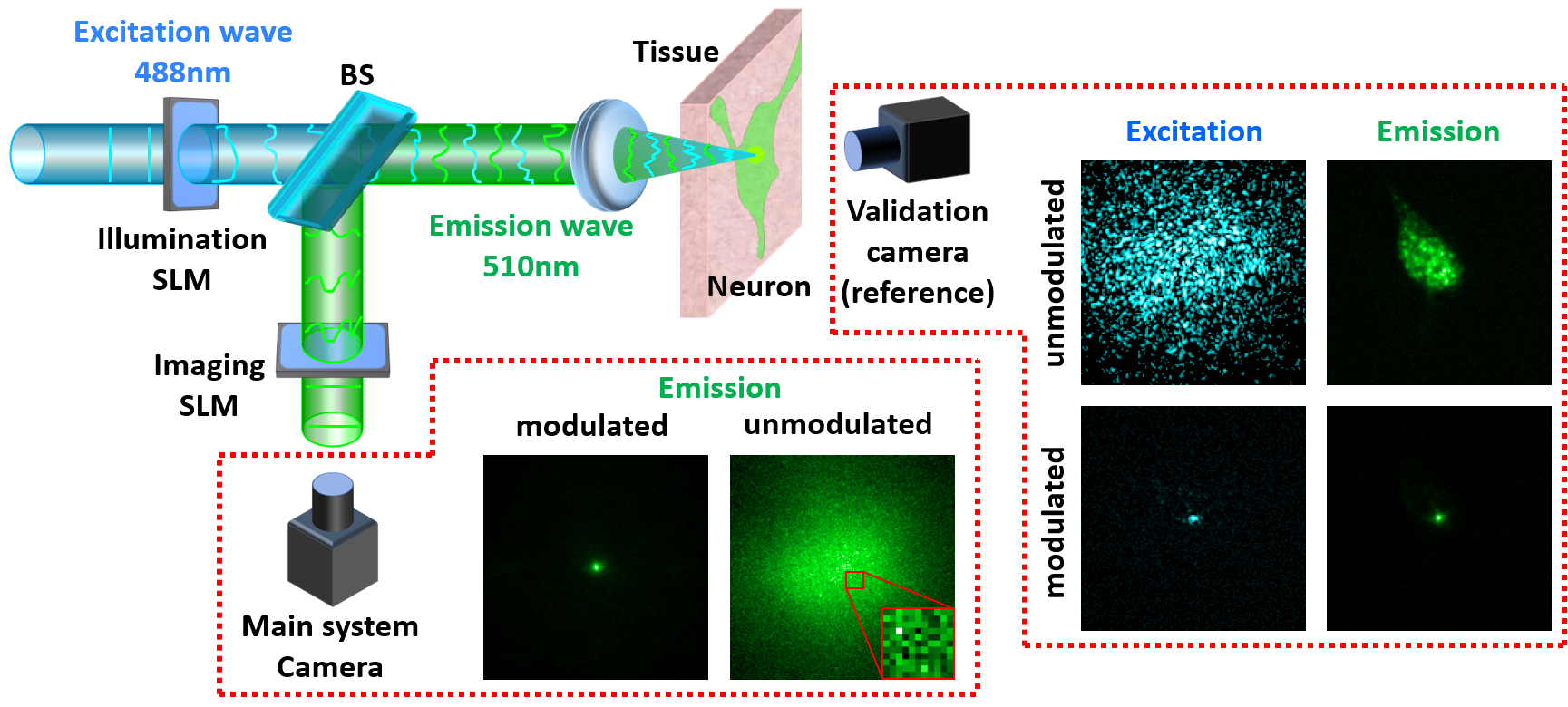}
	\end{center}
	\caption{System schematic: An incoming laser illumination  propagates through a layer of scattering tissue to excite a fluorescent neuron behind it. The emitted light is scattered again through the tissue toward a detector (main camera). In a conventional imaging system, the image of the neuron in the main camera is highly aberrated. Moreover, due to the weak emission, the image is very noisy. By adding two SLMs modulating the excitation and emission wavefronts, we can undo the tissue aberration. This allows us to reshape the incoming illumination into a spot on the neuron, and also to correct the emitted light and focus it into a spot in the main camera. For  reference, we also place a validation camera behind the tissue, which can image the neuron directly and validate that the excitation light has focused into a spot. This camera does not provide any input to our algorithm. We visualize example images from these two cameras with and without the modulation correction.   }\label{fig:setup_simp_1}
\end{figure*}

Earlier proof-of-concept demonstrations have used a validation camera behind the tissue to provide feedback to the algorithm~\cite{Vellekoop:07,Conkey:12,PopoffPhysRevLett2010,Vellekoop2010,Yaqoob2008,chen20203PointTM}, and other approaches have relied on the existence of a guiding star~\cite{Horstmeyer15,Tang2012,Katz:14,Wang20142PAdaptive,Liu2018,Fiolka:12,Jang:13,Xu11,Wang2012,Kong:11,Vellekoop2012}.
In the absence of such a guiding star, and when only non-invasive feedback is available, determining whether a wavefront has focused inside the tissue is not straightforward.  
The difficulty results from the fact that even if we can  find an illumination wavefront that actually focuses into a small spot 
inside the tissue, the light back-scattering from this  spot is aberrated again on its way to the camera, forming yet another scattered pattern. 

\begin{figure*}[t]
	\centering\begin{tabular}{@{}c@{~}c@{~~}c@{~}c@{~~}c@{~}c@{~~}c@{~}c@{}}

%
\begin{tikzpicture}
	\node[inner sep=0pt] (inset-bead) at (0,0){\includegraphics[height=0.18\textwidth]{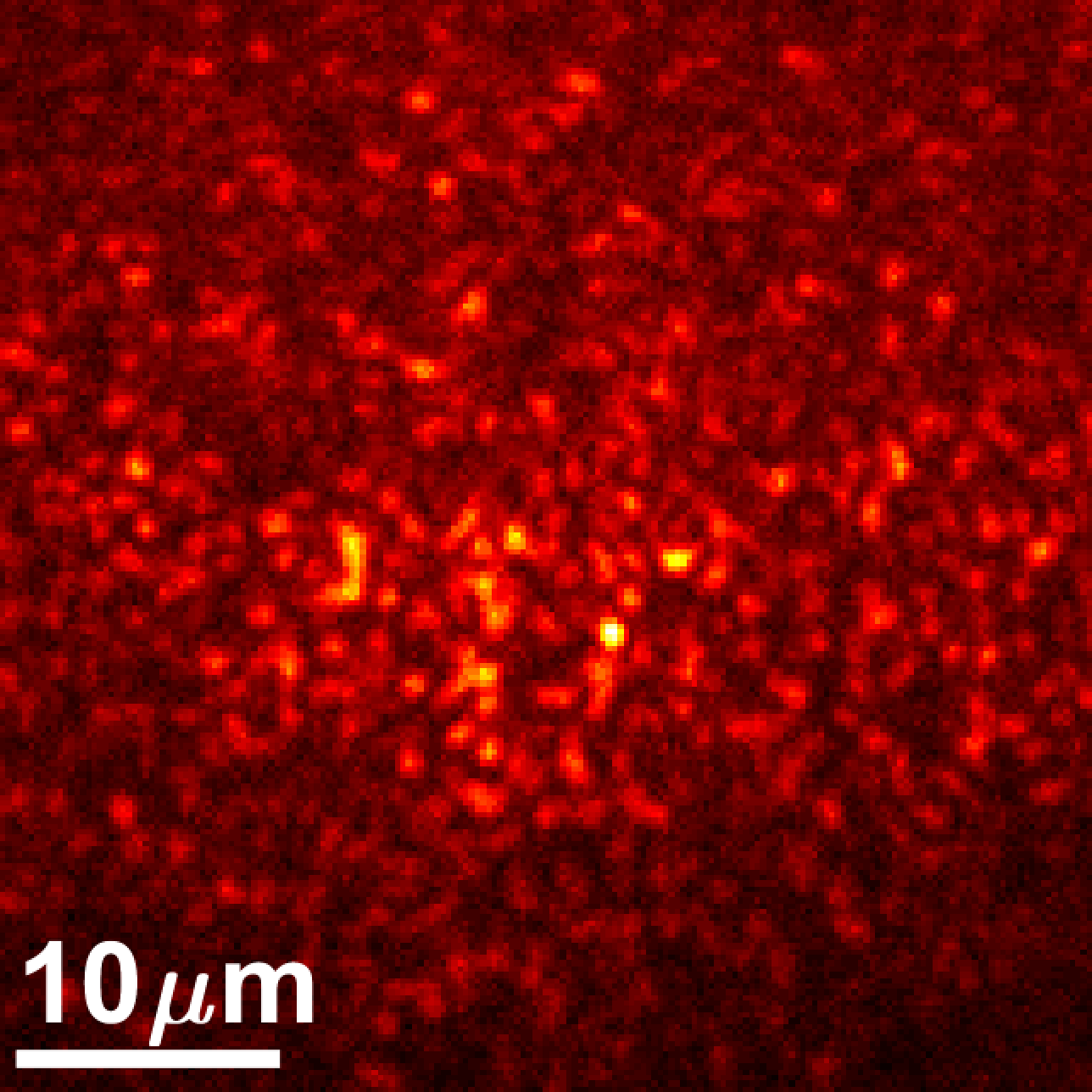}};
\end{tikzpicture}&
\includegraphics[height=0.18\textwidth]{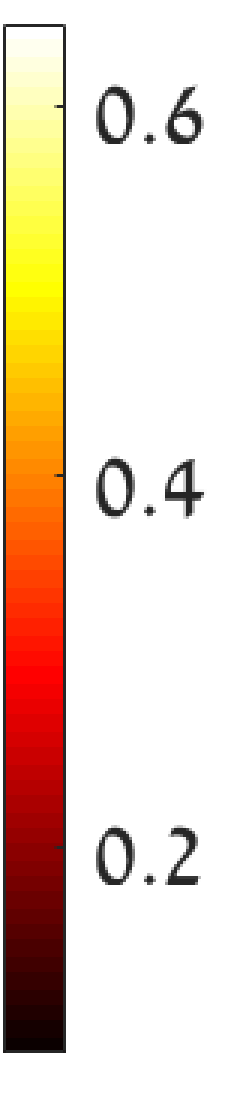}&
\includegraphics[height=0.18\textwidth]{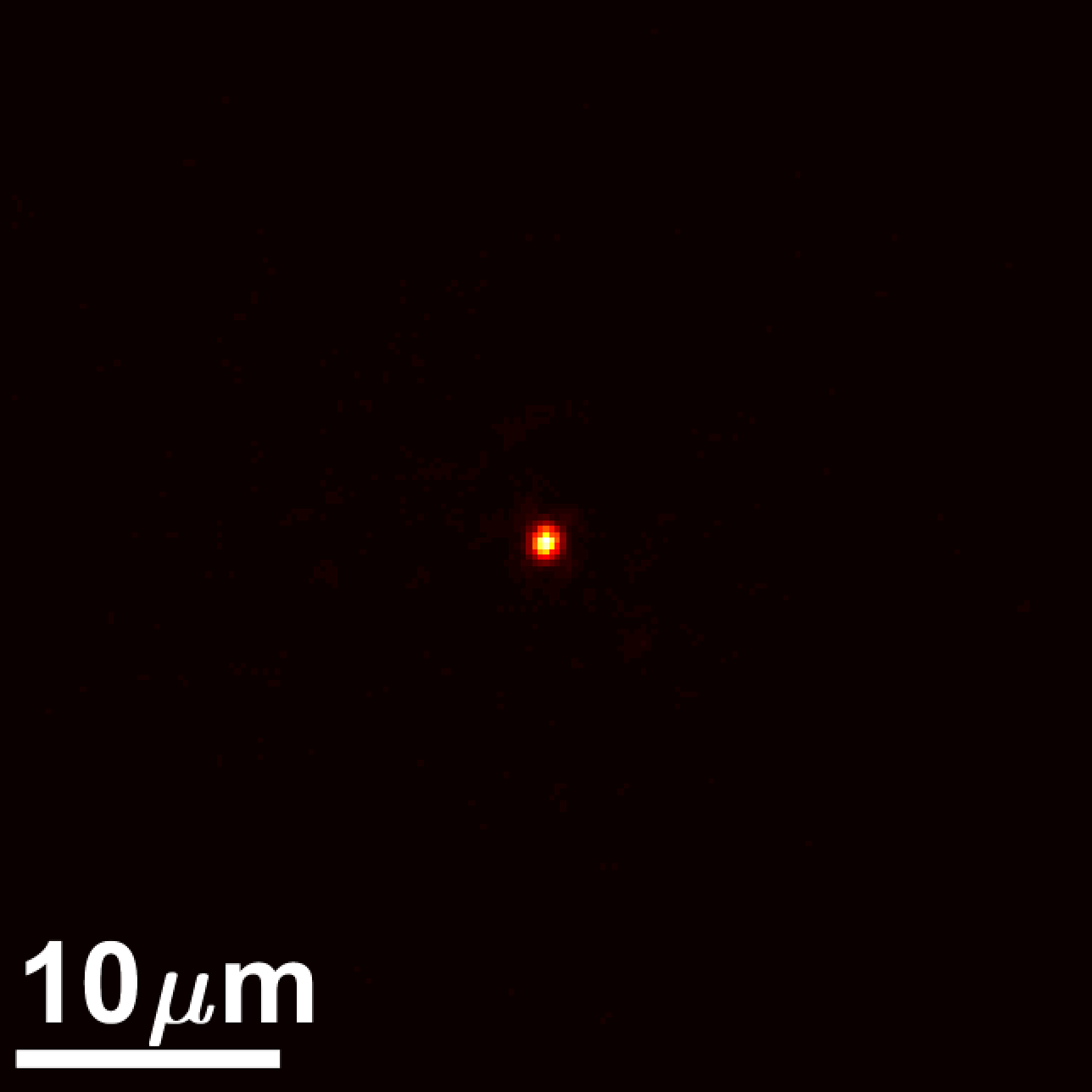}&
\includegraphics[height=0.18\textwidth]{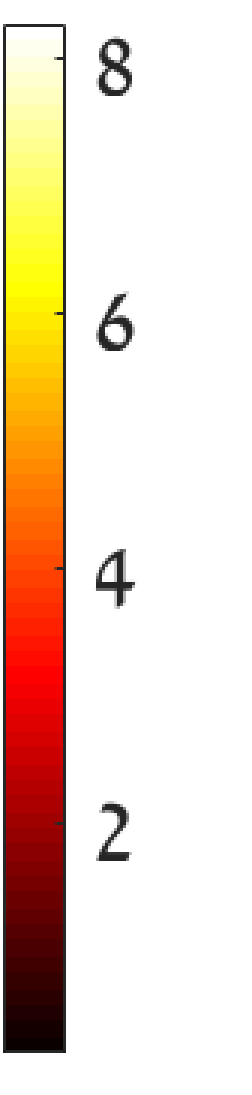}&
\begin{tikzpicture}
	\node[inner sep=0pt] (inset-neuron) at (0,0){\includegraphics[height=0.18\textwidth]{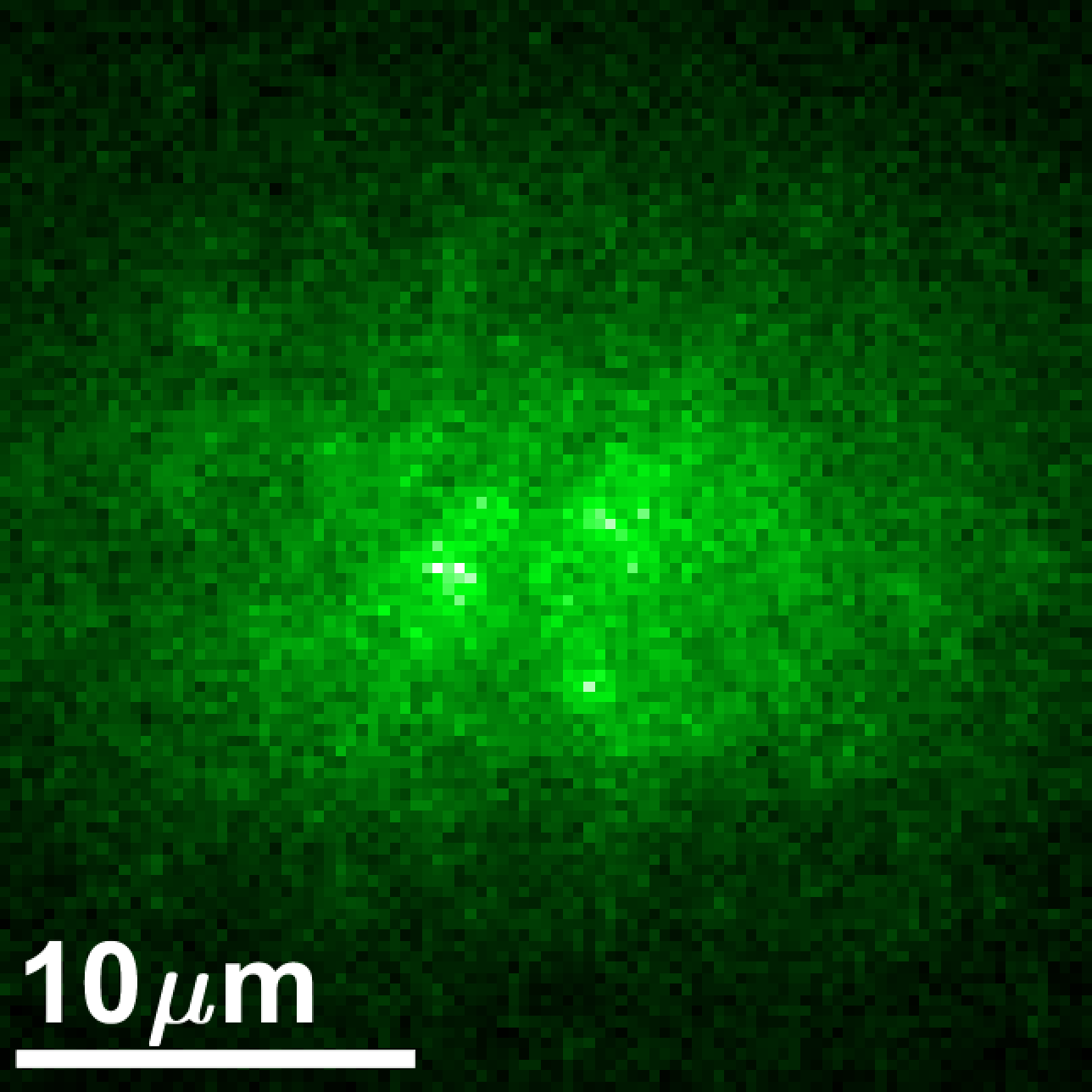}};
\end{tikzpicture}&
\includegraphics[height=0.18\textwidth]{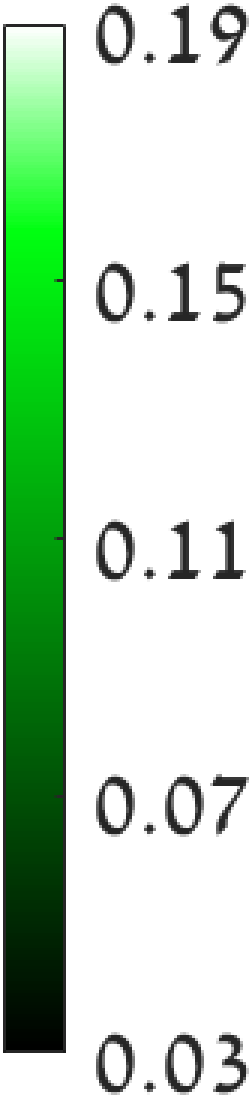}&
\includegraphics[height=0.18\textwidth]{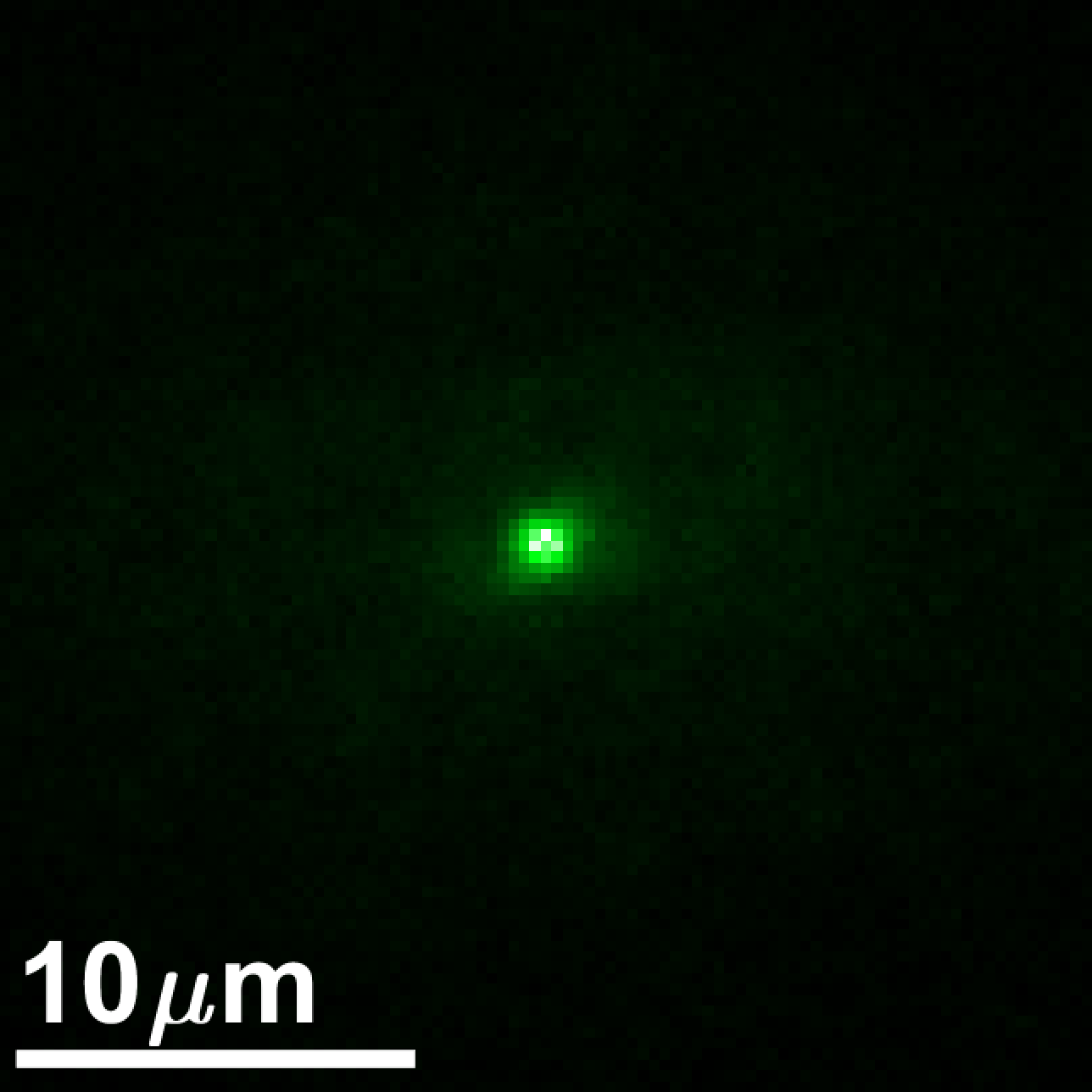}&
\includegraphics[height=0.18\textwidth]{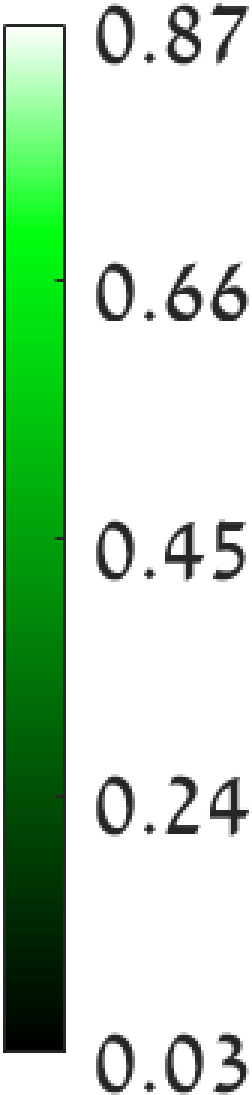}\\		


		\footnotesize{\!\!\!\!\!\!\!\!\!(a) Scattering, bead}&&
		\footnotesize{\!\!\!\!\!\!\!\!\!(b) Focusing, bead}&&\footnotesize{\!\!\!\!\!(c) Scattering, neuron}&	&	\footnotesize{\!\!\!\!\!(d) Focusing, neuron}&\\
	\end{tabular}
	\vspace{-0.1in}
	\caption{Types of fluorescent data: (a,b) emission from invitrogen fluorescent microspheres (excitation/emission at 640/680nm). A single bead is excited and the emitted light scatters through the tissue to generate a wide speckle pattern in (a).  In (b) we use an aberration correction in the imaging arm so that the sensor measures a sharp spot. With such synthetic sources we can image a speckle pattern at high SNR, but this is not always the case with real biological samples. For example, (c,d) demonstrate fluorescent emission from EGFP neurons (excitation/emission at 488/508nm), which is orders of magnitude weaker. In (c) a single fluorescent spot is excited, and the  limited number of photons it emits are spread over multiple pixels. Noise is dominant, and any attempt to measure the variance of this image will evaluate the noise variance rather than the speckle variance. In (d)  aberration correction is applied in the optics. As all photons are collected by a single pixel, SNR is drastically improved. Note that images (c,d) are taken under equal exposure and equal excitation power.}\label{fig:data_type}
	\vspace{-0.05in}
\end{figure*}

Earlier approaches  evaluate whether a wavefront modulation has focused using multi-photon fluorescence feedback. In this way, the light emitted from a fluorescence spot  is a non-linear function of the excitation intensity arriving to it, so when all light is focused into a single spot the total emission  energy  is maximized~\cite{Katz:14,Ji2017review}. 
However, obtaining feedback using single-photon fluorescence is highly desired as the process is significantly simpler and cheaper than the multi-photon one. 
The single-photon case cannot be evaluated using the simple score function  applied in the multi-photon case, since the emission energy is a linear function of the excitation energy and thus the amount of emission energy does not increase when all excitation is focused into a spot.
Recently, progress has been made on non-invasive wavefront shaping using single-photon feedback~\cite{Boniface:19,Dror22}. First, Boniface et al.~\cite{Boniface:19} have suggested that one can evaluate whether an incoming wavefront modulation has focused by computing the variance of the emitted speckle pattern.
More recently, Aizik et al.~\cite{Dror22} have suggested a rapid approach that can find a wavefront shaping modulation using  iterative phase conjugation. 
Both approaches were only demonstrated when the fluorescent feedback was provided by synthetic fluorescent beads, which emit a relatively strong signal. 
However, it is simpler if we could estimate  wavefront shaping modulations using feedback from biological samples,  such as neurons. 
These biological fluorescent samples impose two main differences. First, the targets are not sparse, but have   wide continuous volumes. Thus, an initial excitation pattern is likely to produce a smooth image of emitted light rather than a speckle pattern, as illustrated in  \figref{fig:setup_simp_1}.
 Without speckle variation, the phase-retrieval process of~\cite{Dror22} cannot be carried out. 
A second, more significant  difference is the fact that the signal emitted from such samples is orders of magnitude weaker than the one provided by fluorescent beads, and bleaching is reached much earlier.
Both algorithms~\cite{Boniface:19,Dror22} inherently assume that the speckle pattern emitted from a single fluorescent spot can be measured. However, if the number of fluorescent photons emitted from a neuron spot is  low, and these photons are aberrated and spread over multiple sensor pixels, no speckle pattern can be observed and one can mostly measure noise, see visualization in \figref{fig:data_type}. 
An attempt to measure the variance of this image, as required by~\cite{Boniface:19}, will result in the noise variance rather than the speckle variance. While it is possible to improve the image in \figref{fig:data_type}(c) by using a longer exposure or by increasing laser power, the wavefront shaping optimization process of~\cite{Boniface:19} would require capturing thousands of images of the target, and bleaching happens well before the optimization converges.  

This work proposes a confocal wavefront shaping framework,  which can be applied in low-light scenarios and uses feedback from  biological sources.
To this end, we  propose to use a simultaneous wavefront shaping modulation both on the incoming excitation wavefront and on the outgoing emitted light~\cite{Booth2002,Choi2015}, as illustrated in \figref{fig:setup_simp_1}. 
The advantage is that since scattered photons are corrected in the optical path  and we attempt to bring all photons emitted from a single spot  into a single detector, we can measure them with a much higher signal-to-noise ratio (SNR). 

To quantify the quality of a candidate modulation correction, we do not attempt to maximize the total energy emitted from the target. Rather, we seek to maximize the energy of the corrected wavefront {\em in a single spot}. We show that despite the fact that we use linear single-photon fluorescence, due to the double correction on both the illumination and imaging arms, our score function scales non-linearly with the intensity arriving at the fluorescence target. Thus, the returning energy at a single pixel is maximized by a focusing modulation that manages to bring all light into a single spot. We show that effectively, this score function 
 is  equivalent to the one used by previous two-photon fluorescence wavefront-shaping work~\cite{Katz:14}.

We successfully use our   algorithm to recover wavefront shaping modulations using fluorescent emission from EGFP neurons. By exploiting memory effect correlations, we use the modulation to locally image the shape of these neurons and their thin axons through scattering tissue.

	\section{Results}
\subsection{Definitions} \label{sec:problem_formulation}

\boldstart{Imaging setup:}
Consider the schematic of \figref{fig:setup_simp_1}. A laser beam illuminates a tissue sample, and an SLM can modulate its shape.  
The illumination wavefront propagates through the scattering tissue and excites the fluorescent target behind it. We wish to image that target, but the emitted light is scattered again through the tissue on its way to the camera.  A second phase SLM at the  imaging arm modulates the emitted light. Lastly, the modulated light is measured by the front main camera.
In practice our SLMs are placed at the Fourier planes of the system, see supplementary  for a complete  description.

The setup includes a second validation camera behind the tissue sample 
to assess focusing quality and  capture a clean reference image of the same target.
We emphasize that we only use non-invasive  feedback by the main (front)  camera, and the validation camera {\em does not} provide any input to the algorithm.

\boldstart{Image formation model:} 
Consider a set of $K$ fluorescent particles inside a  sample, and denote their positions by $\ptd_1,\ldots,\ptd_K$.
The illumination SLM   displays a complex 2D electric field  $\bu$.  We  use $\bou$ to denote a $K\times 1$ vector of  the field propagating through the sample at each of the $K$  fluorescent sources. 
We can express $\bou=\TMi \bu$, where  $\TMi$ is the incoming transmission matrix, describing the forward coherent light propagation in the tissue. 
 Likewise, the coherent propagation of light returning from the target to the  SLM of the imaging arm can be described by a  back-propagation transmission matrix $\TMo$.

%

Fluorescent energy emitted from a particle is proportional to  $|\bou_{{k}}|^{2\alpha}$, where $\alpha$ denotes the type of fluorescent excitation. The simplest case  $\alpha=1$ is known as single-photon fluorescence, where the emission is linear in the excitation energy $|\bou_{{k}}|^{2}$. In two-photon fluorescence,  $\alpha=2$, and the emission is proportional to the squared excitation. 

%

Since the laser energy is fixed, for any modulation $\bu$  the energy arriving the fluorescent target is bounded and w.l.o.g. we assume 
 \BE \label{eq:bounded-v} \sum_k |\bou_k|^2\leq 1.\EE
 
	\subsection{Scoring modulations}
\label{sec:metrics}
To estimate  a wavefront shaping modulation, we first need   a score function 
that can evaluate the focusing quality facilitated by a candidate modulation mask, using a noninvasive feedback alone. We start by reviewing scores that were previously introduced in the literature and then propose our new, noise-robust confocal score.

\boldstart{Image quality scores:}
Modulation evaluation is a simpler task when the same modulation can correct a sufficiently large isoplanatic image region. This assumption was made by 
 adaptive optics research~\cite{Booth2014,Ji2017review,HampsonBooth21review,Bonora:13,Antonello:20} and also by  wavefront shaping approaches~\cite{YeminyKatz2021,Stern:19,Daniel:19,Metzler23NeuWS}, who  evaluate the quality of the resulting image, in terms of contrast~\cite{Ji2017review},  sharpness,  variance~\cite{YeminyKatz2021},  or with a neural network regularization~\cite{Metzler23NeuWS,kang2023coordinatebased}. 
However, for thick tissue, wavefront shaping correction can vary quickly between nearby pixels,  and a modulation may only explain a very local region.  This case makes the above image quality scores less applicable, as inherently they evaluate the quality of an image region rather than a pixel. For spatially varying modulations, ideally, 
we need to  evaluate the success of the modulation based on a per-pixel criterion.

\boldstart{The total intensity score:}
Consider  a configuration where we only try to correct the illumination arm, and the SLM in the imaging arm of \figref{fig:setup_simp_1} is not used. 
The simplest score that was considered in the literature~\cite{Katz:14,Ji2017review} is just the total intensity measured over   the entire sensor plane, which is also proportional to the total fluorescent power,
 reducing to \vspace{-0.1cm}
\BE\label{eq:metric-TI}
\Mtric_{\text{TI}}(\bu)\equiv\sum_x I(x)=\sum_k |\bou_k|^{2\alpha}.
\vspace{-0.1cm}\EE
Since   the energy at the target  is bounded (see \equref{eq:bounded-v}),
for the case $\alpha>1$ this score is maximized when $\bou$ is a one-hot vector, which equals 1 at a single entry and zero at all the others. 

Two-photon fluorescence is  expensive and hard to implement, and solutions that can use a single-photon excitation feedback are  highly desirable. However, in the single-photon case where $\alpha=1$, \equref{eq:metric-TI} reduces to the total power in $\bou$, $\Mtric_{\text{TI}}(\bu)= \sum_k |\bou_k|^{2}$, and since this power is fixed, the same amount of energy returns whether we spread the excitation power over multiple fluorescence sources or bring all of it into one spot. 

\boldstart{The variance maximization score:}
Boniface \etal~\cite{Boniface:19} propose an alternative approach to evaluate focusing with linear, single-photon feedback. Their approach also modulates only the illumination wavefront attempting to focus light at a single spot, while the emitted light is scattered to the sensor. To score modulations, they measure the variance of the resulting speckle image and attempt to maximize it.
 The idea is that if we manage to focus all the excitation light at a single spot, the emitted light scattered through the tissue will generate a highly varying speckle pattern on the sensor plane. If the excitation is not focused, multiple sources emit simultaneously. The light emitted by these sources sums incoherently,  and hence the variance of the speckle pattern on the sensor decays. 
They show that as in the two-photon case, speckle variance is  proportional to $\sum_k |\bou_k|^4$.


This score is an important advance; however, it is hard to evaluate for low SNR biological sources. When a low number of photons is spread over multiple sensor pixels, the captured image is very noisy and an attempt to evaluate its variance will result in the noise variance rather than the speckle variance, see \figref{fig:data_type}.

While one can reduce imaging noise by using a longer exposure or by increasing the power of the excitation laser, the optimization of~\cite{Boniface:19} requires capturing many images of the same target, and usually the neuron  bleaches well before  convergence. 

\boldstart{Confocal energy score:}
In this research we suggest a new score for evaluating a wavefront shaping modulation. While the previous scores corrected only the illumination arm, inspired by~\cite{Booth2002,Choi2015,PanJi2023}, we suggest to correct both arms, using two modulations denoted by $\binu,\boutu$.
 As emission is very weak,  the fact that the SLM correction is applied before imaging helps collect all photons at one sensor pixel and improve SNR. 
To score the  focusing quality of each modulation we will use the intensity at the central pixel, rather than the total intensity throughout the  sensor. 
In supplementary we prove that we 
 can express the energy of the central pixel  as: 
\BE\label{eq:inc-confocal} \vspace{-0.1cm}
\Mtric_{\text{Conf}}(\binu,\boutu)\equiv I(0)=
 \sum_k |\boutou_k|^2\cdot  |\binou_k|^2,
\EE
with $\binou=\TMi \binu$ and $\boutou=\boutu^T \TMo$.

As mentioned in \equref{eq:bounded-v}, the energy of $\binou$ is bounded, and due to reciprocity the same applies for $\boutou$. 
It is easy to see that this score is maximized when $\binou,\boutou$ are both one-hot vectors, \blue{which bring all energy to a single joint entry $k$ and have zero energy at all other entries}. That is, the score of \equref{eq:inc-confocal} is maximized when the excitation modulation $\binu$ brings all light to one of the particles $\ptd_k$, and the modulation at the imaging arm $\boutu$ corrects the wavefront emitted from the same particle $\ptd_k$ and brings all of it into the central pixel.


In particular, if the excitation and emission wavelengths are sufficiently similar, we explain in supplementary that it is best to use the same modulation at both the illumination and imaging arms, and the score of \equref{eq:inc-confocal} reduces to $\sum_k |\bou_k|^4$ as in the variance-maximization and two-photon cases.

While the confocal score is equivalent to the variance maximization score above, it is significantly less susceptible to noise. This is due to the fact that the small number of photons we have at hand are collected at a single spot, rather than being spread over multiple pixels.  \figref{fig:data_type}(c-d) shows the images emitted from a single neural spot with and without modulation in the imaging arm, and the significant noise reduction.

 In this work we have  explicitly optimized the confocal score (\equref{eq:inc-confocal}) using standard Hadamard basis optimization~\cite{PopoffPhysRevLett2010}, detailed in Supplementary Sec 2. 	Overall this optimization approach is significantly slower than the iterative phase conjugation of~\cite{Dror22}, but can handle dense targets. \blue{ In our case the excitation and emission wavelengths are similar, yet not identical. While using two different modulations at the excitation and emission arms results in a modestly improved correction, it also doubles the required number of measurements. Alternatively, we can neglect the wavelength difference and use the same correction in both arms. Despite the approximation, the faster optimization is advantageous  in the presence of photobleaching. We compare a single correction to two corrections in supplementary section 5.  }



	\subsection{Experimental results}
\label{sec:results}
\begin{figure*}[t!]
	\begin{center}
		\begin{tabular}{@{}c@{~~}c@{~~}c@{~~}c@{~~~~}c@{~}c@{~~}c@{~}c@{~~}c@{~}c@{}}
			\multicolumn{4}{c}{Validation camera}&\multicolumn{6}{c}{Main camera}\\
			\cmidrule(l{0pt}r{13pt}){1-4} 	\cmidrule(l{0pt}r{0pt}){5-10}

			\vspace{0.1cm}
			\includegraphics[width = 0.115\textwidth]{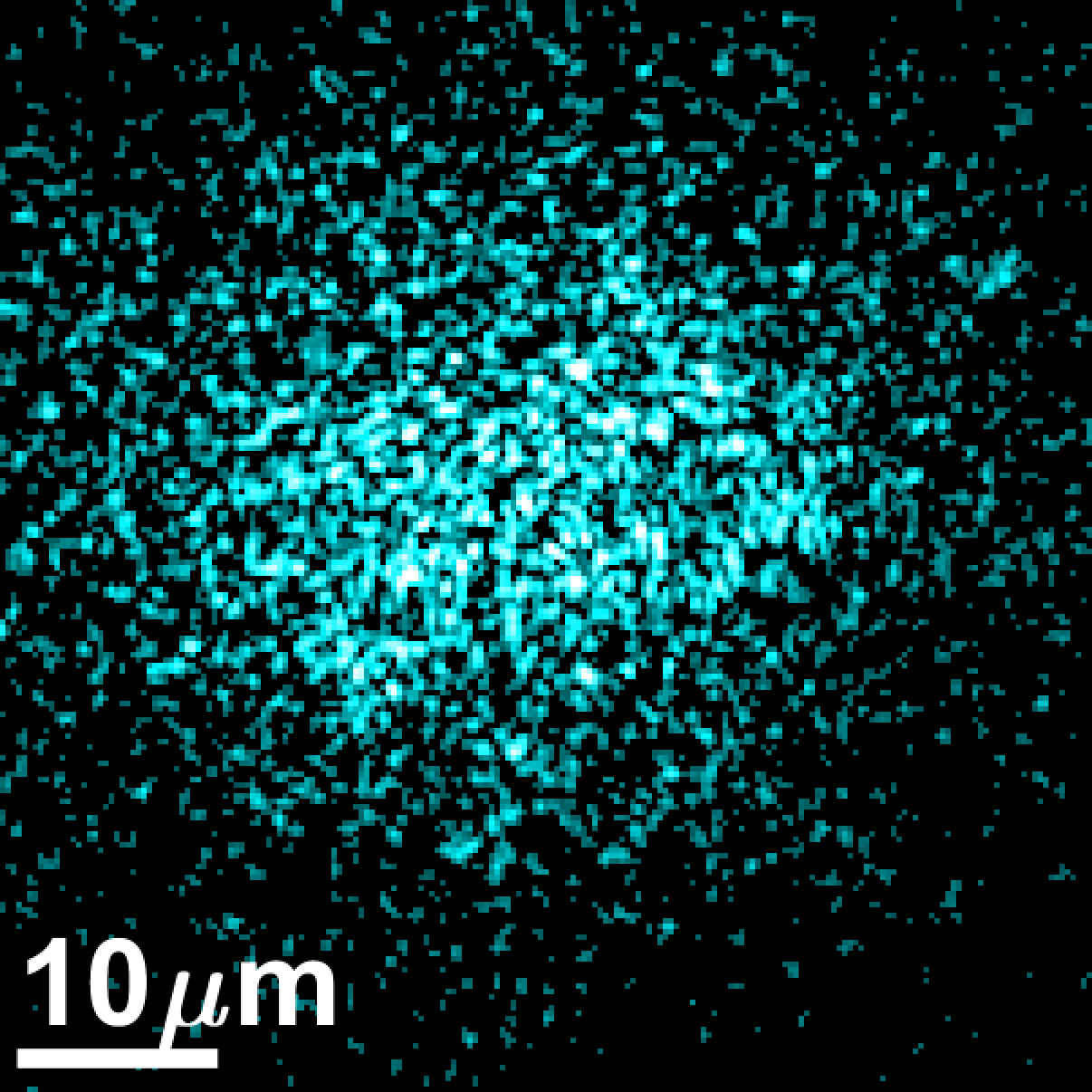}&
			\includegraphics[width = 0.115\textwidth]{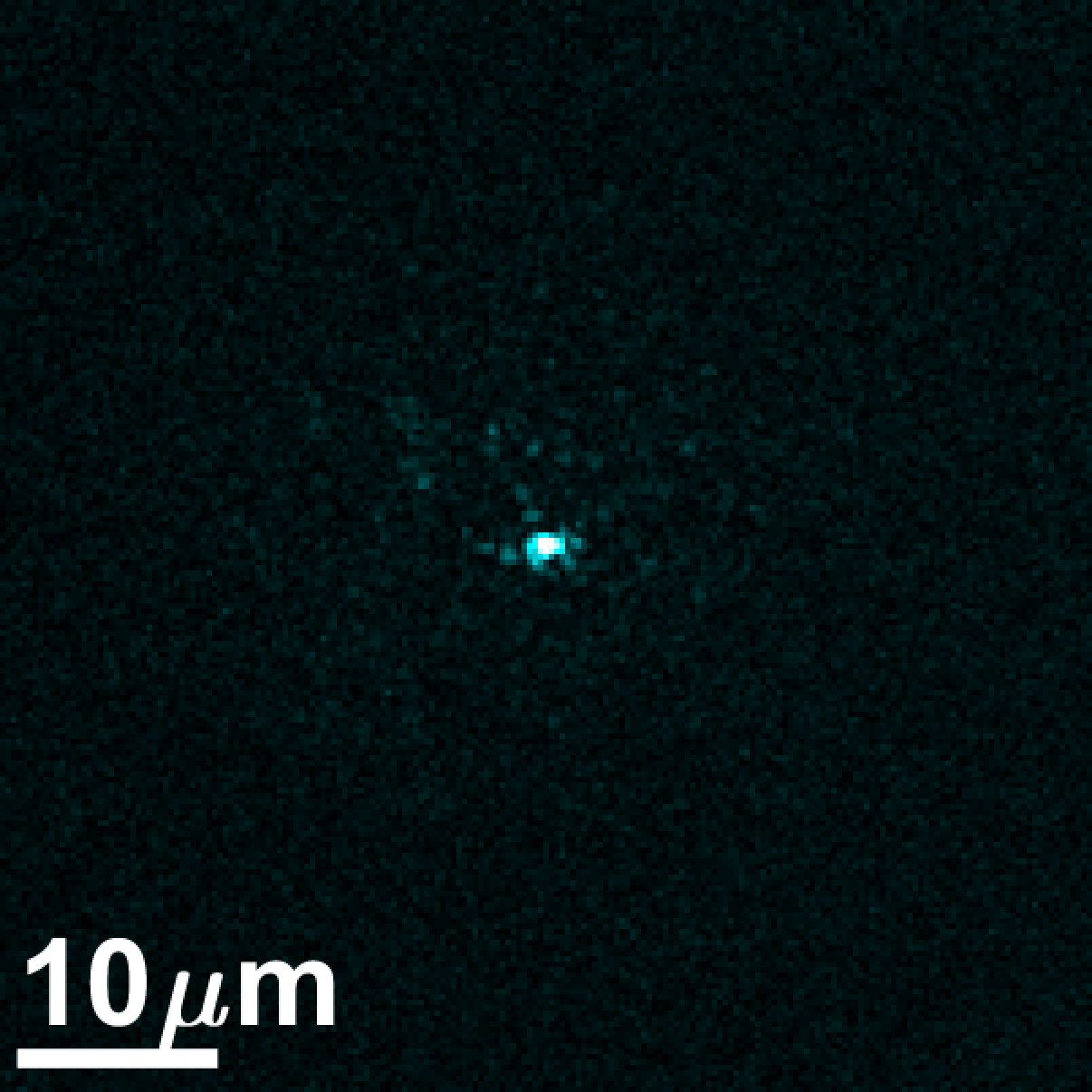}&
			\includegraphics[width = 0.115\textwidth]{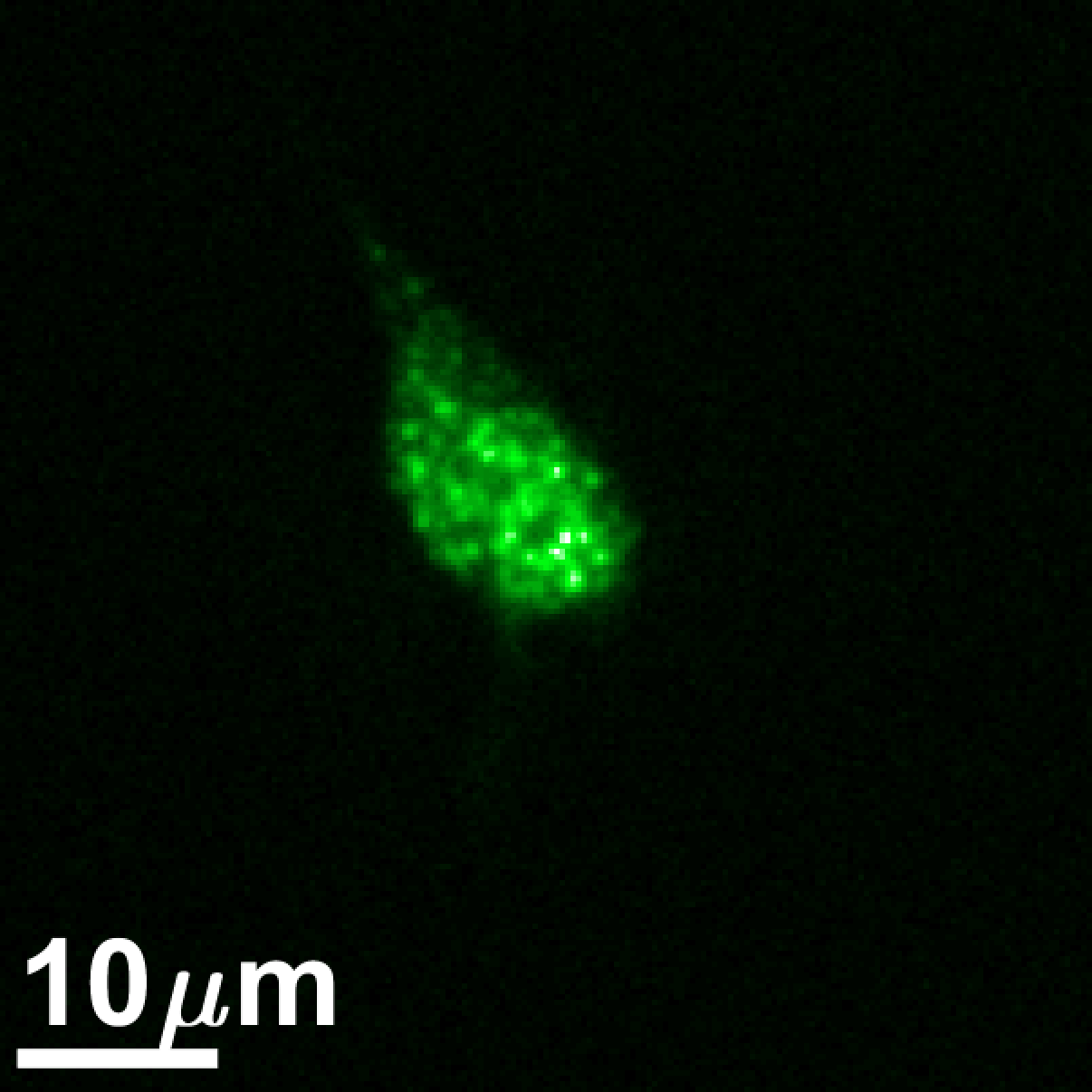}&
			\includegraphics[width = 0.115\textwidth]{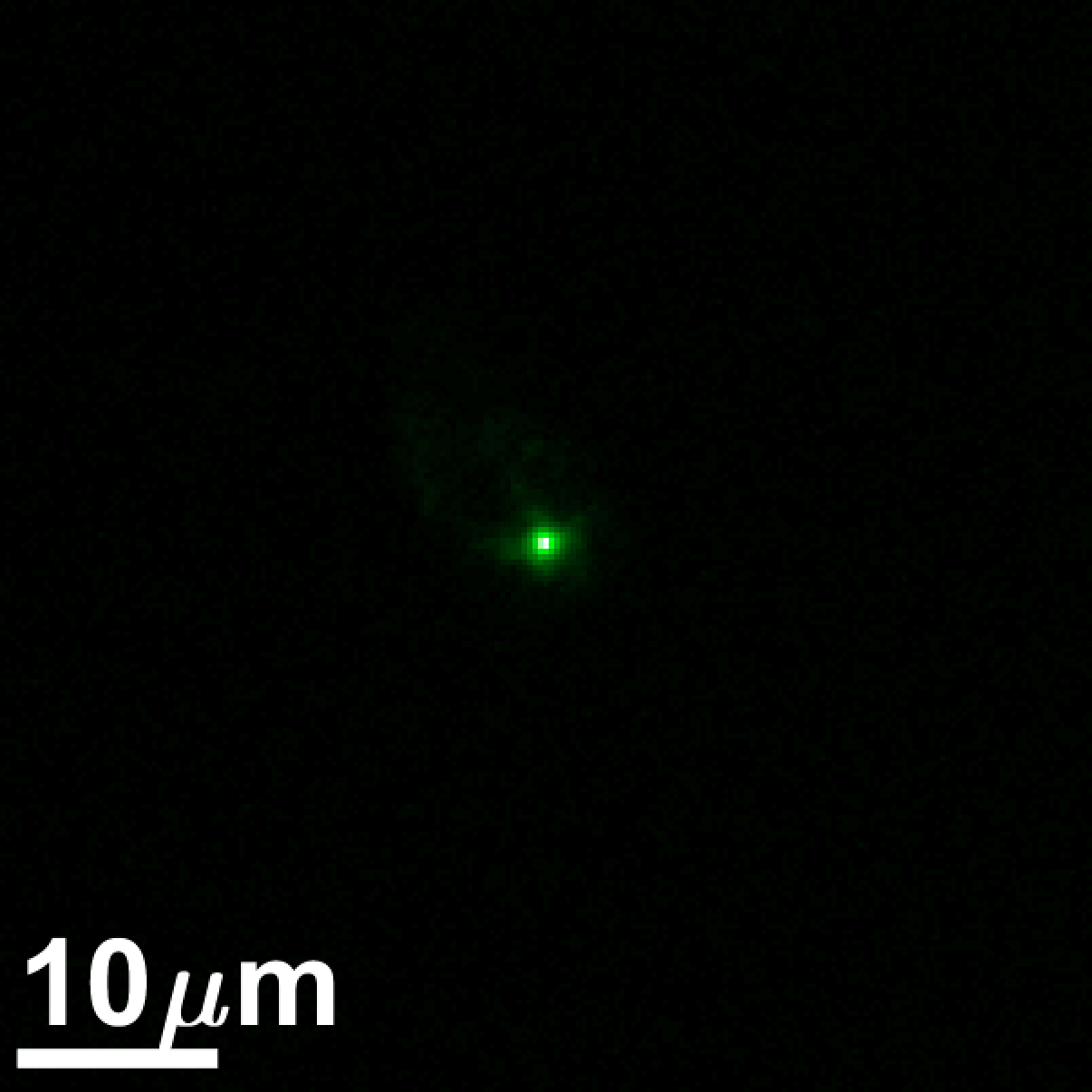}&
			\includegraphics[width = 0.115\textwidth]{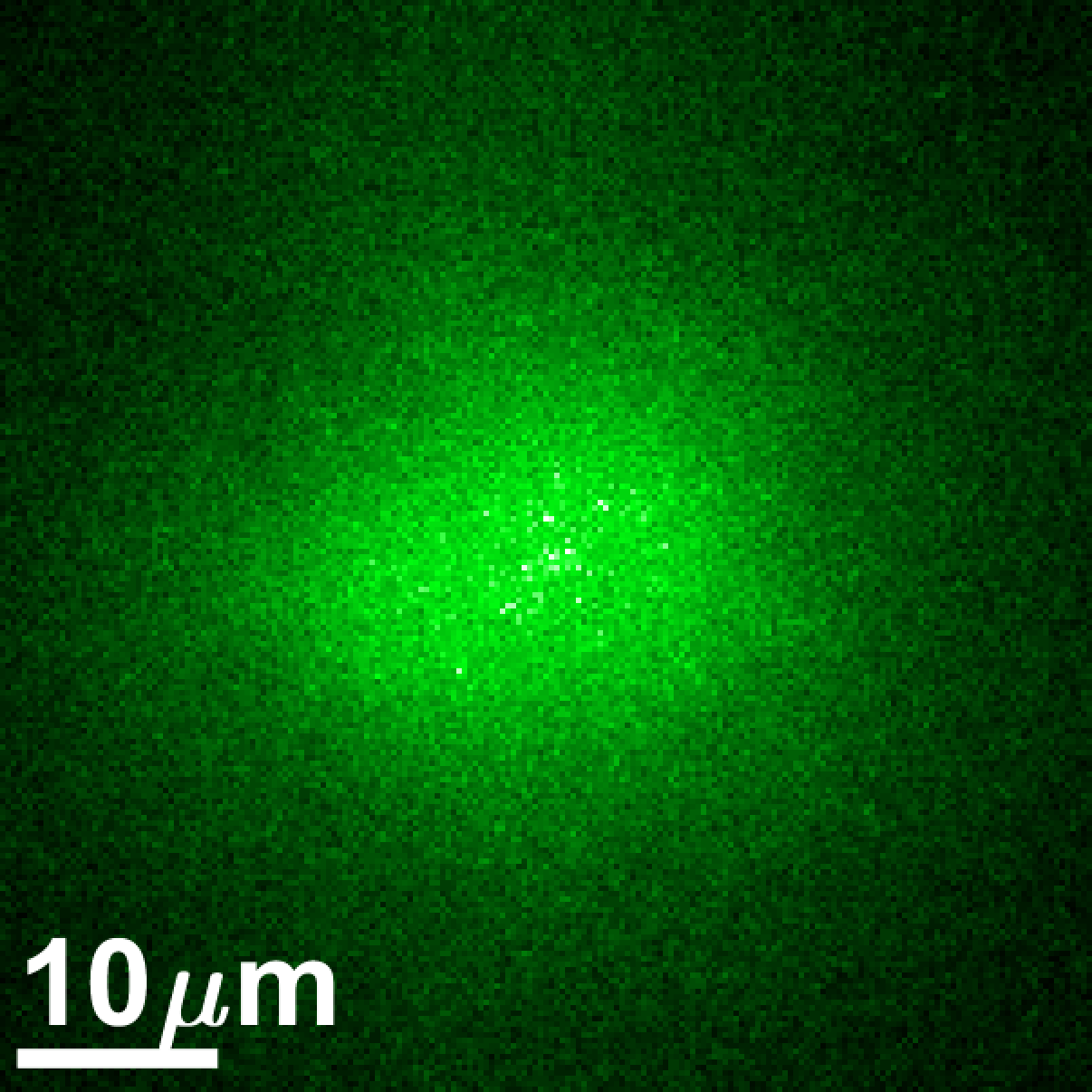}&
			\includegraphics[height = 0.115\textwidth]{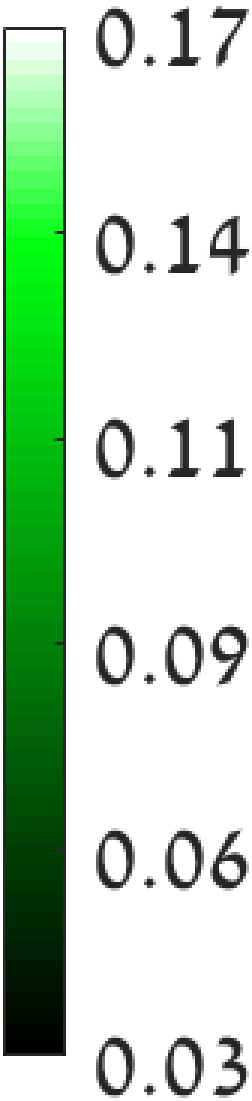}&
			\includegraphics[width = 0.115\textwidth]{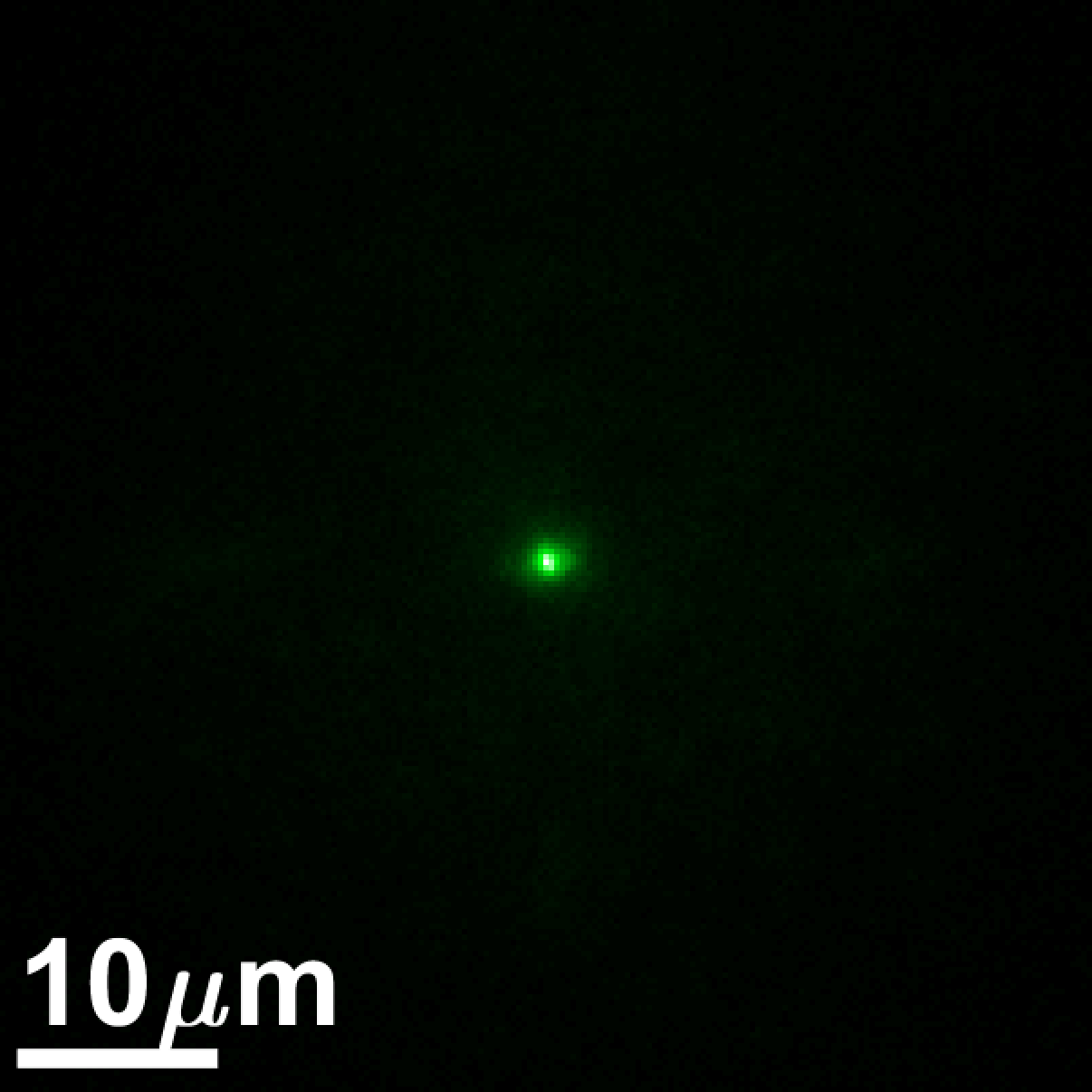}&
			\includegraphics[height = 0.115\textwidth]{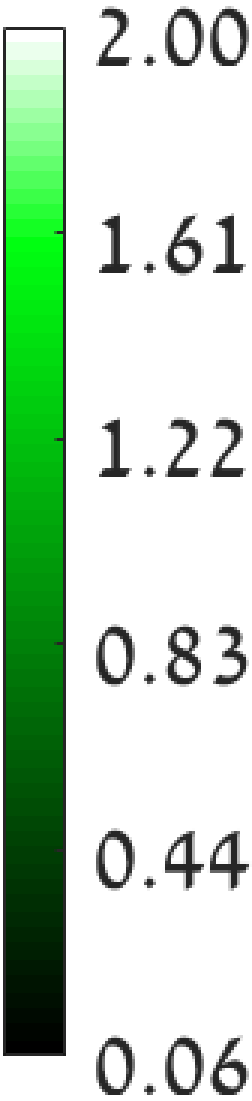}&
			\includegraphics[width = 0.115\textwidth]{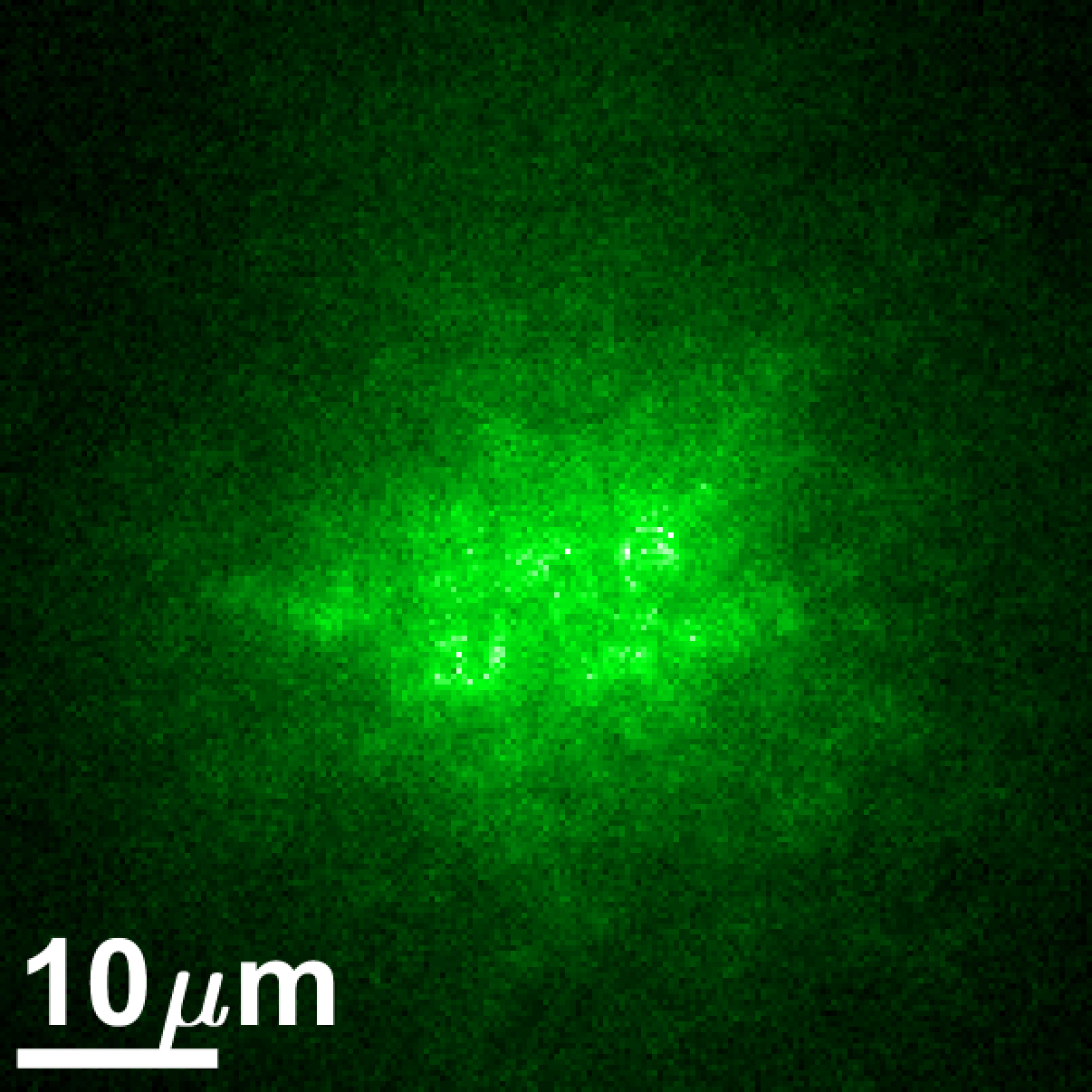}&
			\includegraphics[height = 0.115\textwidth]{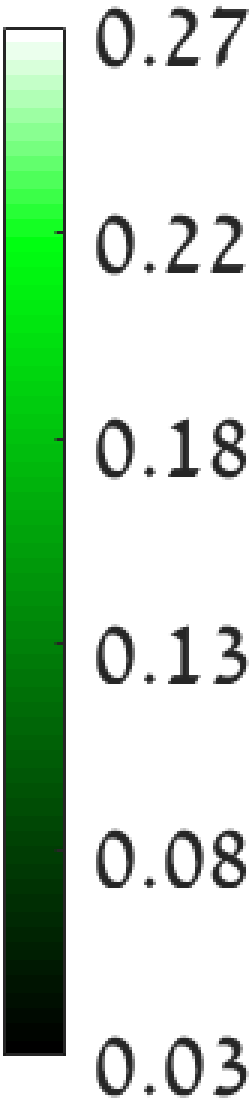}\\


			\includegraphics[width = 0.115\textwidth]{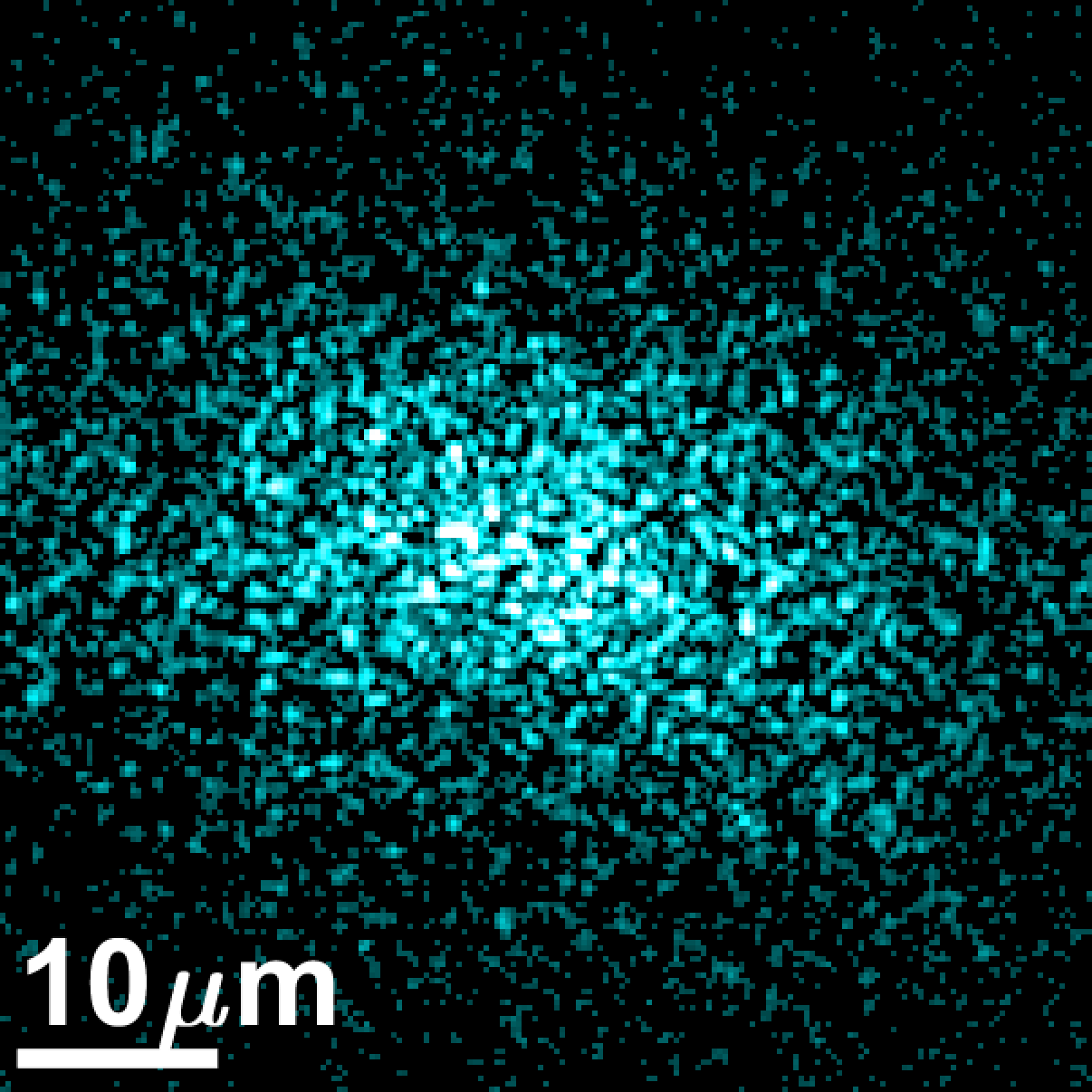}&
			\includegraphics[width = 0.115\textwidth]{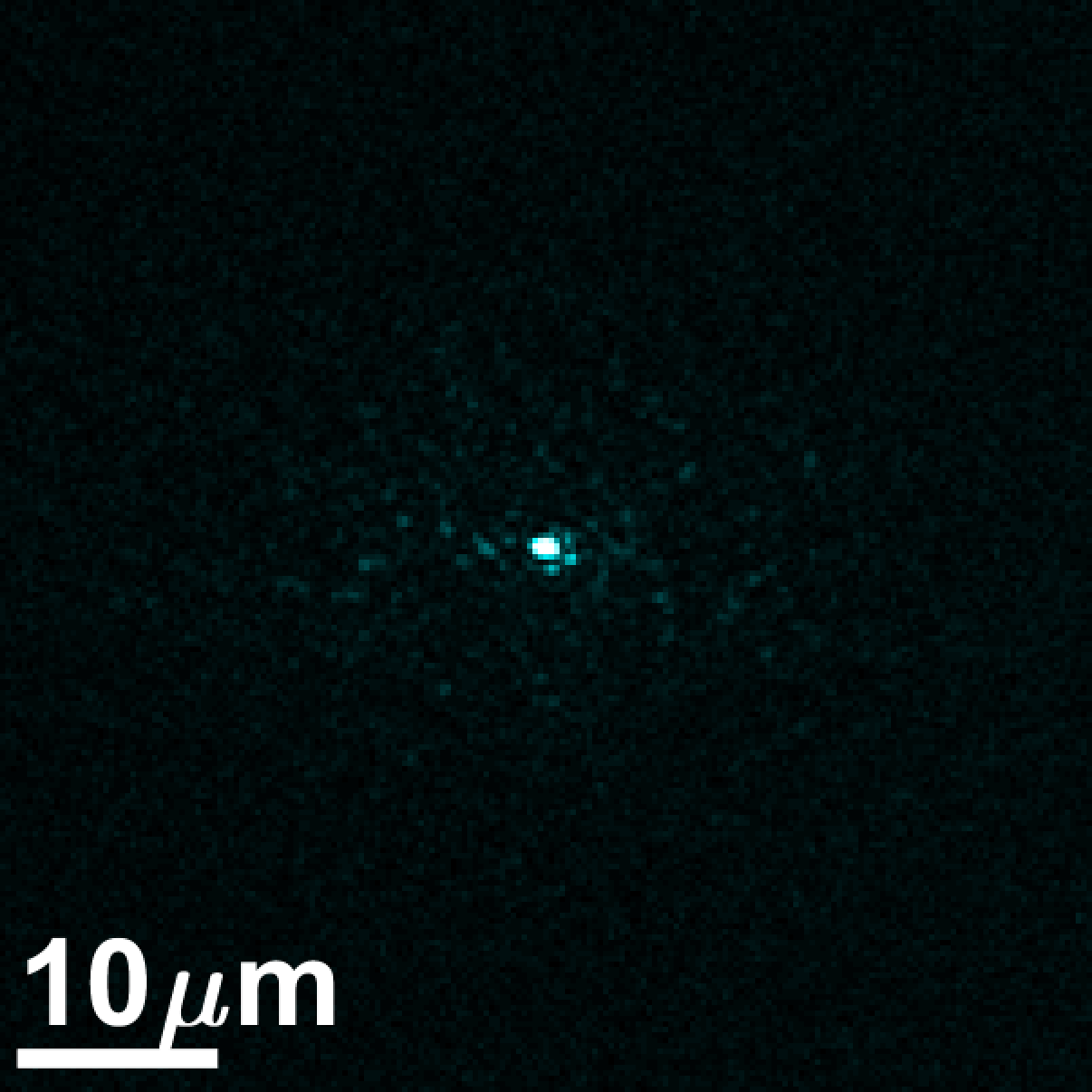}&
			\includegraphics[width = 0.115\textwidth]{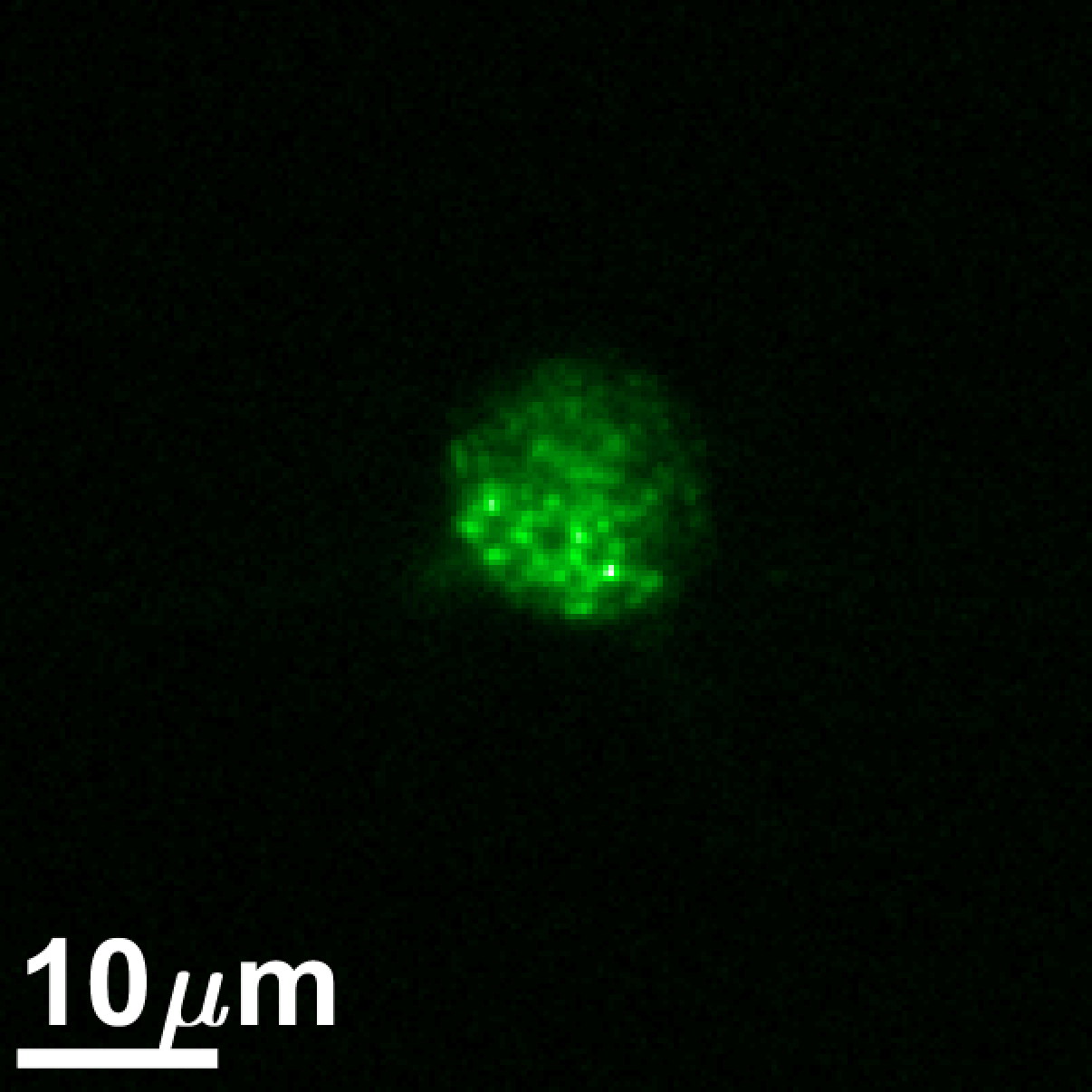}&
			\includegraphics[width = 0.115\textwidth]{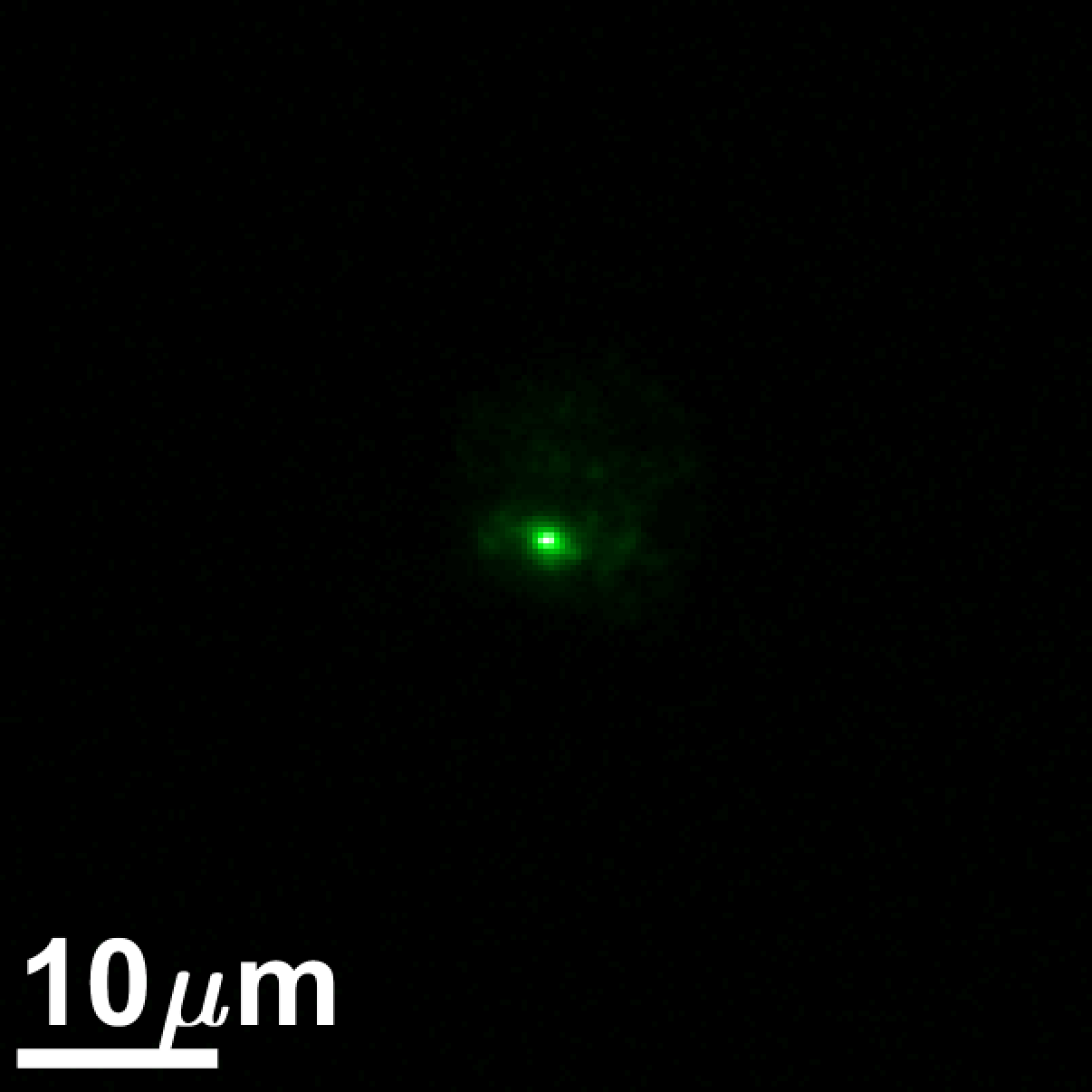}&
			\includegraphics[width = 0.115\textwidth]{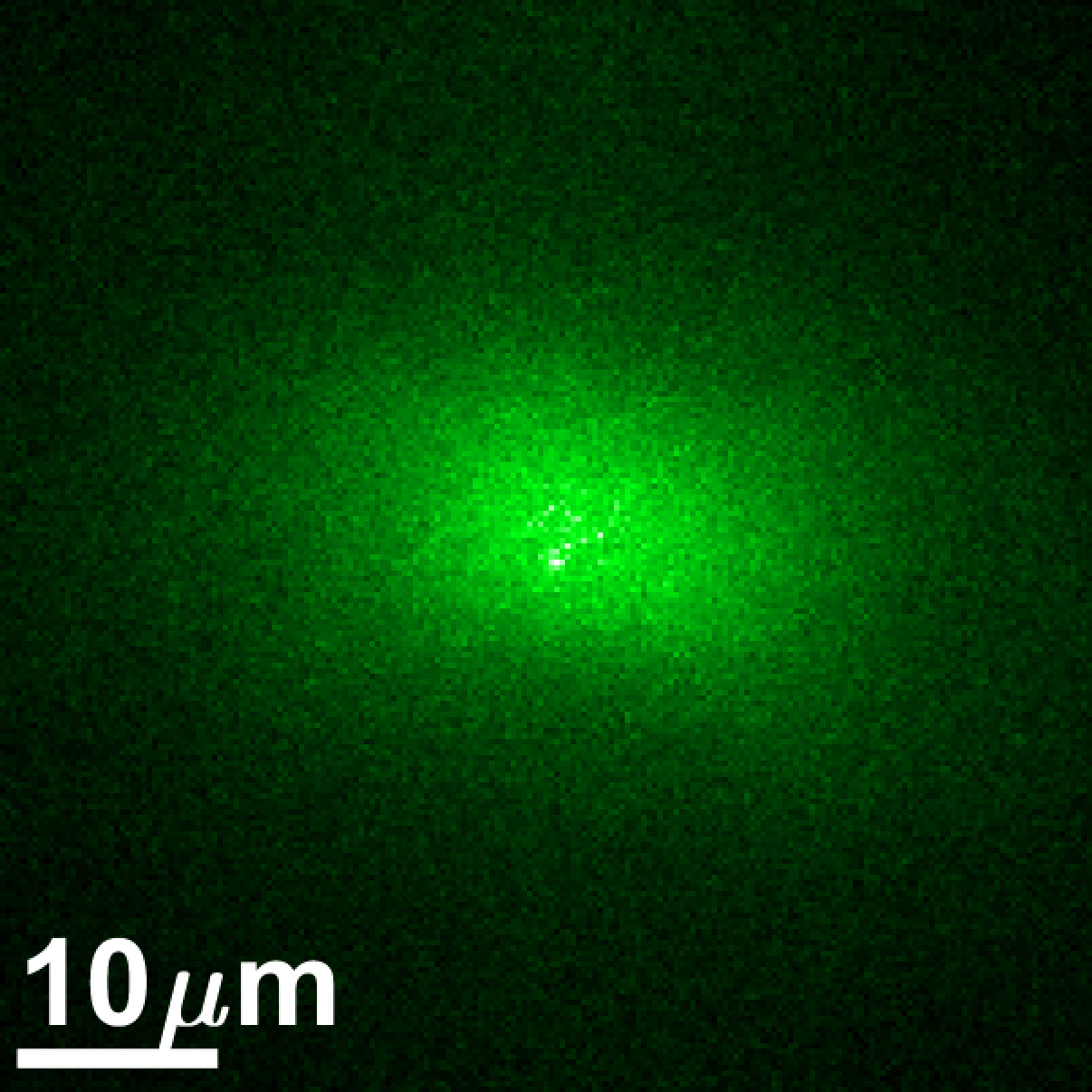}&
			\includegraphics[height = 0.115\textwidth]{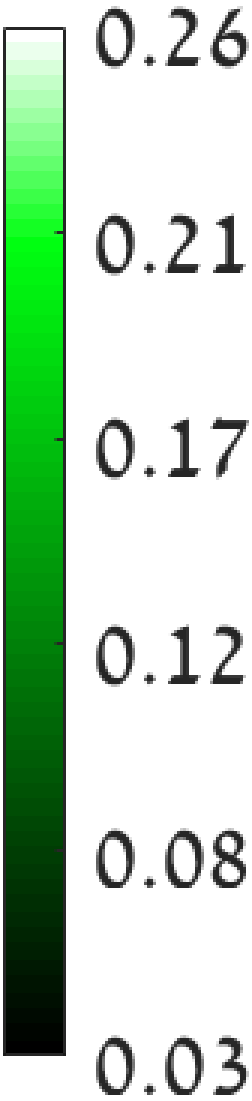}&
			\includegraphics[width = 0.115\textwidth]{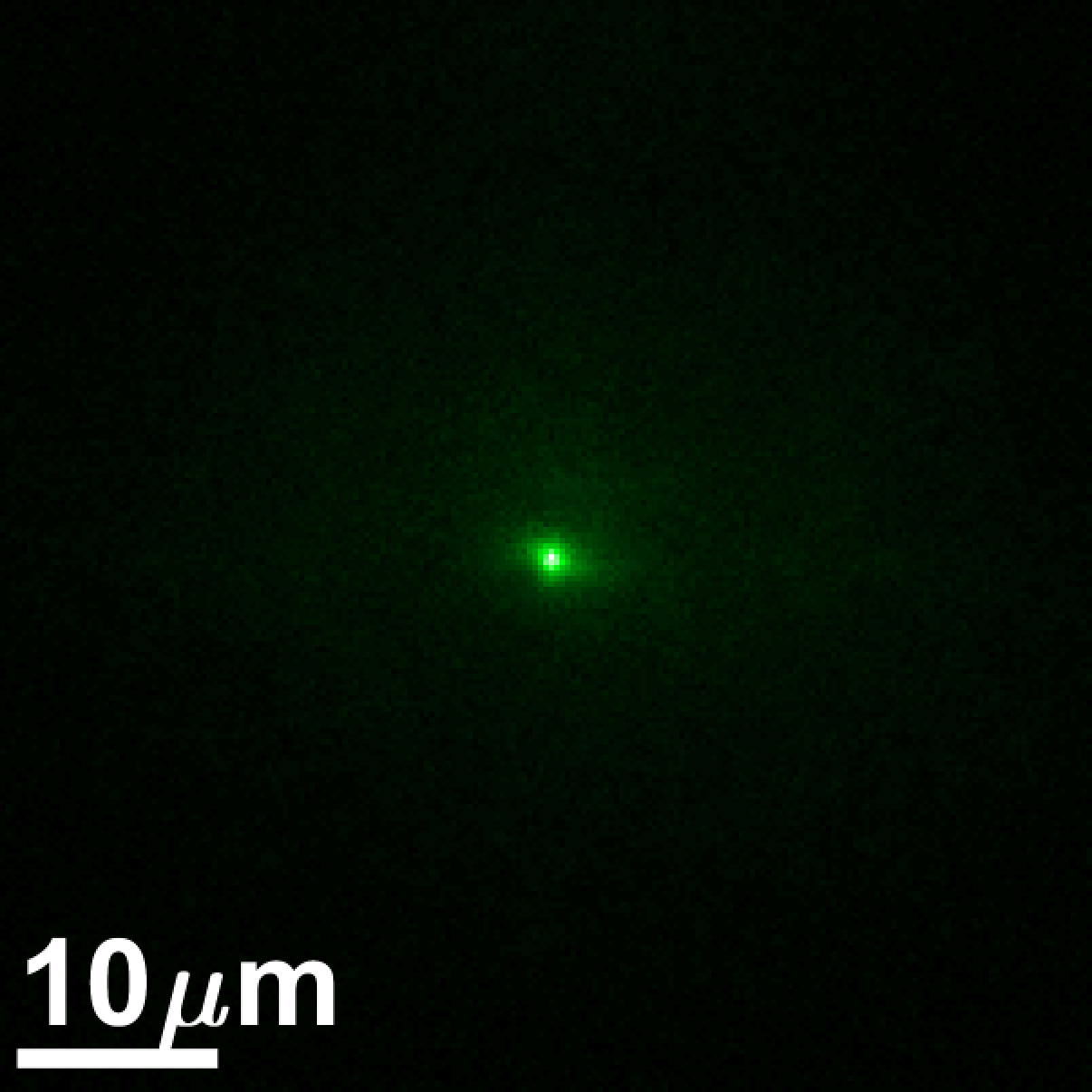}&
			\includegraphics[height = 0.115\textwidth]{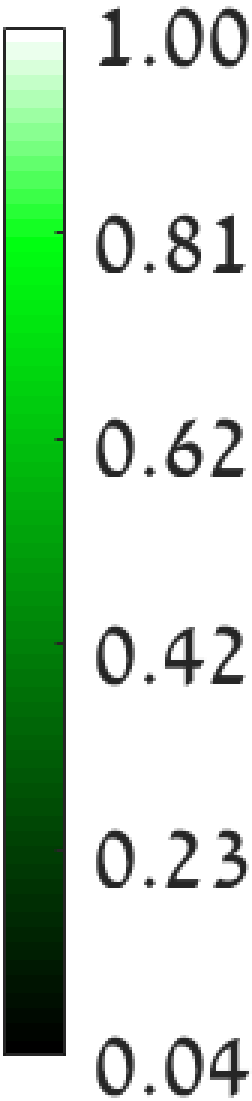}&
			\includegraphics[width = 0.115\textwidth]{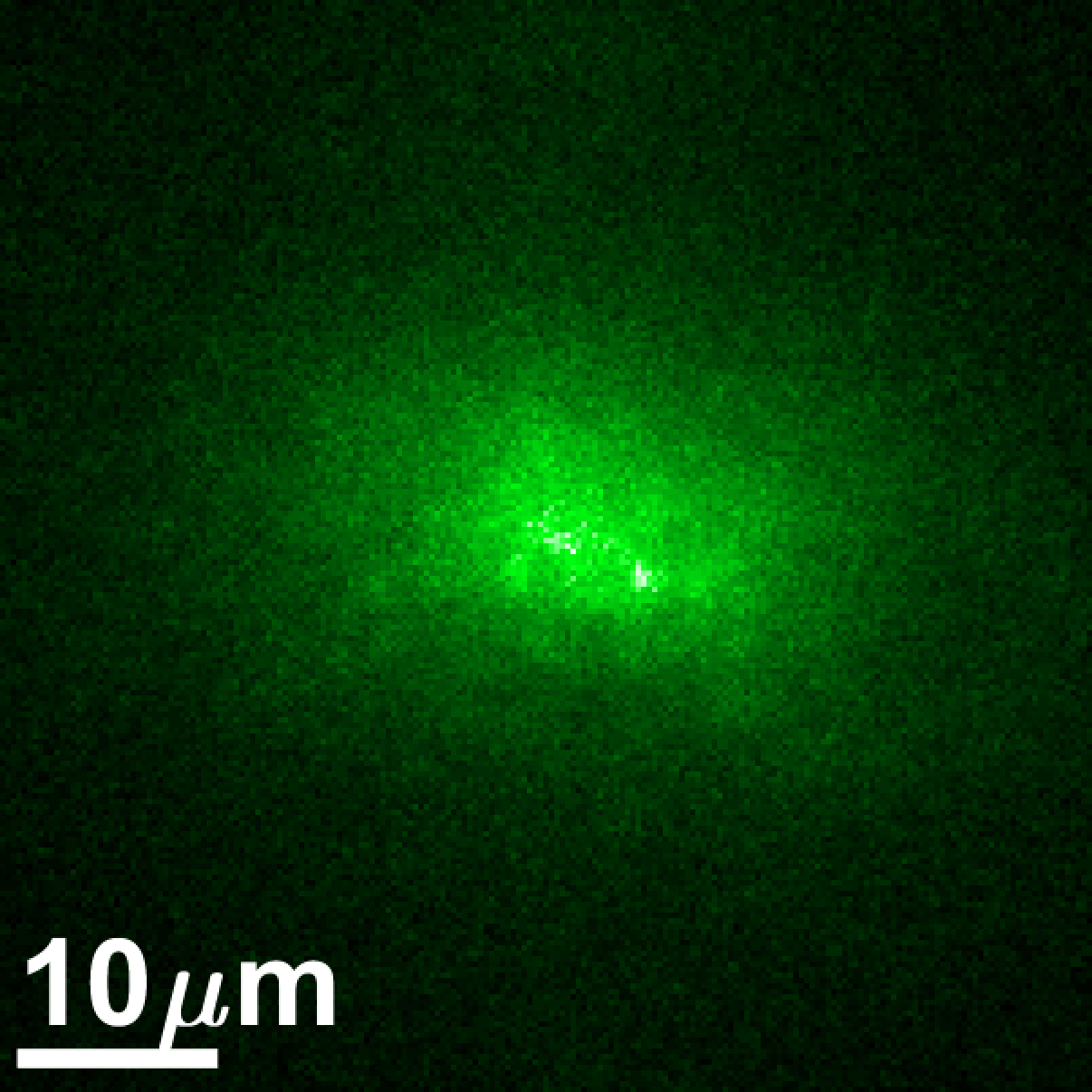}&
			\includegraphics[height = 0.115\textwidth]{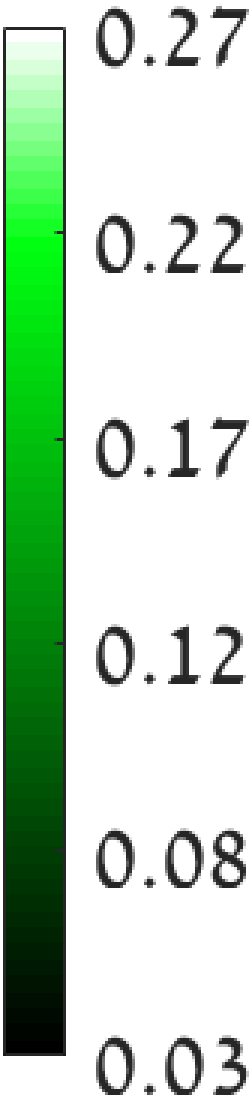}\\
			
			\vspace{0.1cm}
			\includegraphics[height = 0.115\textwidth]{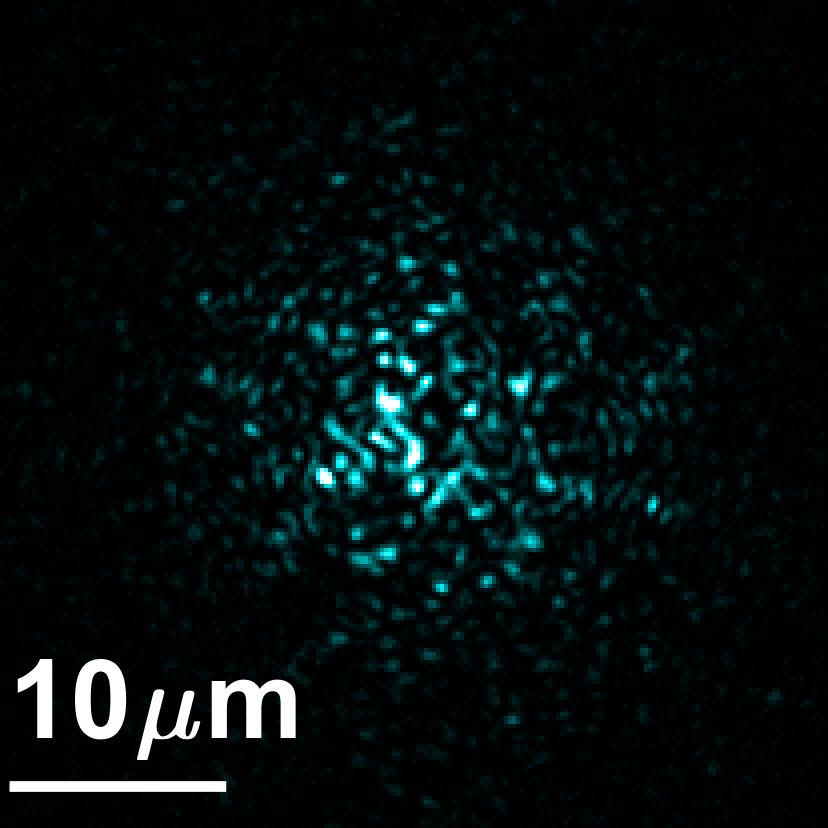}&
			\includegraphics[height = 0.115\textwidth]{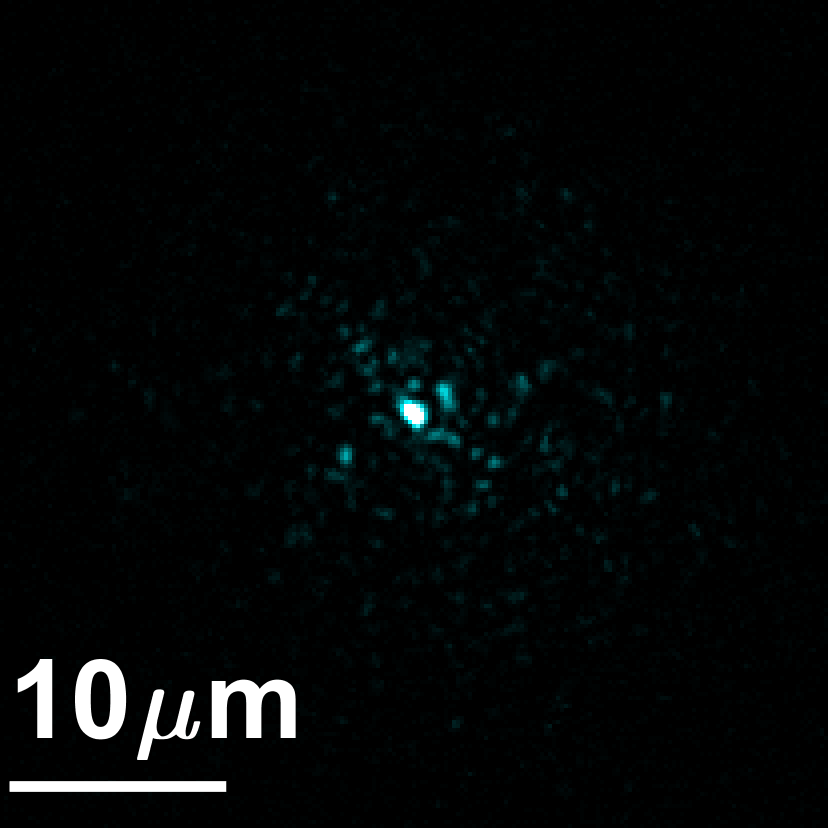}&
			\includegraphics[height = 0.115\textwidth]{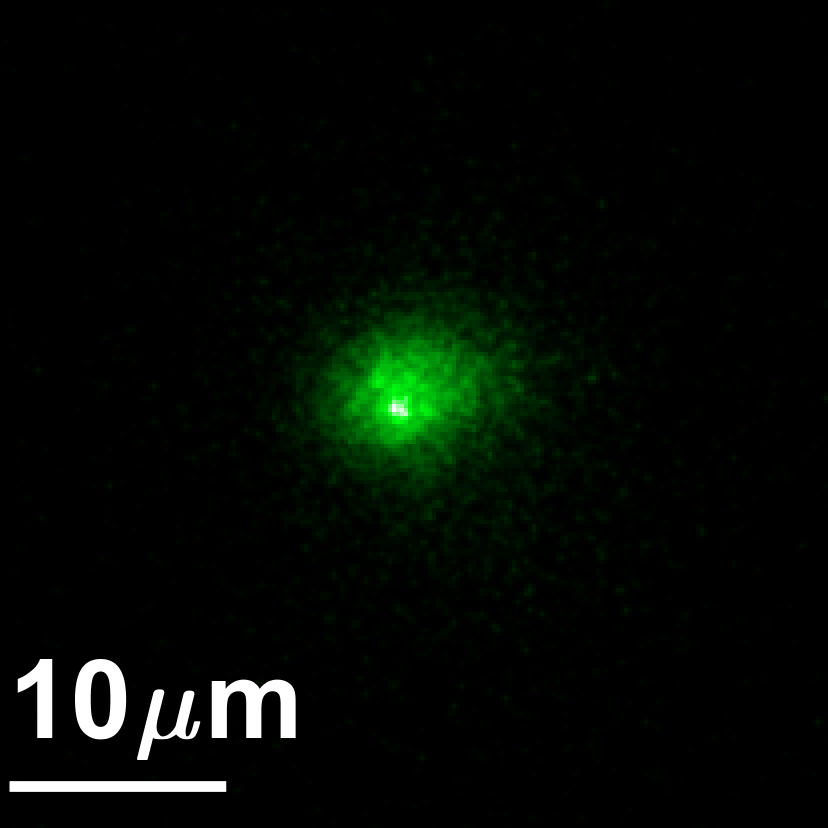}&
			\includegraphics[height = 0.115\textwidth]{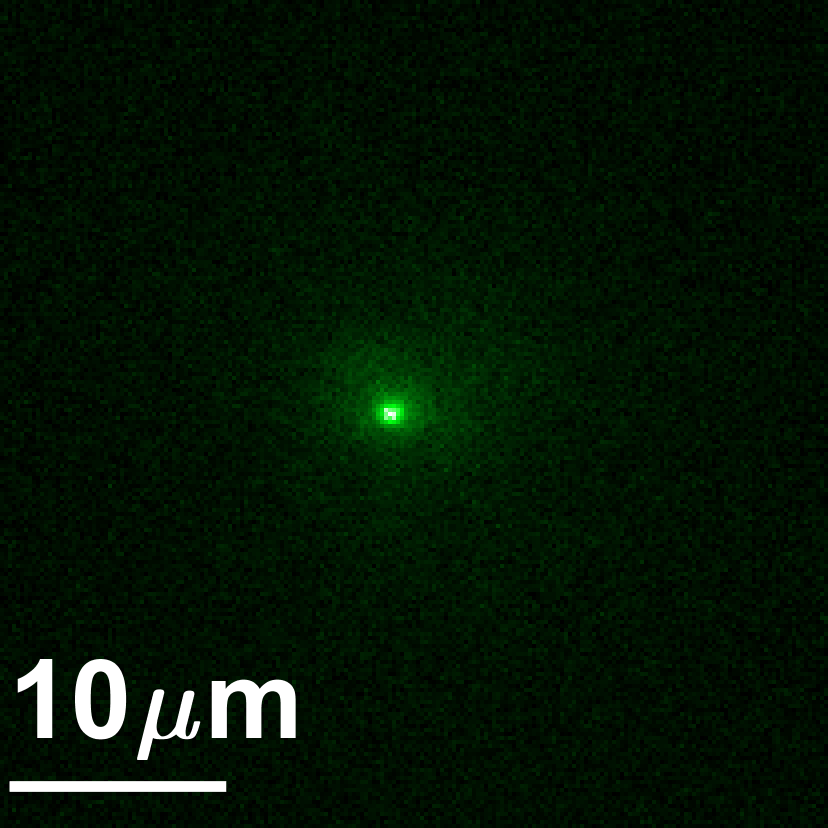}&
			\includegraphics[height = 0.115\textwidth]{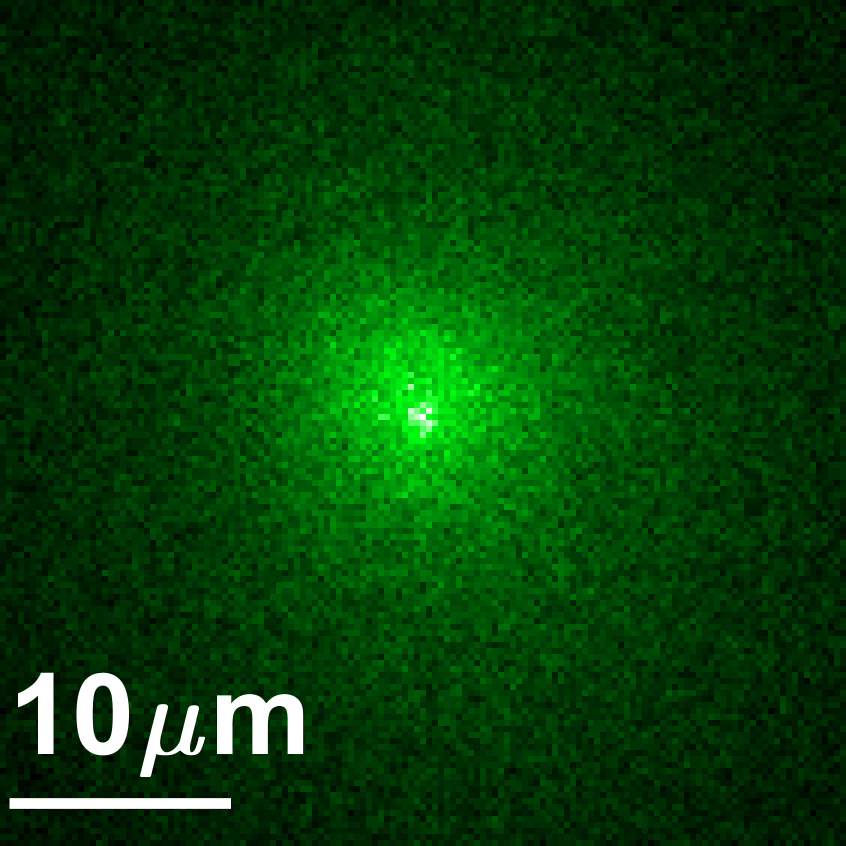}&
			\includegraphics[height = 0.115\textwidth]{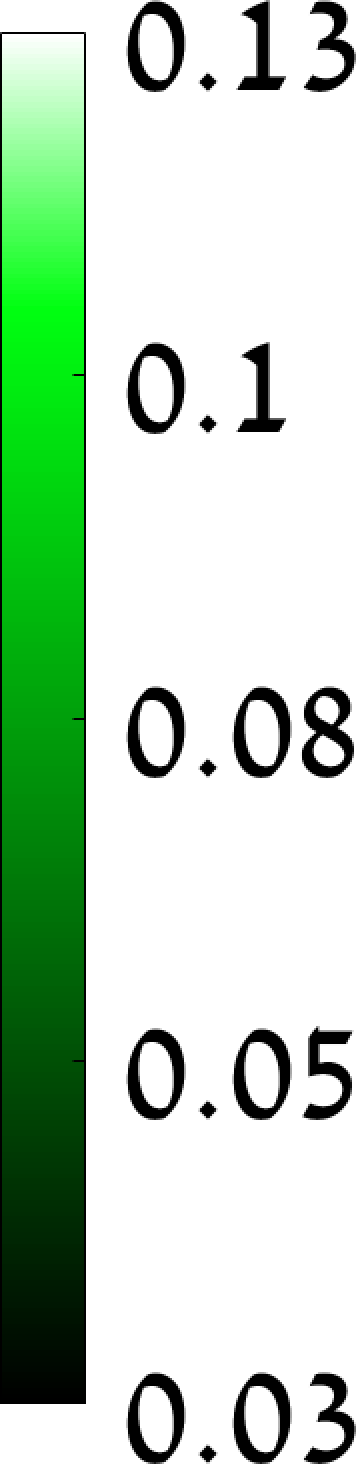}&
			\includegraphics[height = 0.115\textwidth]{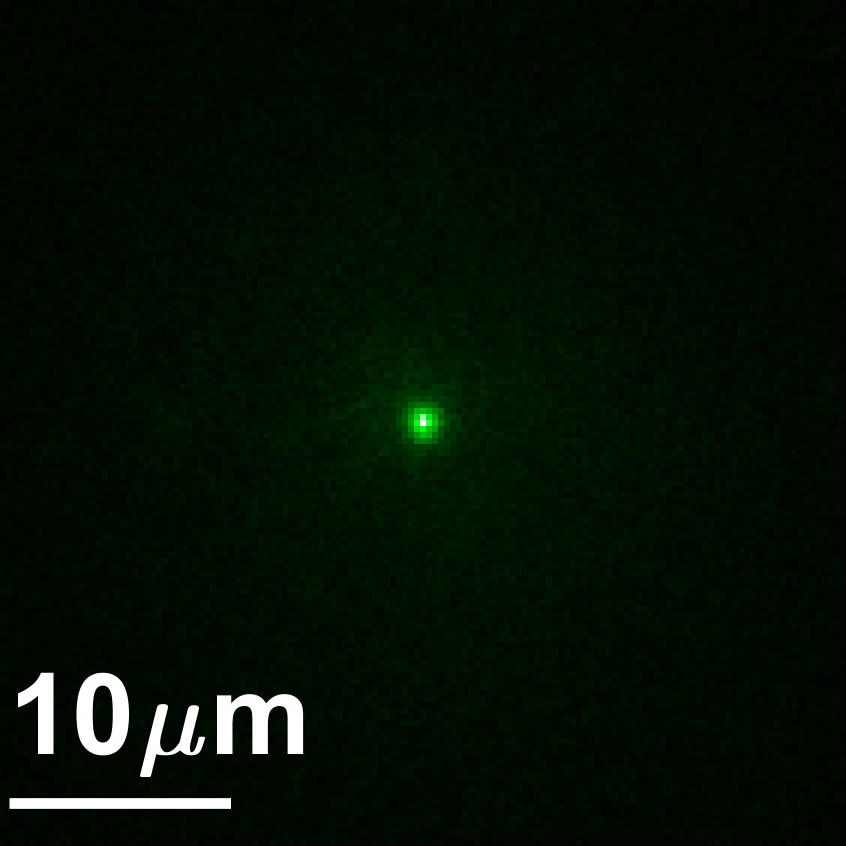}&
			\includegraphics[height = 0.115\textwidth]{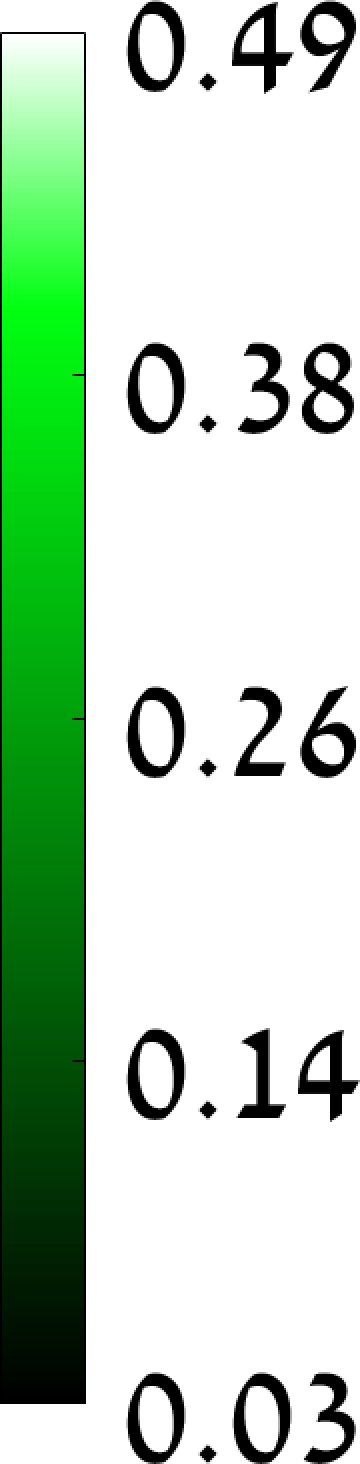}&
			\includegraphics[height = 0.115\textwidth]{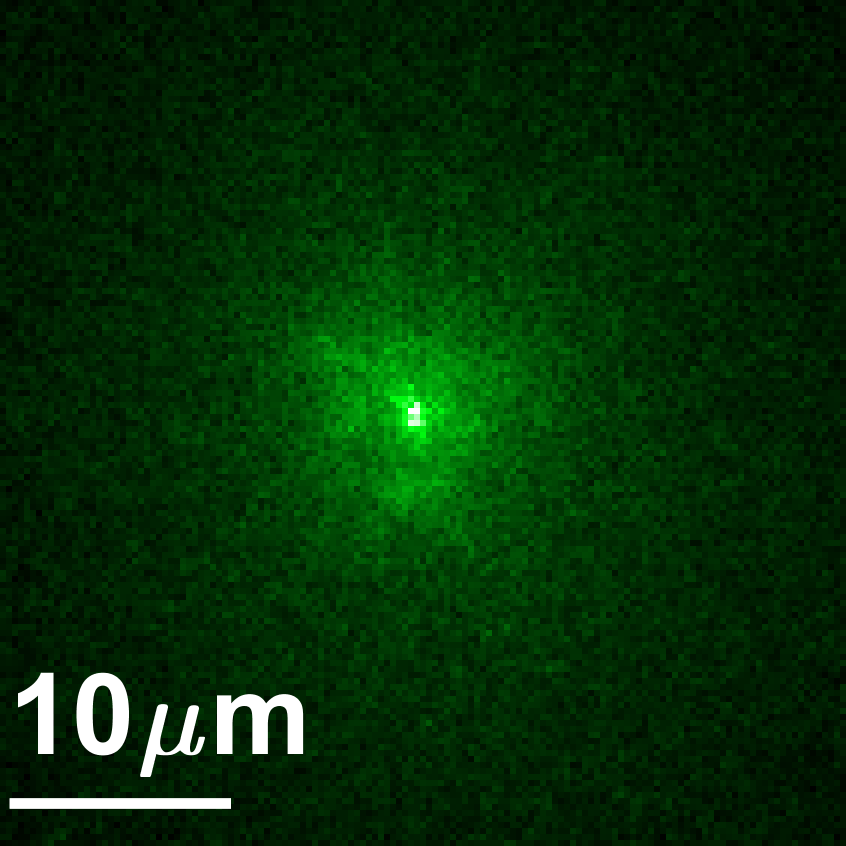}&
			\includegraphics[height = 0.115\textwidth]{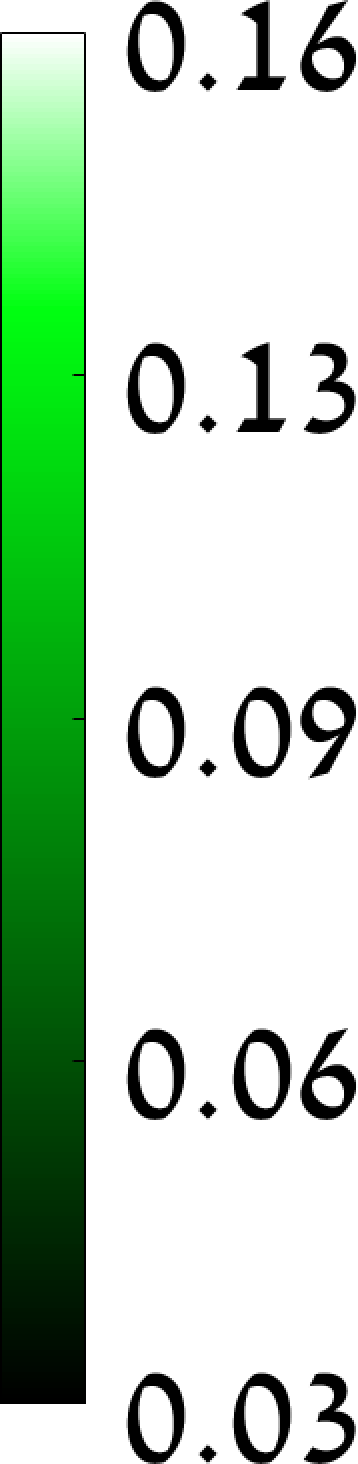}\\
			
			\footnotesize{(a) No mod.} & \footnotesize{(b) With mod.} & \footnotesize{(c) No mod.} & \footnotesize{(d) With mod.} & \footnotesize{(e) No mod.} &   & \footnotesize{(f) With mod.} &   & \footnotesize{(g) PSF}\\
			\multicolumn{2}{c}{Excitation}&\multicolumn{2}{c}{Emission}&\multicolumn{6}{c}{Emission}
		\end{tabular}
		\caption{Wavefront shaping results: we visualize views from the validation and main cameras, each row demonstrates a different tissue sample. 
			(a-b) The excitation light as viewed by the validation camera at the back of the tissue. Due to significant scattering, at the beginning of the algorithm when no modulation (mod.) is available, a wide speckle pattern is generated.  After  optimization, the modulated wavefront is brought into a single spot.
			(c-d)  By placing a band-pass filter on the validation camera, we visualize the emitted light with and without the modulation correction.
			(e-f) Views of the emitted light at the main front camera  with and without the modulation correction. Note that this is the only input used by our algorithm. Without modulation, light is scattered over a wide image area and the image is noisy.  A sharp clean spot can be imaged when the limited number of photons is  brought into a single sensor pixel.
			(g) By correcting the emission such that a single spot is excited and leaving the imaging path uncorrected, we can visualize the actual aberration of a single fluorescent point source. \blue{The top two examples demonstrate a thin brain layer behind parafilm, and the lower one is a thick brain slice. }
		}\label{fig:confocal}
	\end{center}
\end{figure*}


%
We image slices of mice brain with EGFP  neurons, excited at $488nm$ and imaged at $508nm$.
\blue{ We use two types of aberrations. In the first case we use thin  brain slices of thickness $50\mu m$, which are almost aberration-free. We generate scattering by placing these slices behind a layer of chicken breast tissue ($200-300\mu m$ thick) or parafilm whose optical properties were measured in~\cite{Boniface:19}. The advantage is that since the fluorescence is present only in a thin 2D layer we can obtain a clean reference from a validation camera. In a second experiment we image through $400\mu m$-thick brain slices. Since the target is 3D it is not always possible to capture clear aberration-free references. All experiments were	approved by Institutional Animal Care and Use Committee (IACUC) at the Technion (IL-149-10-2021), as well as  the Hebrew University of Jerusalem (MD-20-16065-4). More details about the mices are included in supplementary file.}

In \figref{fig:confocal}  we visualize some results of our algorithm.
 \figref{fig:confocal}(a) shows an image of the initial excitation pattern from the validation camera behind the tissue.  As can be observed, the tissue exhibits significant scattering. 
In  \figref{fig:confocal}(b), we visualize the excitation light after  optimizing the wavefront shaping modulation, which is nicely focused into a sharp spot. 
In  \figref{fig:confocal}(c-d), we also show the emitted light.
Before optimization a wide area is excited and we can see the neuron shape. At the end of the optimization a single point is excited.
In  \figref{fig:confocal}(e-f), we visualize the views of the front main camera, providing the actual input to our algorithm.  
Before optimization the emitted light is scattered over a wide sensor area.  
As a low number of photons is spread over multiple sensor pixels, the captured imaged is noisy. 
At the end of the optimization the aberration is corrected and all the photons are brought into a single sensor pixel.
In \figref{fig:confocal}(g), we demonstrate the actual point spread function of the tissue aberration. For that we have used the correction only at the illumination arm and focused  the illumination to excite a single spot.  We used a blank SLM at the imaging arm so the emitted light is not corrected. One can see that the aberration of a single fluorescent spot is rather wide. 
Each of the images in \figref{fig:confocal} is normalized so that its maximum is 1, but as indicated by the colorbar, the spot at the focused images received a much higher number of photons than the wide scattering images of unfocused light, despite the fact that all images were captured under equal exposure and equal excitation power. 
\begin{figure*}[t!]
	\begin{center}
		\begin{tabular}{@{}c@{~}c@{~~}c@{~}c@{~~}c@{~}c@{~~}c@{~}c@{}}

			\includegraphics[height= 0.18\textwidth]{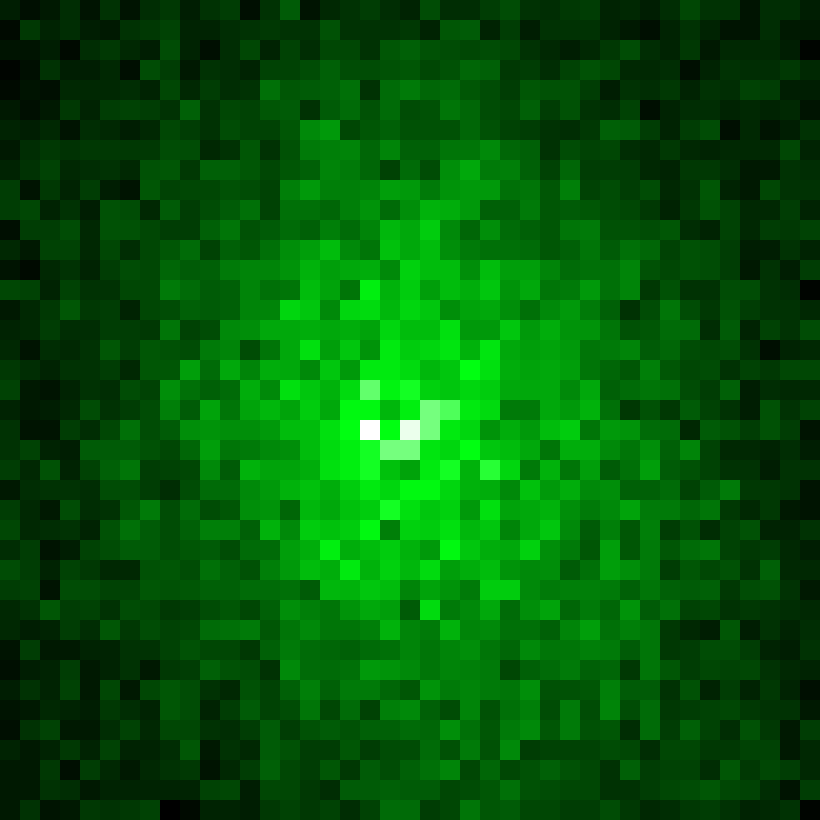}&
			\includegraphics[height= 0.18\textwidth]{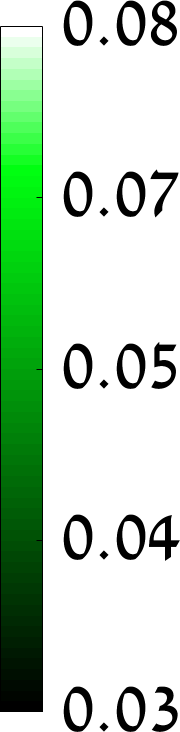}&
			\includegraphics[height= 0.18\textwidth]{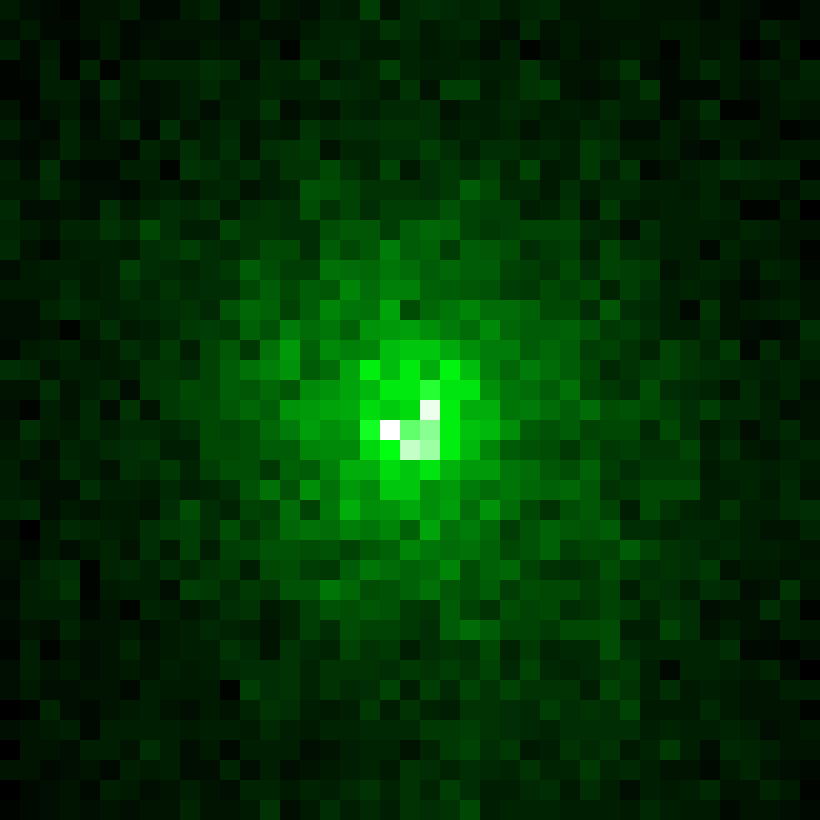}&			
			\includegraphics[height= 0.18\textwidth]{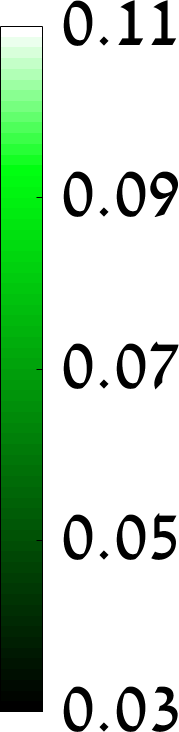}&
			\includegraphics[height= 0.18\textwidth]{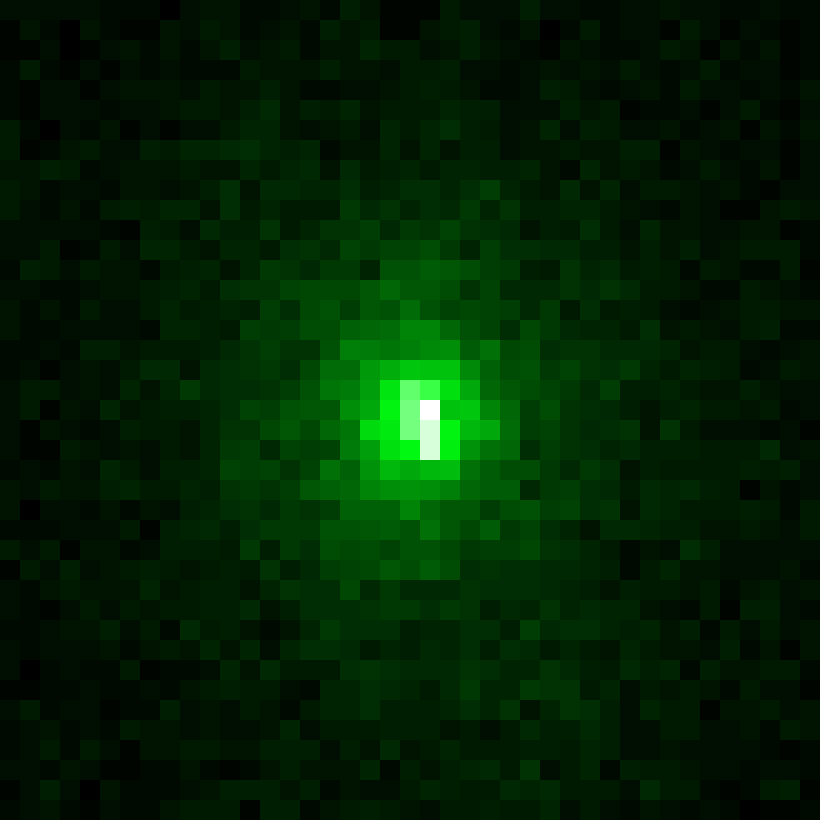}&			
			\includegraphics[height= 0.18\textwidth]{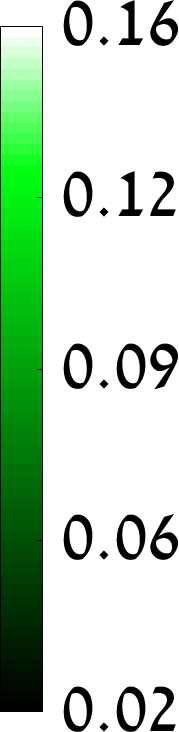}&
			\includegraphics[height= 0.18\textwidth]{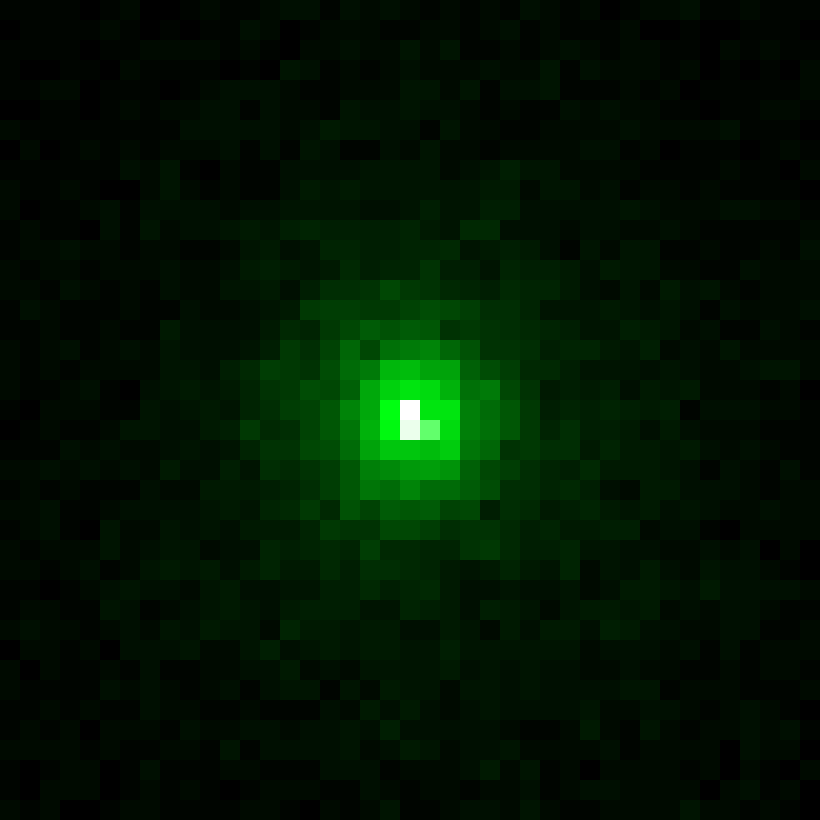}&			
			\includegraphics[height= 0.18\textwidth]{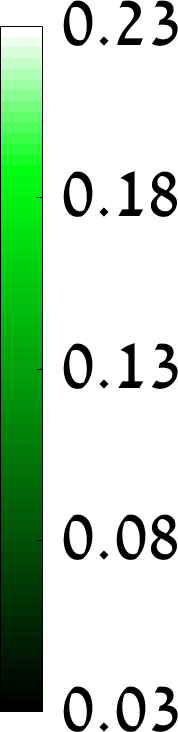}	
			\\
Itr 0&&Itr 200&&Itr 400&&Itr 1200&\\
\blue{SNR 76.89}&&\blue{SNR 223.97}&&\blue{SNR 343.28}&&\blue{SNR 537.15}&
		\end{tabular}
		\caption{Convergence with noise.
		Views from the main camera at 
		 a few iterations of our algorithm. In the beginning a small number of photons is spread over multiple sensor pixels and the resulting image is very noisy. As the algorithm proceeds and a wavefront shaping modulation is recovered, the low number of photons is brought to a single sensor spot and the measured image has a higher SNR. \blue{SNR is calculated by capturing 40 images with the same modulation.} \Anat{Are you sure about SNR number? aren't they high?}}
	 \label{fig:noise_conv}
	\end{center}
\end{figure*}

To better appreciate the noise handled by our algorithm, in \figref{fig:noise_conv} we visualize a $41\times 41$-pixel window captured by the main camera, at  a few iterations of our algorithm. In the beginning this image is very noisy, because a small number of emitted photons are spread over multiple sensor pixels. However, when the optimization proceeds, it finds a better modulation correction.
As a result, all the laser power is brought to excite one spot and
all the emitted photons are collected to one sensor spot and measured with better SNR.

\begin{figure*}[t!]
	\begin{center}

		\begin{tabular}{@{}c@{~~}c@{~~}c@{}}

			\includegraphics[width= 0.2\textwidth]{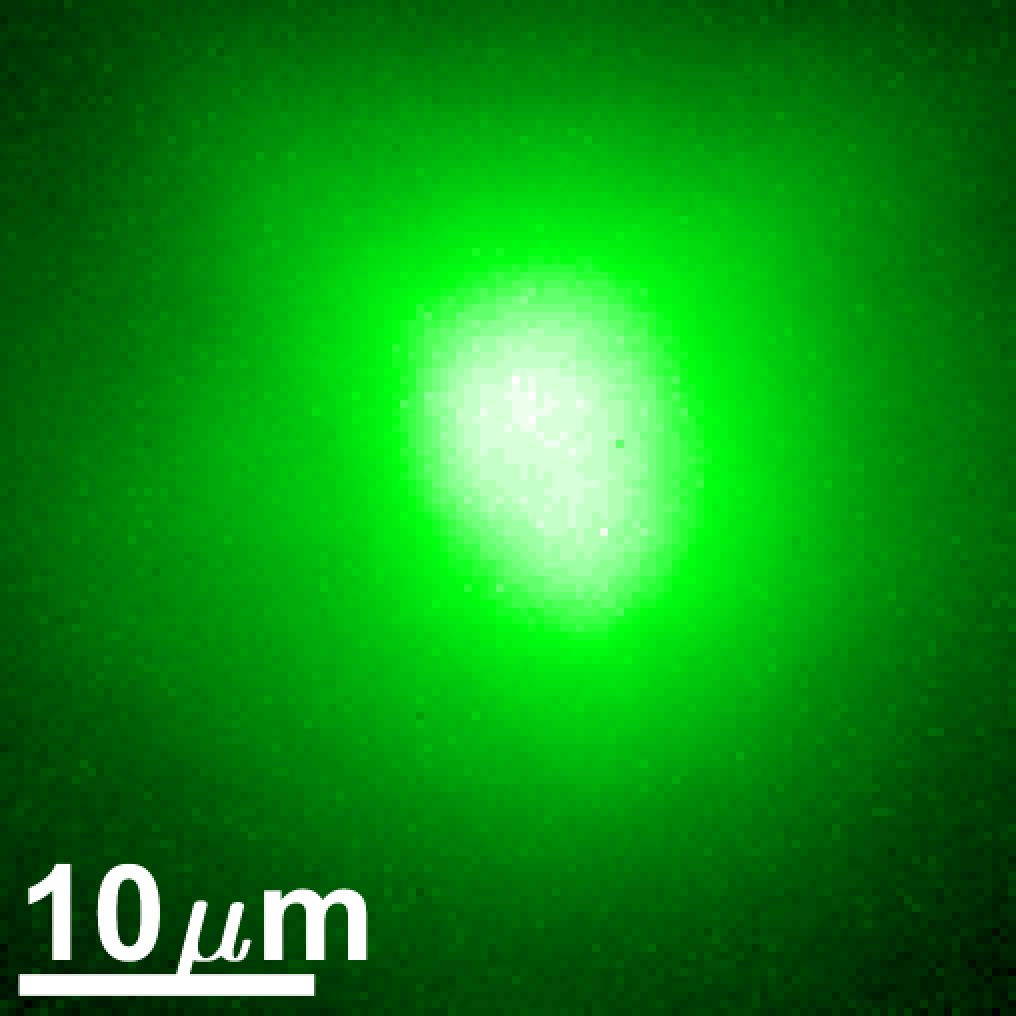}&
			\includegraphics[width= 0.2\textwidth]{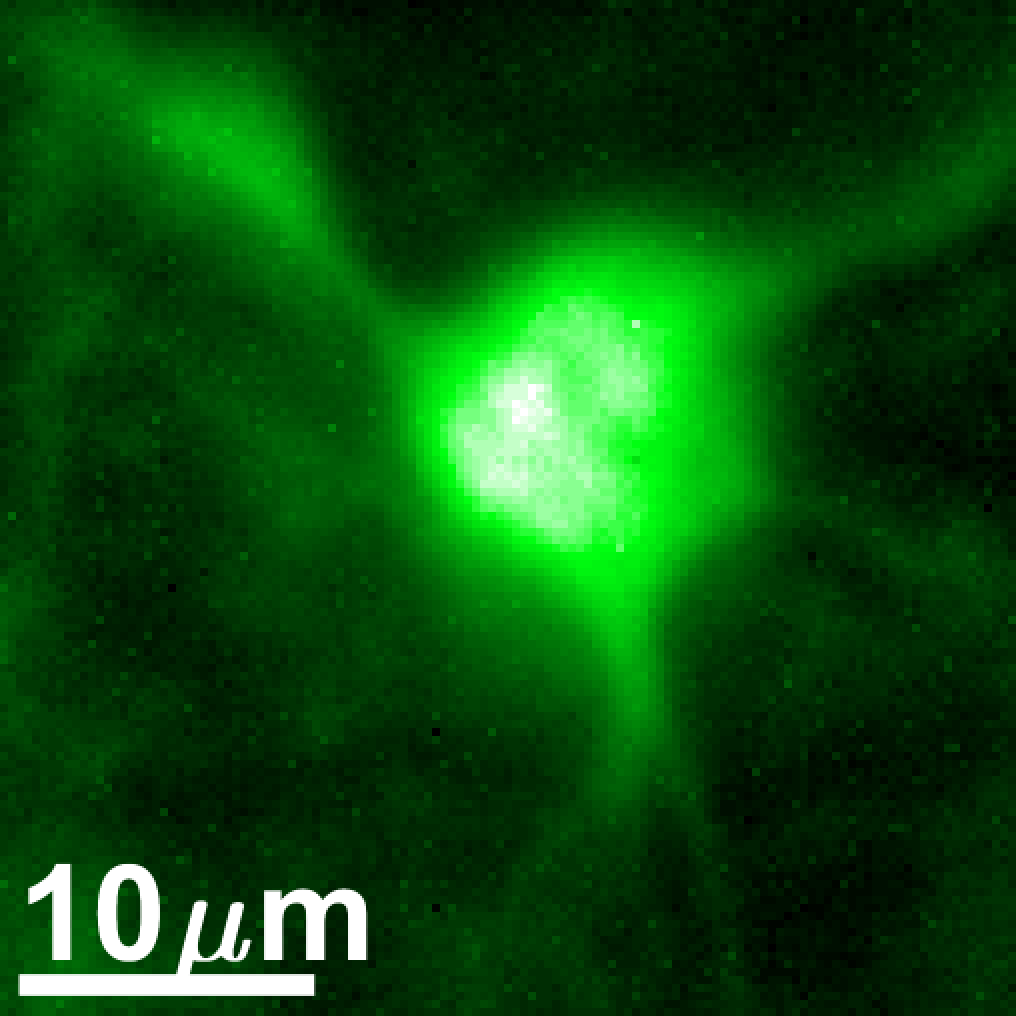}&
			\includegraphics[width= 0.2\textwidth]{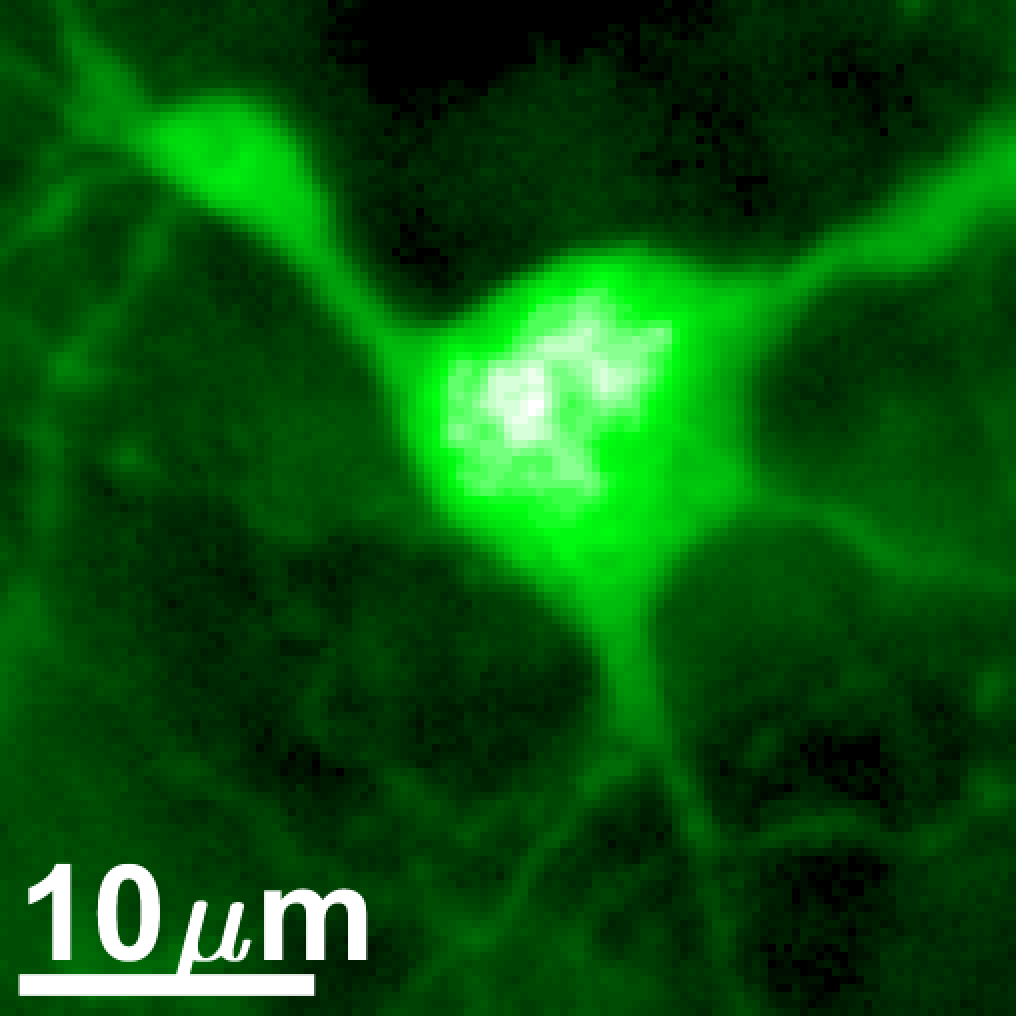}\\
			
			\includegraphics[width= 0.2\textwidth]{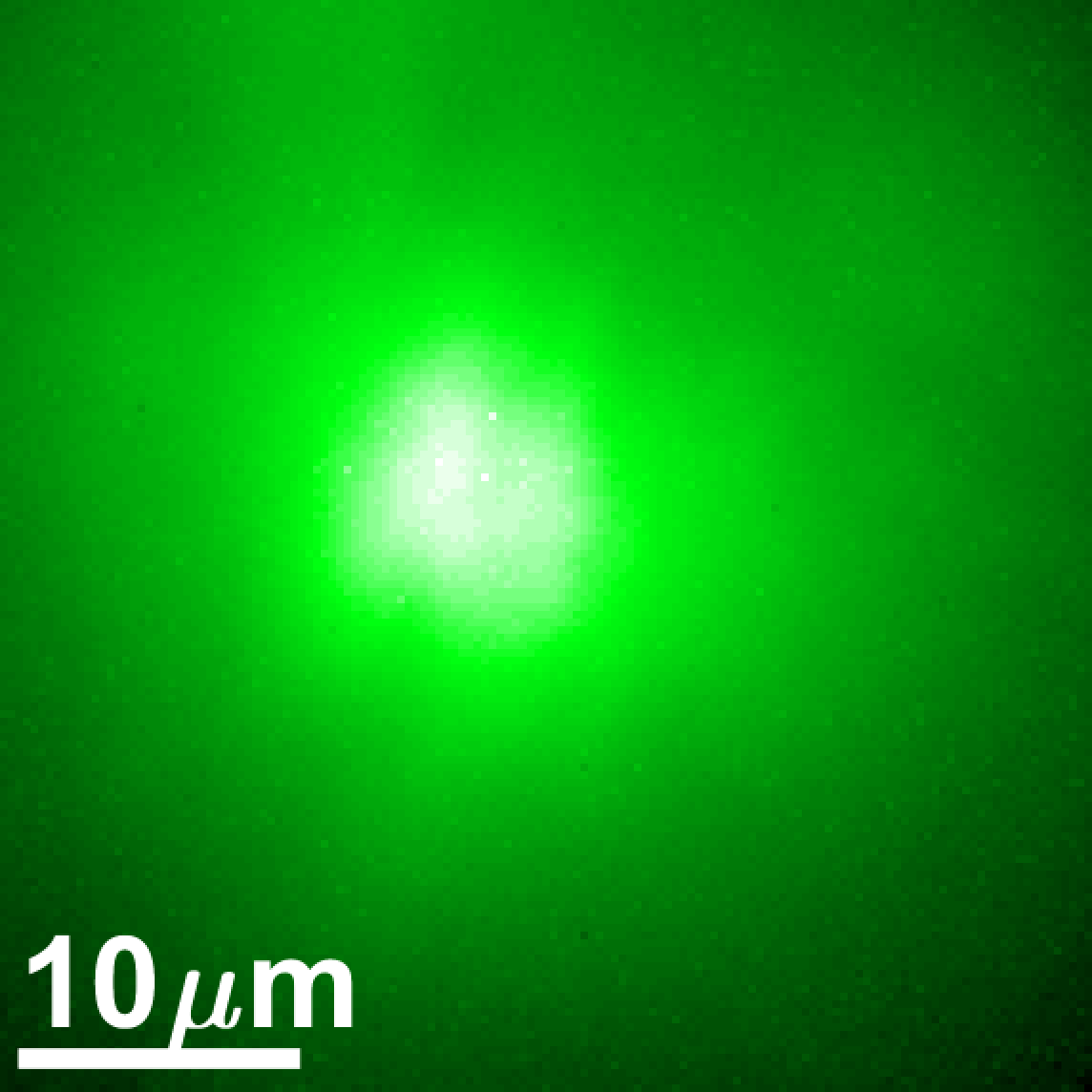}&
			\includegraphics[width= 0.2\textwidth]{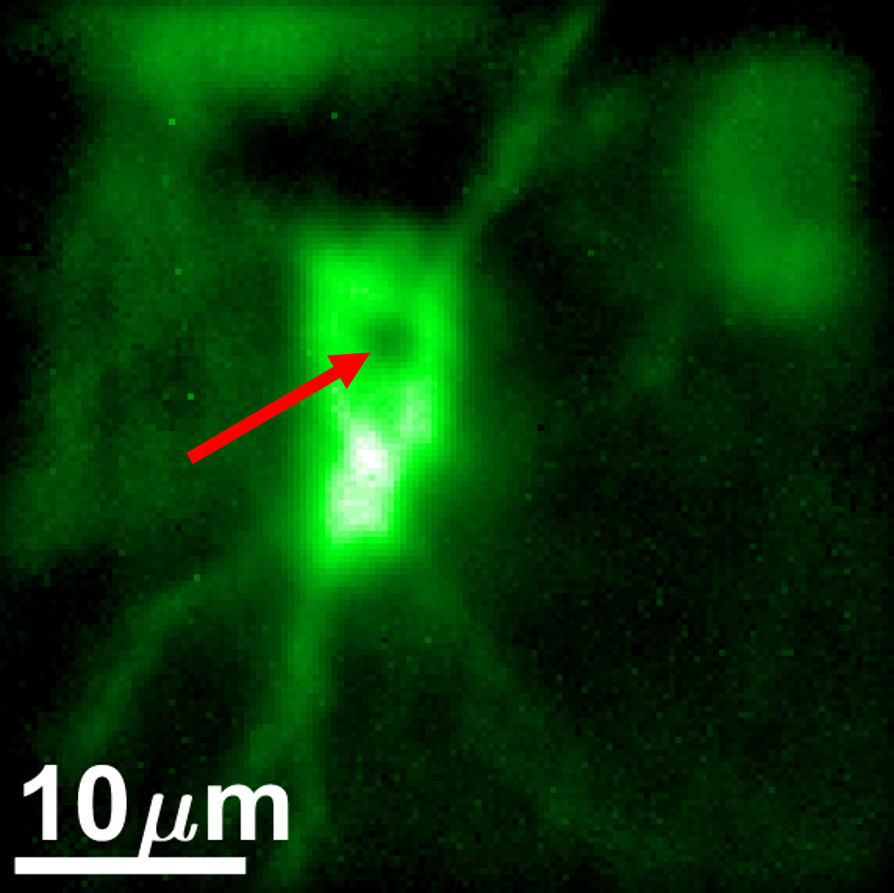}&
			\includegraphics[width= 0.2\textwidth]{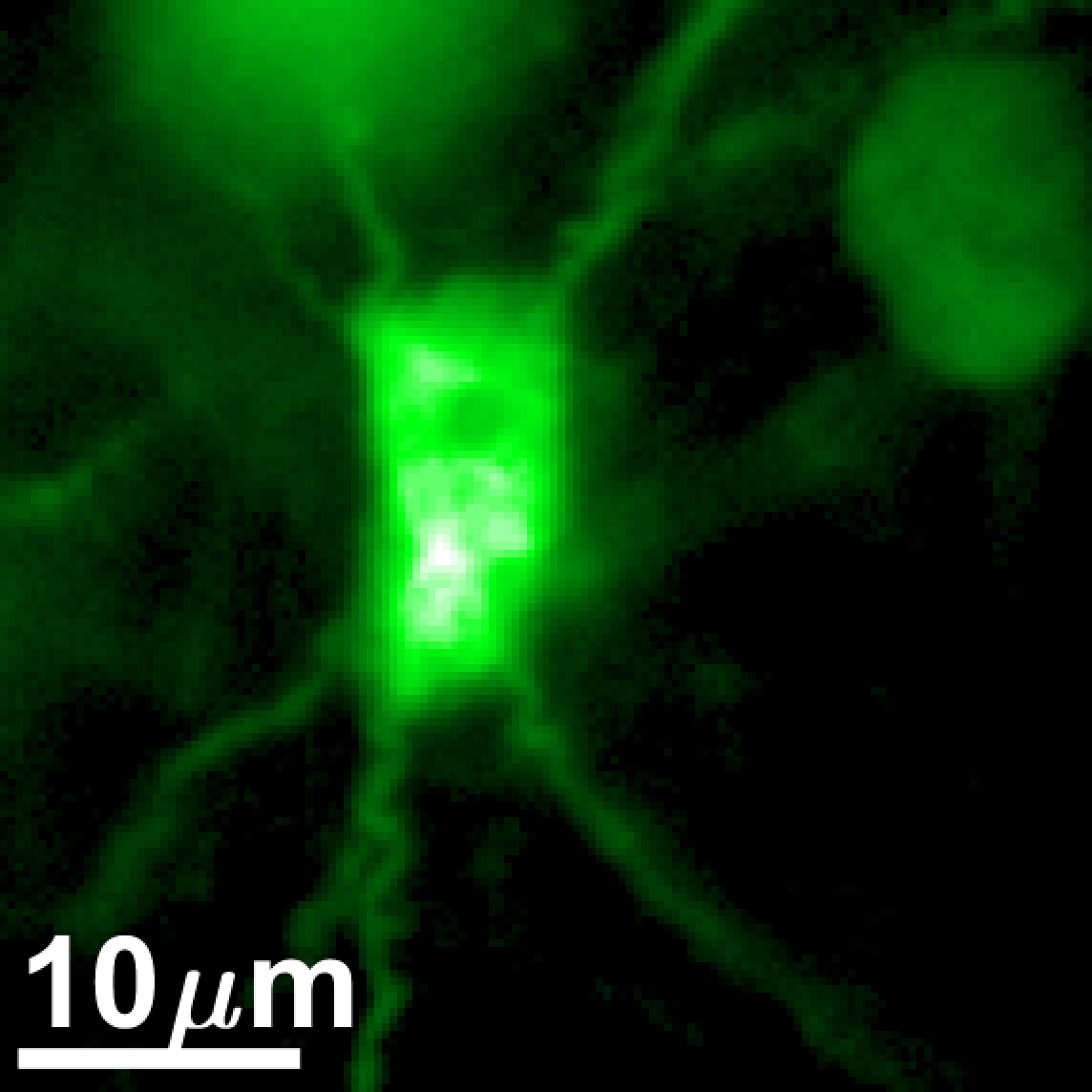}\\
			
			\includegraphics[width= 0.2\textwidth]{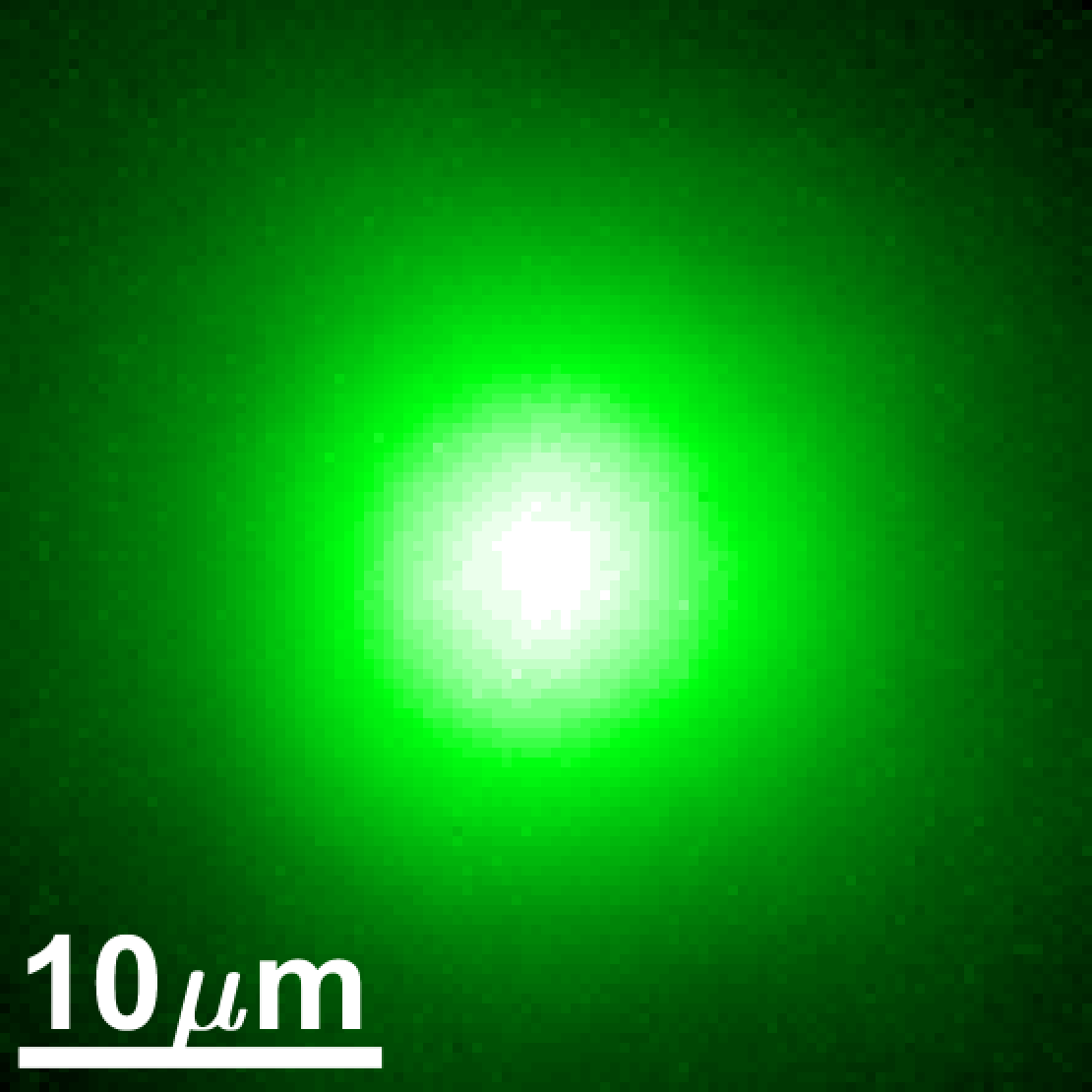}&
			\includegraphics[width= 0.2\textwidth]{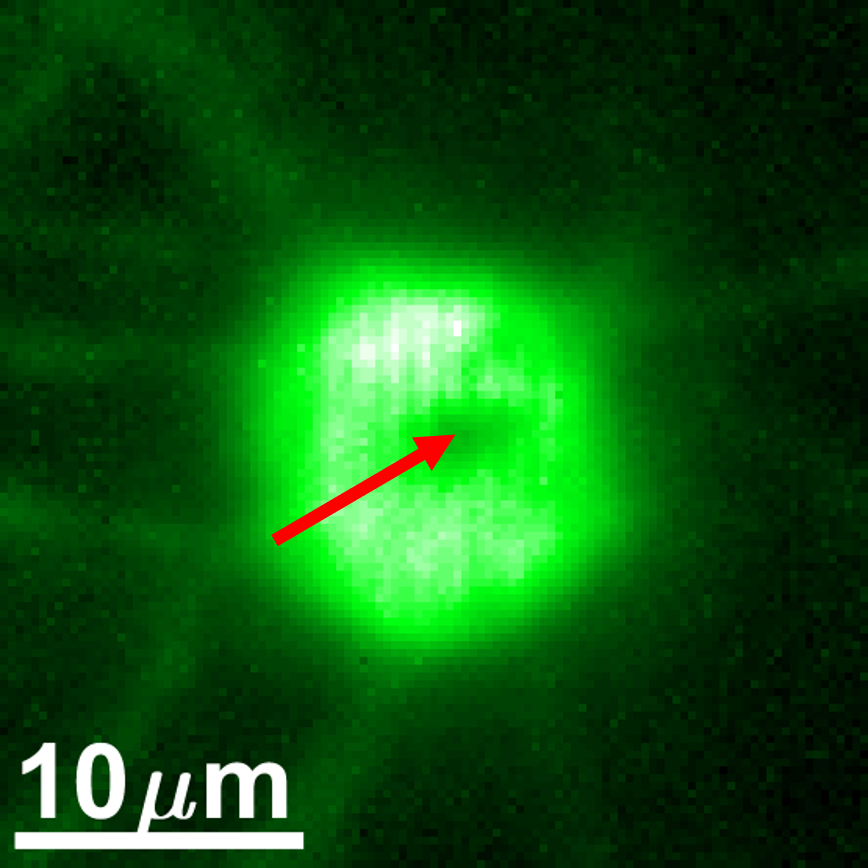}&
			\includegraphics[width= 0.2\textwidth]{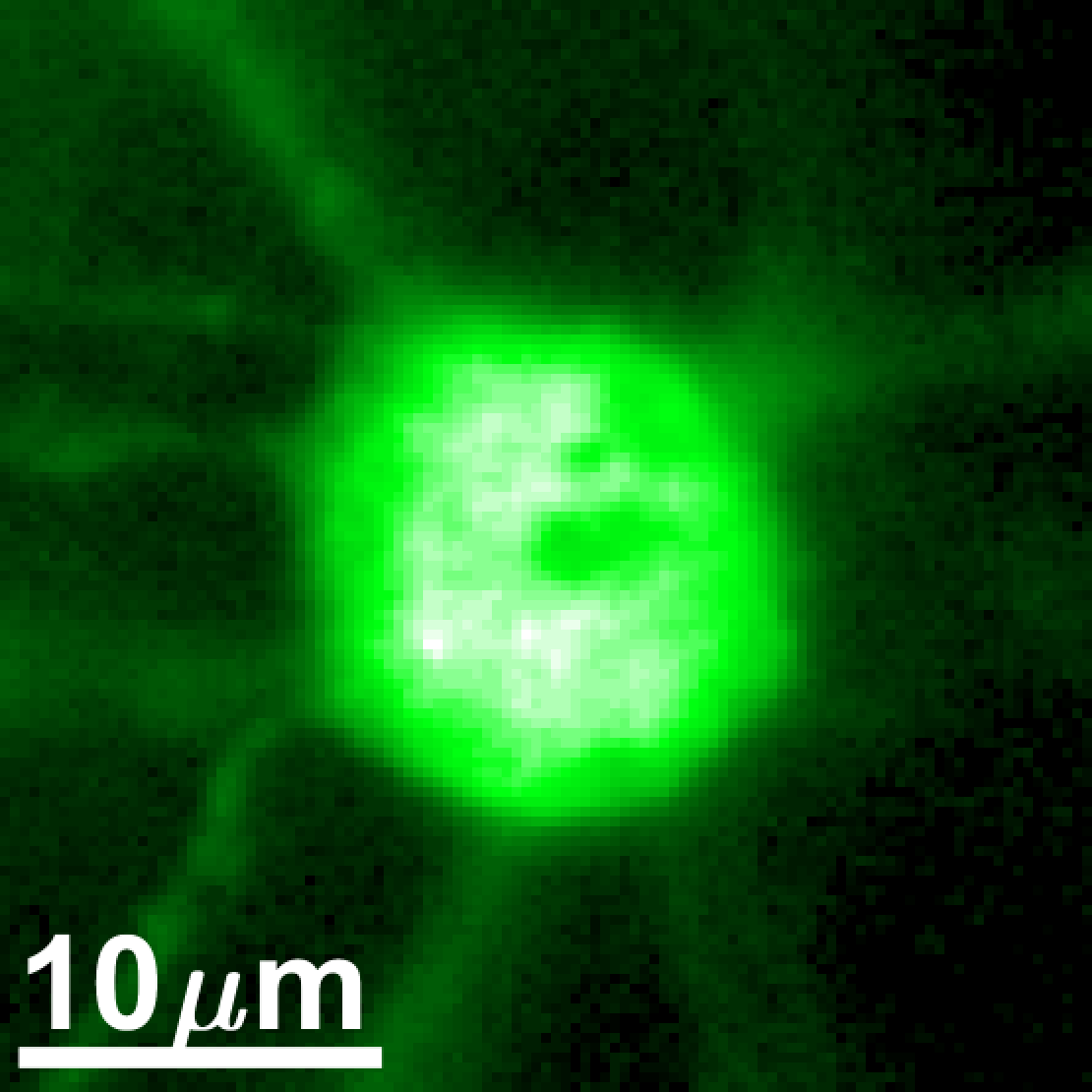}\\			
			
			\includegraphics[width= 0.2\textwidth]{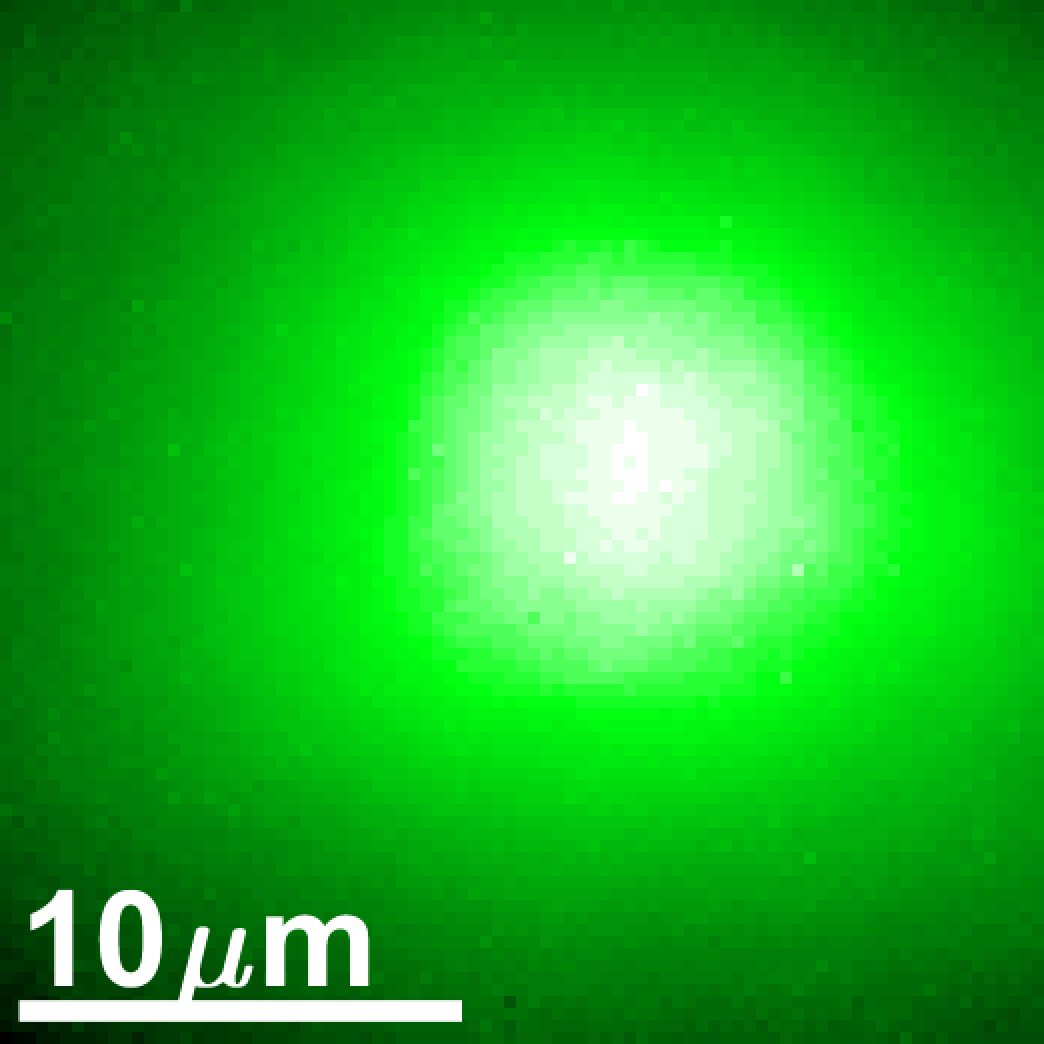}&
			\includegraphics[width= 0.2\textwidth]{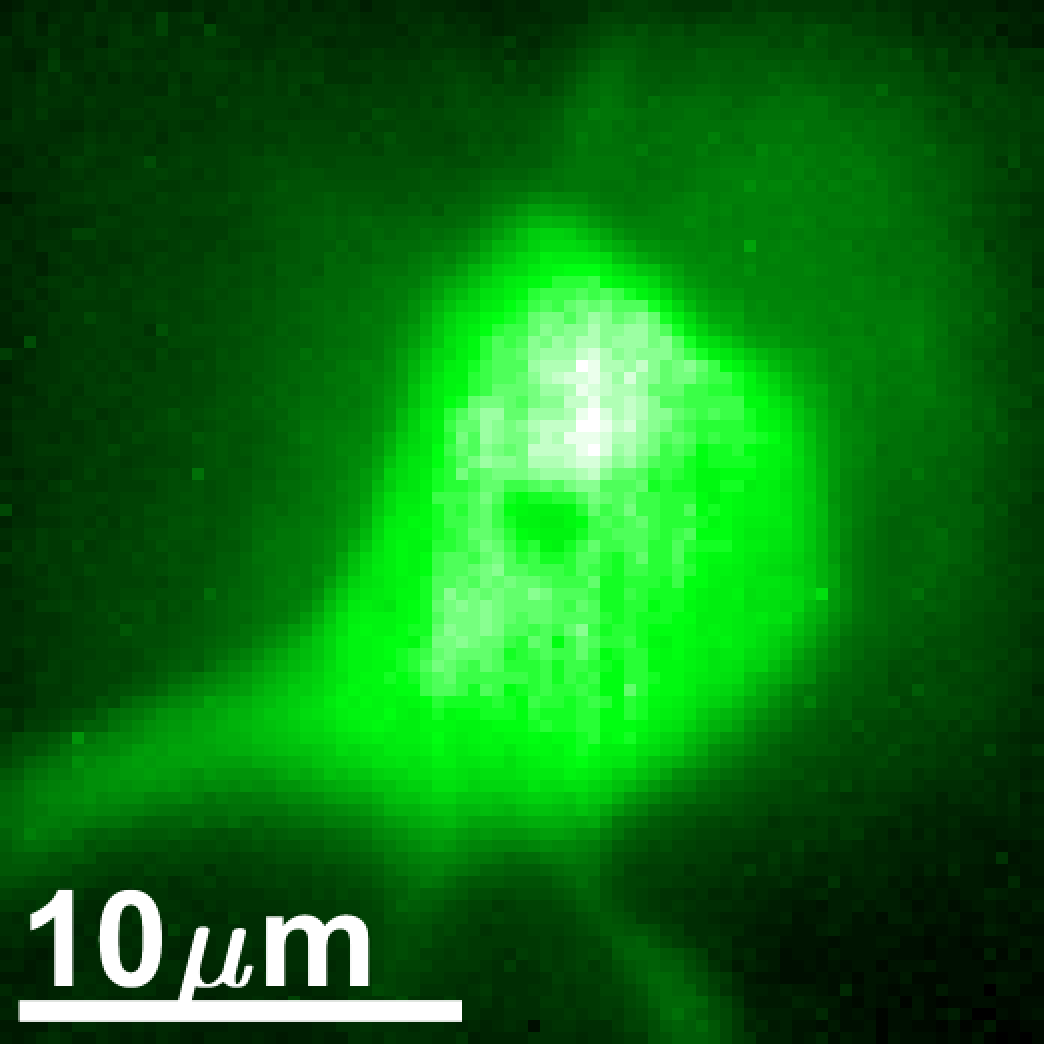}&
			\includegraphics[width= 0.2\textwidth]{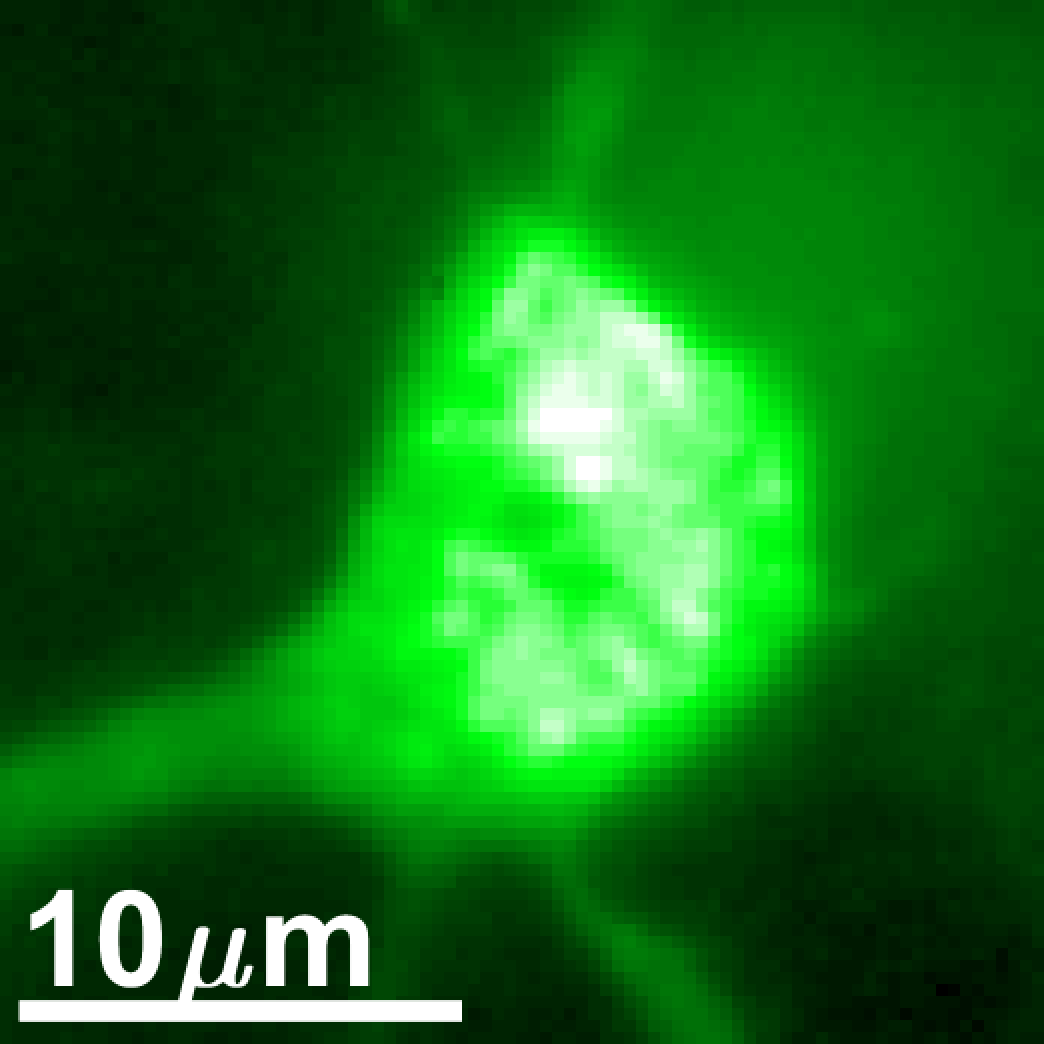}\\

		\end{tabular}
		\caption{Imaging a wide area.  We image a thin fluorescent brain slice behind a scattering layer. The top two images correct scattering through chicken breast tissue and the lower ones through parafilm.   (a) Image of the neuron from the main camera with no correction, strong scattering is present and the neuron structure is lost. (b) Image with our modulation correction, the neuron shape as well as some of the axons are revealed. (c) A  reference  image of the same neuron, from the validation camera. The arrow marks a spot at which the  optimization has converged. This spot is darker as it bleached during the optimization.
		}\label{fig:big_area}
	\end{center}
\end{figure*}

We use the recovered wavefront shaping modulation to image a wide area rather than a single spot.  In \figref{fig:big_area} we demonstrate results for a thin brain slice beyond chicken breast and parafilm. 
For that, we excite a wide area and use a correction only at the imaging arm. Due to the memory effect~\cite{Judkewitz14,osnabrugge2017generalized}, the same modulation can allow us to image a small local patch rather than a single spot. 
With the correction, the neuron is observed with a much higher contrast and even the axons (thin lines around the neuron) emerging from it, whose emission is much weaker, can be partially observed. \textbf{ Additional results are provided in supplementary file}. 
We note that noise is less visible in the aberrated images of \figref{fig:big_area}(a), because these images are captured with a much longer exposure compared to the optimization images in \figref{fig:noise_conv}. While it is possible to capture a few noise-free images of such targets, it is not possible to do that for all the optimization iterations without bleaching. 
In \figref{fig:big_area} we mark with an arrow some points at which the algorithm has converged. One could see a darker spot as such points have bleached during optimization.

\begin{figure*}[t!]
	\begin{center}
		\begin{tabular}{@{}c@{~~}c@{~~}c@{}}
			
			\includegraphics[height= 0.21\textwidth]{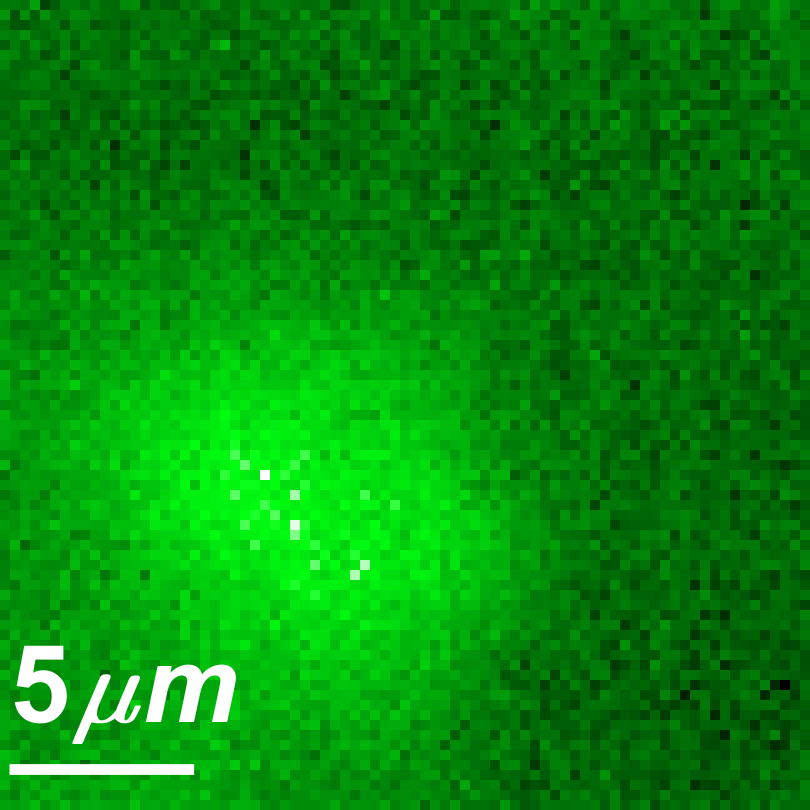}&
			\includegraphics[height= 0.21\textwidth]{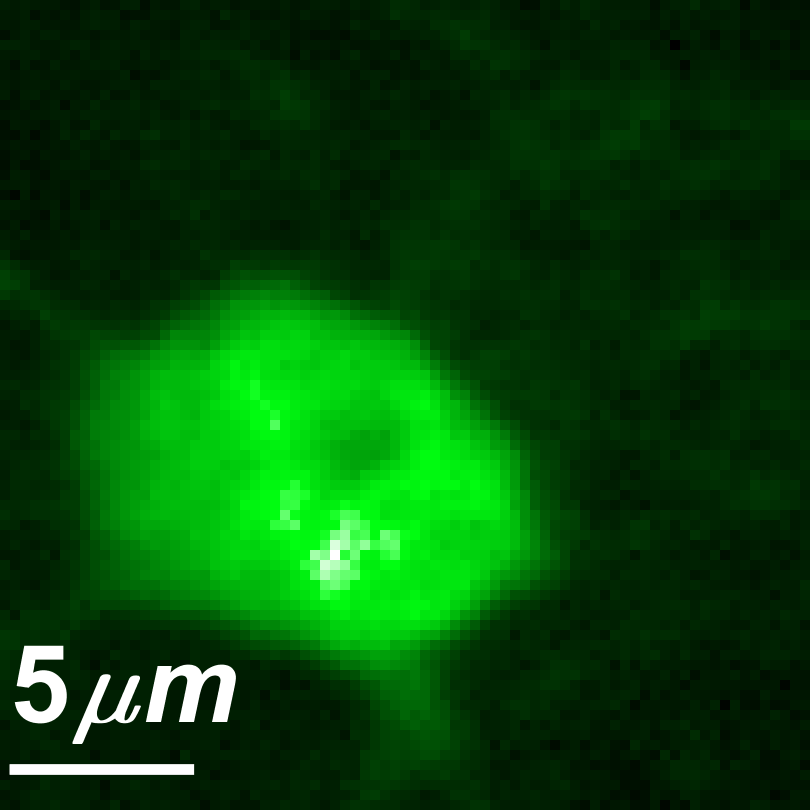}&
			\includegraphics[height= 0.21\textwidth]{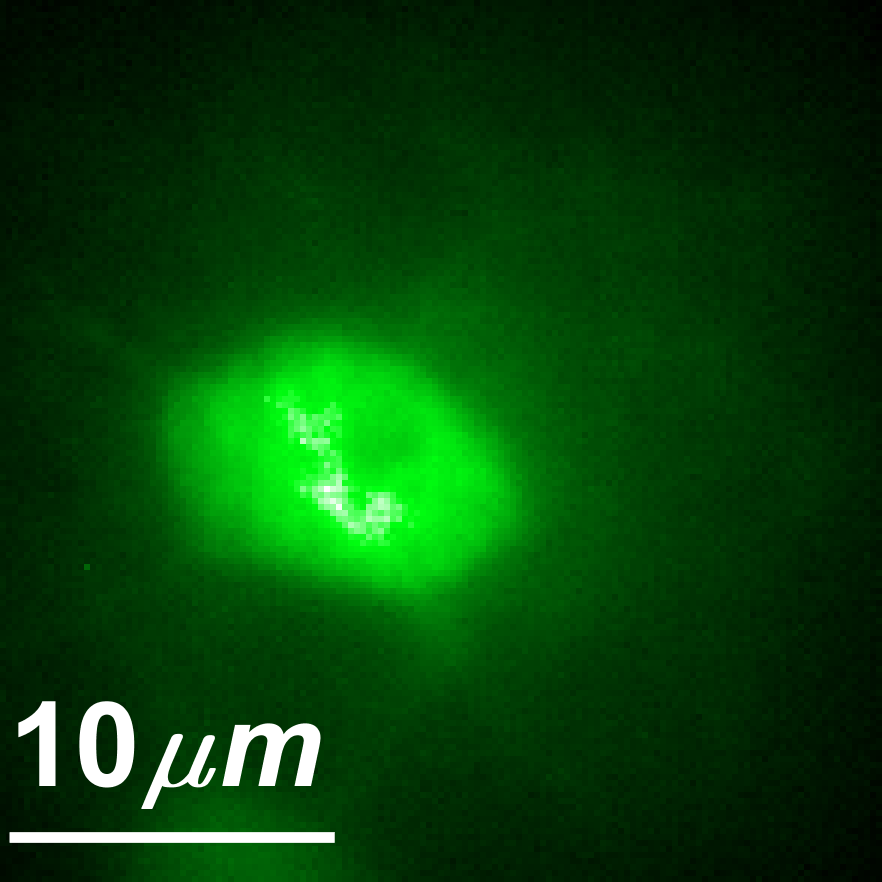}\\
         	\includegraphics[height= 0.21\textwidth]{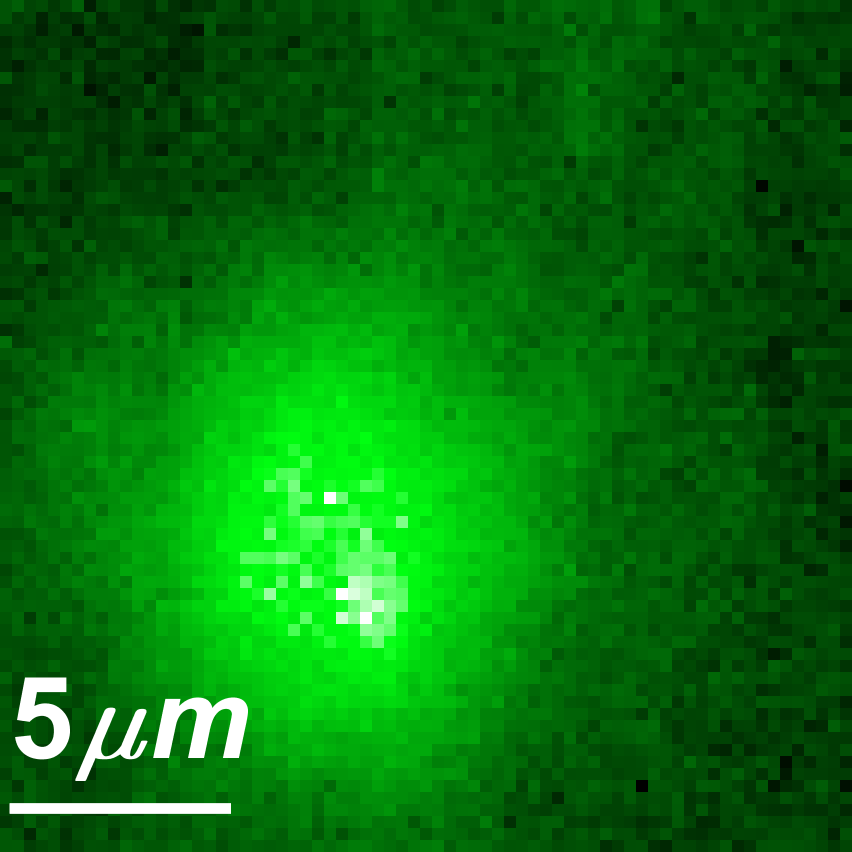}&
            \includegraphics[height= 0.21\textwidth]{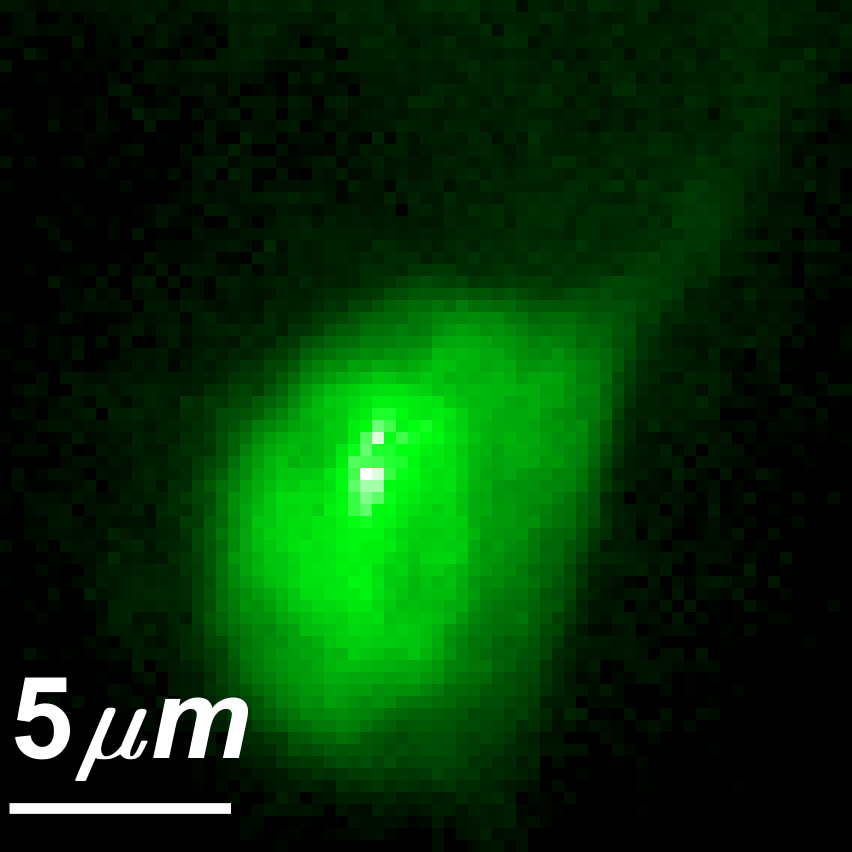}&
            \includegraphics[height= 0.21\textwidth]{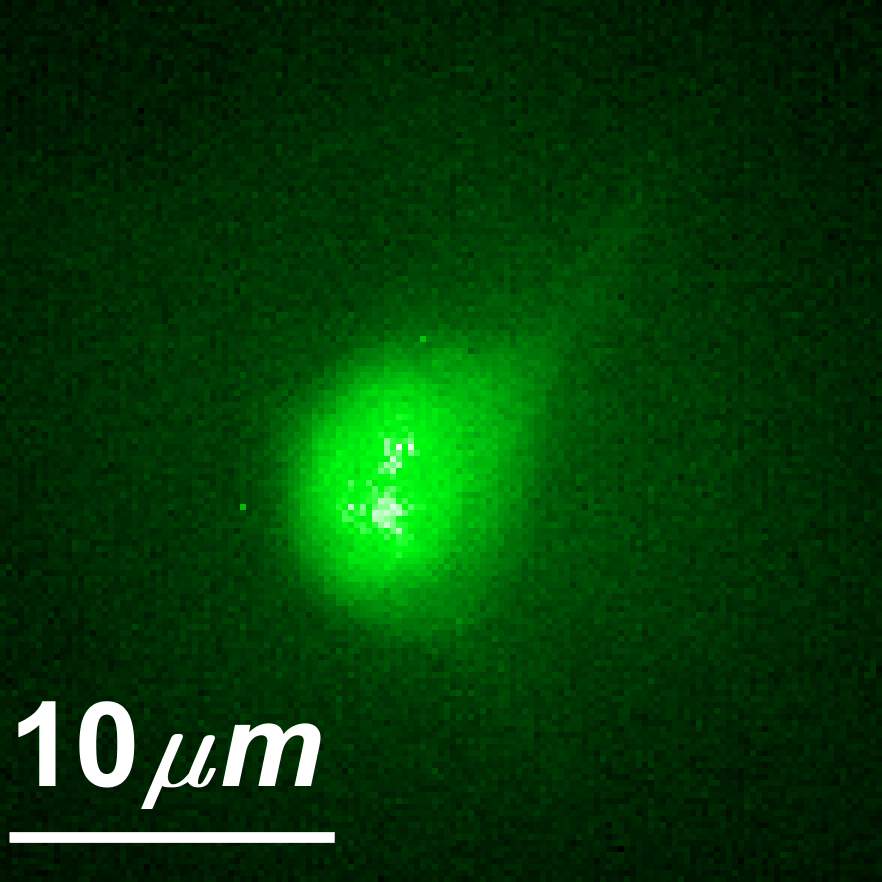}\\
			\includegraphics[height= 0.21\textwidth]{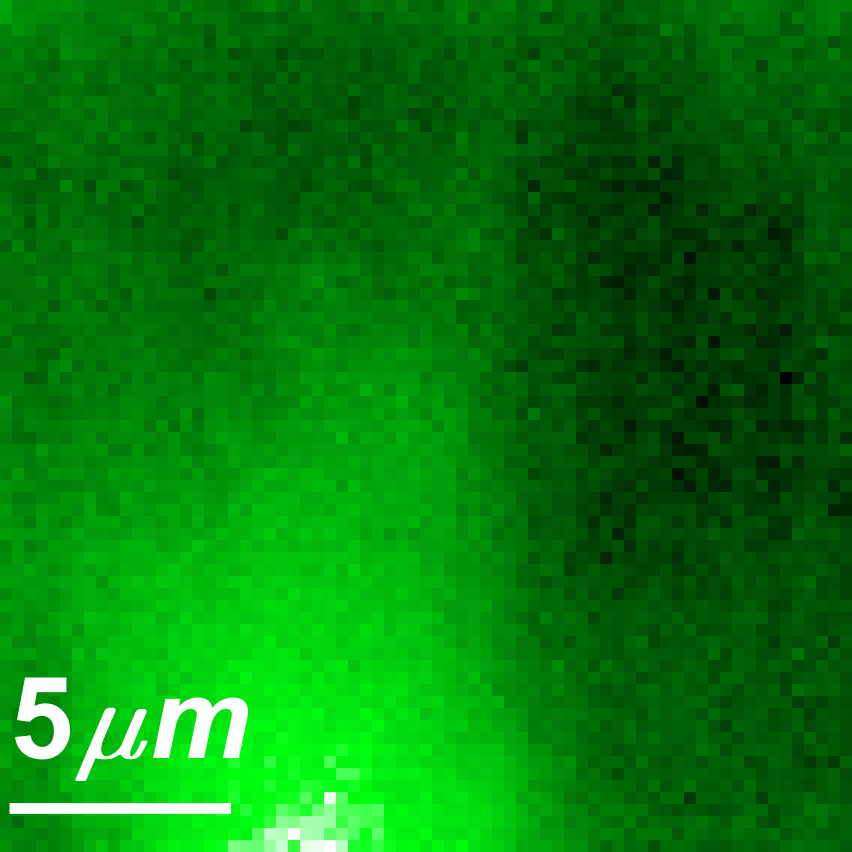}&
			\includegraphics[height= 0.21\textwidth]{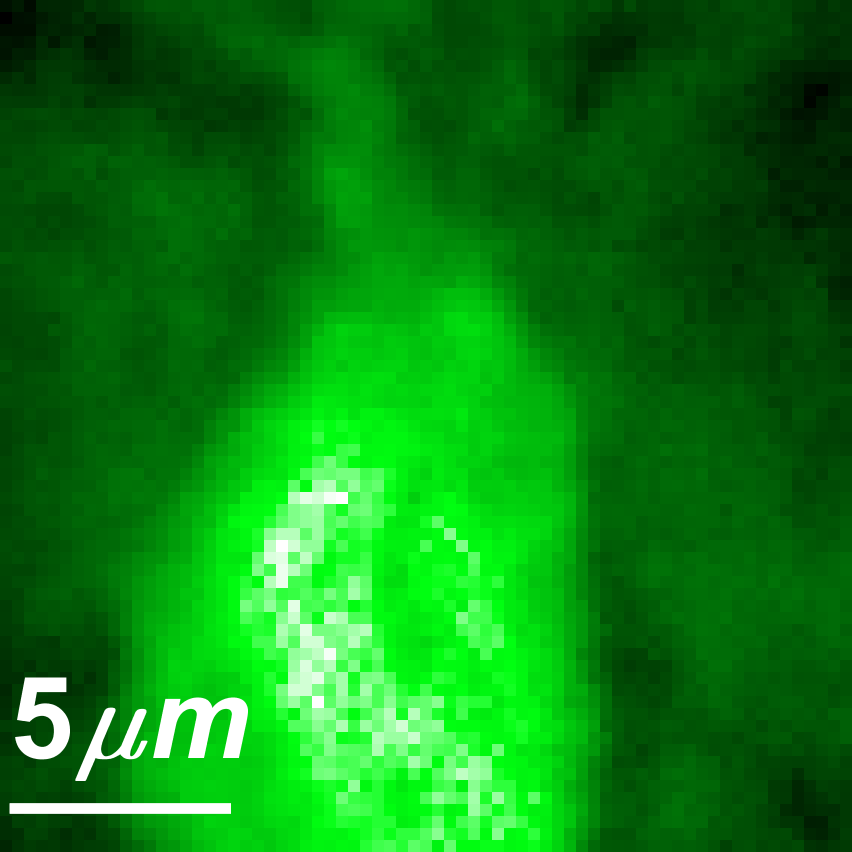}&
			\includegraphics[height= 0.2\textwidth]{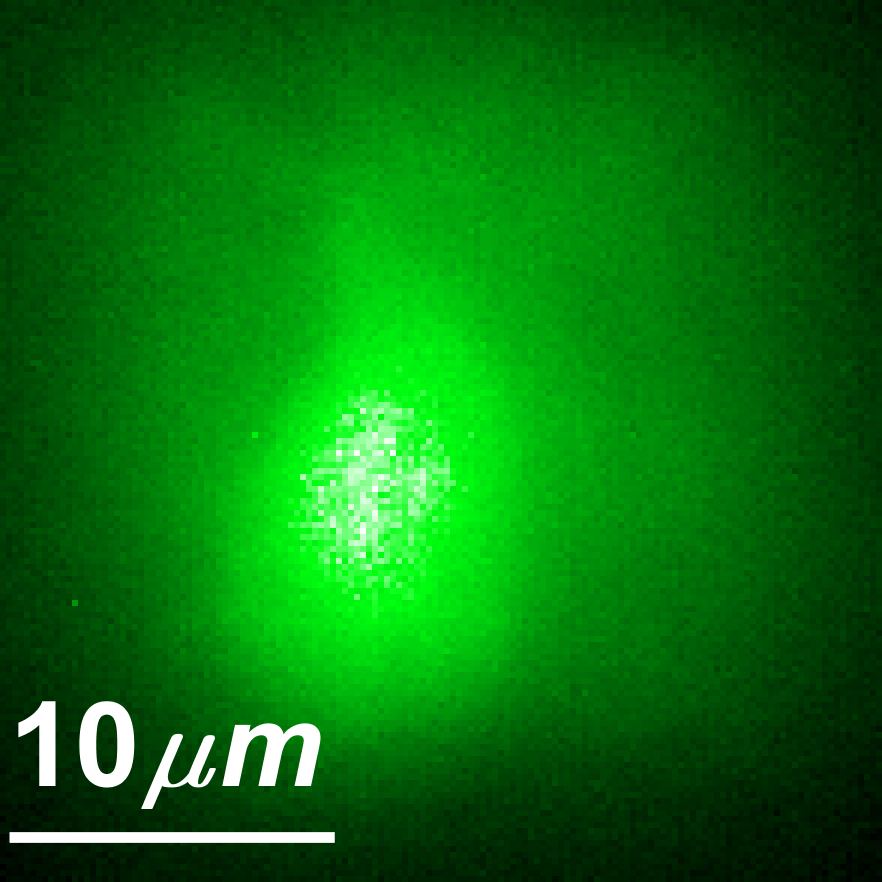}\\

	{\small (a) Uncorrected confocal } &	{\small  (b) Corrected confocal} & 	{\small (c) Reference}\\
	{\small Main camera} & 	{\small  Main camera }&	{\small   Validation camera}\\
		\end{tabular}
		\caption{\blue{Imaging a wide 3D target, through a $400\mu m$ thick fluorescent brain slice.   (a) A confocal image of the neuron from the main camera with no correction, strong scattering is present and the neuron structure is lost. (b) A confocal image with our modulation correction, the neuron shape as well as some of the axons are revealed. (c) A  reference  image of the same neuron, from the validation camera. Due to the 3D spreading of the fluorescent components, the validation camera cannot always capture an aberration-free image of the target. }	}\label{fig:confocal_images}

	\end{center}
\end{figure*}

\blue{In \figref{fig:confocal_images} we show imaging through a $400\mu m$   brain slice. Due to the 3D structure of the target, to isolate a neuron at a single-depth plane we have to use a slow confocal scanning, where the modulation is placed on  both arms and is tilt-shifted to excite and image different spots of the target. We compare this with an uncorrected confocal scan that is significantly aberrated. In some cases there is a small shift between the corrected and uncorrected confocal images, because the recovered modulation has also shifted the focal spot inside the target. Note that our confocal scanning is currently implemented by tilting and shifting the modulation pattern on the SLM and not with a proper galvo mirror. Since this approach is very slow we could only scan small windows. We also include a full frame image from the validation camera behind the tissue, but due to the 3D  fluorescence structure, in some cases this does not provide a clear ground truth. \textbf{ Additional results are provided in the supplementary file.}}

\blue{	In Sec. 6 of the supplementary file we compare our confocal score with the variance maximization approach of~\cite{Boniface:19}, showing that our approach can converge using a significantly smaller number of photons. We also compare against  one of the non-local approaches by~\cite{YeminyKatz2021}. This approach assumes that a single modulation can correct a wide image region rather than a single spot. Our evaluation shows that when memory-effect (ME) exists over a wide extent, this algorithm can indeed recover good modulations, but the quality of the results degrades for a small ME, where the size of the iso-planatic patches that can be corrected with a single modulation is small. }

	\section{Discussion}
\label{sec:disc}

In this research, we have analyzed score functions for wavefront shaping correction, using non-invasive feedback at the absence of a guiding star.
To assess focusing quality, we seek a score function that can measure a non-linear function of the light emitted by different sources. This is naturally achieved when using two-photon fluorescent feedback, but is harder to achieve with linear fluorescence. We show that by using a confocal correction at both the illumination and imaging arms we can measure such a non-linear feedback, which is maximized when all excitation light is brought into one spot.
Moreover, the fact that our system uses a correction of the emitted light as part of the optical path allows us to bring the limited number of emitted photons into a single sensor spot, facilitating a high SNR measurement. 

It is worth noting that while our approach can recover focusing modulations, it can converge to different spots on the fluorescent target, depending on initialization.
While we cannot know where it has focused, we can use memory effect correlations to image the surrounding window.
Another drawback of the approach is that in our current implementation it takes about 30 min to optimize for one modulation pattern. 
Some of this can be largely optimized with better hardware, such as a faster SLM.
However, the iterative optimization is inherently slower than the power iterations of~\cite{Dror22}.
Beyond better hardware, we are exploring algorithmic alternatives which can accelerate optimization.


	\subsection*{Data availability} The data that support the findings of this study are available from the corresponding author upon reasonable request.
\subsection*{Code availability} All code that was used to acquire the data and process it is available in ~\cite{Aizik_code:23}. 
\subsection*{Acknowledgments} We thank the following people in their help  preparing neural samples: Amit Parizat, Etay Aloni, Zhige Lin and Amit Zeisel from the from the Technion department of Biotechnology and Food Engineering, as well as Ariel Gilad and Odeya Marmor-Levin from the Medical Neurobiology department at the Hebrew University.  
This research was funded by the European Research Council [101043471],  United States-Israel Binational Science Foundation [2008123/2019758] and Israel Science Foundation [1947/20]. 
\subsection*{Author contributions} Both authors contributed extensively to the work presented in this paper.
\subsection*{Competing interests} The authors declare no competing interests.

\nocite{DrorArxiv}


	\twocolumn[
	\begin{@twocolumnfalse}
		\huge{Supplementary Material}


		\vspace*{1cm}
	\end{@twocolumnfalse}

	]

	\setcounter{section}{0}
	\setcounter{figure}{0}
	\section{Mathematical derivation}\label{sec:methods}

We provide a derivation of an image formation model using a transmission matrix formulation, and use it to prove the non-linear properties of our confocal  score. This derivation explains why the confocal score can favor modulations that focus at a single spot despite the fact that it collects linear fluorescence feedback.
\subsection{Image formation model} 
Consider a set of $K$ fluorescent particles inside a  sample, and denote their positions by $\ptd_1,\ldots,\ptd_K$.
We assume the SLM in the illumination arm is illuminated with a spatially uniform plane wave and use the SLM to display a complex 2D electric field that we denote by  $\bu$.  Although $\bu$ is a 2D field, we reshape it as a 1D vector. We also use $\bou$ to denote a $K\times 1$ vector of  the field propagating through the sample at each of the $K$  fluorescent sources.

The relation between $\bu$ and $\bou$ is linear and can be described as a  multiplication by a (very large)  matrix  \BE\label{eq:bou-Tbu}\bou=\TMi \bu,\EE where  $\TMi$ is the incoming transmission matrix, describing the forward coherent light propagation in the tissue. We note that $\TMi$   is specific to the tissue sample being tested, and different tissue samples are described by very different transmission matrices. For thick tissue, $\TMi$ can be an arbitrarily complex matrix incorporating multiple scattering events in the tissue.
Likewise, if the light returning from the target is coherent, its propagation  to the SLM of the imaging arm can be described as $\TMo\bou$, where $\TMo$ is the coherent back-propagation transmission matrix.

We denote by $\CamSLMVect$ the wavefront placed on the SLM of the imaging arm, and by $\DiagM(\CamSLMVect)$ a diagonal matrix with  $\CamSLMVect$ on its diagonal. Our SLMs are placed at the Fourier planes of the imaging system, and we denote by $\Fr$ the Fourier transform of the wavefront from the SLM to the camera sensor. With this notation, coherent light propagating from the target particles to the sensor, through the SLM, can be expressed as
\BE
\Fr \DiagM(\CamSLMVect)\TMo\bou.
\EE

In the fluorescent case, the emissions from different points are incoherent, hence the recorded intensity is the sum of emitted intensity from each fluorescent bead:
\BE \label{eq:img-form-incoherent} 
I=\sum_k |\Fr \DiagM(\CamSLMVect)\ckTMo|^2 |\bou_{{k}}|^{2\alpha}, 
\EE
where  $|\bou_{{k}}|^2$ is the energy of the excitation light arriving at particle $\ptd_k$, and $\ckTMo$ is the $k$-th column of $\TMo$, so that $\Fr \DiagM(\CamSLMVect) \ckTMo$ is the wavefront arriving to the sensor from  $\ptd_k$.  
In \equref{eq:img-form-incoherent} $\alpha$ denotes the type of fluorescent excitation. The simplest case  $\alpha=1$ is known as single-photon fluorescence where the emission is linear in the excitation energy $|\bou_{{k}}|^{2}$. In two-photon fluorescence,  $\alpha=2$, namely the emission is proportional to the squared excitation. 

\boldstart{Phase conjugation: }
When the incoming and outgoing light have the same wavelength, the Helmholtz  reciprocity principle leads to wave conjugation. Namely, if we record the wavefront emitted from a source point inside the tissue and illuminate with the conjugate wavefront,  light will focus at the same point.  This implies that the returning transmission matrix is the transpose of the incoming one ~\cite{RevModPhys.89.015005}: \BE \label{eq:conj-transmission-M} \TMo={\TMi}^\top. \EE 
In the fluorescent case,  $\TMi$ and $\TMo$ describe propagation at two different excitation and emission wavelengths. However, for linear single-photon fluorescence the excitation and emission wavelengths are relatively similar and 
$\TMo\approx {\TMi}^\top$. While our algorithm {\em does not} require the incoming and outgoing transmission matrices to be the same, this similarity will help us draw intuition on what is being optimized.

%
%
%
%
%

\boldstart{Normalization:}
we assume for simplicity that our transmission matrices are normalized such that every column or row has a unit energy, that is for every $k$ \vspace{-0.1cm}
\BE\label{eq:bounded-rows}
\sum_x |\TMo_{x,k}|^2= 1,\quad \sum_x |\TMi_{k,x}|^2=1. \vspace{-0.1cm}
\EE
This means that the total amount of energy that can arrive to particle $\ptd_k$ or emerge from it is fixed.
As the laser energy is fixed, we also assume w.l.o.g. that all illumination vectors have a unit norm $\|\bu\|=1$.
As propagation through the tissue does not generate new energy,  every incoming vector $\bu$ should satisfy  \vspace{-0.1cm}
$\|\TMi \bu\|\leq1$ and thus the energy at the target is also bounded: \BE \label{eq:bounded-v-methods} \sum_k |\bou_k|^2\leq 1.\EE

\subsection{Score functions}
We provide a longer derivation of the modulation scores mentioned in the main paper and explain how they can evaluate focusing.

\boldstart{The total intensity score:}
Consider  a configuration where we only try to correct the illumination arm, and the SLM in the imaging arm is not used (equivalently, $ \DiagM(\CamSLMVect)$ in \equref{eq:img-form-incoherent} is the identity matrix). One of the earlier scores that were considered in the literature~\cite{Katz:14,Ji2017review} is just the total intensity measured over   the entire sensor plane. 
Using \equpref{eq:img-form-incoherent} {eq:bounded-rows} it is easy to show that this total intensity score reduces to \vspace{-0.1cm}
\BE\label{eq:metric-TI}
\Mtric_{\text{TI}}(\bu)\equiv\sum_x I(x)=\sum_k |\bou_k|^{2\alpha}.
\vspace{-0.1cm}\EE
Since   the energy at the target  is bounded (see \equref{eq:bounded-v-methods}),
for the case $\alpha>1$ this score is maximized when $\bou$ is a one-hot vector, which equals 1 at a single entry and zero at all the others. 
However, in the single-photon case where $\alpha=1$, \equref{eq:metric-TI} reduces to the total power in $\bou$, $\Mtric_{\text{TI}}(\bu)= \sum_k |\bou_k|^{2}$, and since this power is fixed, the same amount of energy returns whether we spread the excitation power over multiple fluorescence sources or bring all of it into one spot. 

\boldstart{The variance maximization score:}
Boniface \etal~\cite{Boniface:19} have recently suggested that to evaluate focusing with linear single-photon feedback, one should  maximize the variance of the intensity measured by the sensor. The idea is that if we manage to focus all the excitation light at a single spot, the emitted light scattered through the tissue will generate a highly varying speckle pattern on the sensor plane. If the excitation is not focused, multiple sources emit simultaneously. The light emitted by these sources is summed incoherently,  and hence the variance of the speckle pattern on the sensor decays. 
A short calculation provided in~\cite{Boniface:19}  shows 
\vspace{-0.3cm}
\BEA 
\Mtric_{\text{Var}}(\bu)\!&\!\equiv\!& \!{\text{Var}}[I]\equiv\frac1{n}\sum_x |I(x)|^2 -\left(\frac1{n}\sum_x I(x)\right)^2 \nonumber
\\
\!&\!\propto\!&\! \sum_k |\bou_k|^4,
\vspace{-0.1cm}
\EEA
where $n$ is the number of image pixels.
Hence, as before, the score is a non-linear function of the power at different fluorescent particles and is maximized by a one-hot vector.

\boldstart{Confocal score:}
Below we derive the relation between the energy at the central pixel and the vector of fluorescent power $\bou$, with the goal of showing that the confocal score is a non-linear function of the power of the excitation vector.  

Denoting by $\Fr_{0,\rightarrow}$ the central row of the Fourier transformation from the SLM plane to the image plane, 
the contribution of the particle $\ptd_k$ to our measurement at the central pixel is
\BE\label{eq:ctrl-pixel}\Fr_{0,\rightarrow} \DiagM(\boutu)\ckTMo.\EE
Assuming w.l.o.g., that the central pixel is measuring the DC component of the Fourier transformation, corresponding to simple averaging, we can express the value at the central pixel  in \equref{eq:ctrl-pixel} as the product of the SLM modulation $\boutu$ at the imaging arm, with the corresponding column of the outgoing transmission matrix:
\BE\Fr_{0,\rightarrow} \DiagM(\boutu)\ckTMo = {\boutu}^\top \ckTMo.\EE

By modulating both illumination and imaging arms,  we can express the energy of the central pixel  as: 
\BEA\label{eq:inc-confocal-methods} \vspace{-0.1cm}
\!\Mtric_{\text{Conf}}(\binu,\boutu)\!&\!\equiv\!&\! I(0)
=\sum_k |{\boutu}^{\top} \ckTMo|^2 |\rkTMi \binu|^2\nonumber\\\!&\!=\!&\! \sum_k |\boutou_k|^2\cdot  |\binou_k|^2.
\EEA

with $\binou=\TMi \binu$ and $\boutou=\boutu^T \TMo$.
As mentioned in \equref{eq:bounded-v-methods}, the energy of $\binou$ is bounded, and due to reciprocity the same applies for $\boutou$. 
It is easy to see that this score is maximized when $\binou,\boutou$ are both one-hot vectors at the same entry $k$. That is, the score of \equref{eq:inc-confocal-methods} is maximized when the excitation modulation $\binu$ brings all light to one of the particles $\ptd_k$, and the modulation at the imaging arm $\boutu$ corrects the wavefront emitted from the same particle $\ptd_k$ and brings all of it into the central pixel.


In particular, if the excitation and emission wavelengths are sufficiently similar so that we can approximate $\TMo \approx {\TMi}^\top$, it is best to use the same modulation at both the illumination and imaging arms, and the score of \equref{eq:inc-confocal-methods} reduces to $\sum_k |\bou_k|^4$ as in the variance-maximization and two-photon cases.

\boldstart{Optimization:} 
In this work we have  explicitly optimized the confocal score (\equref{eq:inc-confocal-methods}) using standard Hadamard basis optimization~\cite{PopoffPhysRevLett2010}, detailed in \secref{sec:optalg}.
Overall the Hadamard optimization is significantly slower than~\cite{Dror22}. 
We note that when all fluorescent sources $\ptd_k$ have the same power, there are $K$ different solutions that can maximize  \equref{eq:inc-confocal-methods} and an optimization may converge to any of them.
Also note that we constrain the solution such that regardless of the position of $\ptd_k$, it will be imaged in the central pixel. If  $\ptd_k$ is not at the center of the frame, the wavefront $\boutou$, placed at the Fourier plane of the imaging system will contain a tilt, shifting $\ptd_k$ to the central pixel.

\section{Optimization algorithm}
\label{sec:optalg}

This section provides details on our optimization algorithm. 
In \secref{sec:one-vs-two-modul} below we compare two different correction models. (i) Assumes that as our excitation and emission wavefronts are sufficiently similar, and use the same modulation  $\bu$ in both the illumination and imaging arms. (ii) Solve for a different modulation in each arm.
Here we provide the mathematical details for the optimization algorithm. 
\subsection{Same modulation in both arms}
Assuming the excitation an emission wavelengths are sufficiently similar $\TMo\approx {\TMi}^\top$ the confocal score derived in   \equref{eq:inc-confocal-methods} reduces to:
\BE \label{eq:conf-score-sup}
\begin{gathered}
	\Mtric_{\text{Conf}}(\bu) =\eta\sum_k |\bu^T \TMo_k|^2 \cdot |\TMi_k \bu|^2 = \\= \eta\sum_k|\TMi_k \bu|^4.
\end{gathered}
\EE
where $\eta$ is fluorescence efficiency, $\TMi_k$ is the k'th row of the input transmission matrix from the illumination SLM to the particles,$\TMo_k$ is the k'th column of the output transmission matrix from the particles to the imaging SLM and $\bu$ is the wavefront correction vector on both SLMs.


Our optimization scans a set of binary phase masks and for each of them finds the phase that maximizes the score function. We show below that if only the phase of this mask can be adjusted, the score can be expressed as the sum of two sinusoidal functions, which can be measured using 5 samples.
By capturing 5 shots we can select the optimal value for this phase and proceed to the subsequent mask.

 
\paragraph{Dictionary representation:}

In our optimization, we mark  the phase function on the SLM in the $m$th iteration as $\psiomegasub{(m)}$. We express the phase mask as a superposition of a dictionary of $N$ binary masks, weighted by phases $\phi_n$:
\BE
\psiomegasub{m}=\sum_{n=1}^{N}\phi_n\cdot\muellomega.
\EE
Here $\phi_n$ is a scalar coefficient, $\muellomega$ is a binary vector whose size is equivalent to  the number of entries in the SLM,  and $\omega$ is the index of an entry in the SLM plane. Since we use phase-only SLMs, the amplitude in each pixel remains constant and the wavefront correction vector is expressed as:
\BE\label{eq-bu-m-itr}
\bu=\exp\left\{{ i\psiomegasub{m}}\right\}=\exp\left\{i\sum_{n=1}^{N}\phi_n\cdot\muellomega\right\}.
\EE

In the $m$th iteration of our optimization,  we choose a single dictionary element $n$  and update $\phi_n$ to obtain:
\BE
\psiomegasub{m}= \psiomegasub{m-1} + \phi_n\cdot  \muellomega,
\EE
while the values of the  coefficients $\phi_1,\ldots,\phi_{n-1},\phi_{n+1},\ldots,\phi_{N}$ are  fixed to the value they had in iteration $m-1$.

\paragraph{Sinusoidal model:}
\begin{claim}
If we only vary the scalar $\phi_n$	the confocal score of \equref{eq:conf-score-sup} can be expressed as a sum of two sinusoids 
	\BE \label{eq:cost_abc}
	\Mtric_{\text{Conf}}(\phi_n)=|A|+|B|\cos(\phi_n+\angle{B})+|C| \cos(2\phi_n+\angle{C}),
\EE
where $A,B,C$ are complex scalars, and we denote their amplitude and phase by $|A|,|B|,|C| $,  $\angle{A},\angle{B}, \angle{C} $  respectively.
\end{claim}
\begin{proof}
	We consider a binary  mask $\muellomega$,  and denote its complimentary by $\overline{\muellomega}$. With this notation we can express the modulation $\bu$ of \equref{eq-bu-m-itr} as
	\BE
	\begin{gathered}
		\bu(\phi)=e^{i\psiomegasub{m}}=e^{i\left(\psiomegasub{m-1} + \phi \cdot\muellomega\right)}\\=e^{i\psiomegasub{m-1}}\overline{\muellomega}+ e^{i\phi}e^{i\psiomegasub{m-1}}{\muellomega}. 
	\end{gathered}
	\EE
	
	We mark
	\BE
	\begin{gathered}
		\bu^{(1)}=e^{i\psiomegasub{m-1}}\cdot \overline{\muellomega},\\ \bu^{(2)}=e^{i\psiomegasub{m-1}}\cdot \muellomega,
	\end{gathered}
	\EE
	and so, $\bu(\phi)$ simplifies to:
	\BE
	\bu(\phi)=\bu^{(1)} + \bu^{(2)}\cdot e^{i\phi}.
	\EE
	With this notation we can express 
	\BE \bou_k=\TMi_k \bu(\phi)=v^{(1)}_k+v^{(2)}_k e^{i\phi},\EE
	with $v^{(1)}_k=\TMi_k\bu^{(1)} $ and   $v^{(2)}_k=\TMi_k\bu^{(2)} $. A short calculation shows that $|\bou_k|^4$ follows the form of \equref{eq:cost_abc}.  The confocal score of \equref{eq:conf-score-sup} is equivalent to $\sum_k|\bou_k|^4$.
\end{proof}
\begin{figure}[t!]
	\begin{center}
		\begin{tabular}{@{}c@{~}c@{~}c@{}}
			\includegraphics[width= 0.15\textwidth]{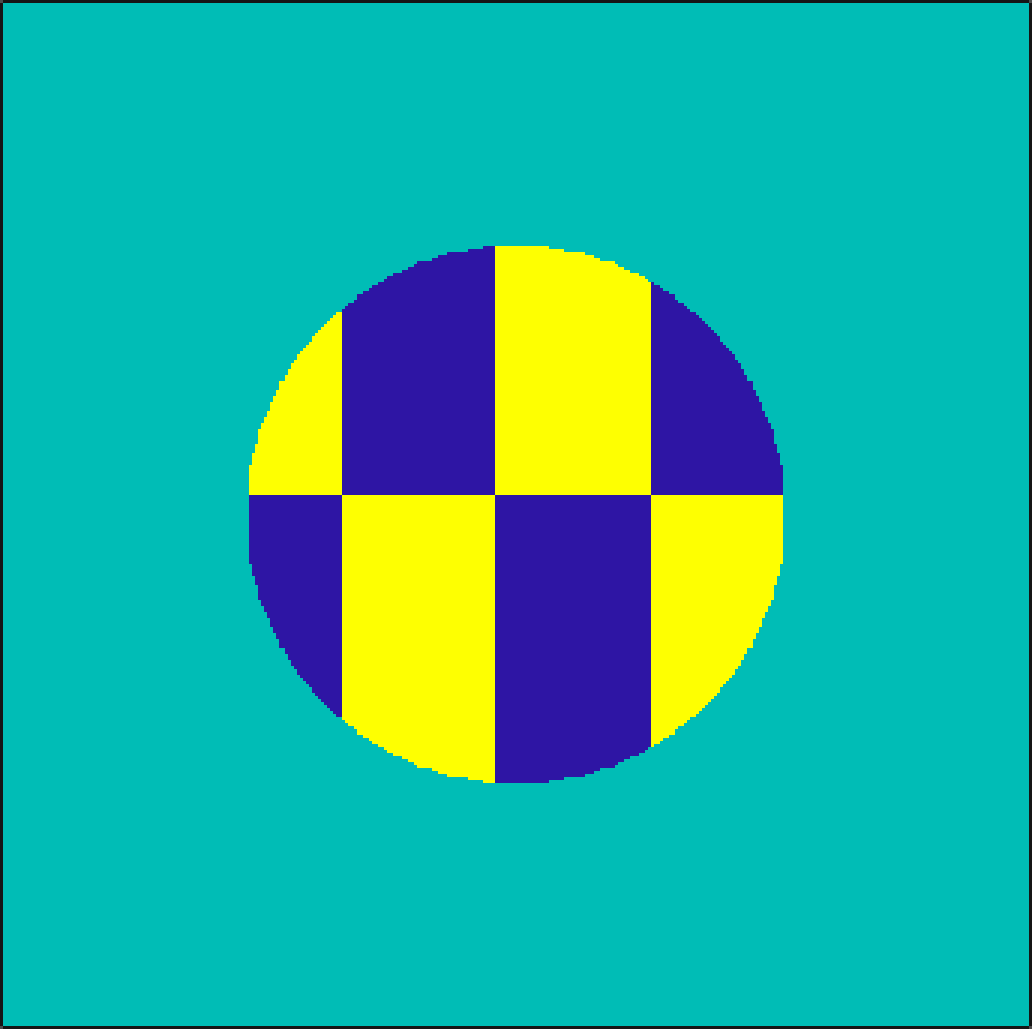} &
			\includegraphics[width= 0.15\textwidth]{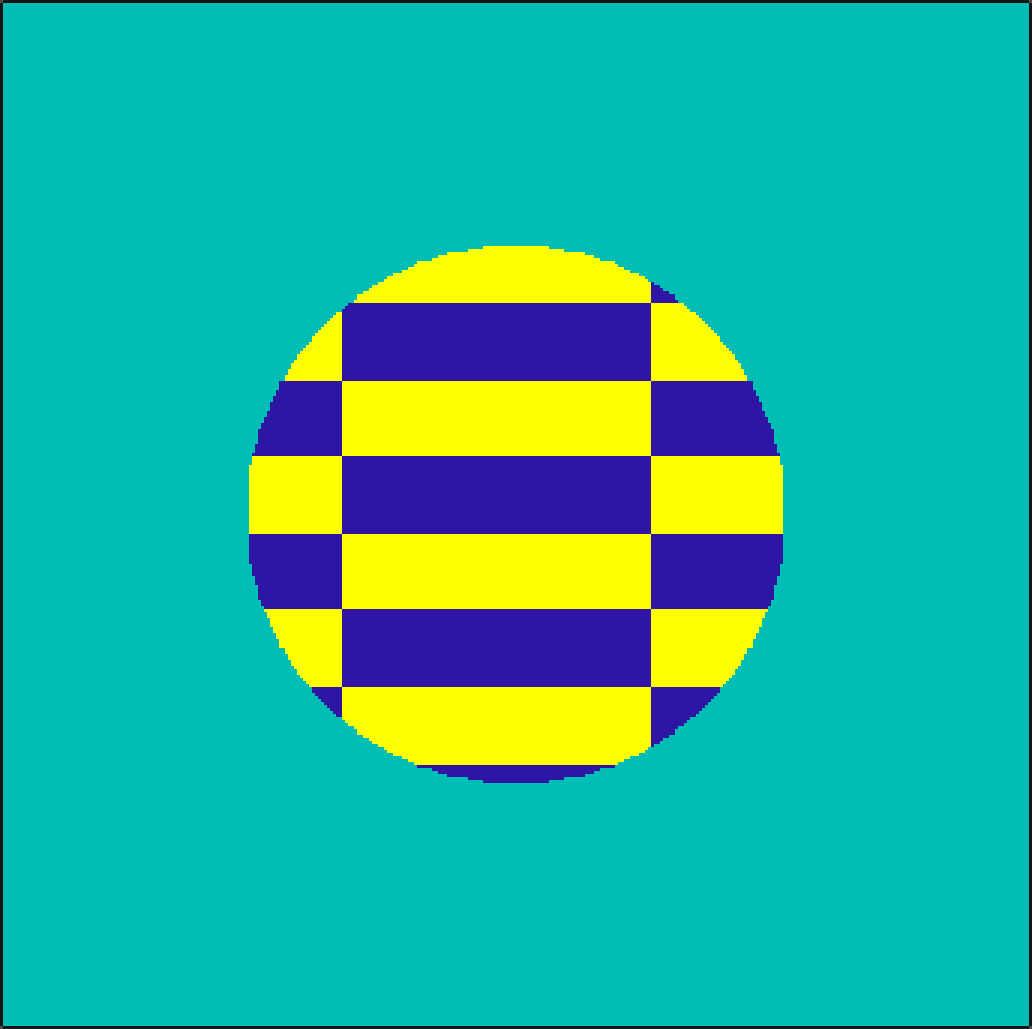} &
			\includegraphics[width= 0.15\textwidth]{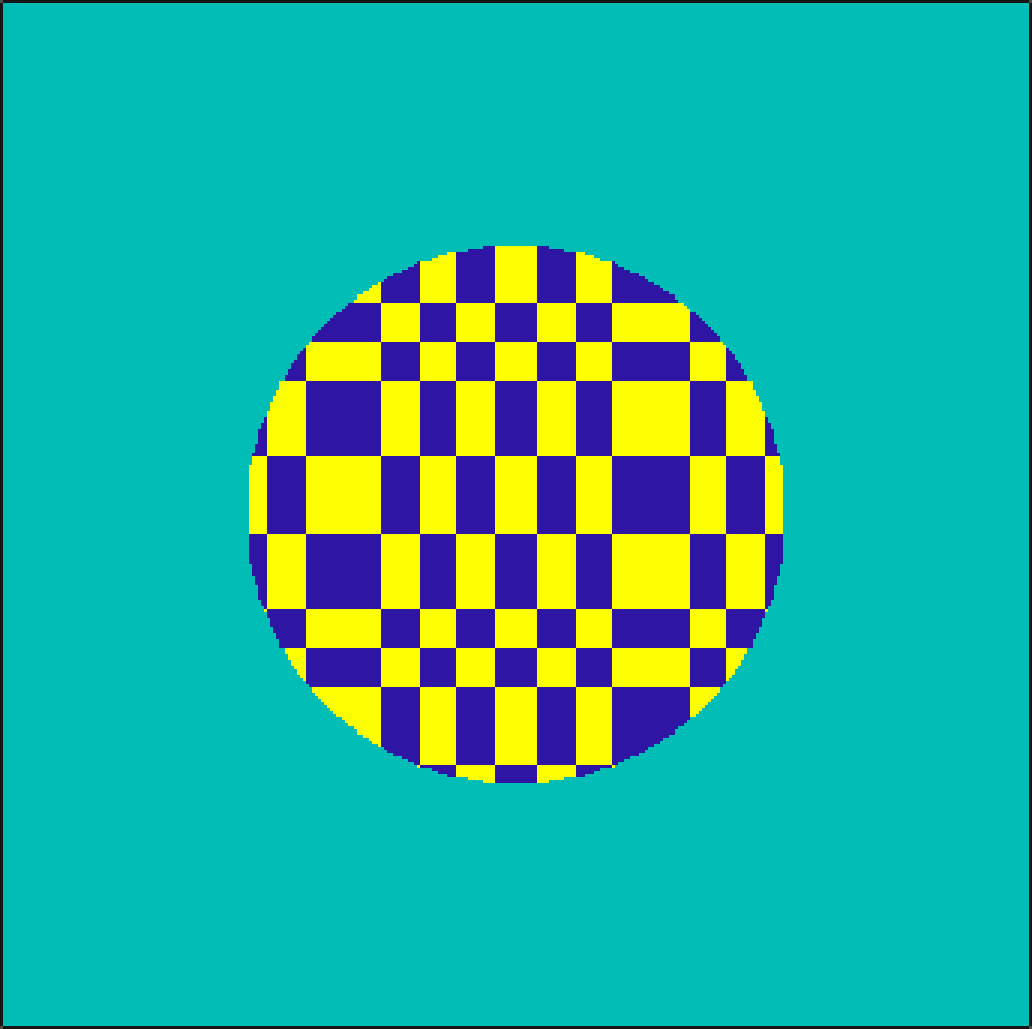} 

		\end{tabular}
	\captionof{figure}
		{ Examples of Hadamard masks used in our optimization. The masks are placed in the Fourier plane and they are all cropped to the circular aperture area. In the figure, the leftmost mask has low frequencies,  while the rightmost one corresponds to higher frequencies.   }
		\label{fig:hadamard}
	\end{center}
\end{figure}

\paragraph{Optimization details:}
Using the above claim, we can consider  $J\geq 5$ equally spaced phases  $ \phi^j=[1\ldots J]\frac{2\pi}{J}$, generate the wavefronts  $\psiomegasub{m}^j= \psiomegasub{m-1} + \phi^j\cdot  \muellomega$ and place them on both SLMs. For each phase we measure the intensity at the central pixel. We use these $J$ intensity measurements to fit a sinusoid and find the value $\phi_n$ maximizing \equref{eq:cost_abc}.
 
In practice we notice that $|C|$ is usually smaller than $|A|,|B|$ and it is enough to fit the score function as a single sinusoid, hence we can also reduce the number of samples to $J=3$ or $J=4$.

 \begin{figure*}[!t]
	\begin{center}
		\includegraphics[width= 1\textwidth]{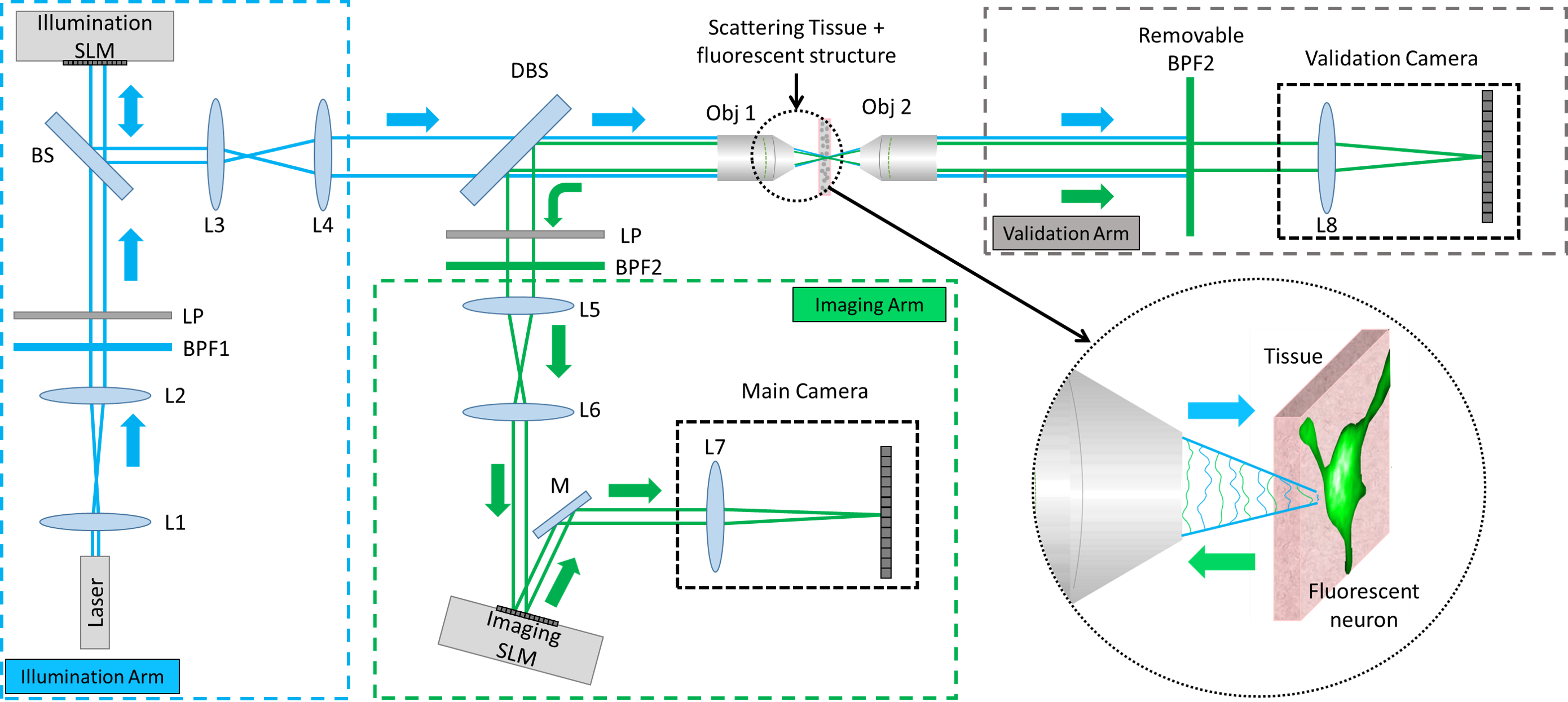}
	\end{center}
	

\caption{Imaging setup: A laser beam is exciting a fluorescent target at the back of a tissue layer, and fluorescent emission is scattered again through the tissue, reflects at a dichroic beam-splitter and is collected by a main (front) camera. We place two SLMs in the Fourier planes of both illumination and imaging arms to allow reshaping these wavefronts. A validation camera views the fluorescent target at the back of the tissue directly. This camera is  not actually used by the algorithm, and is only  assessing  its success. LP=linear polarizer, BS=beam-splitter, DBS=dichroic beam-splitter, BPF=bandpass filter, M=mirror, $L1\ldots L8$=lenses, Obj=Objective. \blue{ Components are listed in Table \ref{fig:components}}.\\
}\label{fig:setup}
\end{figure*}

 \begin{table*}[!t]
	\centering    
	\begin{tabular}{lcl}
		 Laser&& Coherent Sapphire 488-200CW\\
		 LP && Linear polarizer, Thorlabs LPVISE200-A\\
		 BS && Beam-splitter, Edmund Optics 68413\\
		 M&& Mirror, Thorlabs PFR10-P01\\
		 DBS&& Dichroic beam-splitter, Thorlabs DMSP490\\
		 BPF1&& Illumination bandpass filter, Edmund Optics 65147\\
		 BPF2 && Imaging bandpass filters, Edmund Optics 67030 + 65151\\
		 L1&& Plano convex, 4cm\\
		 L2&& Plano convex, 40cm\\
		 L3&& Plano convex, 10cm\\
		 L4,L5,L8&& Tube lense 20cm, Thorlabs TTL200\\
		 L6&&Plano convex, 12.5cm\\
		 L7&&Plano convex, 15cm\\
		 Objectives &&Thorlabs N20X-PF\\
		 Main camera&&Teledyne Photometrics Prime BSI Express\\
		 Validation camera&&Flea3 FL3-G3-28S4M\\
		 SLM&&Holoeye Pluto	
	\end{tabular}
	\caption{\blue{List of components.}\Dror{fixed. let me know what you think about L1,1 and L1,2. should i change to something else?}\Anat{It looks good. why won't you just call them L1 and L2 and increase all other lens indices by 1? About the laser, maybe you should draw a collimated output, not a fiber} }
	
	\label{fig:components}
\end{table*}

\paragraph{Binary masks: }
We divide the SLM to $\sqrt{N}\times\sqrt{N}$ super-pixels, and choose to use binary Hadamard masks as our dictionary.
We show several such Hadamard masks in \figref{fig:hadamard}.
The advantage of this dictionary is that the $\phi_n$ we adjust in each iteration is displayed on half of the SLM area, rather than on a single pixel. Thus, it has more impact on the intensity we measure and the measurement is less sensitive to noise.  

Since the speckles from a single source have a compact support, as demonstrated e.g. in Fig. 3  of the main paper, we conclude that the correction mask has more content at the lower frequencies and less content on the higher ones. To account for that, we first scan the Hadamard basis elements that correspond to low frequencies and later the Hadamard elements with higher frequency content. We  return to the same dictionary element more than once during optimization, and we invest more optimization iterations in the low frequency elements. 


\subsection{Different modulations in illumination and imaging arms}
So far  we considered a model constraining the modulations on both SLMs to be the same. This is only an approximation to the desired modulation, because the illumination and excitation wavelengths can differ. To solve for two different modulations we alternate between updating the $\phi_n$ values in the excitation modulation, and updating the $\phi_n$ values of the emission modulation. If we fix one modulation and vary the other, the confocal score $\Mtric_{\text{Conf}}(\phi_n)$ is a single sinusoid rather than a sum of two sinusoids as in  \equref{eq:cost_abc}, and its phase can be fitted with $J=3$ samples. 
In \secref{sec:one-vs-two-modul} below we compare two different modulations to a single modulation in both arms. We find out that two different modulations can lead into a somewhat better correction, but this optimization also doubles the number of exposures. Therefore, in the presence of photo-bleaching, a single modulation usually leads to better results.  

\section{Setup}
In \figref{fig:setup} we visualize the full  imaging setup for wavefront-shaping correction. \blue{All components are listed in Table \ref{fig:components}.}  A laser beam illuminates a tissue sample via  a microscope objective. A phase SLM in the illumination arm modulates the illumination pattern. The illumination wavefront propagates through the scattering tissue and excites the fluorescent target behind it. We wish to image that target, but the emitted light is scattered again through the tissue on its way to the objective. Scattered light is collected via the same objective,
and reflected at a dichroic  beam-splitter. A second phase SLM  at the  imaging arm modulates the emitted light. Lastly, the modulated light is measured by the front main camera.
In our setup the SLMs  are placed in the Fourier plane of the system.
\blue{We use a $10nm$ bandpass filter in the imaging arm to image a relatively  monochromatic light.}

As we want to correct the scattering of the sample itself rather then aberrations in the optical path, before starting the optimization  we place the sample so the fluorescent light has smallest support in the main camera, then focus the objective of the validation camera (Obj2 in \figref{fig:setup}) such that the neuron is in focus. 
We elaborate on the construction and alignment of this setup in \secref{sec:calibration}.

We image slices of mice brain with EGFP neurons, excited at $488nm$ and imaged at $508nm$.
\blue{ We used two types of aberrations. In the first case we used thin,  brain slices of thickness $50\mu m$ which are almost aberration free, and generate scattering by placing these slices behind a layer of chicken breast tissue ($200-300\mu m$ thick) or parafilm whose optical properties were measured in~\cite{Boniface:19}. The advantage is that since the fluorescence is present only in a thin 2D layer we can obtain a clean reference from a validation camera.   In a second experiment we image through thick brain slices. The slice were originally cut to be  $400\mu m$ thick, though while squeezing between two cover-glass some of the water evacuated and the resulting slice is somewhat thinner.  Since the target is 3D it is not always possible to capture clear aberration free references even with the help of a validation camera. The thin slices contain Betz neurons coming from a 6w female Strain:C57BL6 mouse.  The thick slices contain pyramidal cells of layer 2-3 in the cortex, from a triple transgenic mouse, Rasgrf2-2A-dCre;CamK2a-tTA;TITL-GCaMP6f line 7m old.}  


%
\begin{figure*}[t!]
	\begin{center}
		\begin{tabular}{@{}c@{~~}c@{~~}c@{~~}c@{~~~~}c@{~}c@{~~}c@{~}c@{~~}c@{~}c@{}}
			\multicolumn{4}{c}{Validation camera}&\multicolumn{6}{c}{Main camera}\\
			\cmidrule(l{0pt}r{13pt}){1-4} 	\cmidrule(l{0pt}r{0pt}){5-10}

			\vspace{0.1cm}
			\includegraphics[width = 0.115\textwidth]{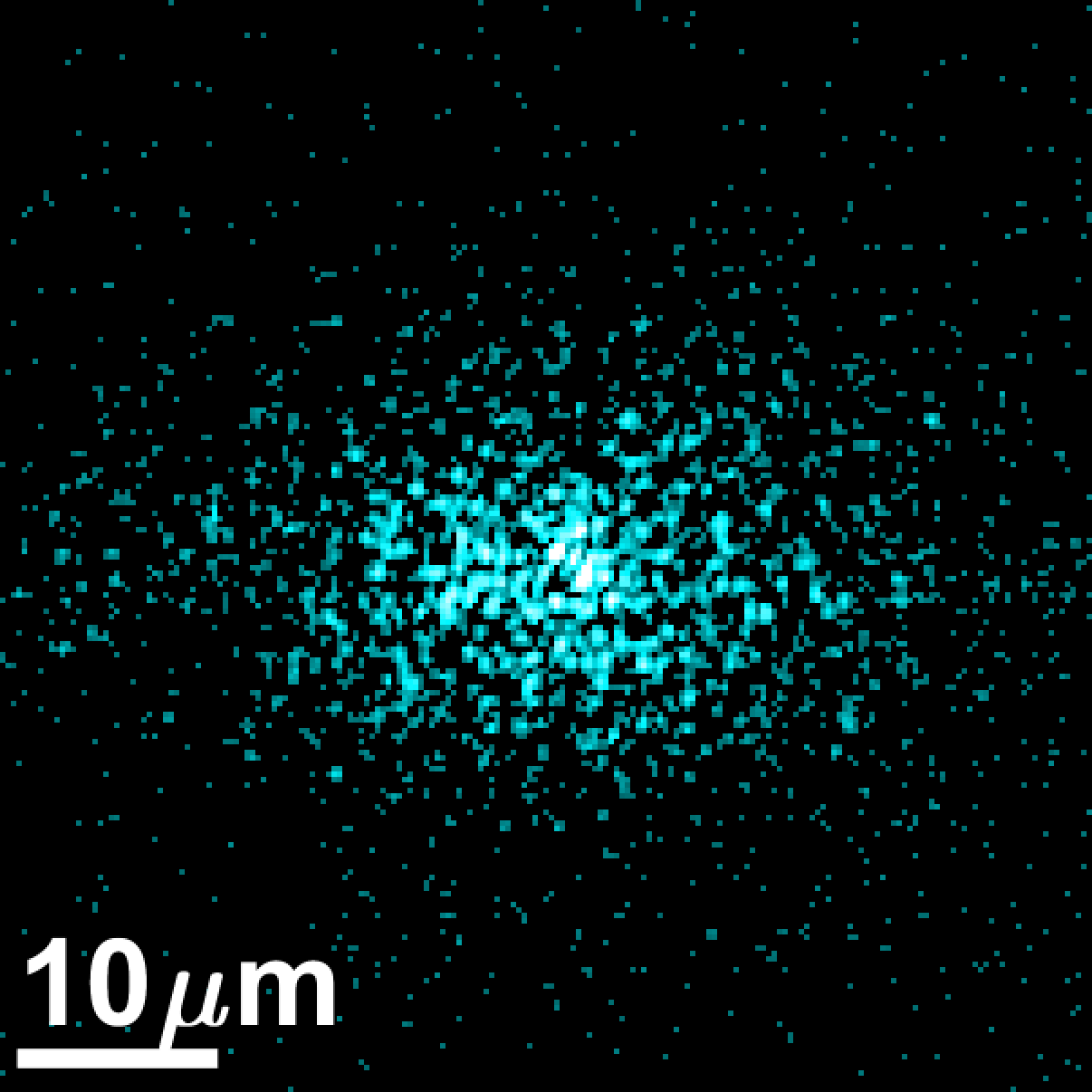}&
			\includegraphics[width = 0.115\textwidth]{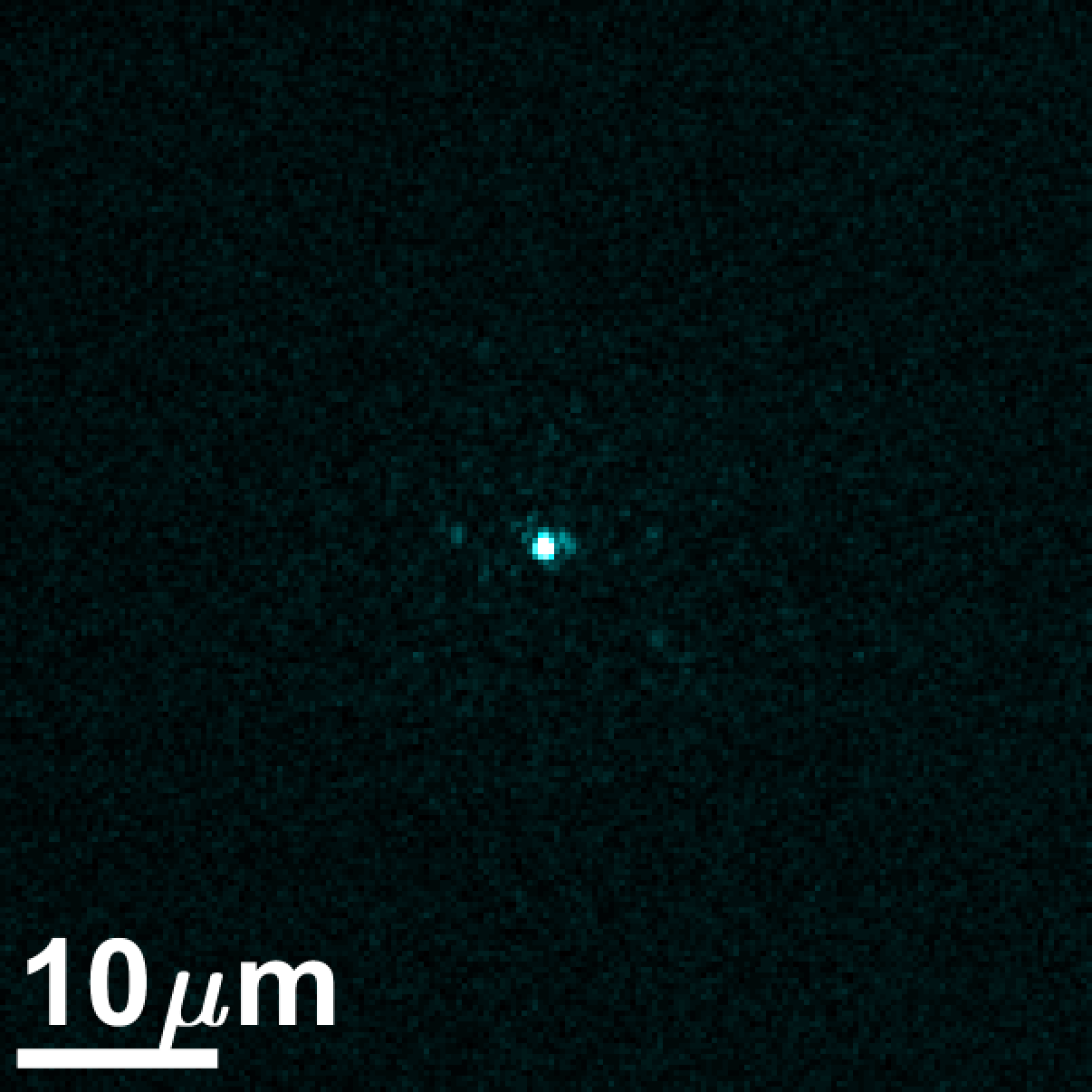}&
			\includegraphics[width = 0.115\textwidth]{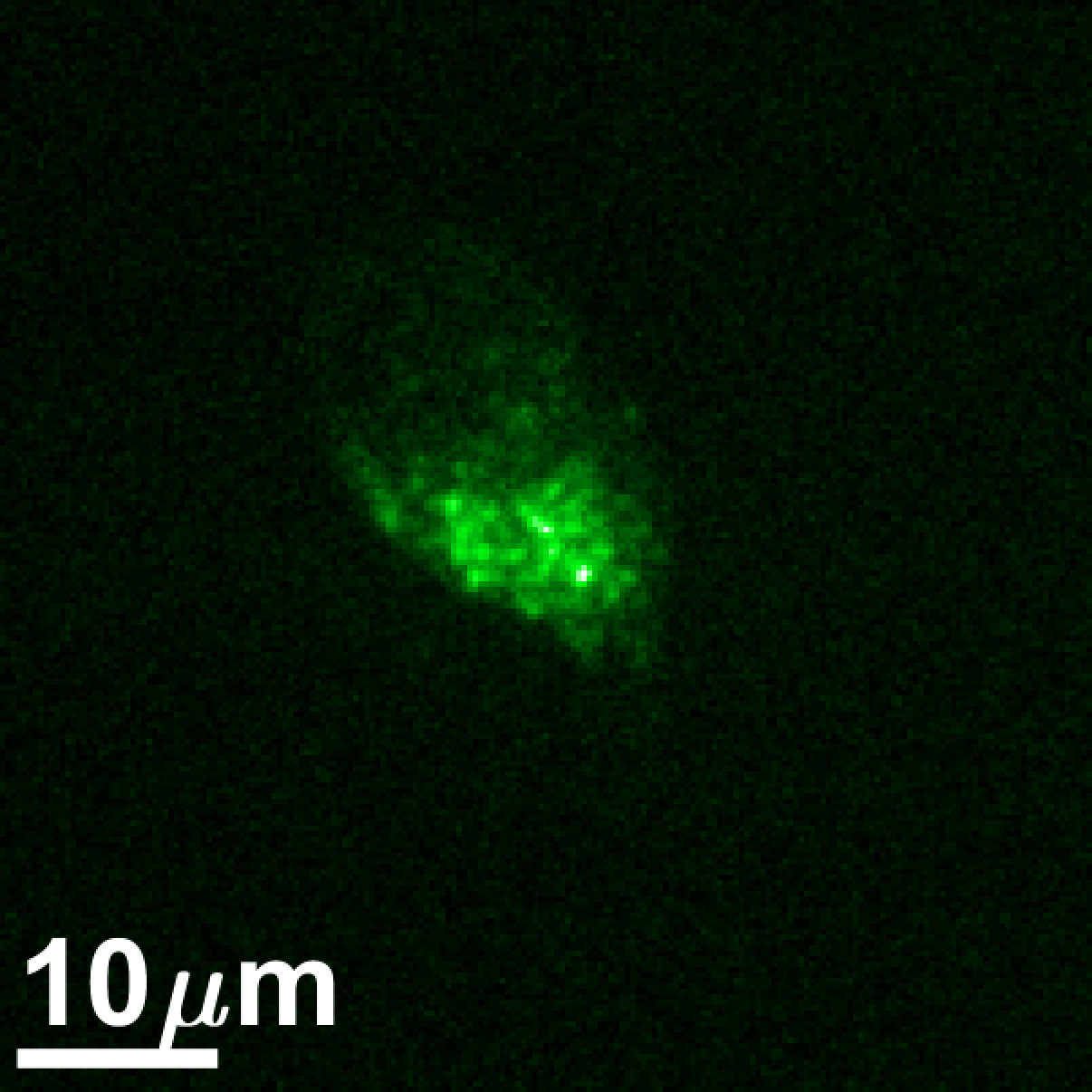}&
			\includegraphics[width = 0.115\textwidth]{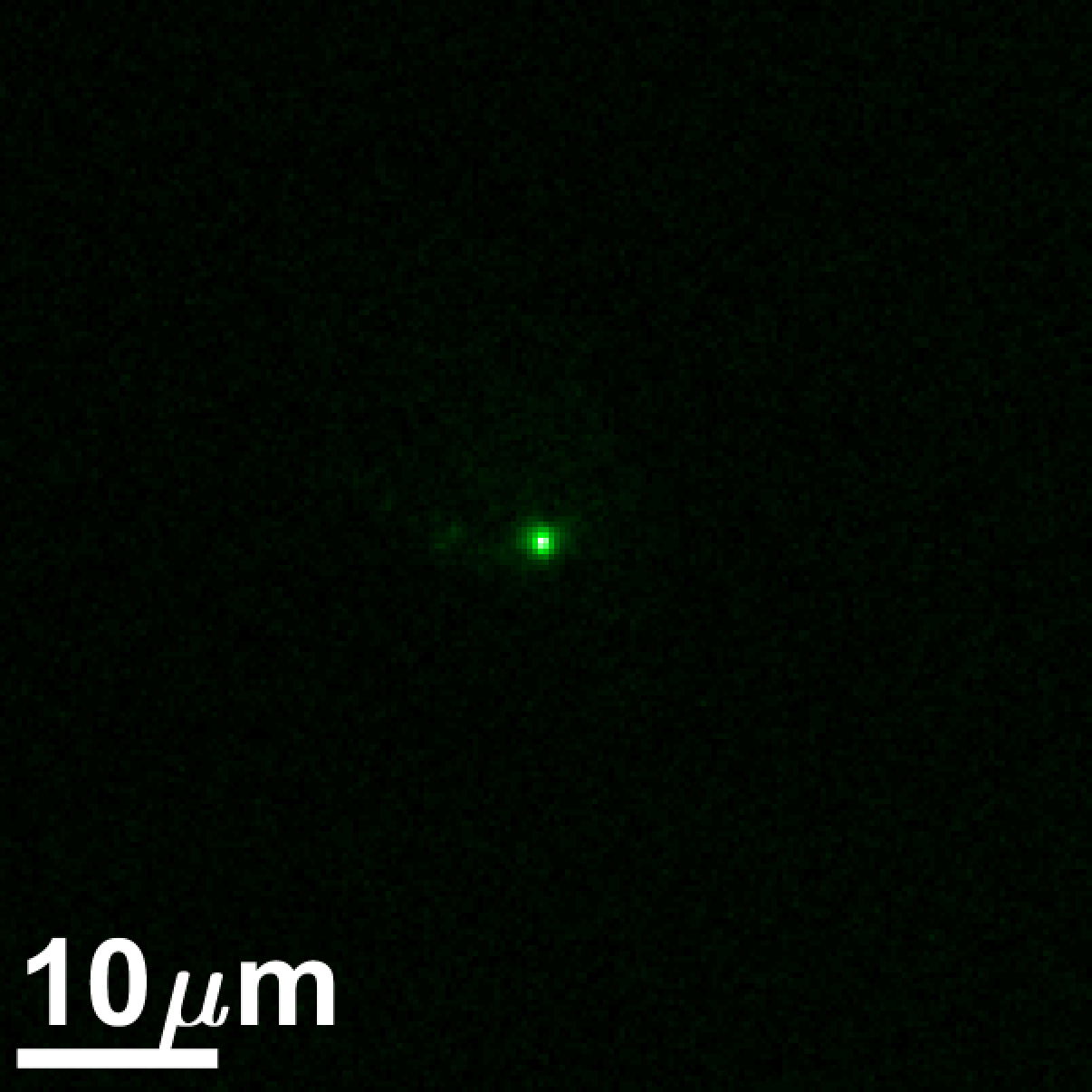}&
			\includegraphics[width = 0.115\textwidth]{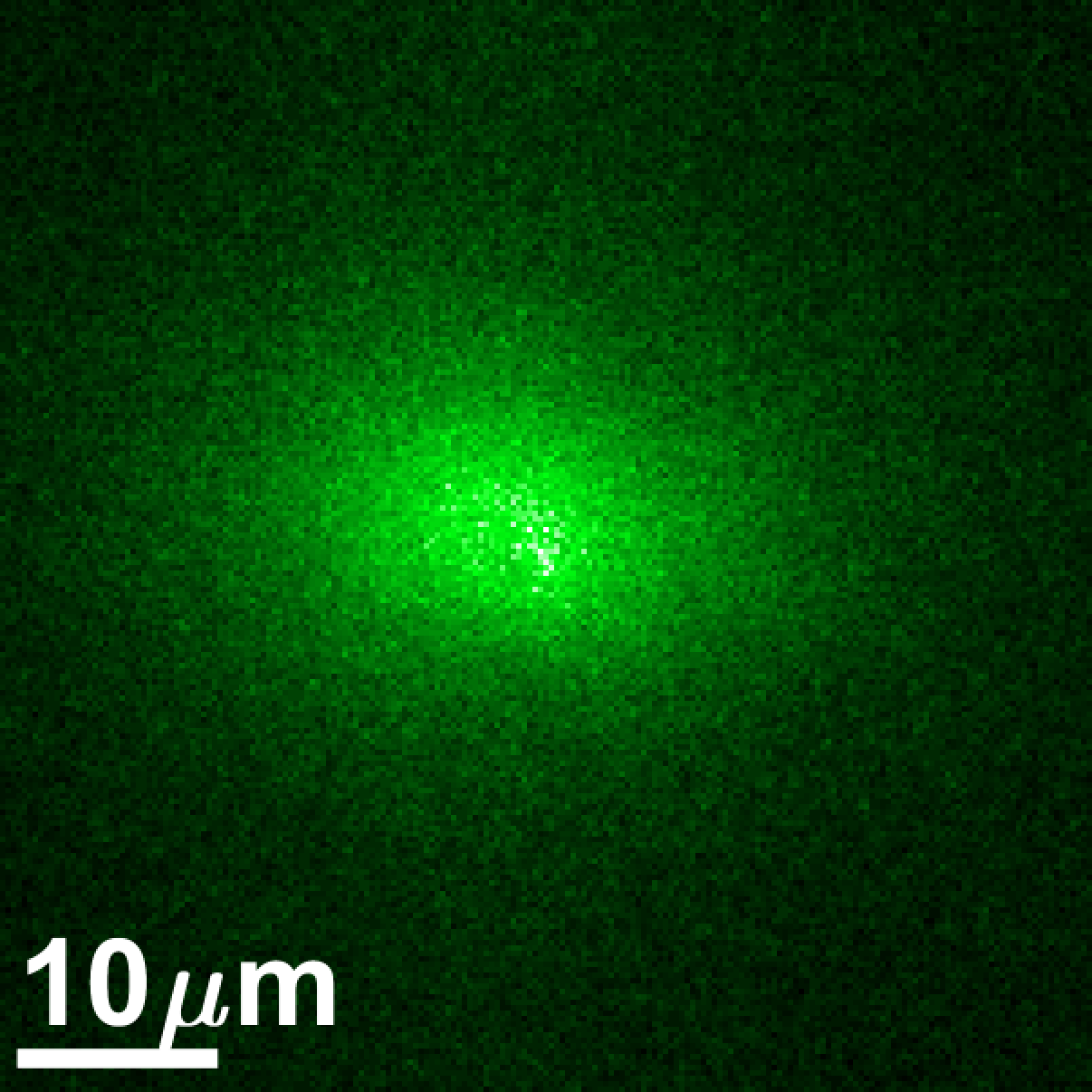}&
			\includegraphics[height = 0.115\textwidth]{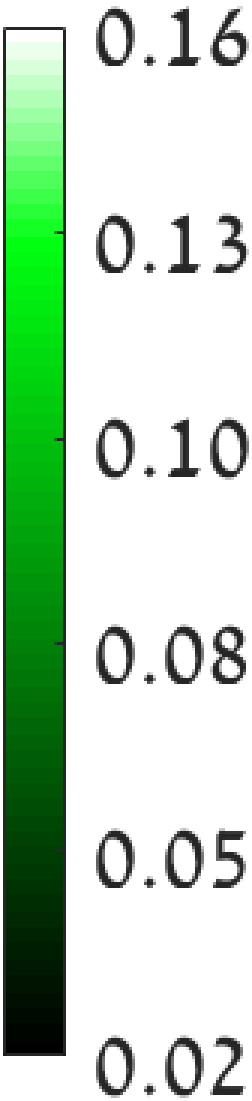}&
			\includegraphics[width = 0.115\textwidth]{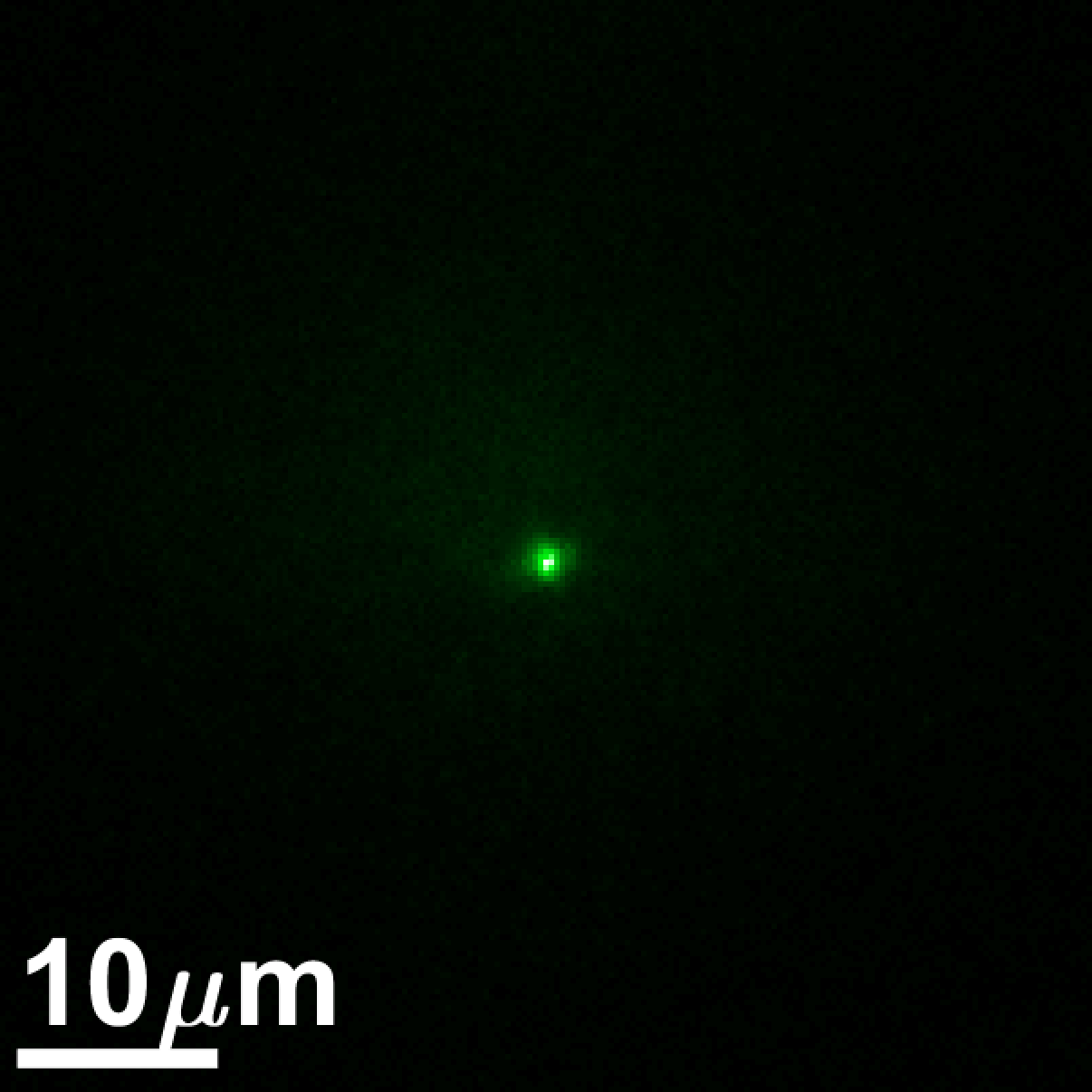}&
			\includegraphics[height = 0.115\textwidth]{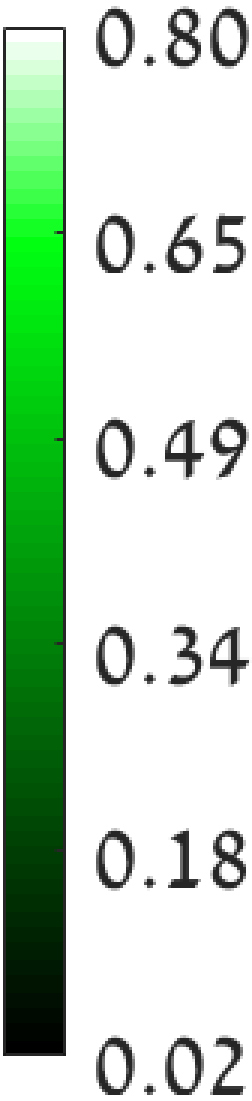}&
			\includegraphics[width = 0.115\textwidth]{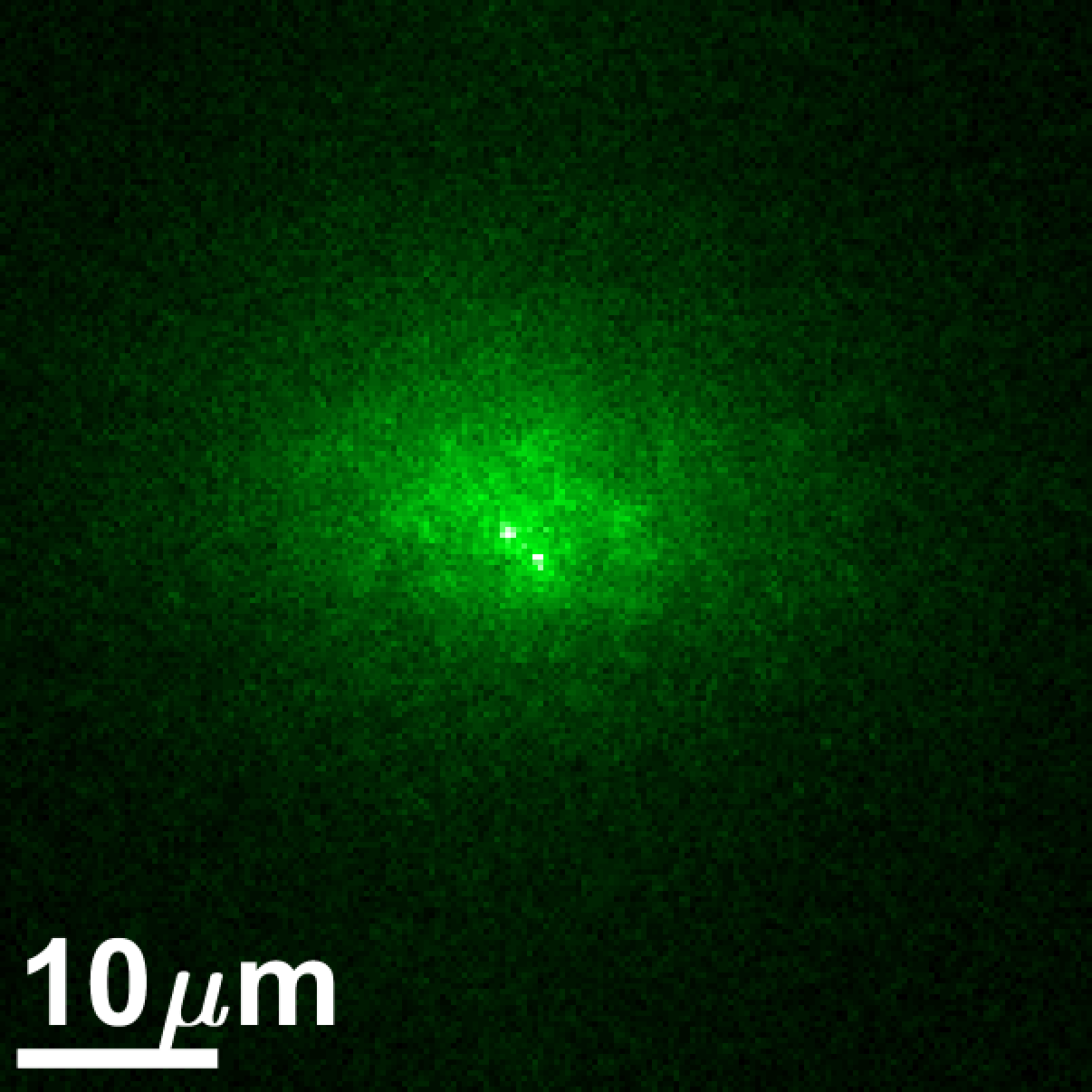}&
			\includegraphics[height = 0.115\textwidth]{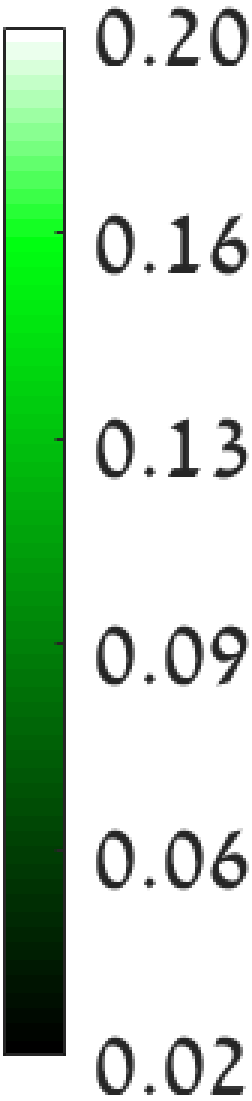}\\

			\vspace{0.1cm}
			\includegraphics[height = 0.115\textwidth]{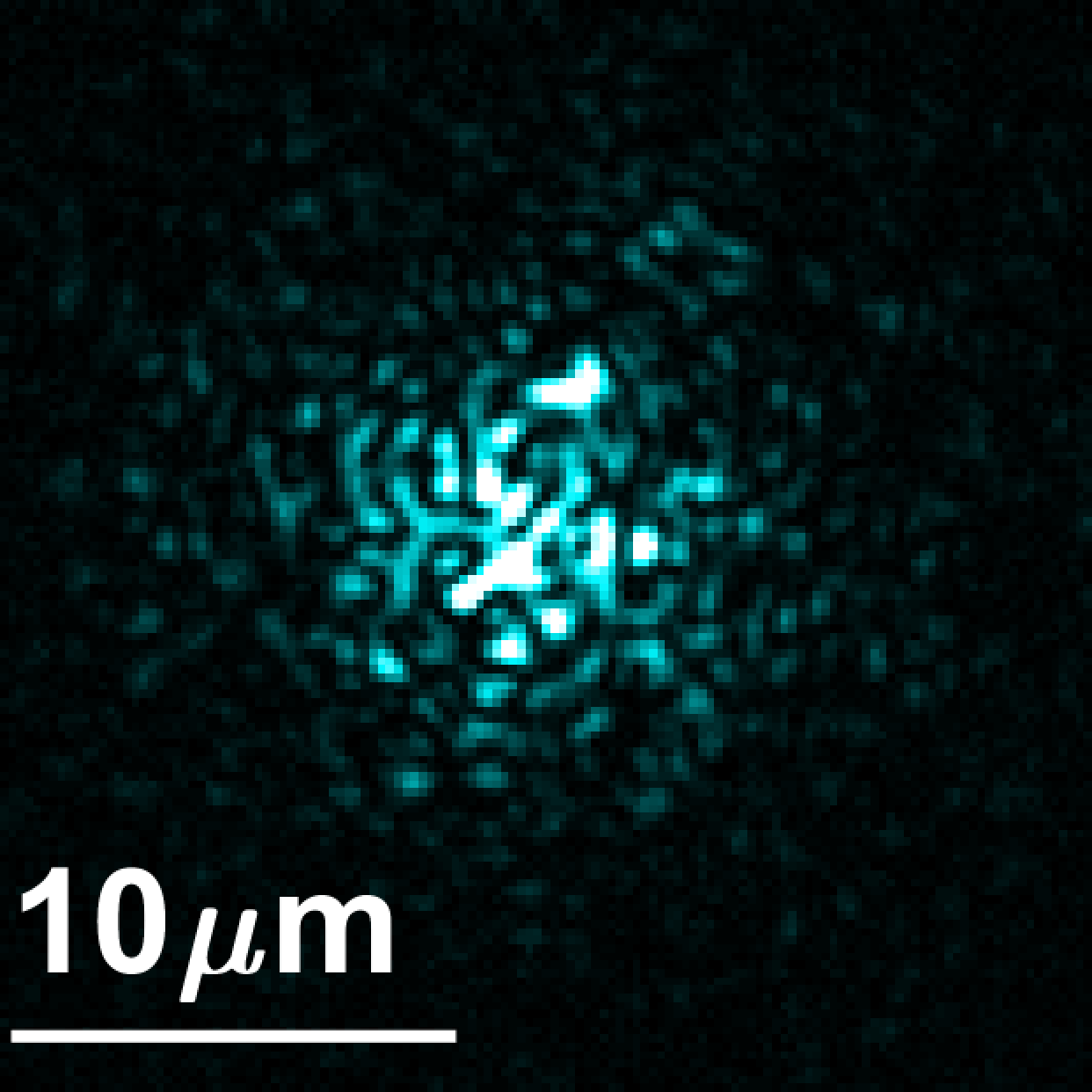}&
			\includegraphics[height = 0.115\textwidth]{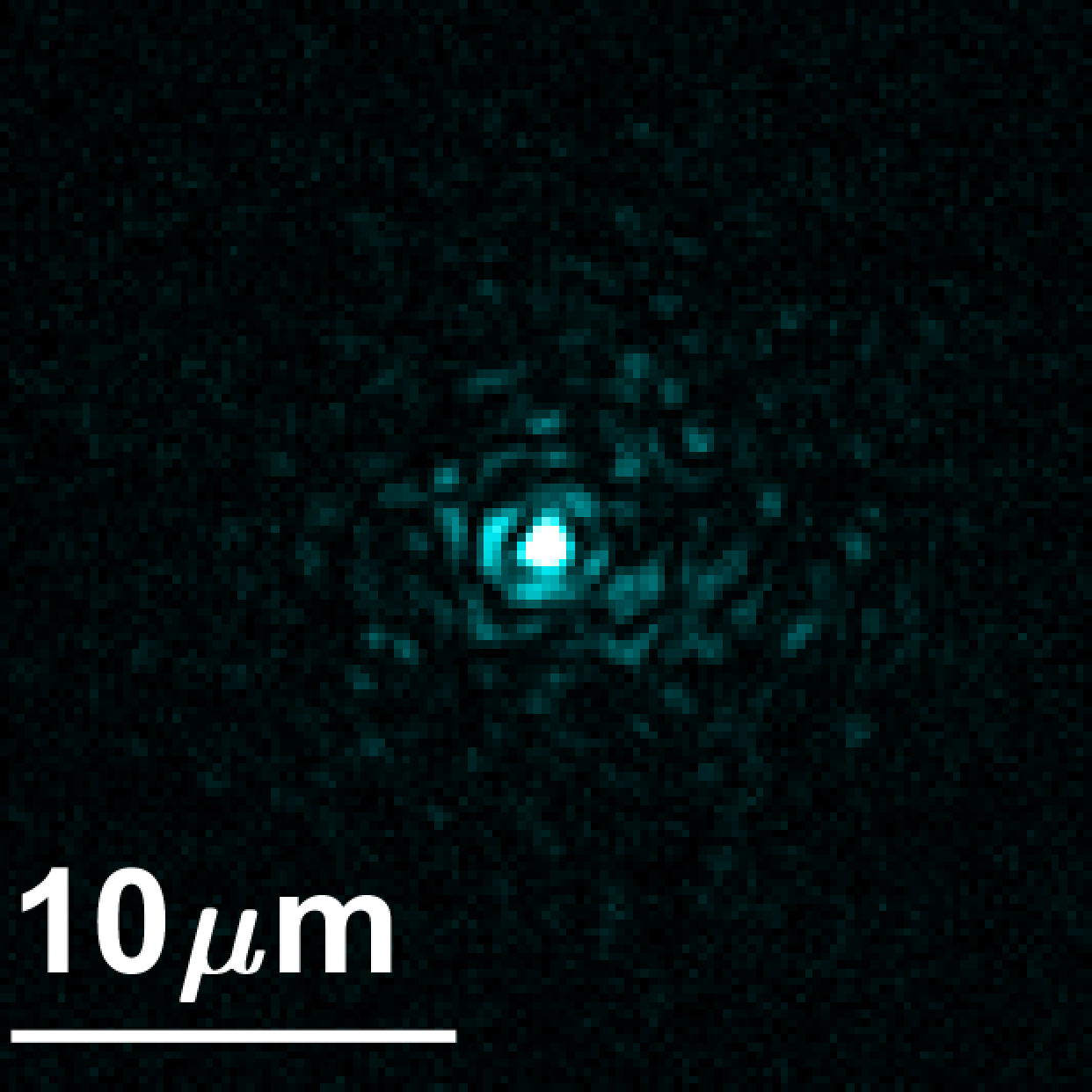}&
			\includegraphics[height = 0.115\textwidth]{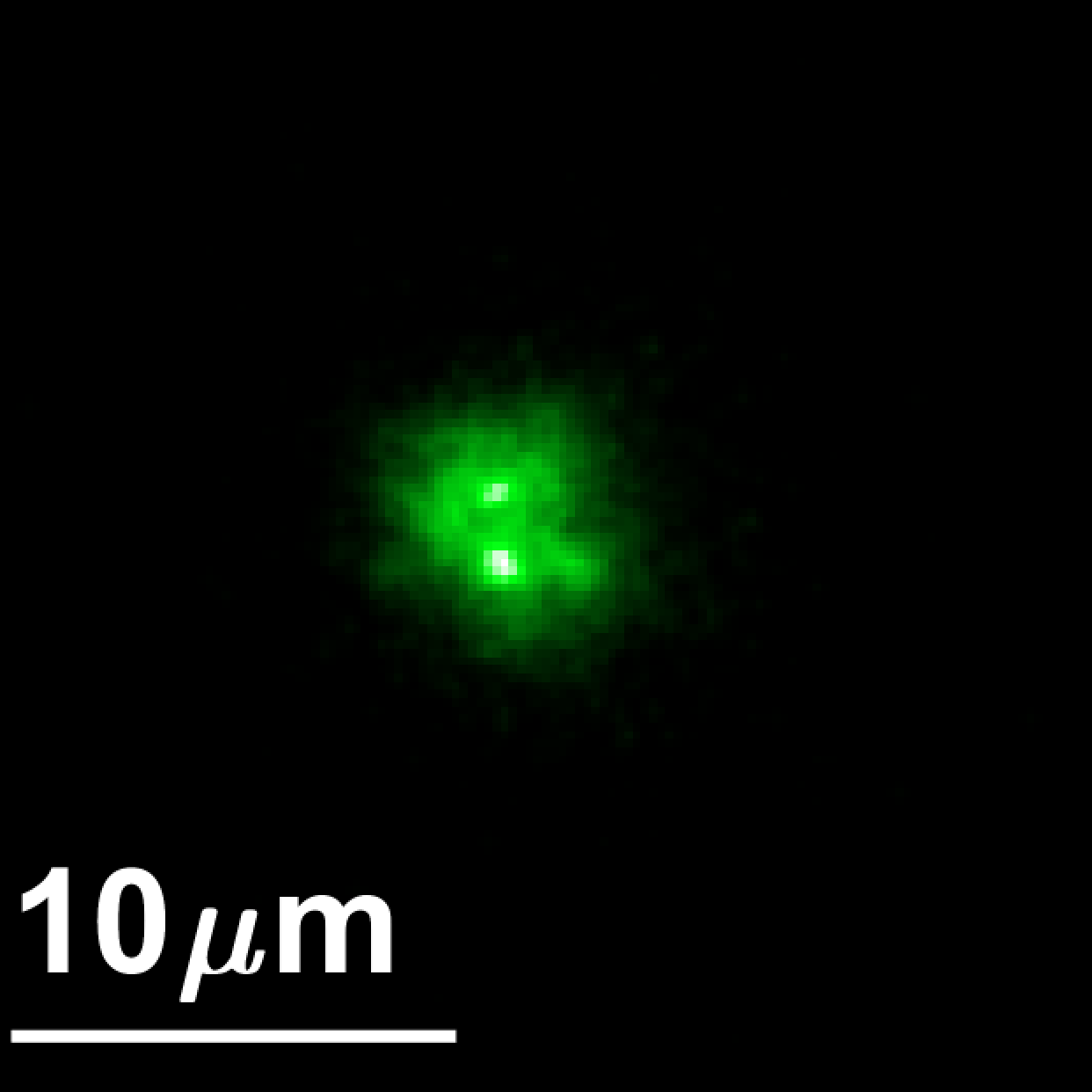}&
			\includegraphics[height = 0.115\textwidth]{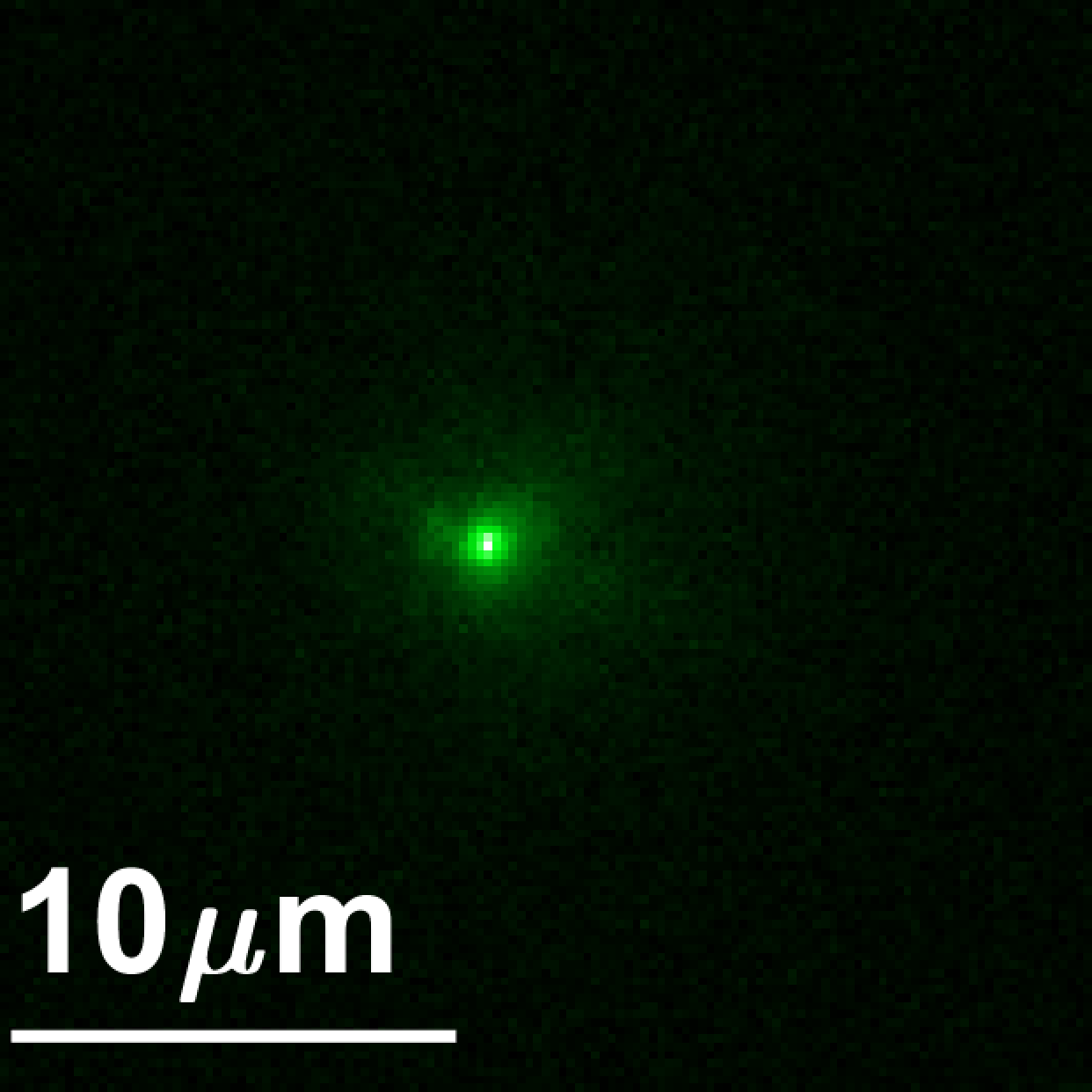}&
			\includegraphics[height = 0.115\textwidth]{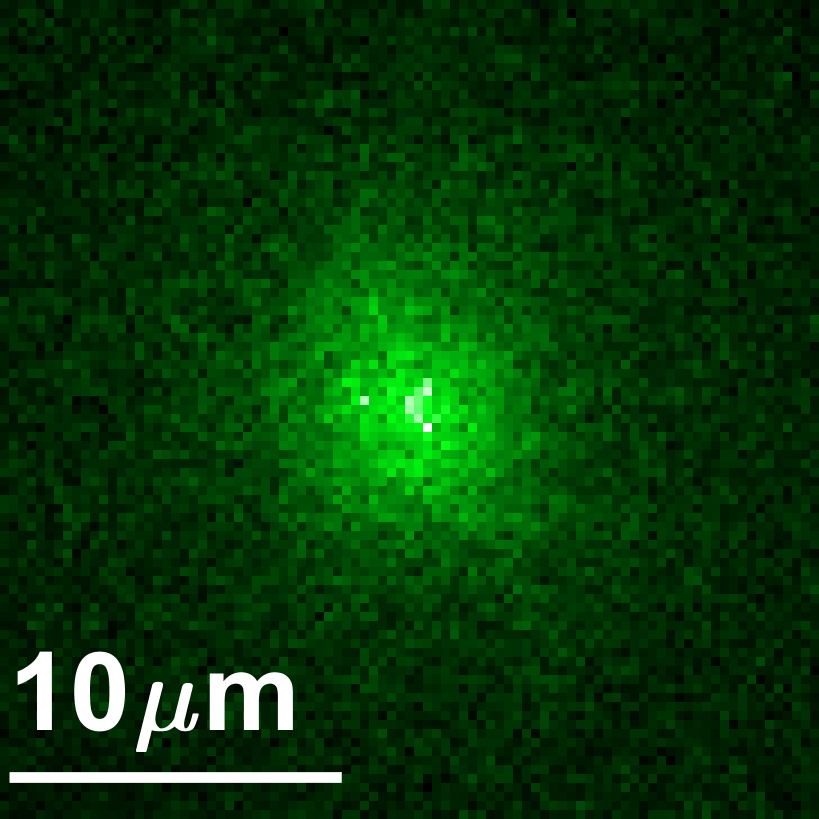}&
			\includegraphics[height = 0.115\textwidth]{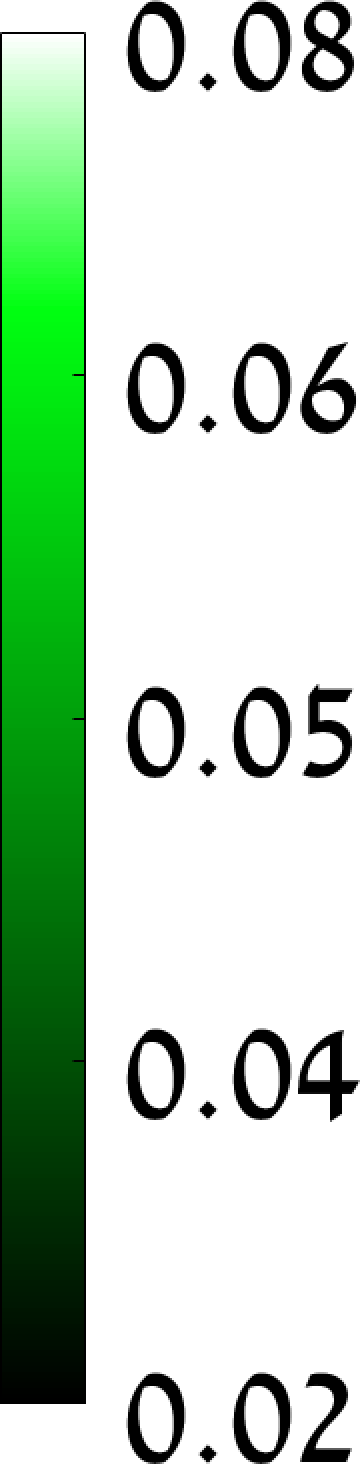}&
			\includegraphics[height = 0.115\textwidth]{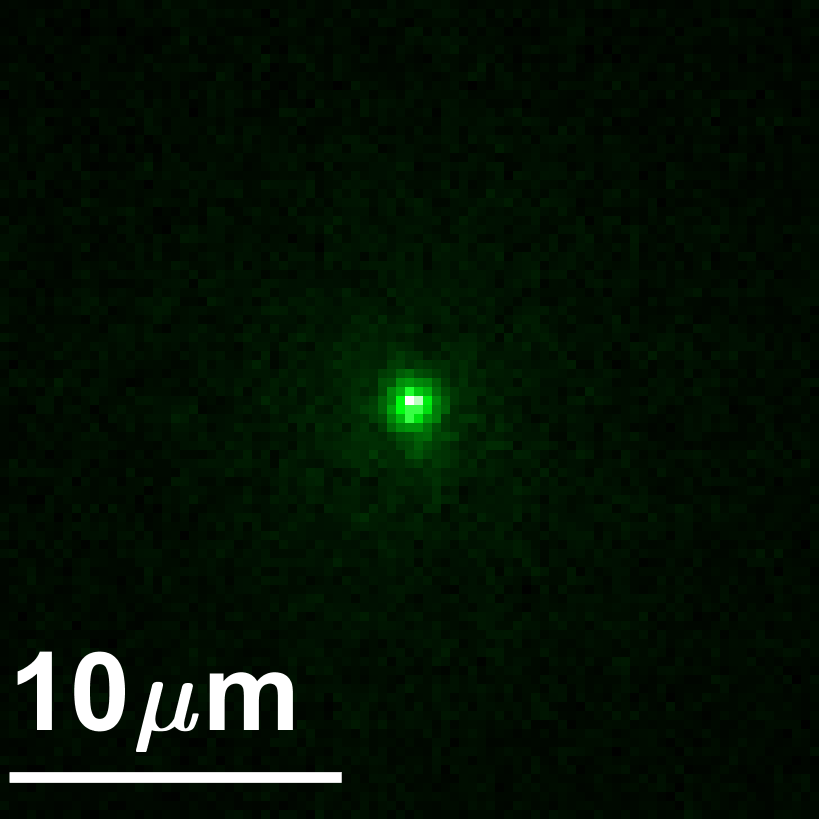}&
			\includegraphics[height = 0.115\textwidth]{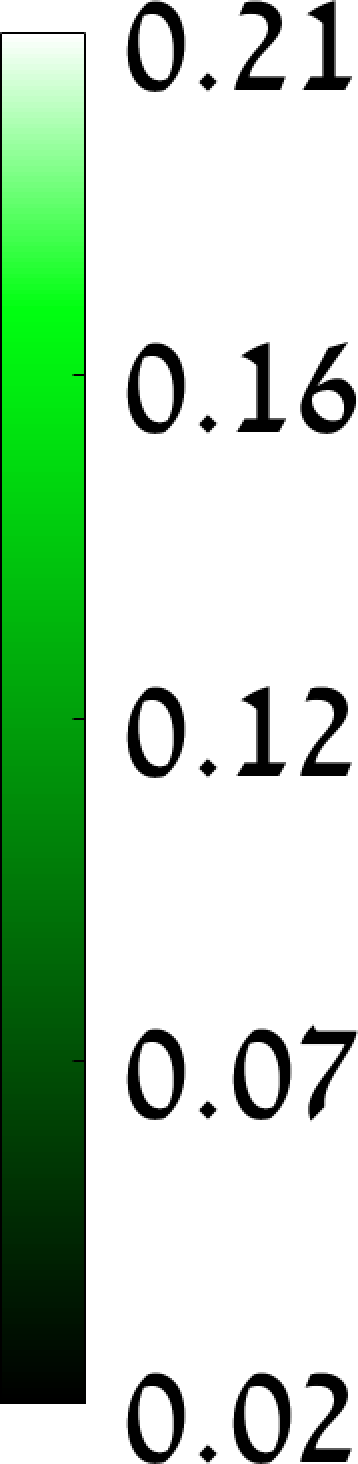}&
			\includegraphics[height = 0.115\textwidth]{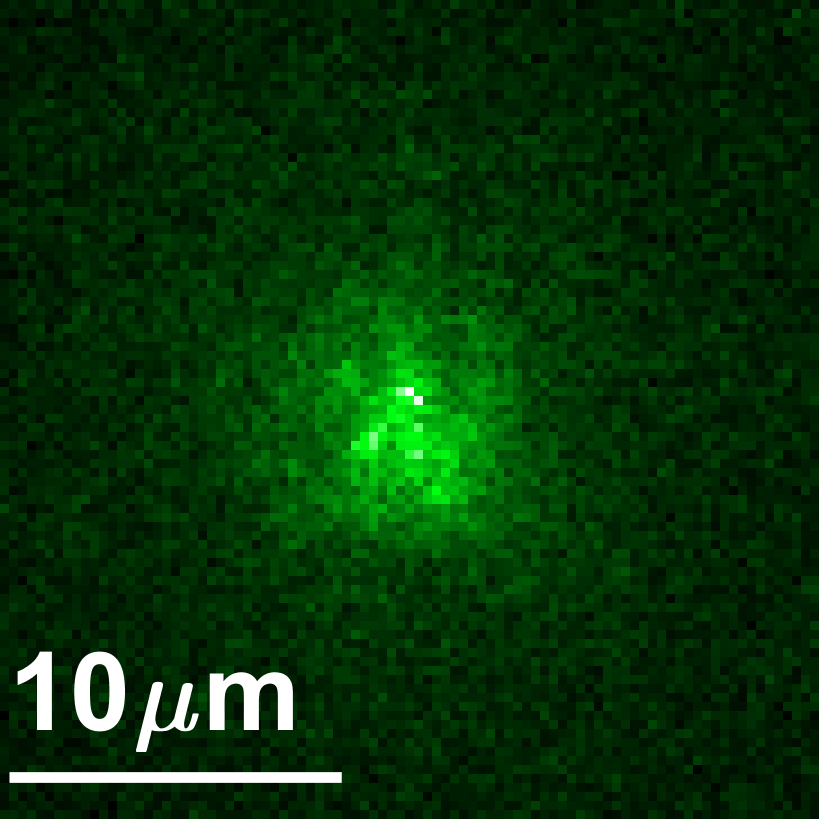}&
			\includegraphics[height = 0.115\textwidth]{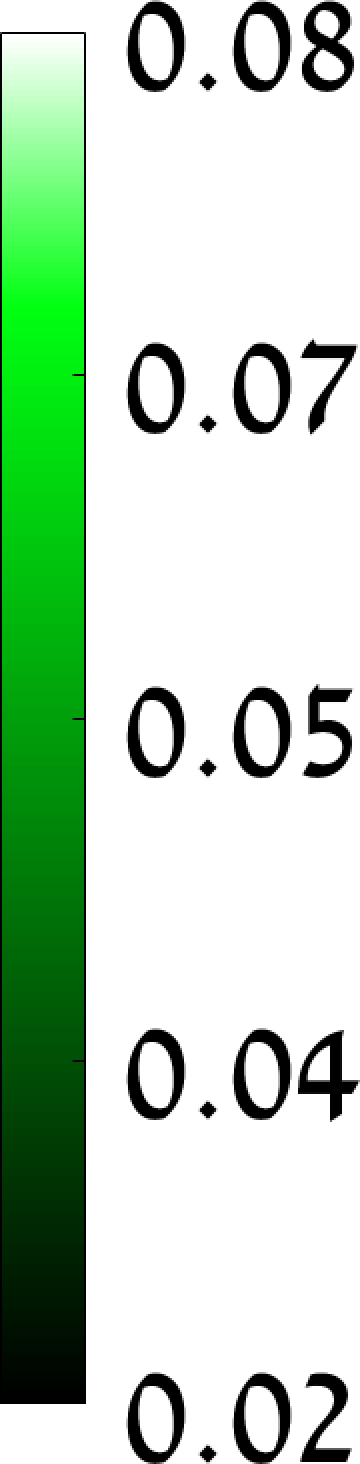}\\

			\vspace{0.1cm}
			\includegraphics[height = 0.115\textwidth]{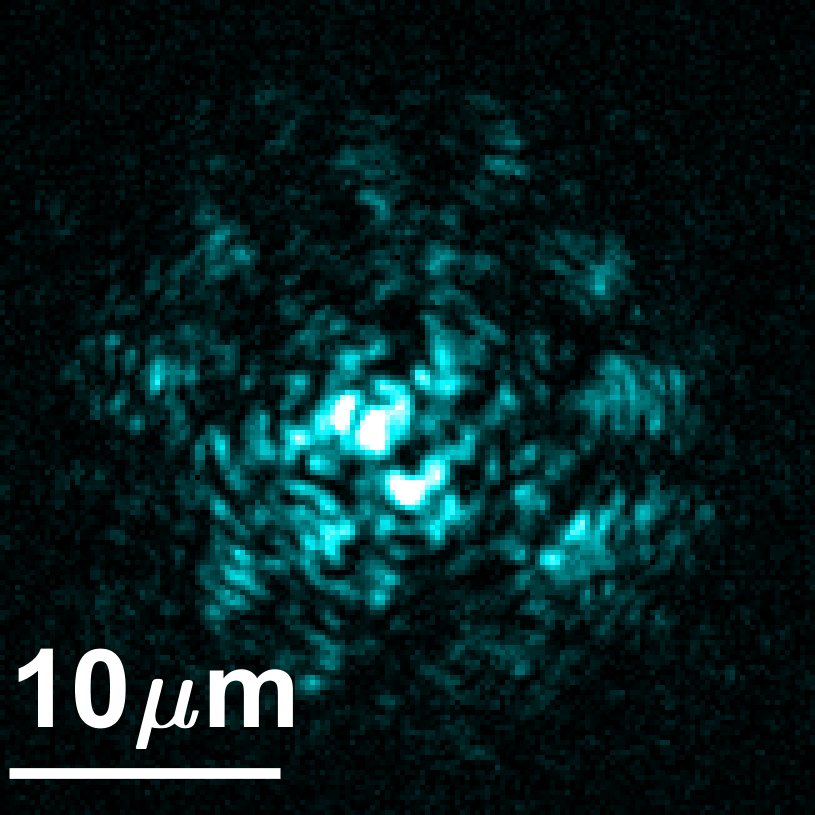}&
			\includegraphics[height = 0.115\textwidth]{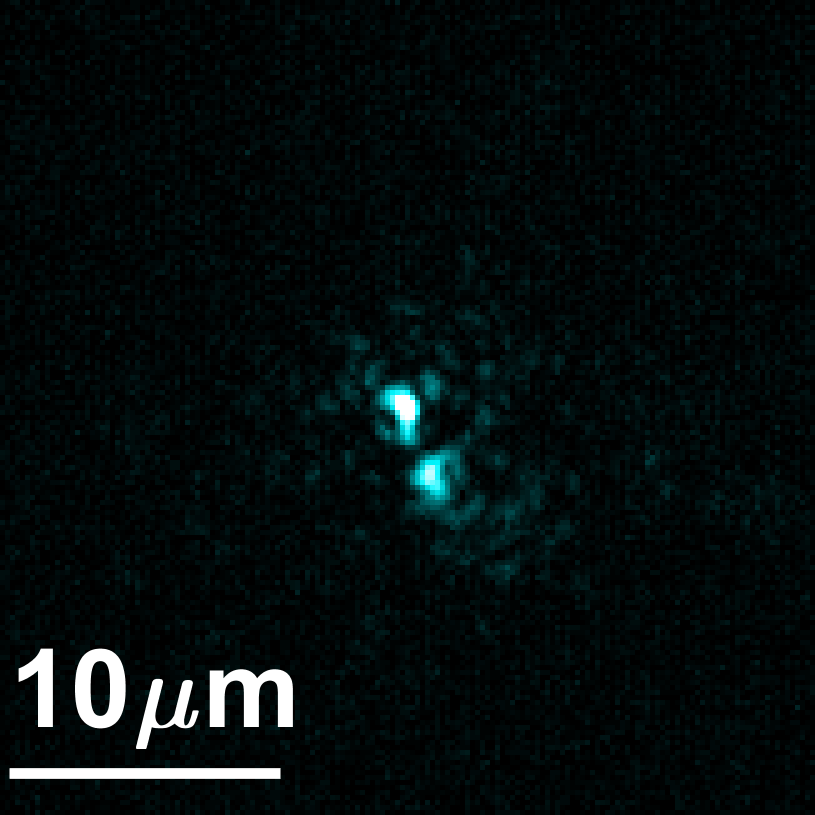}&
			\includegraphics[height = 0.115\textwidth]{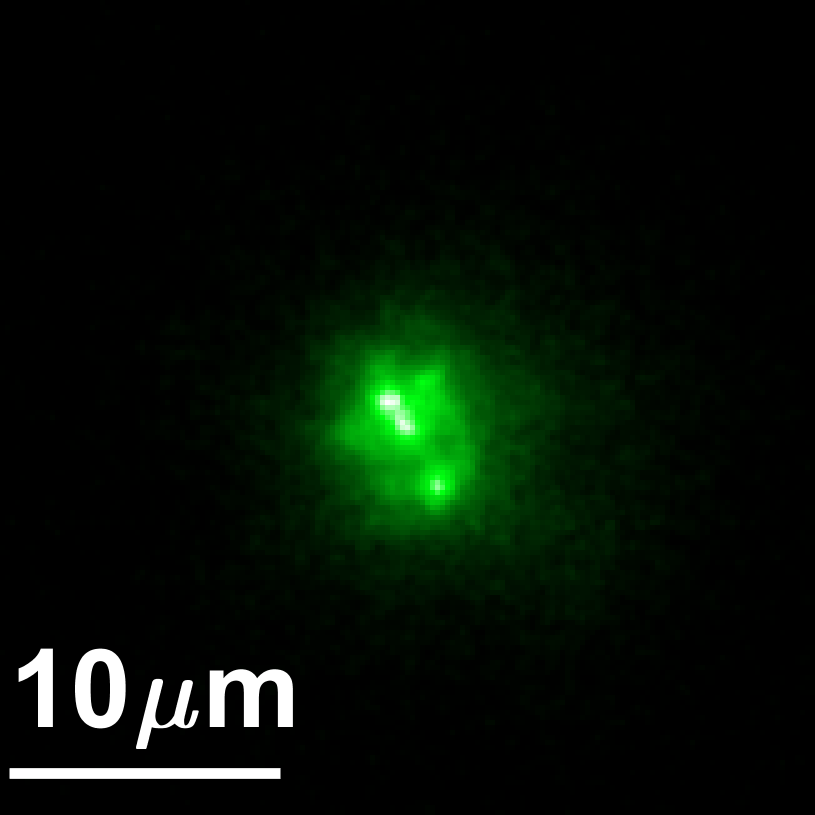}&
			\includegraphics[height = 0.115\textwidth]{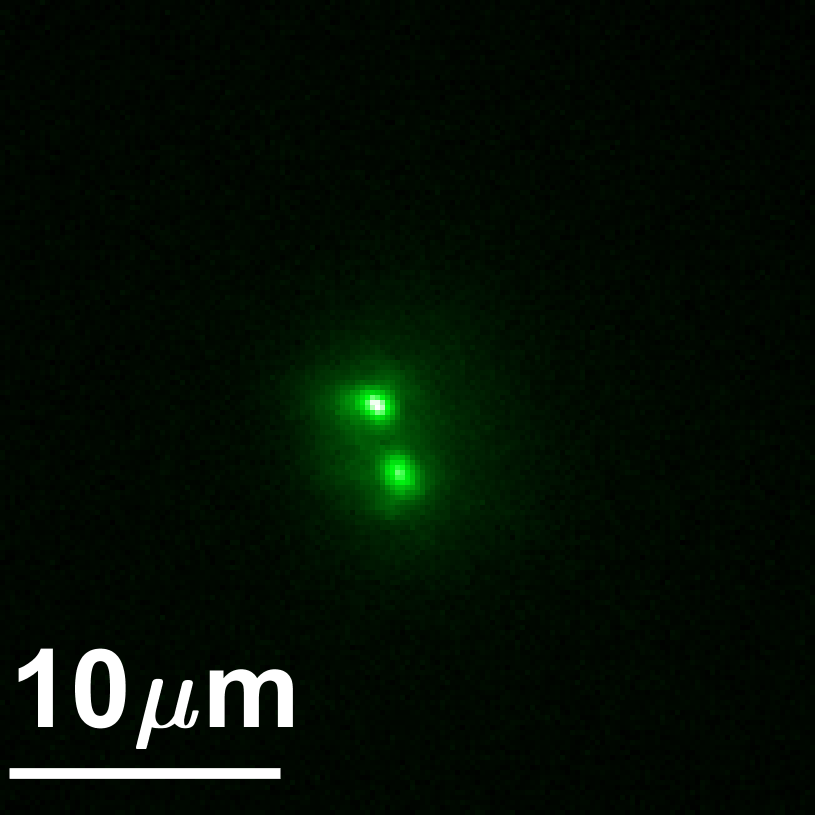}&
			\includegraphics[height = 0.115\textwidth]{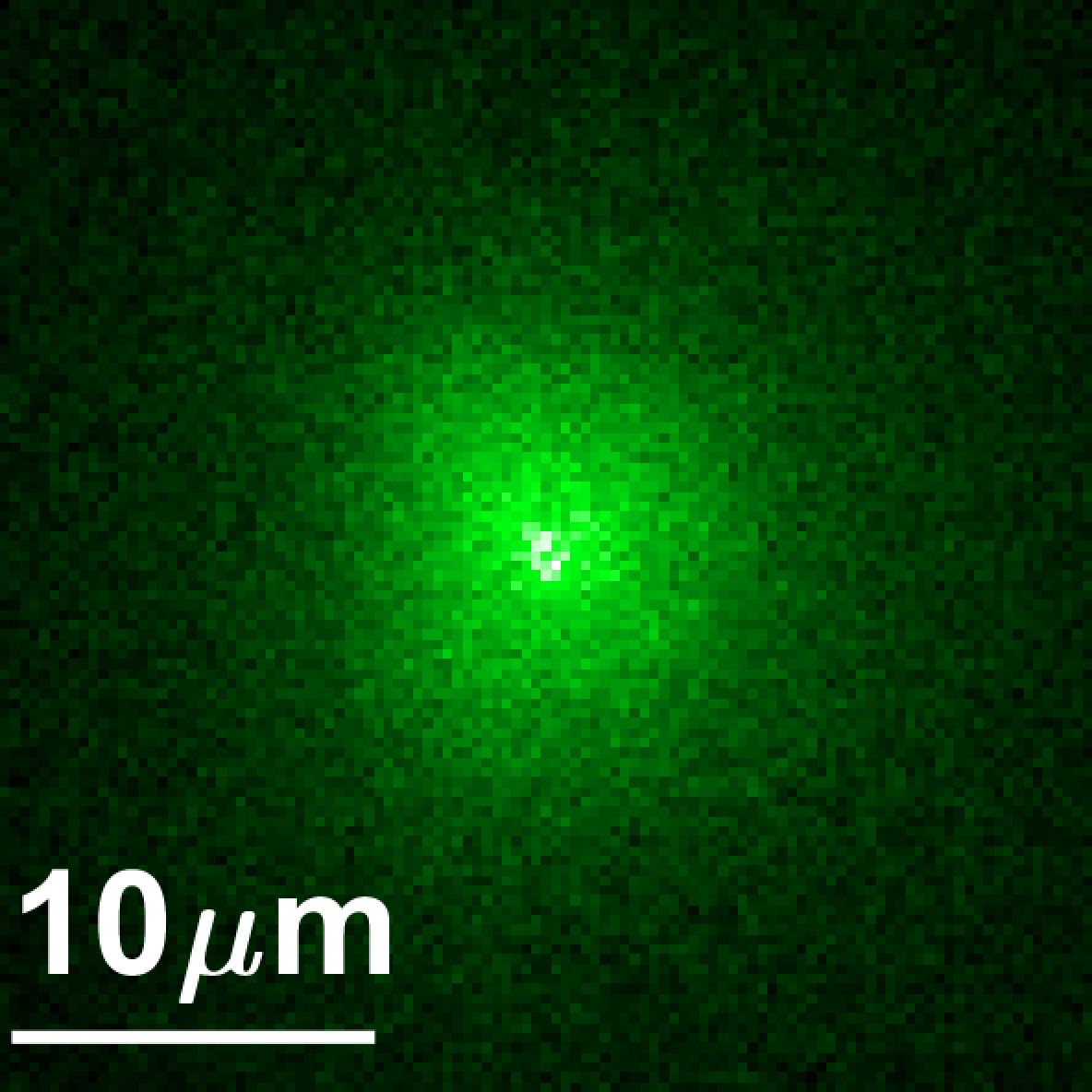}&
			\includegraphics[height = 0.115\textwidth]{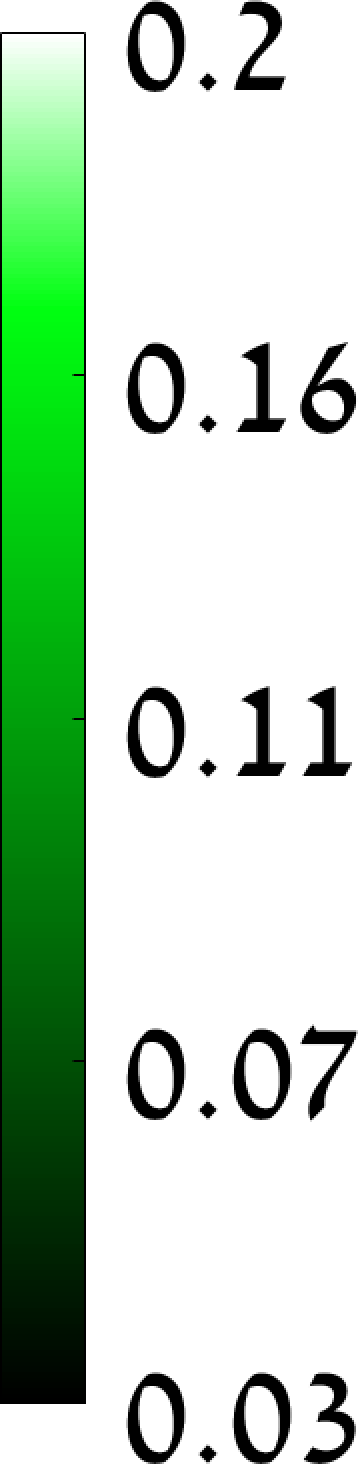}&
			\includegraphics[height = 0.115\textwidth]{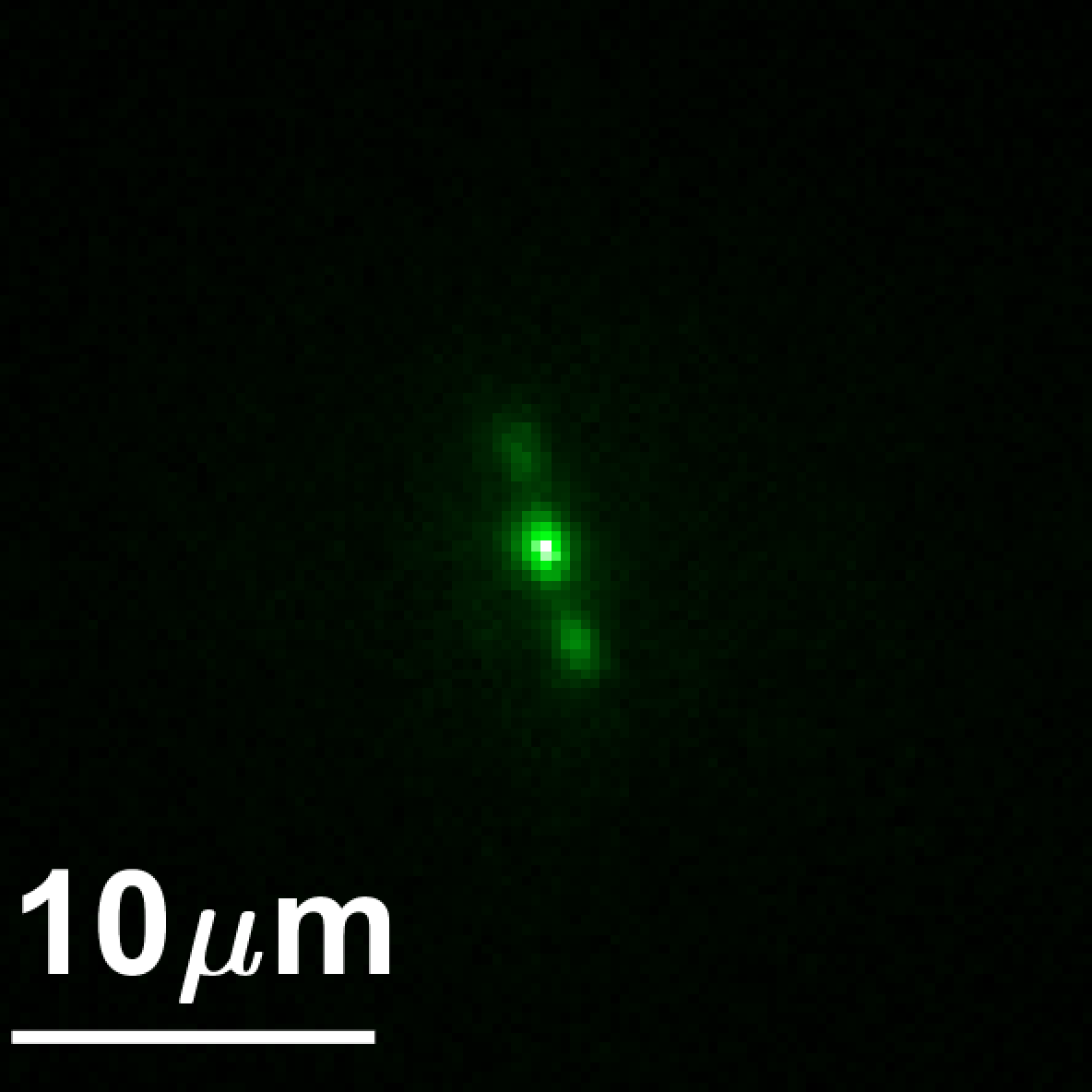}&
			\includegraphics[height = 0.115\textwidth]{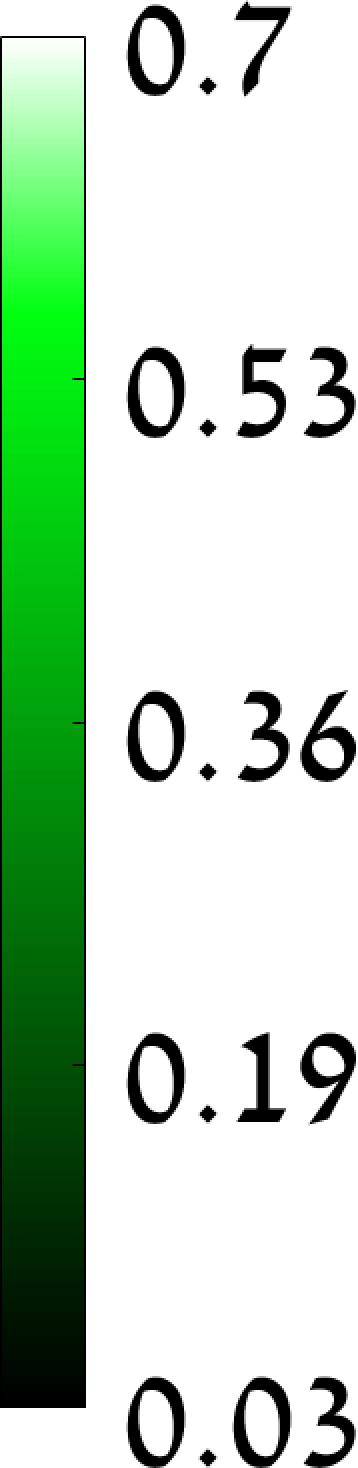}&
			\includegraphics[height = 0.115\textwidth]{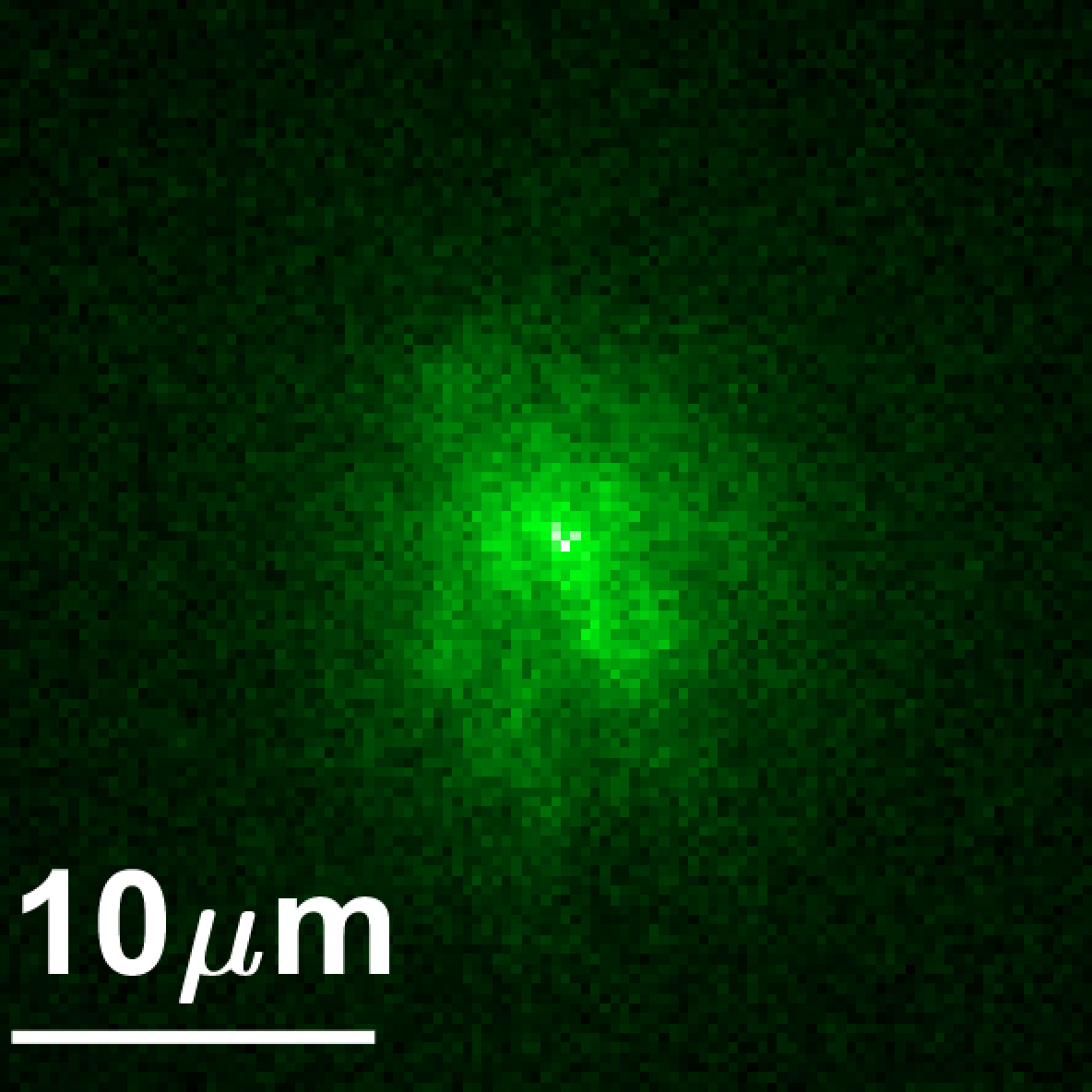}&
			\includegraphics[height = 0.115\textwidth]{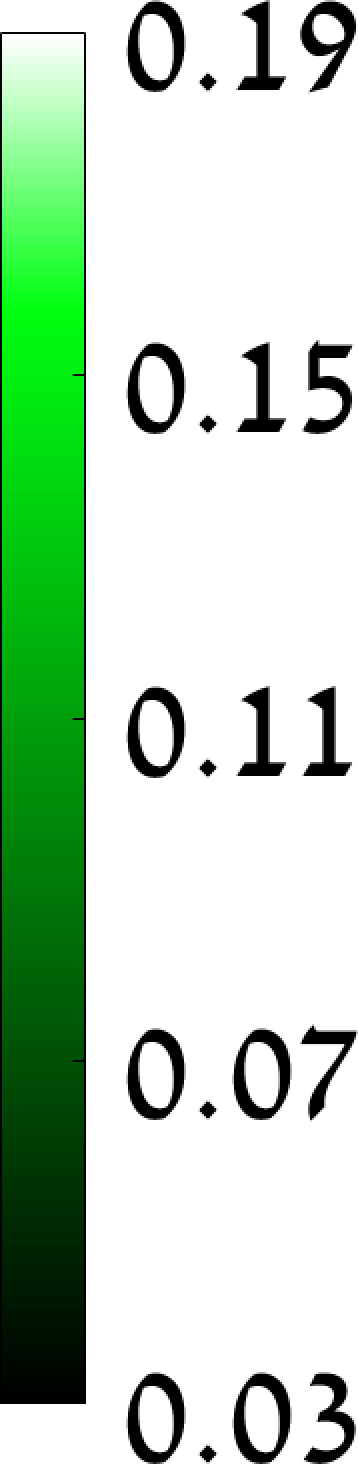}\\

			
			\footnotesize{(a) No mod.} & \footnotesize{(b) With mod.} & \footnotesize{(c) No mod.} & \footnotesize{(d) With mod.} & \footnotesize{(e) No mod.} &   & \footnotesize{(f) With mod.} &   & \footnotesize{(g) PSF}\\
			\multicolumn{2}{c}{Excitation}&\multicolumn{2}{c}{Emission}&\multicolumn{6}{c}{Emission}
		\end{tabular}

		\caption{\blue{Additional wavefront shaping results: we visualize views from the validation and main cameras.
			(a-b) The excitation light as viewed by the validation camera at the back of the tissue. Due to significant scattering, at the beginning of the algorithm when no modulation (mod.) is available, a wide speckle pattern is generated.  After  optimization, the modulated wavefront is brought into a single spot.
			(c-d)  By placing a band-pass filter on the validation camera, we visualize the emitted light with and without the modulation correction.
			(e-f) Views of the emitted light at the main front camera  with and without the modulation correction. Note that this is the only input used by our algorithm. Without modulation, light is scattered over a wide image area and the image is noisy.  A sharp clean spot can be imaged when the limited number of photons is  brought into a single sensor pixel.
			(g) By correcting the emission such that a single spot is excited and leaving the imaging path uncorrected, we can visualize the actual aberration of a single fluorescent point source. Each row demonstrates a different tissue sample.  The top  example demonstrates a thin brain layer behind parafilm, and the two lower ones are from a thick brain slice. The lowest row demonstrates a failure example where the optimization converged at two spots rather than one.} }\label{fig:converging2}

	\end{center}
\end{figure*}

\begin{figure*}[t!]
	\begin{center}
		
		\begin{tabular}{@{}c@{~~}c@{~~}c@{}}

			\includegraphics[width= 0.2\textwidth]{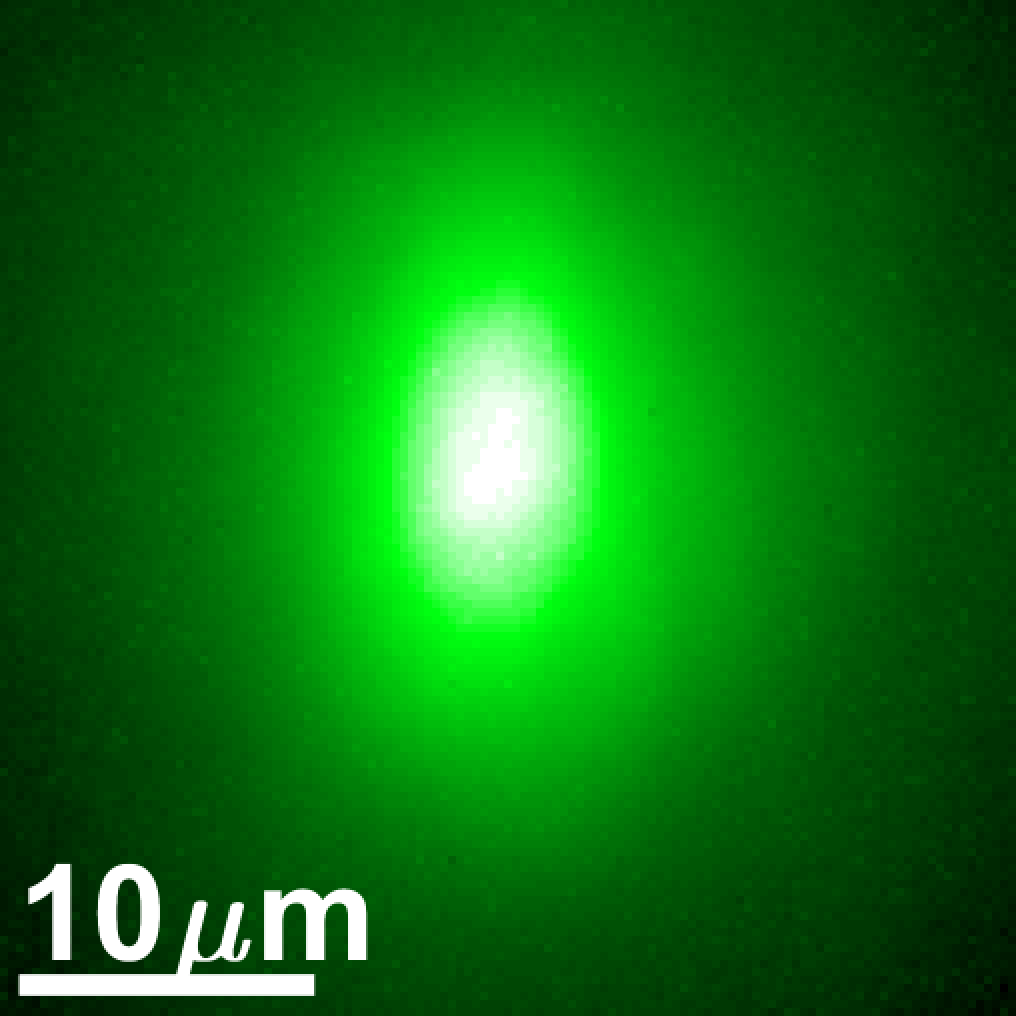}&
			\includegraphics[width= 0.2\textwidth]{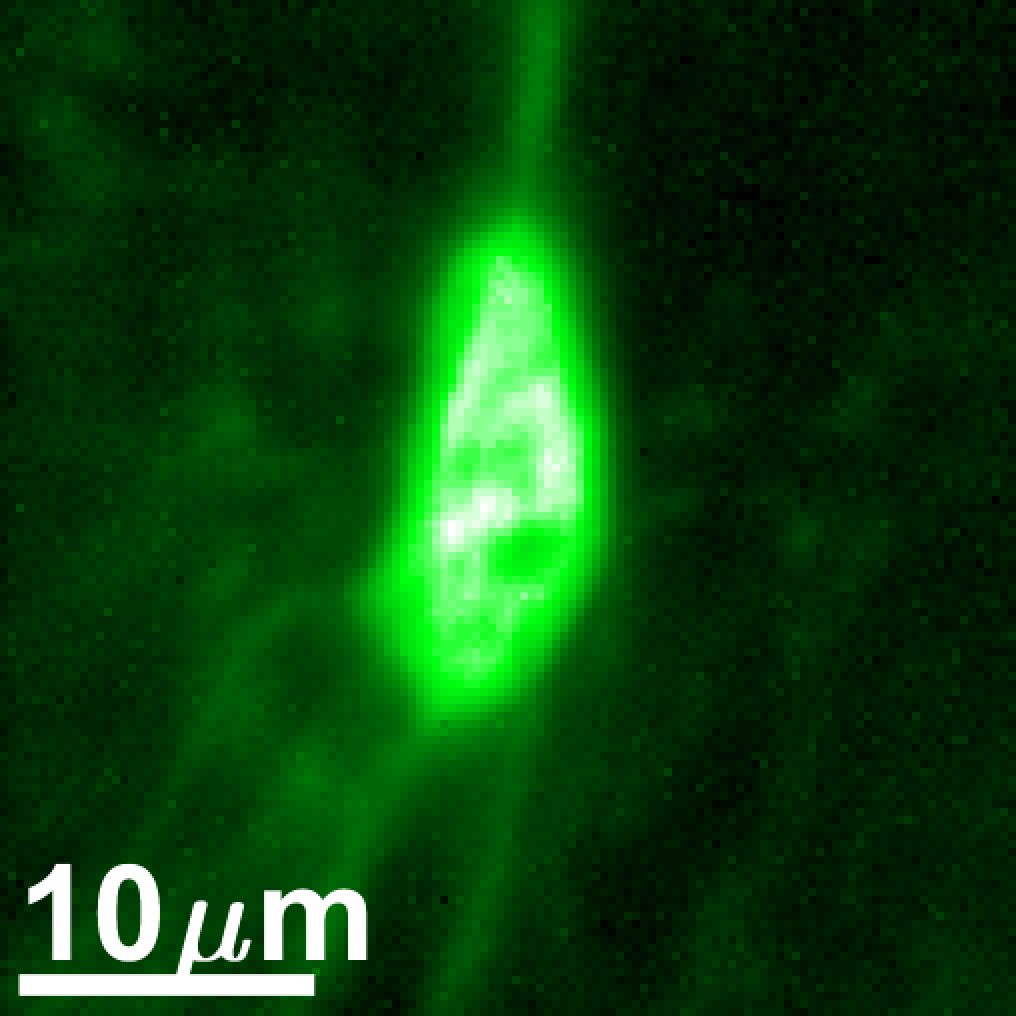}&
			\includegraphics[width= 0.2\textwidth]{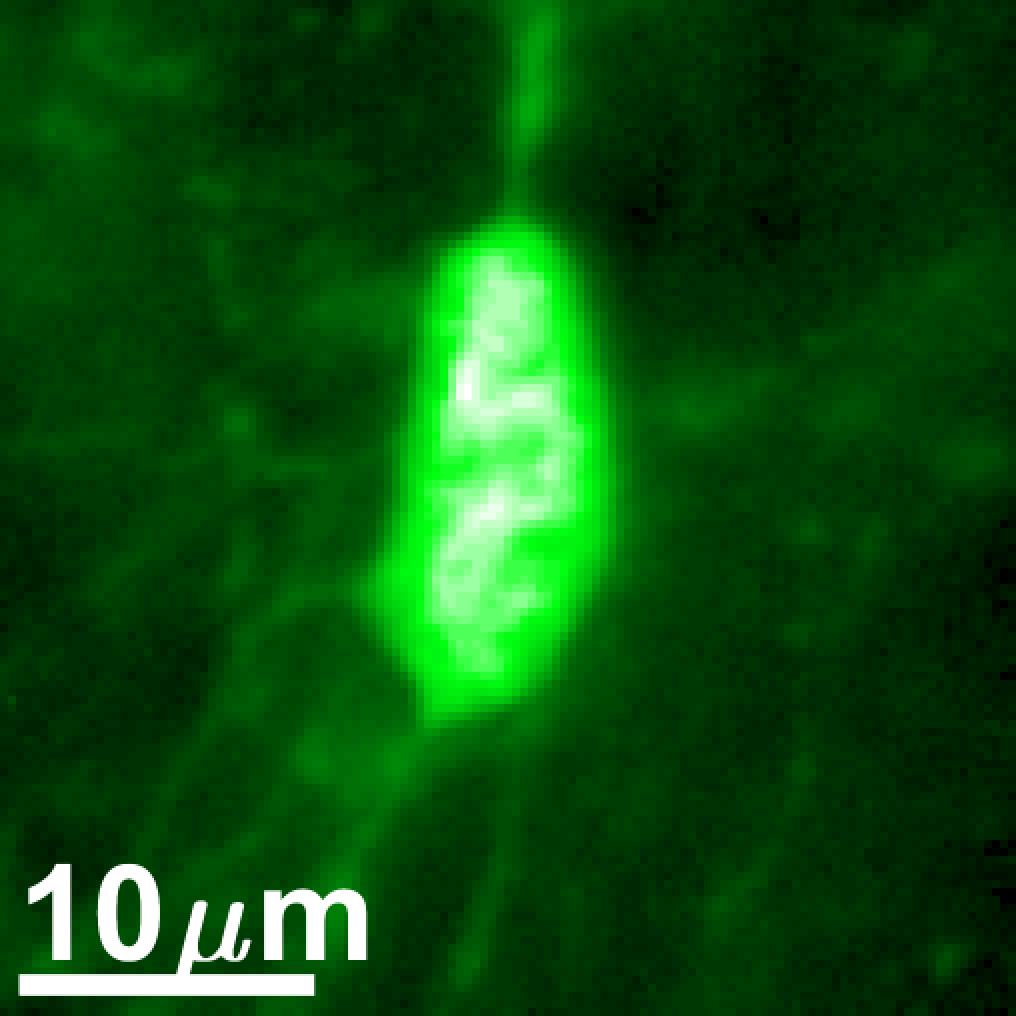}\\			
			
			\includegraphics[width= 0.2\textwidth]{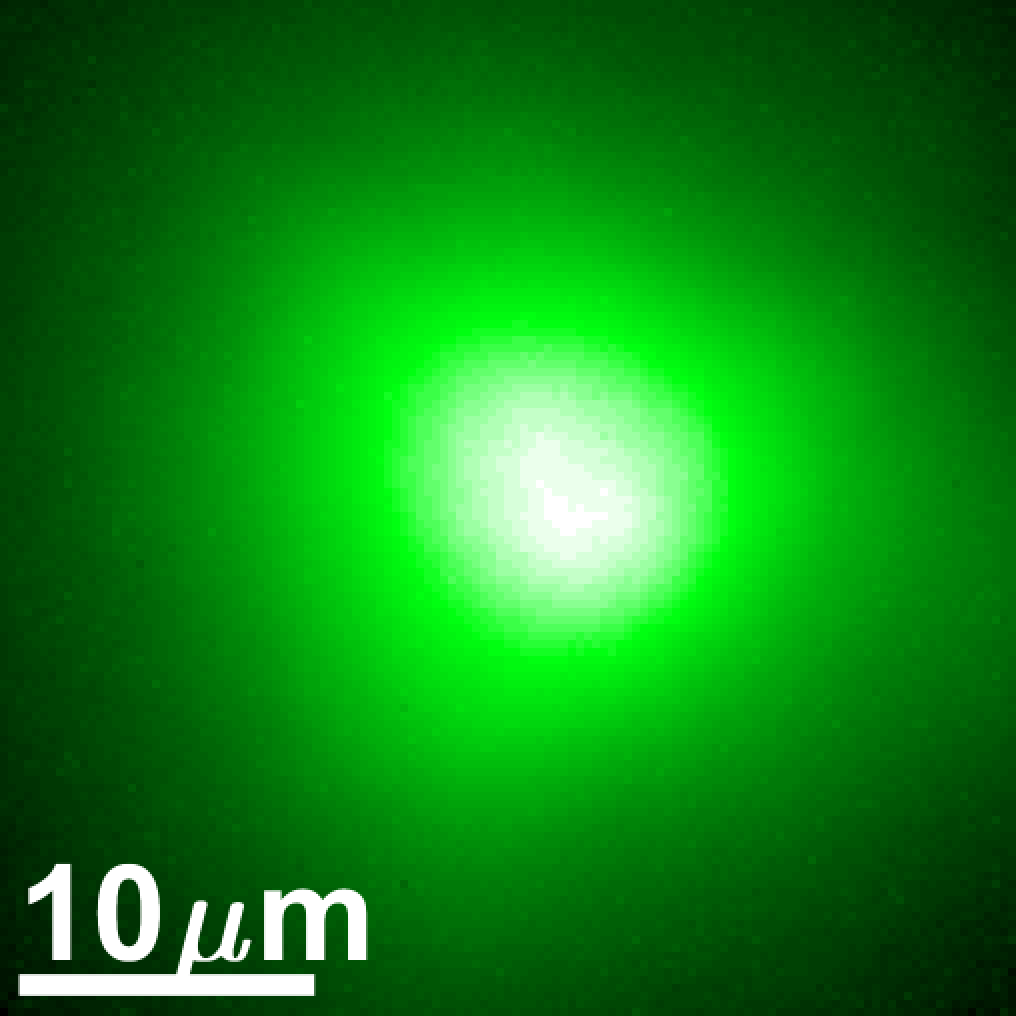}&
			\includegraphics[width= 0.2\textwidth]{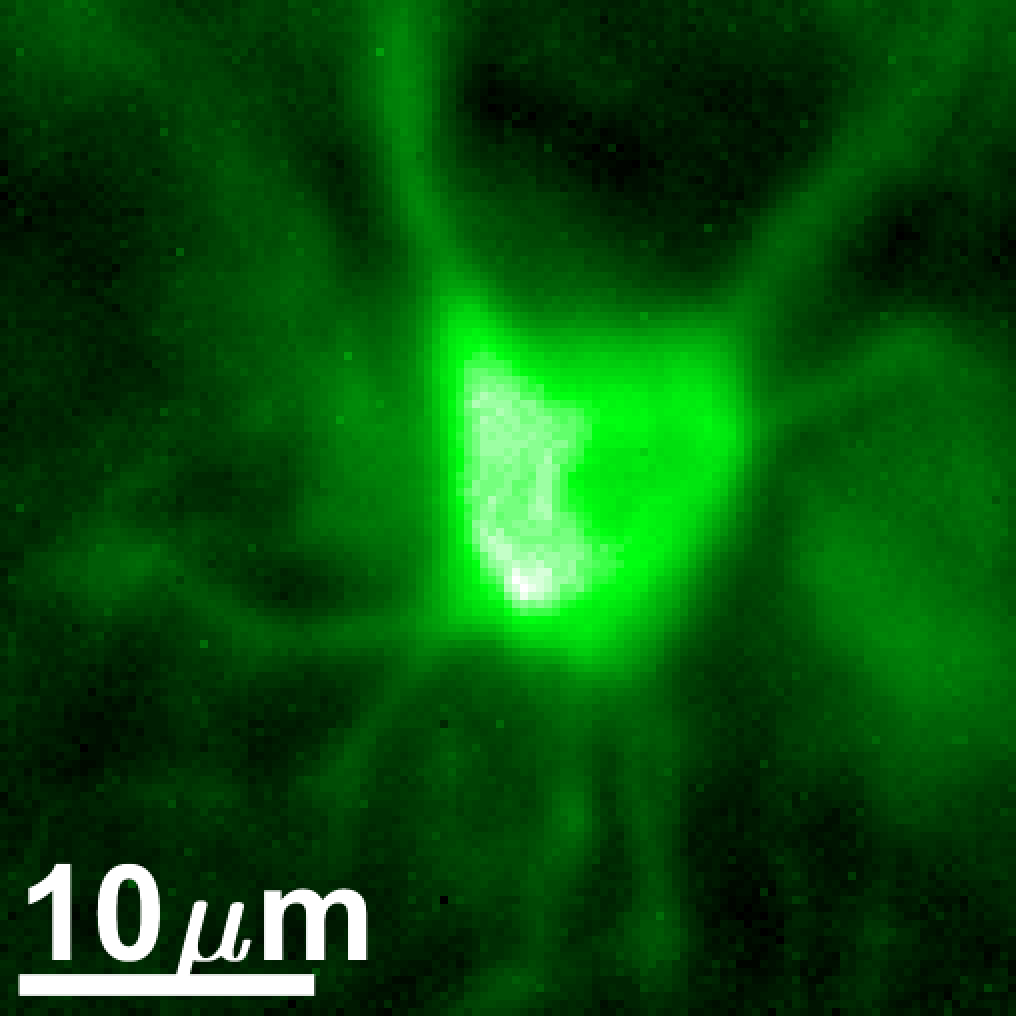}&
			\includegraphics[width= 0.2\textwidth]{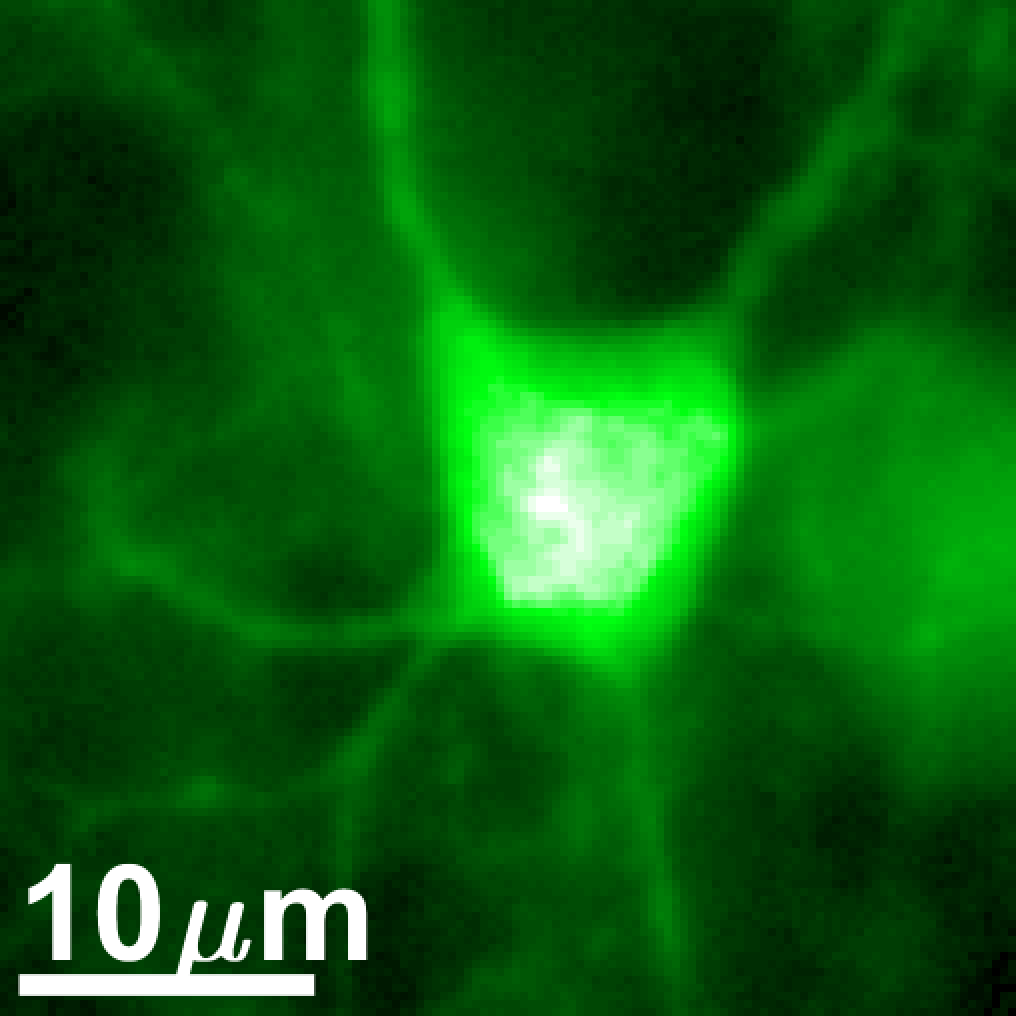}\\

			(a) Uncorrected  & (b) Corrected & (c) Reference  \\
			Main camera   & Main camera   & Validation camera             \\

		\end{tabular}
		\captionof{figure}
		{Additional results, a thin brain layer behind chicken breast tissue:  (a) Image of the neuron from the main camera with no correction, strong scattering is present and the neuron structure is lost. (b) Image with our modulation correction, the neuron shape as well as some of the axons are revealed. (c) A clean reference   image of the same neuron, from the validation camera.} 
		\label{fig:big_area_sup_chicken}
	\end{center}
\end{figure*}

\begin{figure*}[t!]
	\begin{center}
		
		\begin{tabular}{@{}c@{~~}c@{~~}c@{}}
			
			\includegraphics[width= 0.18\textwidth]{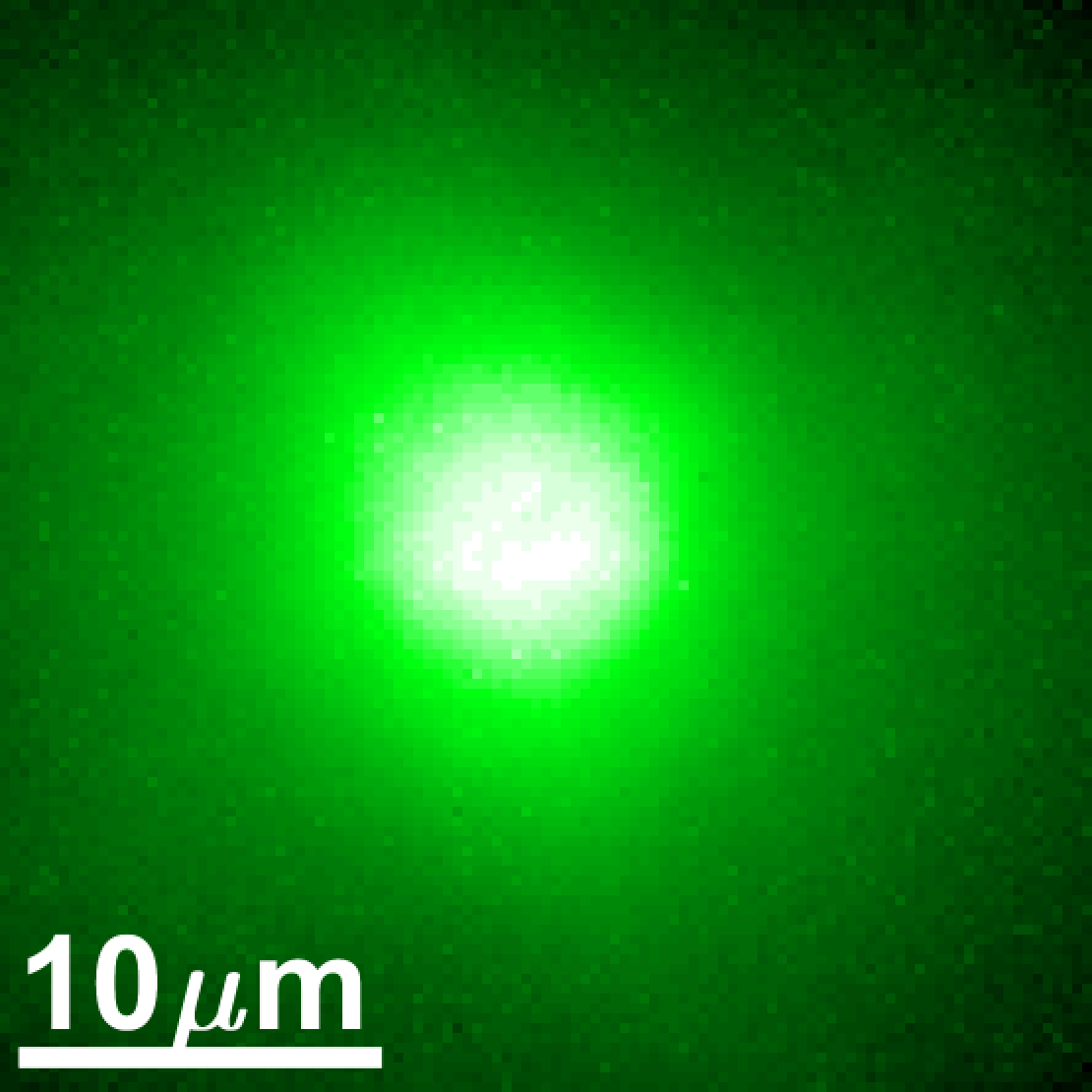}&
			\includegraphics[width= 0.18\textwidth]{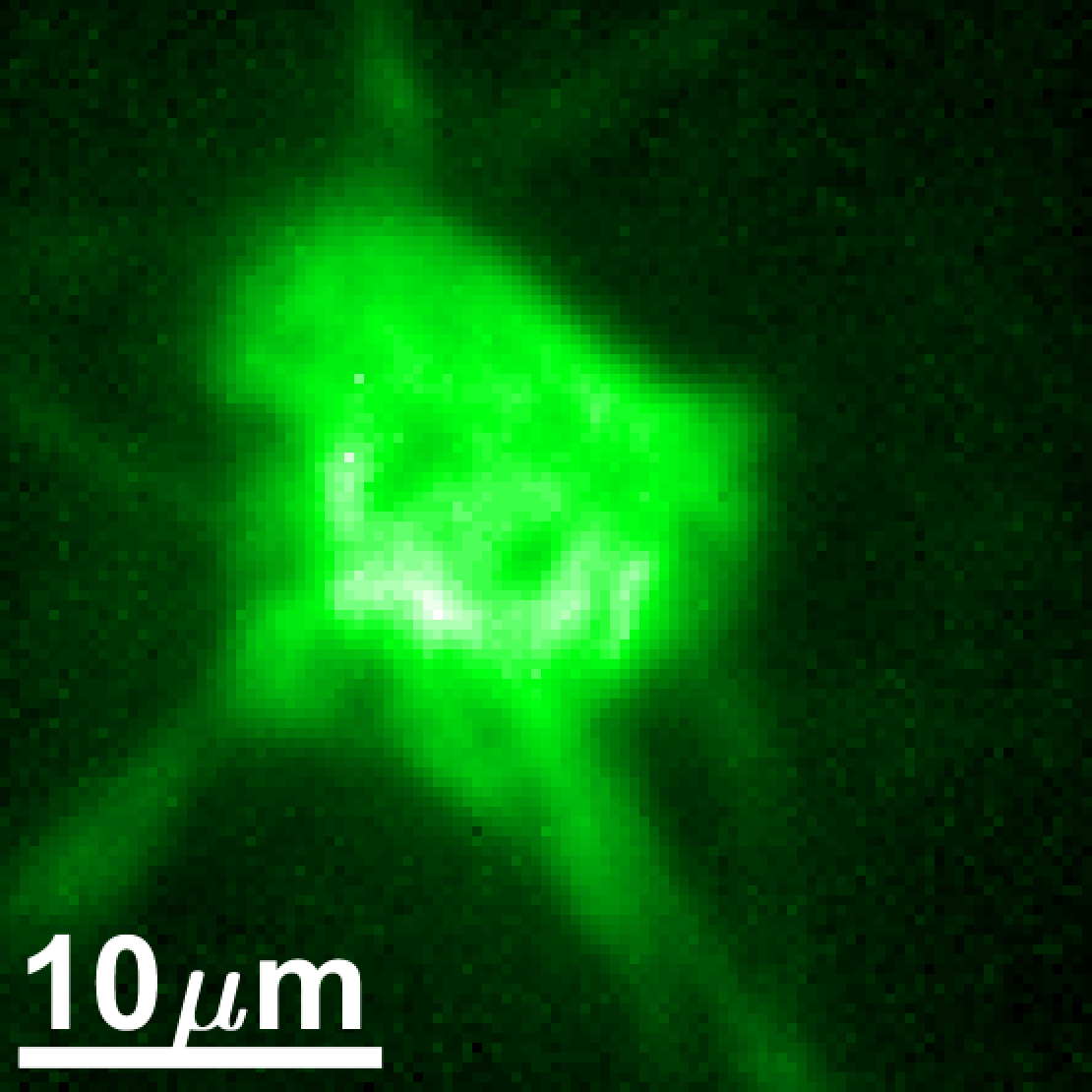}&
			\includegraphics[width= 0.18\textwidth]{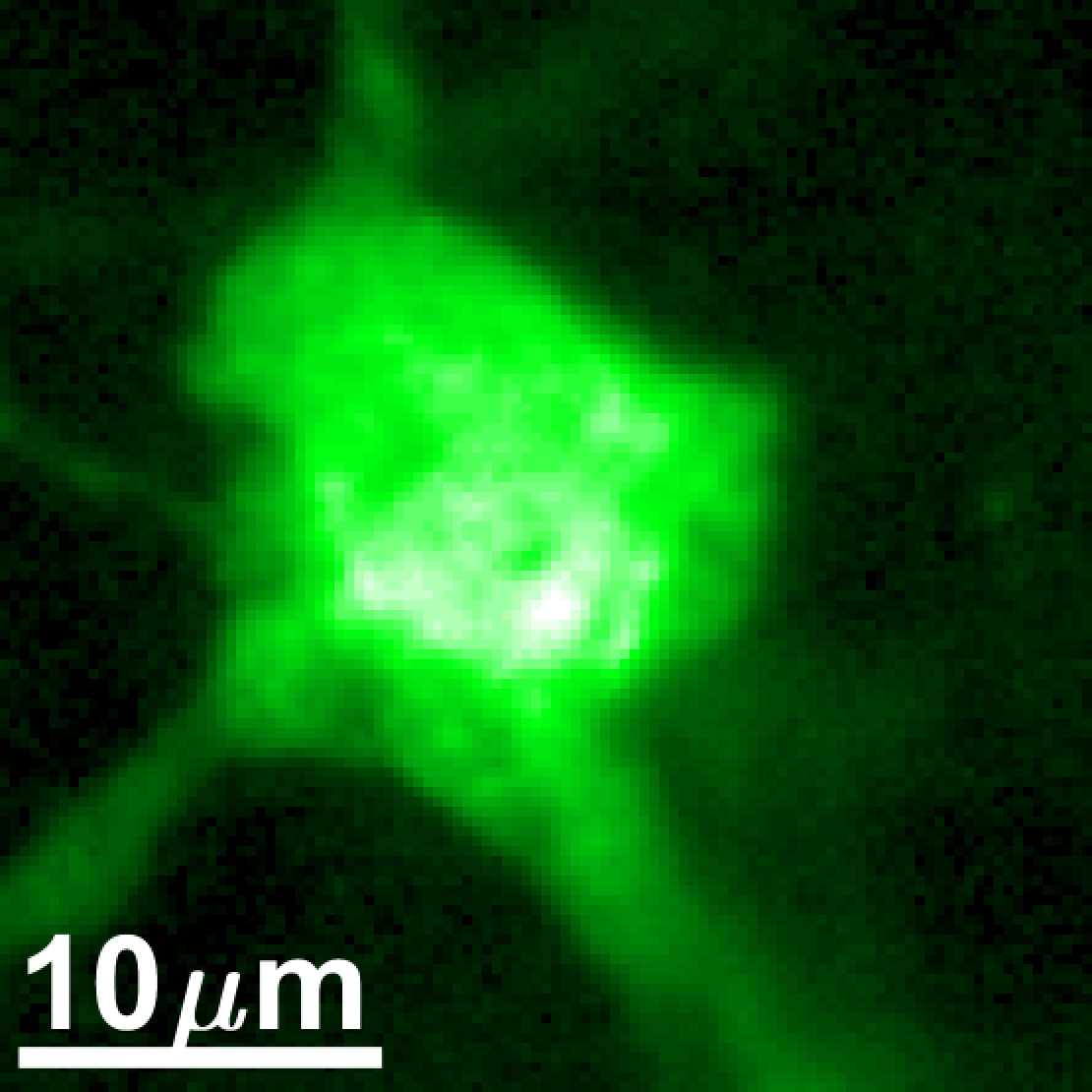}\\						
			
			\includegraphics[width= 0.18\textwidth]{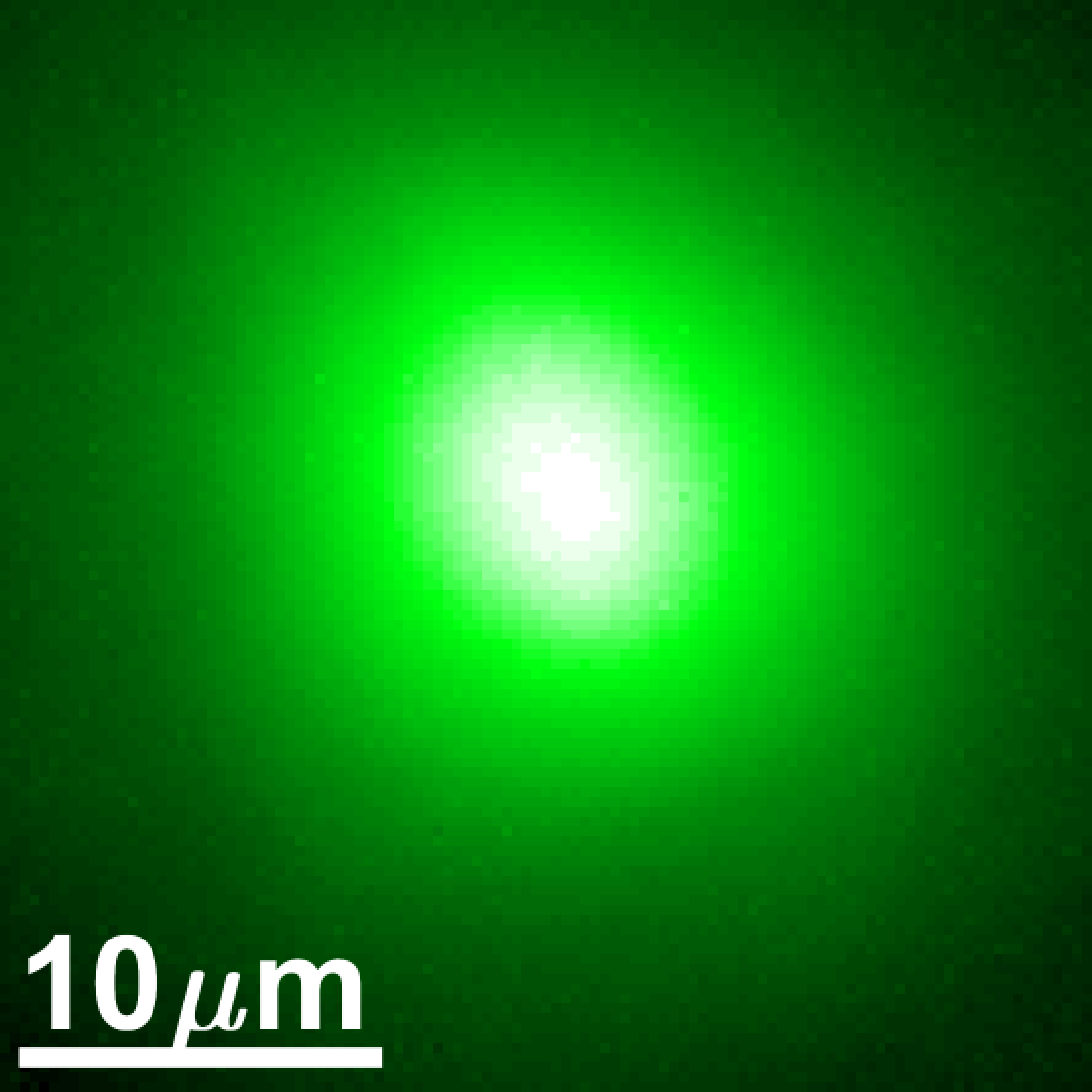}&
			\includegraphics[width= 0.18\textwidth]{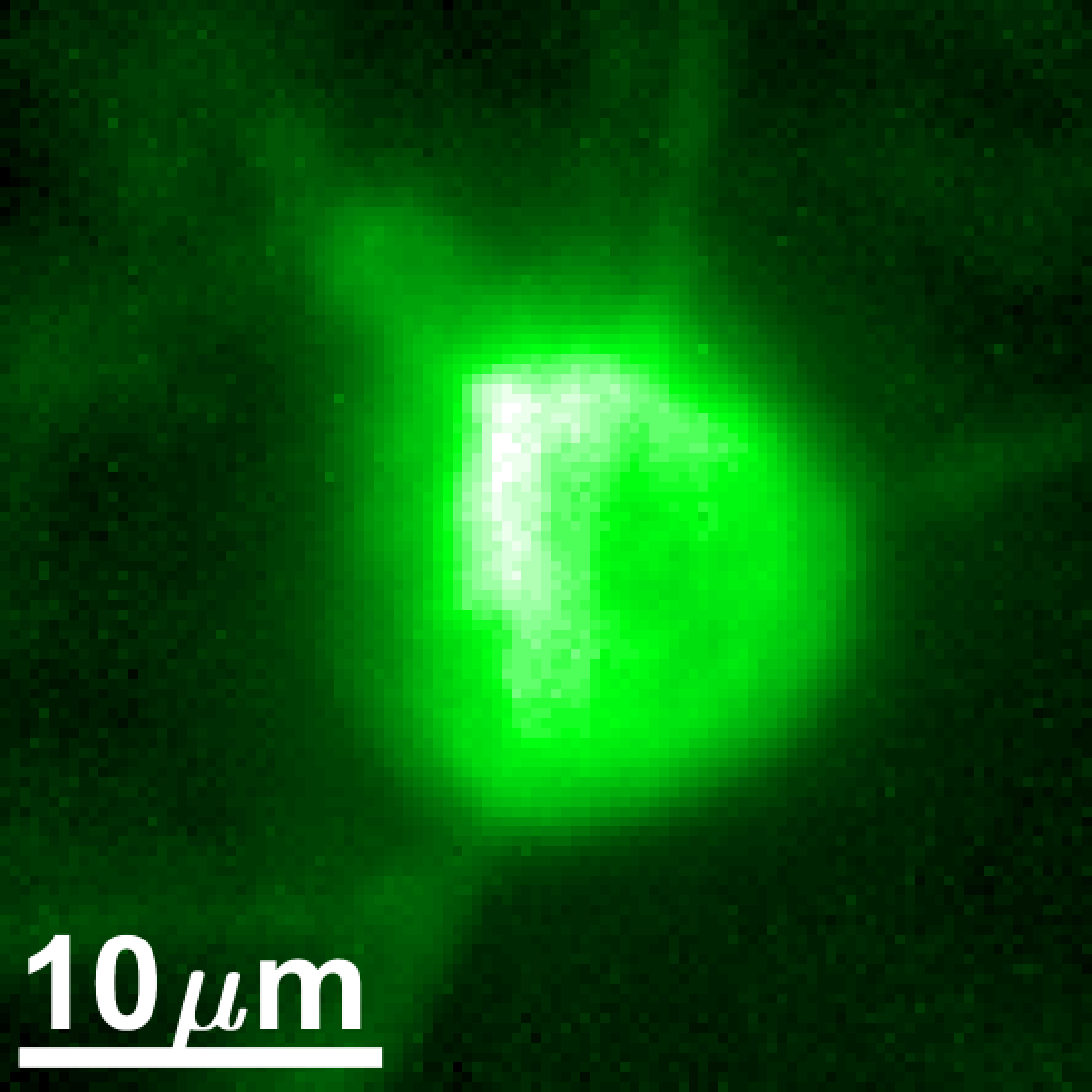}&
			\includegraphics[width= 0.18\textwidth]{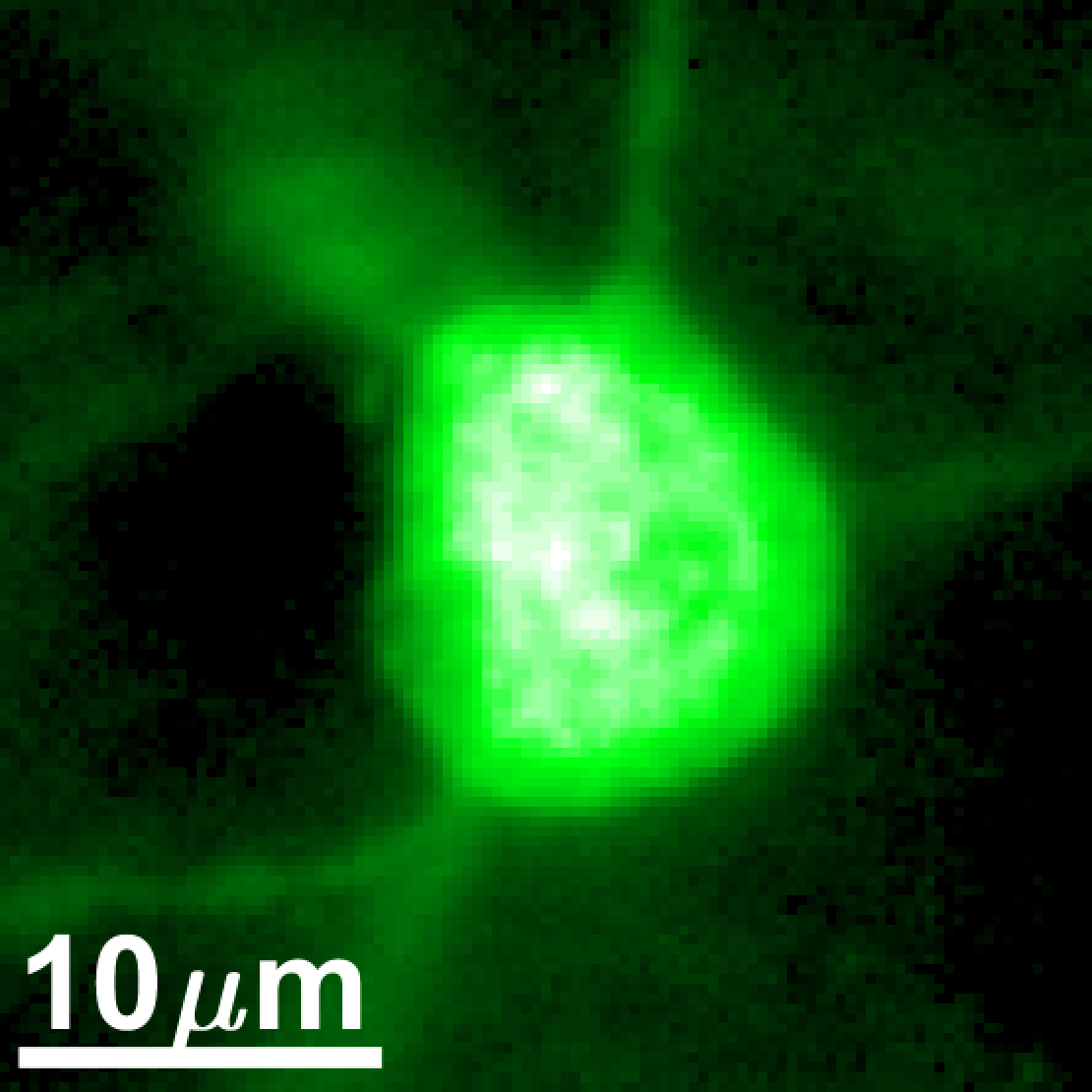}\\			
			
			\includegraphics[width= 0.18\textwidth]{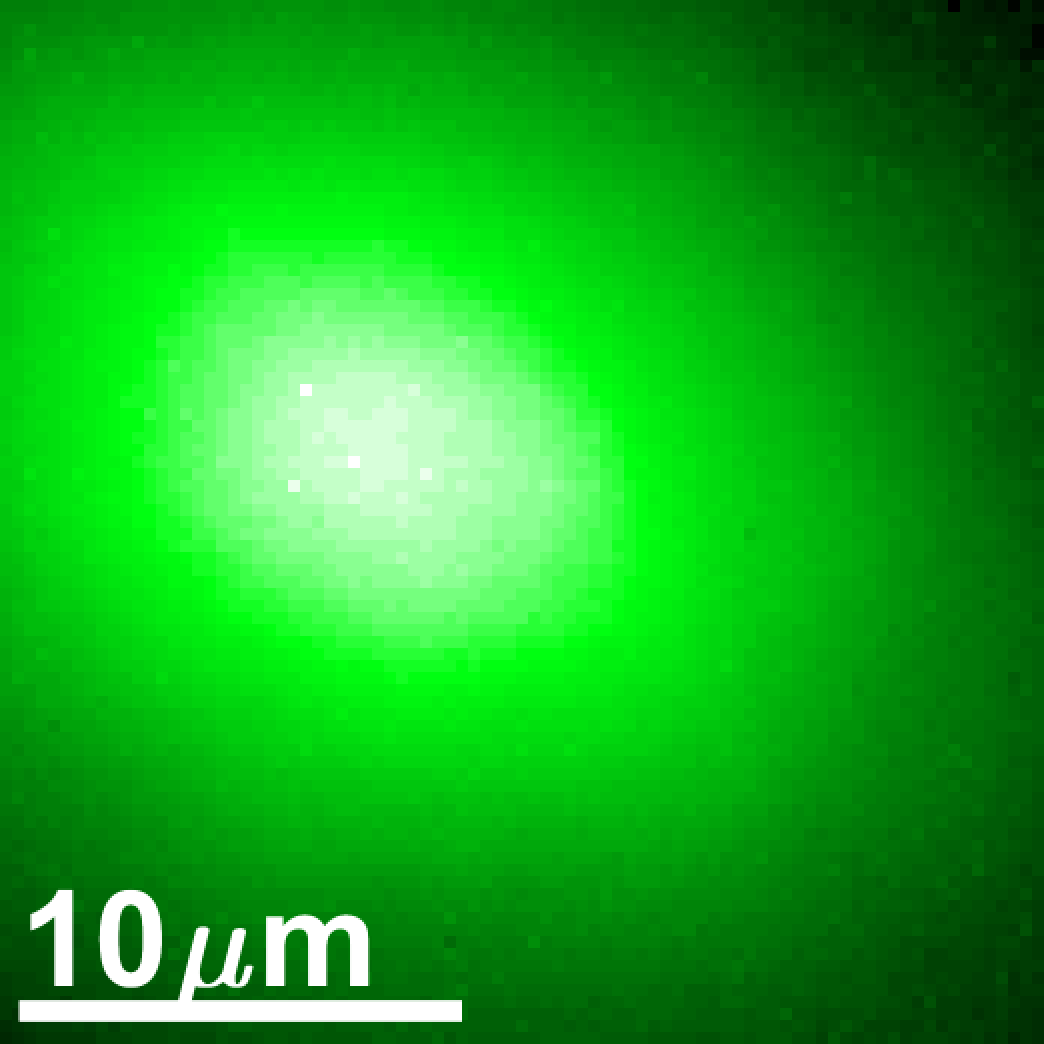}&
			\includegraphics[width= 0.18\textwidth]{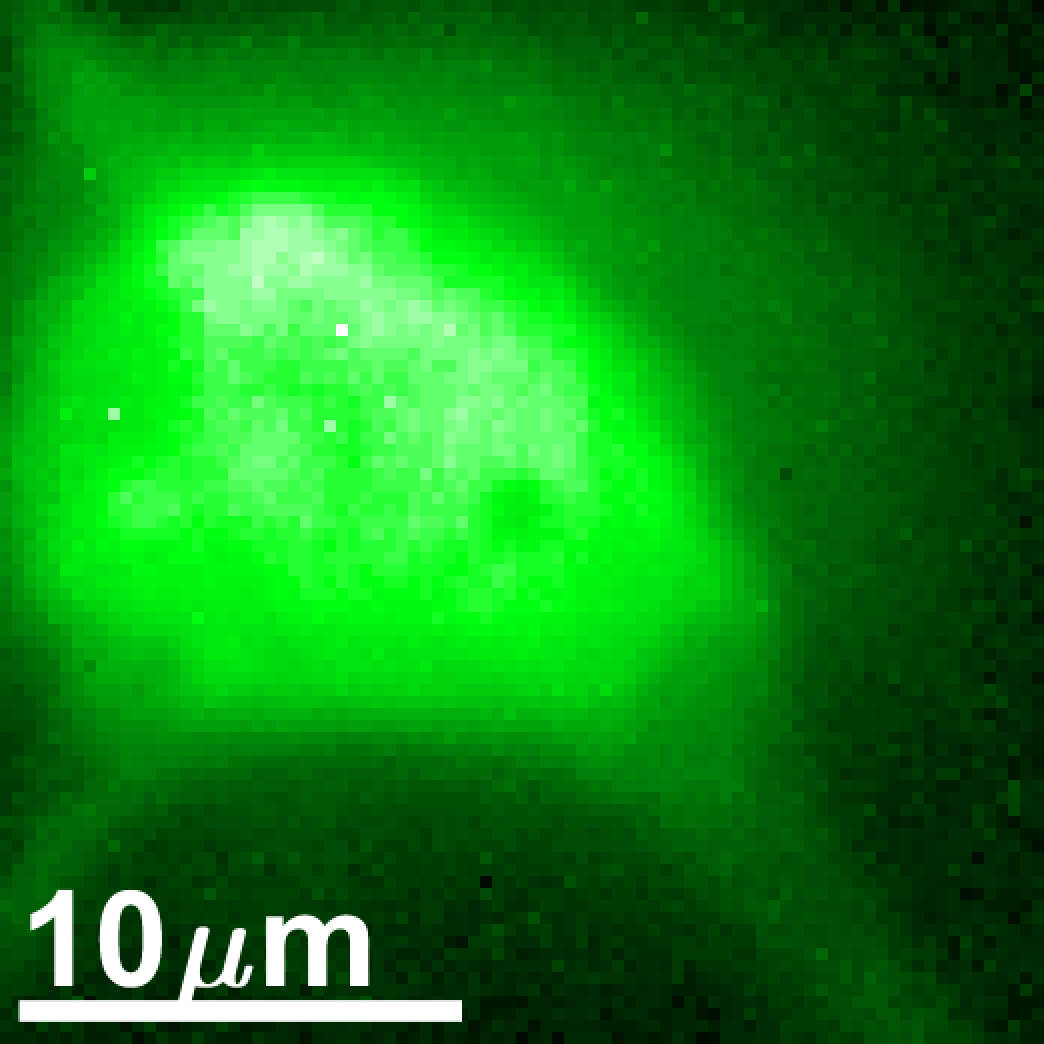}&
			\includegraphics[width= 0.18\textwidth]{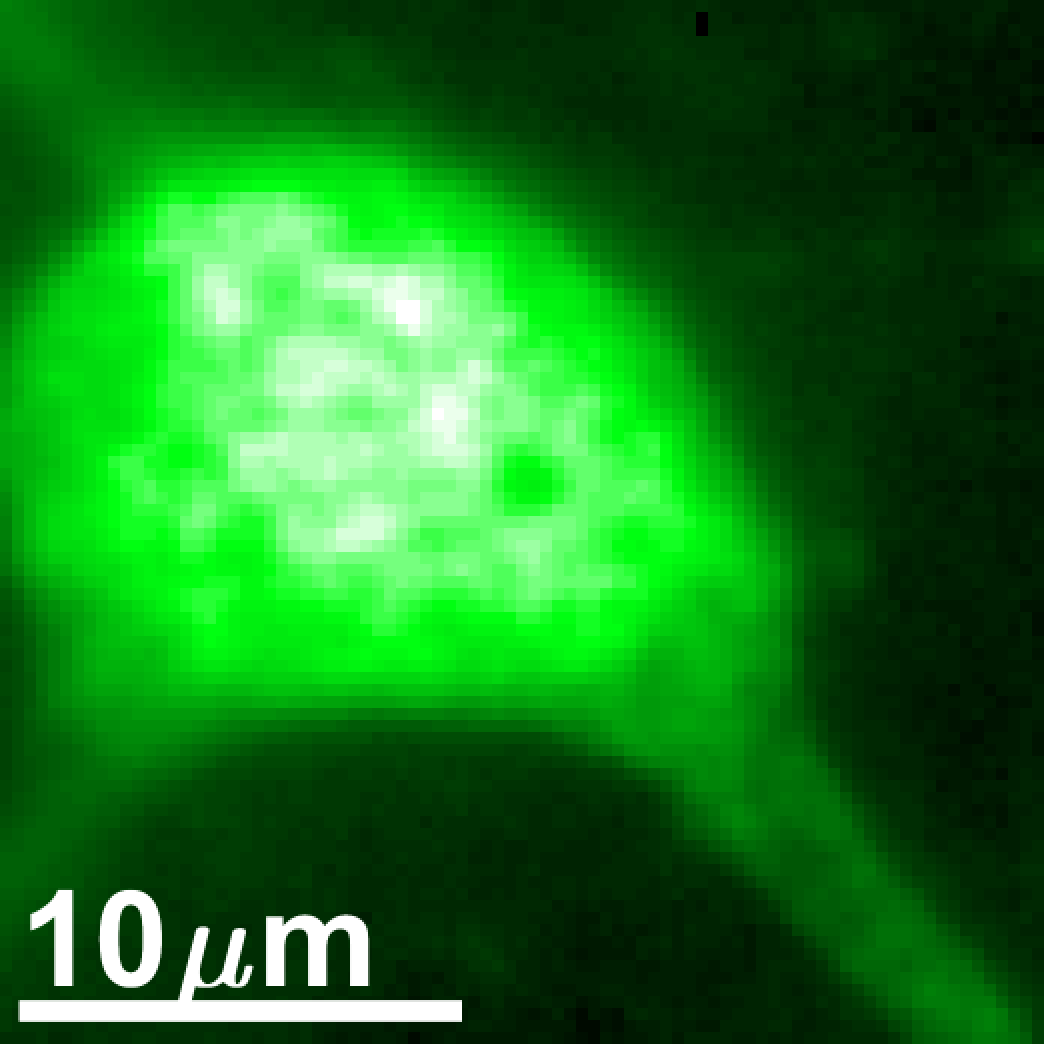}\\			
			
			\includegraphics[width= 0.18\textwidth]{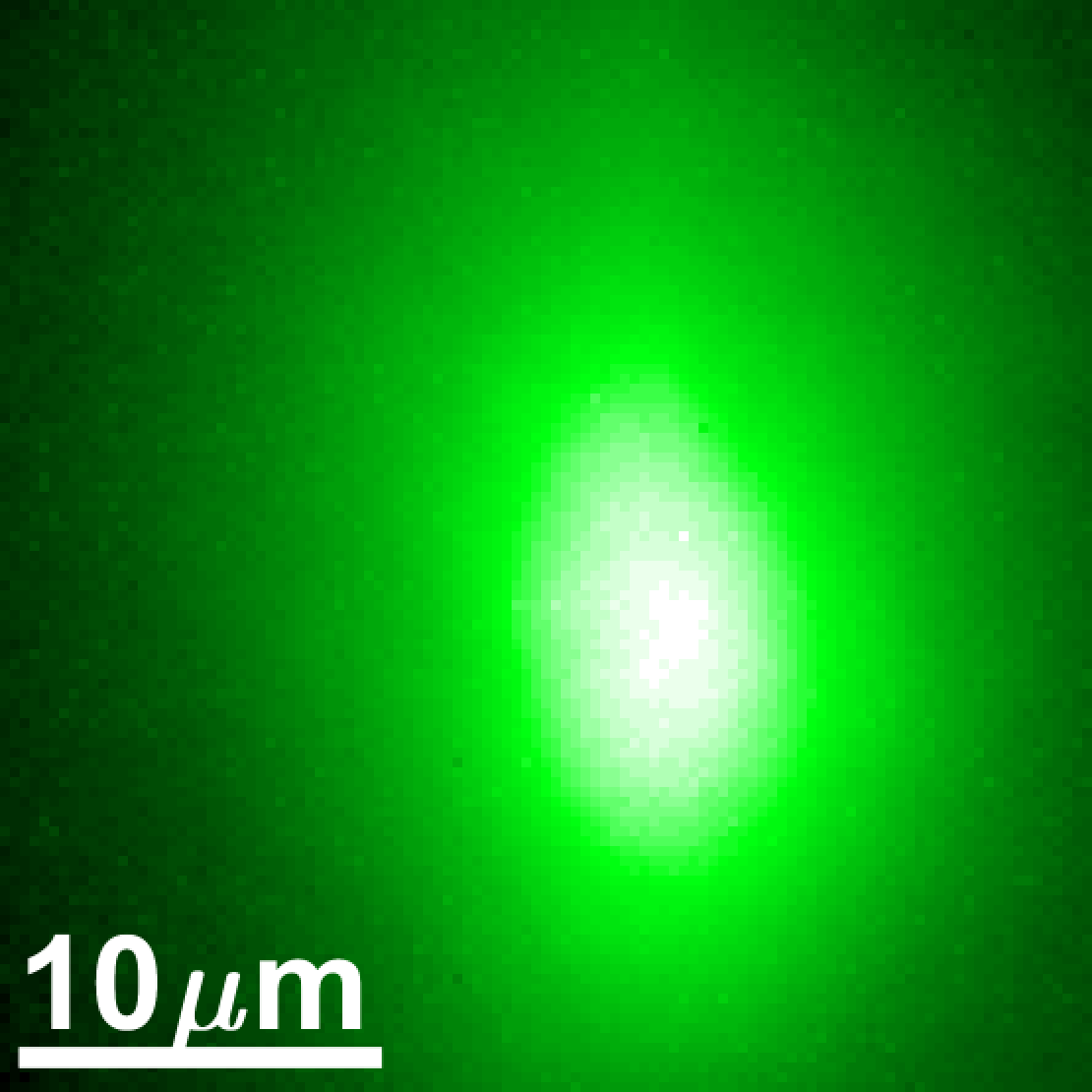}&
			\includegraphics[width= 0.18\textwidth]{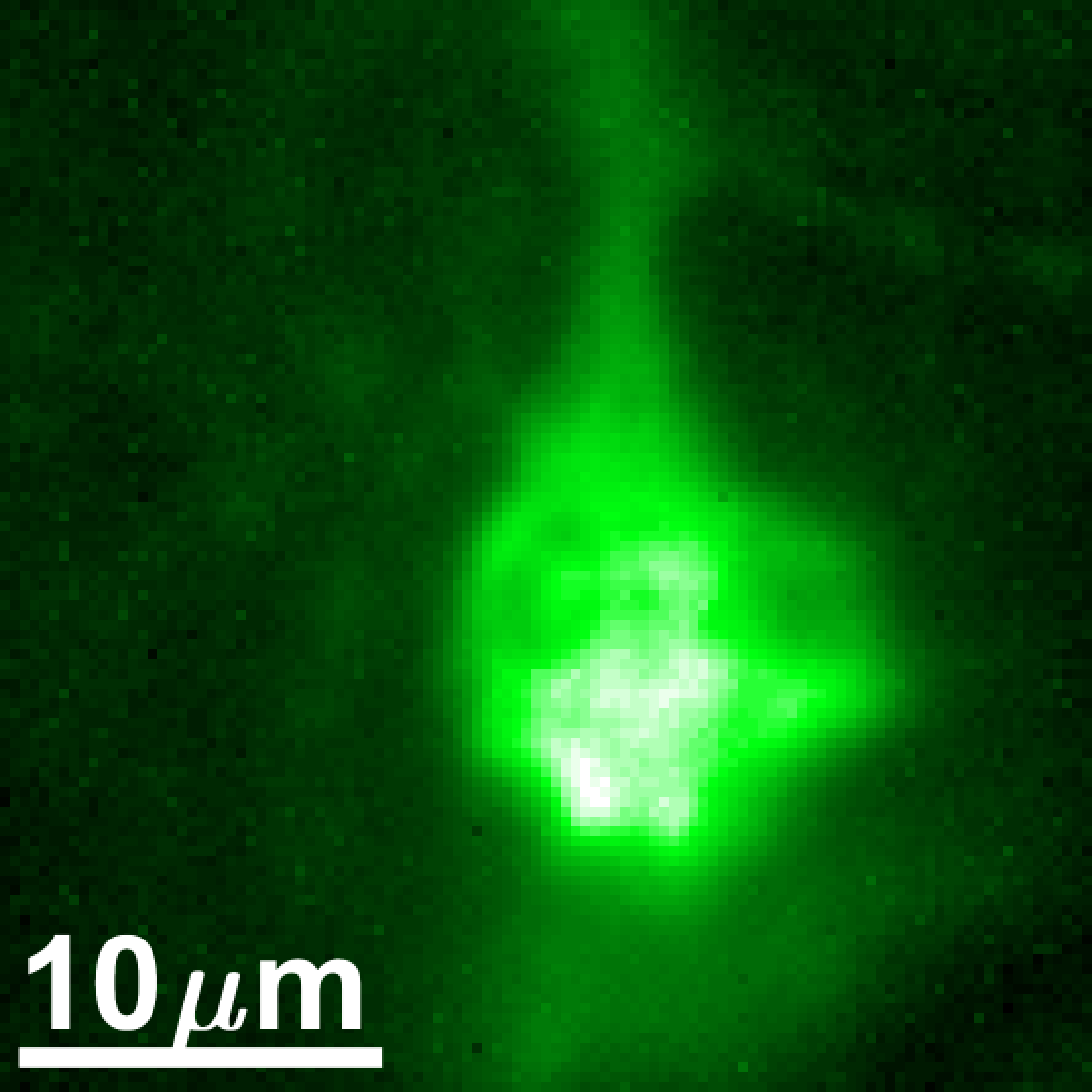}&
			\includegraphics[width= 0.18\textwidth]{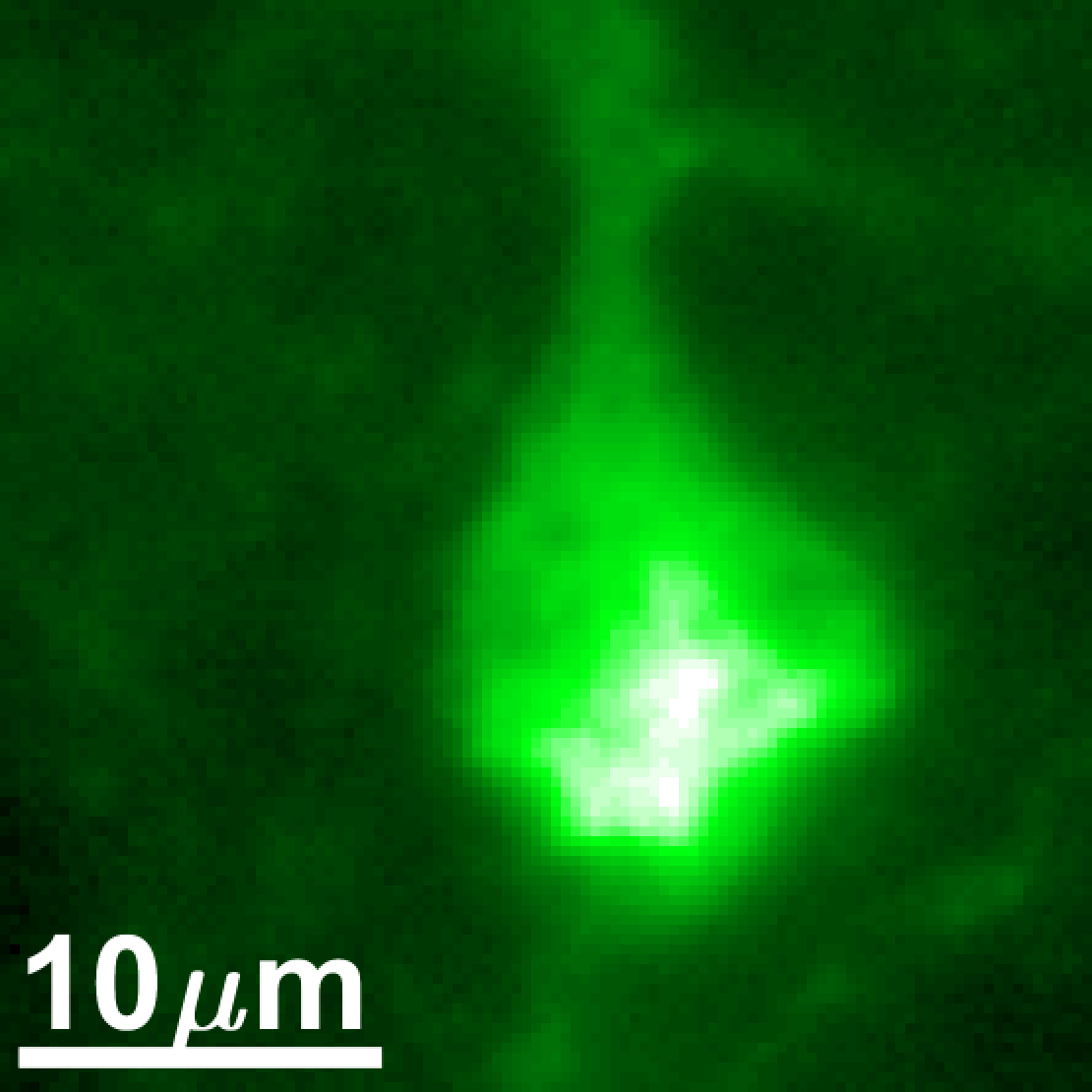}\\			
			
			\includegraphics[width= 0.18\textwidth]{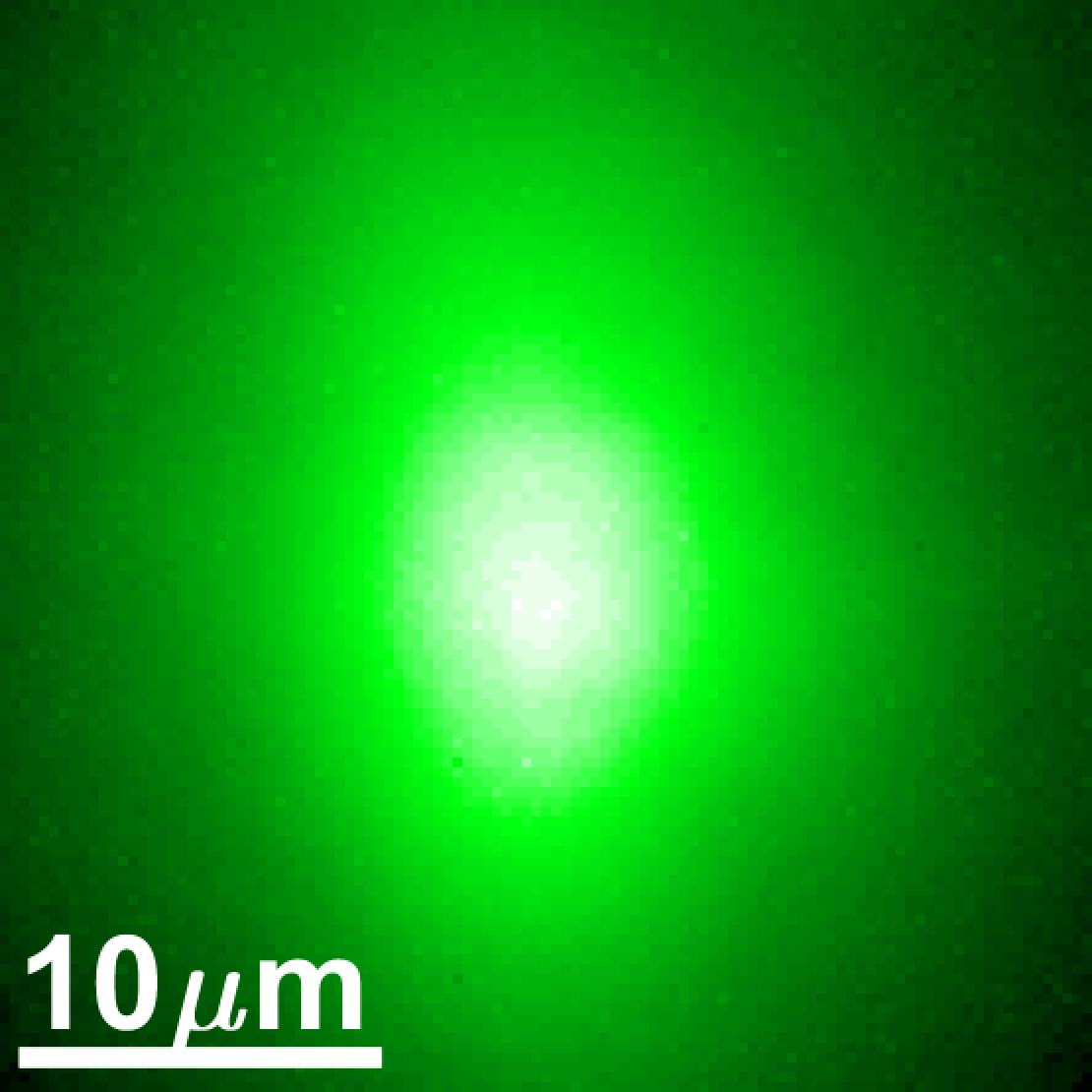}&
			\includegraphics[width= 0.18\textwidth]{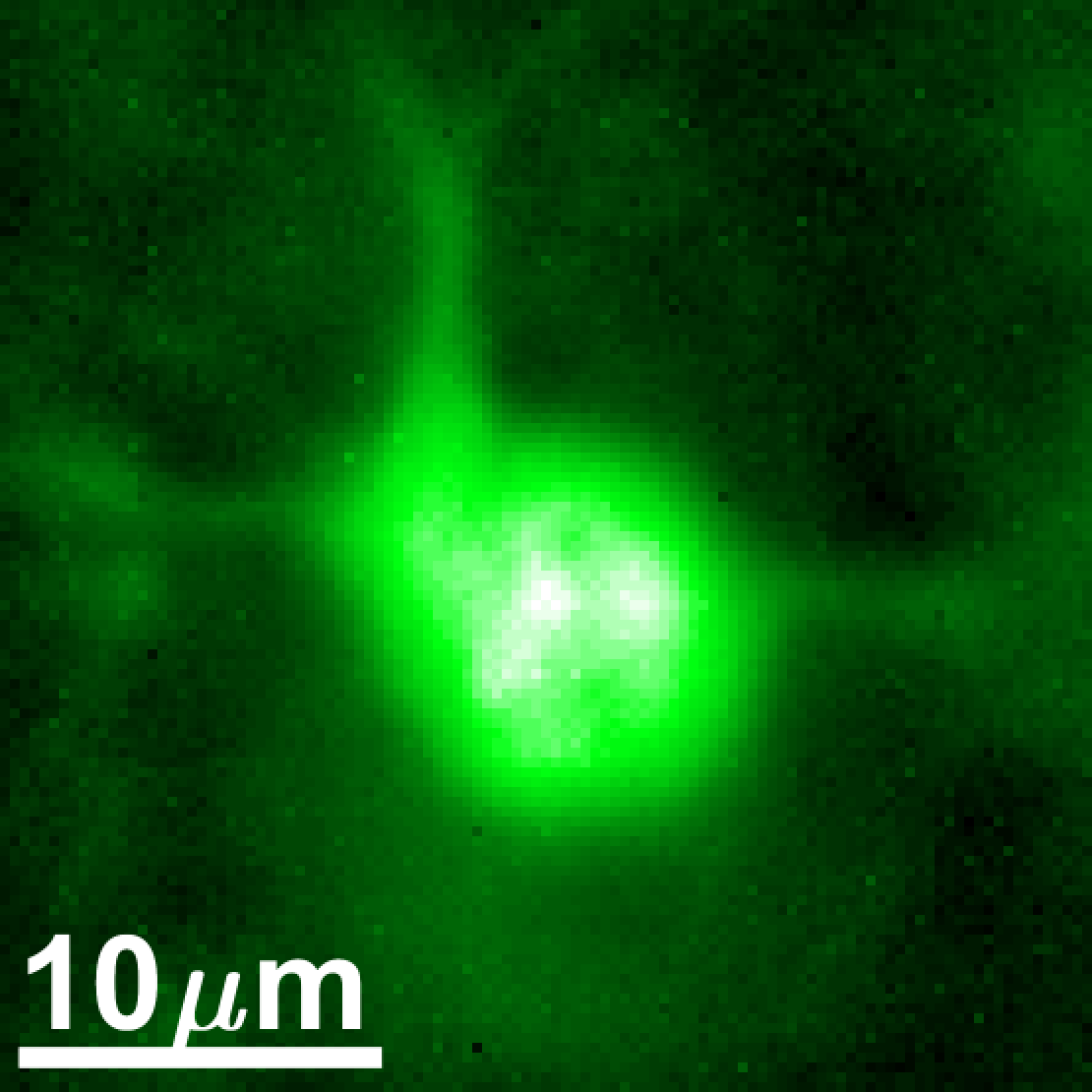}&
			\includegraphics[width= 0.18\textwidth]{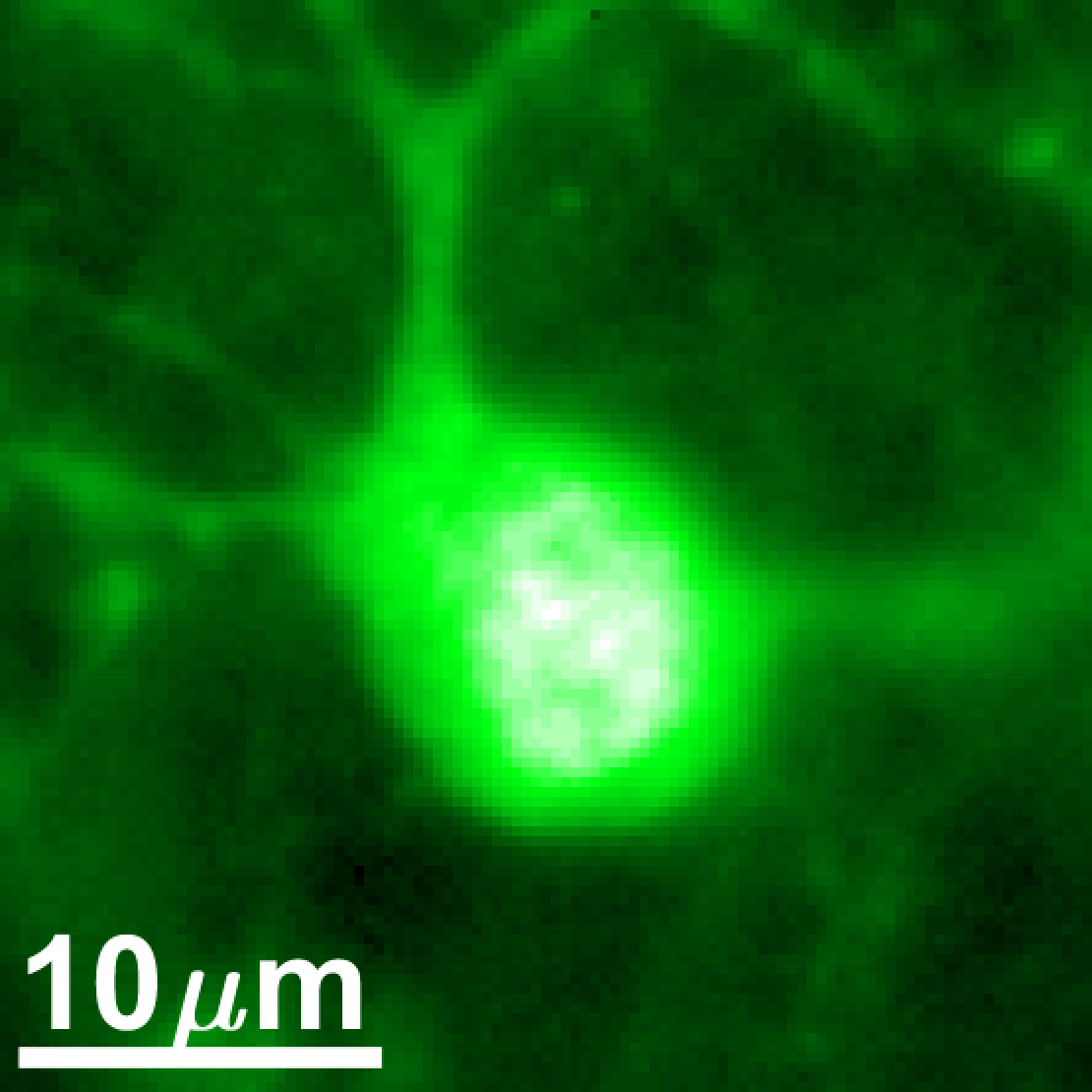}\\			
			
			\includegraphics[width= 0.18\textwidth]{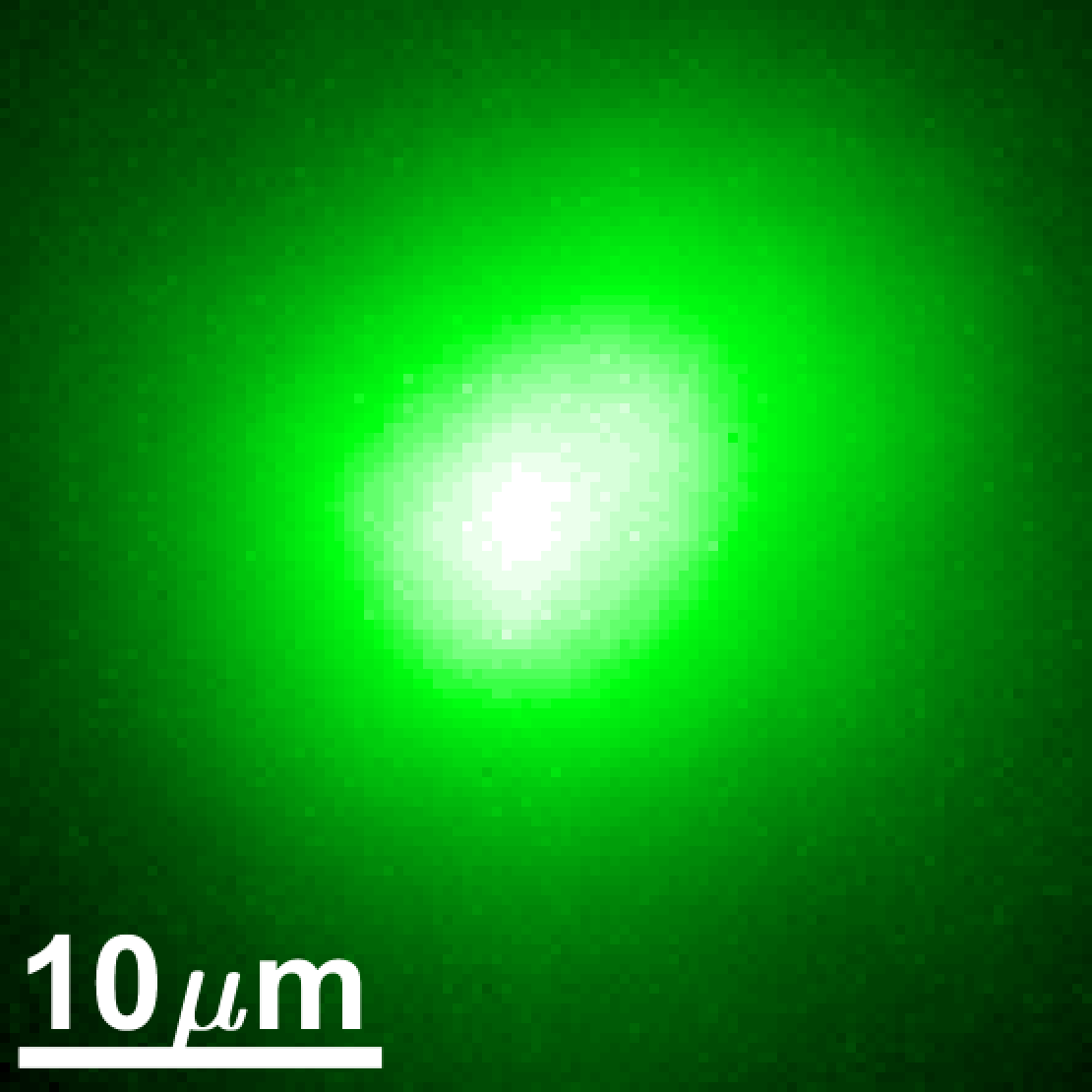}&
			\includegraphics[width= 0.18\textwidth]{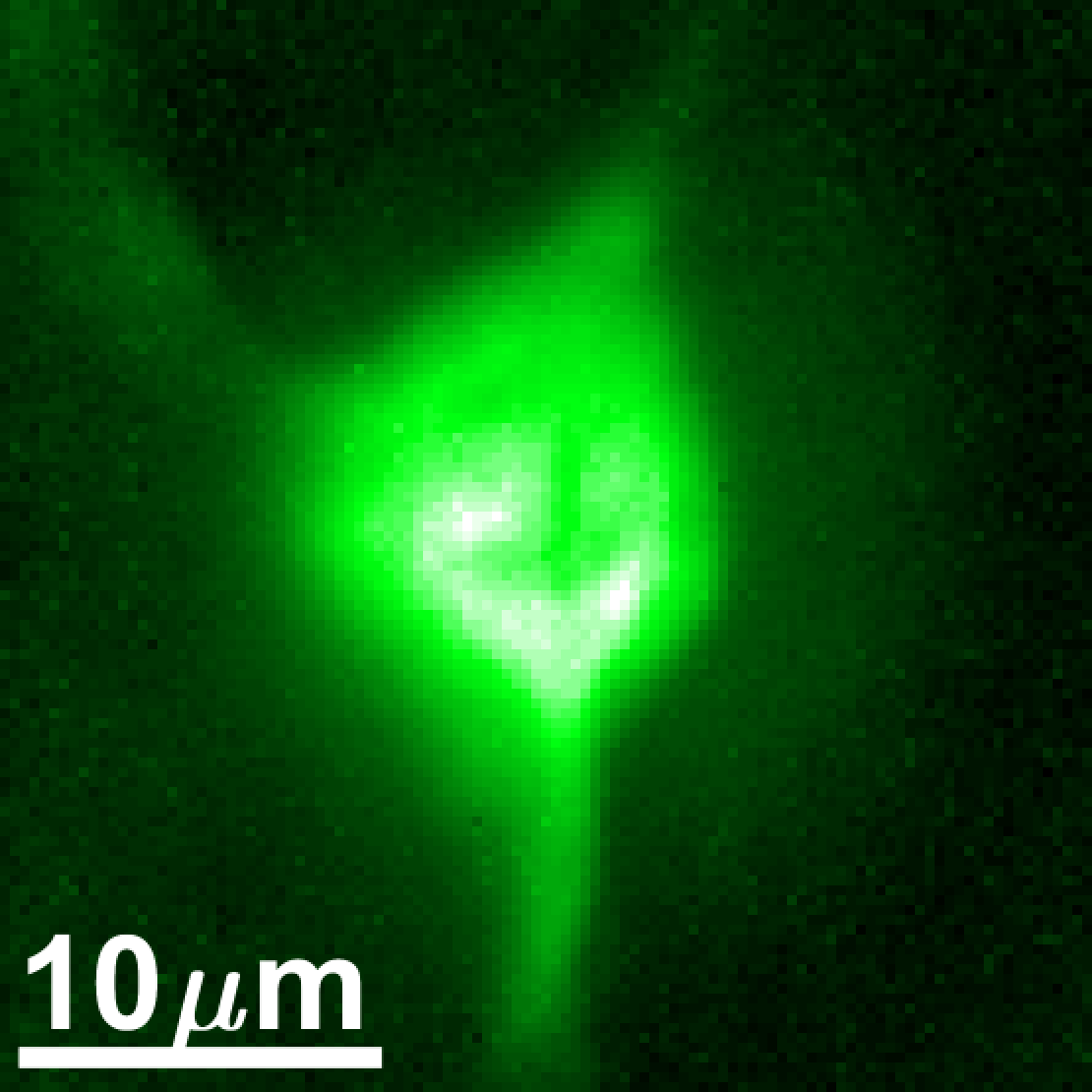}&
			\includegraphics[width= 0.18\textwidth]{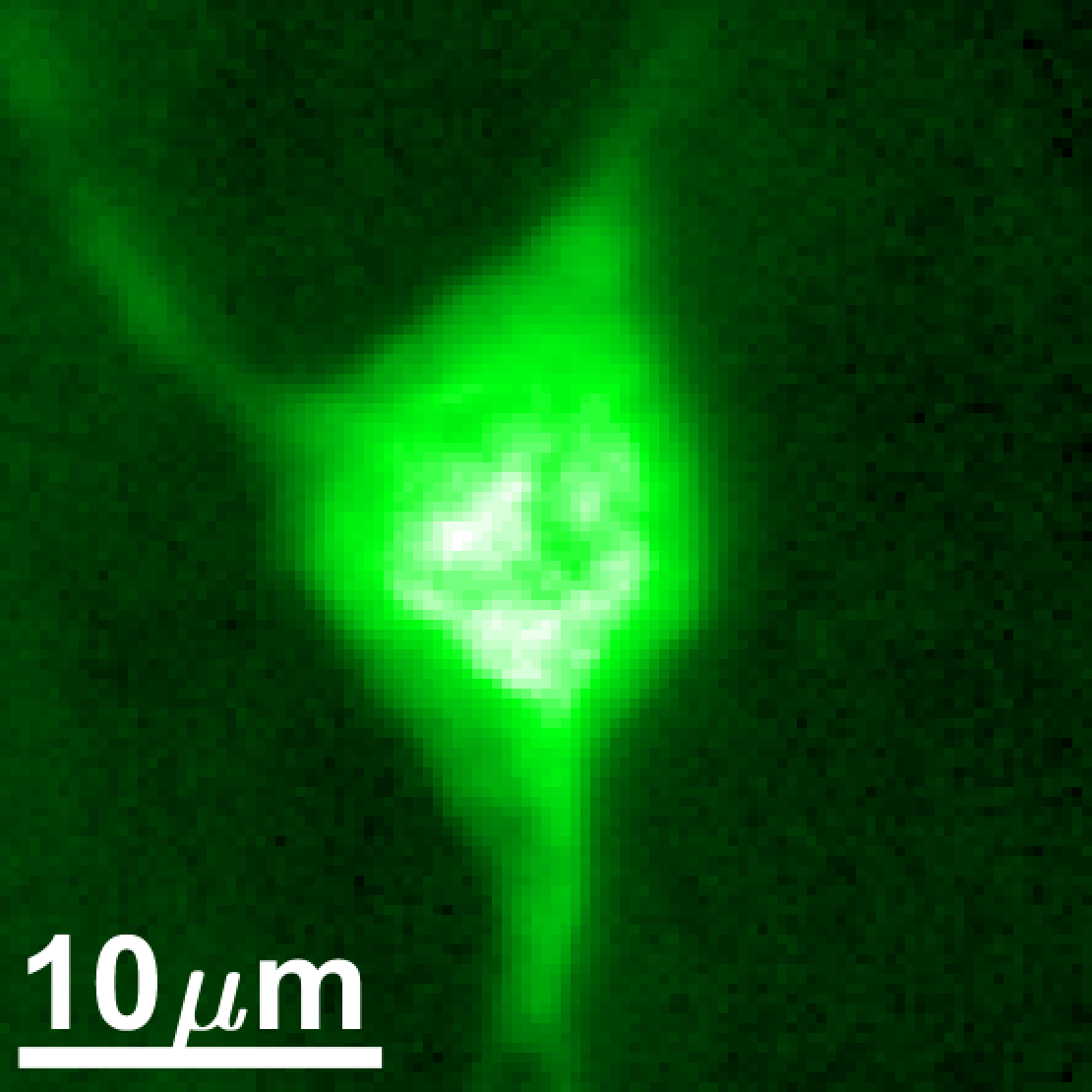}\\

			(a) Uncorrected  & (b) Corrected & (c) Reference  \\
			Main camera   & Main camera   & Validation camera             \\

		\end{tabular}
		\captionof{figure}
		{Additional results, a thin brain layer behind parafilm:  (a) Image of the neuron from the main camera with no correction, strong scattering is present and the neuron structure is lost. (b) Image with our modulation correction, the neuron shape as well as some of the axons are revealed. (c) An undistorted  image of the same neuron, from the validation camera.}
		\label{fig:big_area_sup_parafilm}
	\end{center}
\end{figure*}

\begin{figure*}[t!]
	\begin{center}
		\begin{tabular}{@{}c@{~~}c@{~~}c@{}}

			\includegraphics[height= 0.22\textwidth]{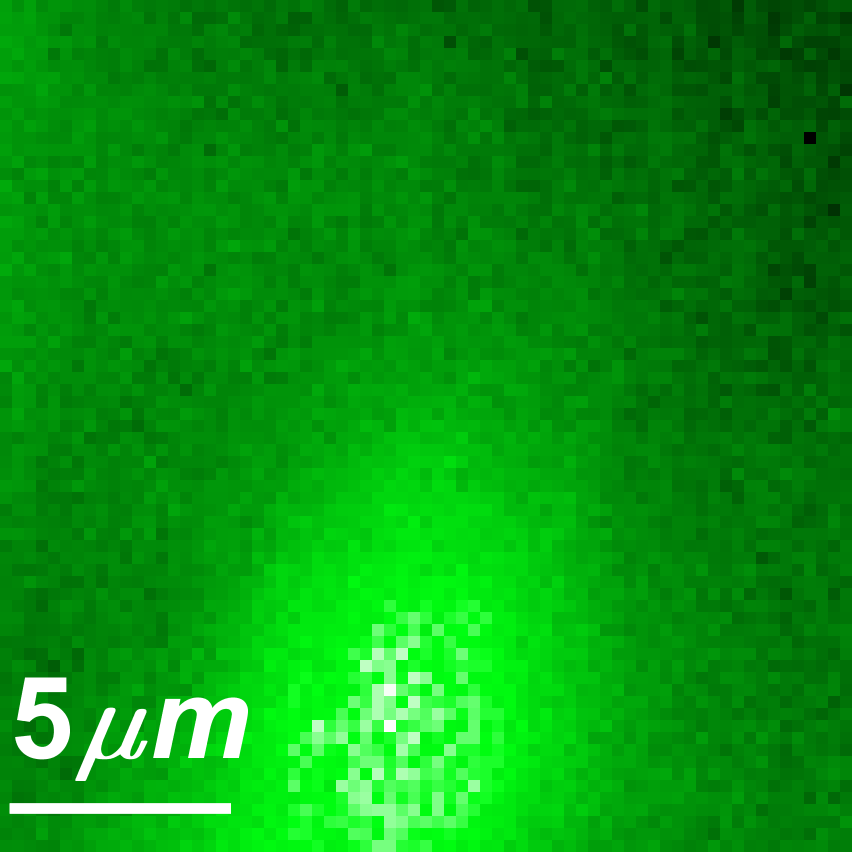}&
			\includegraphics[height= 0.22\textwidth]{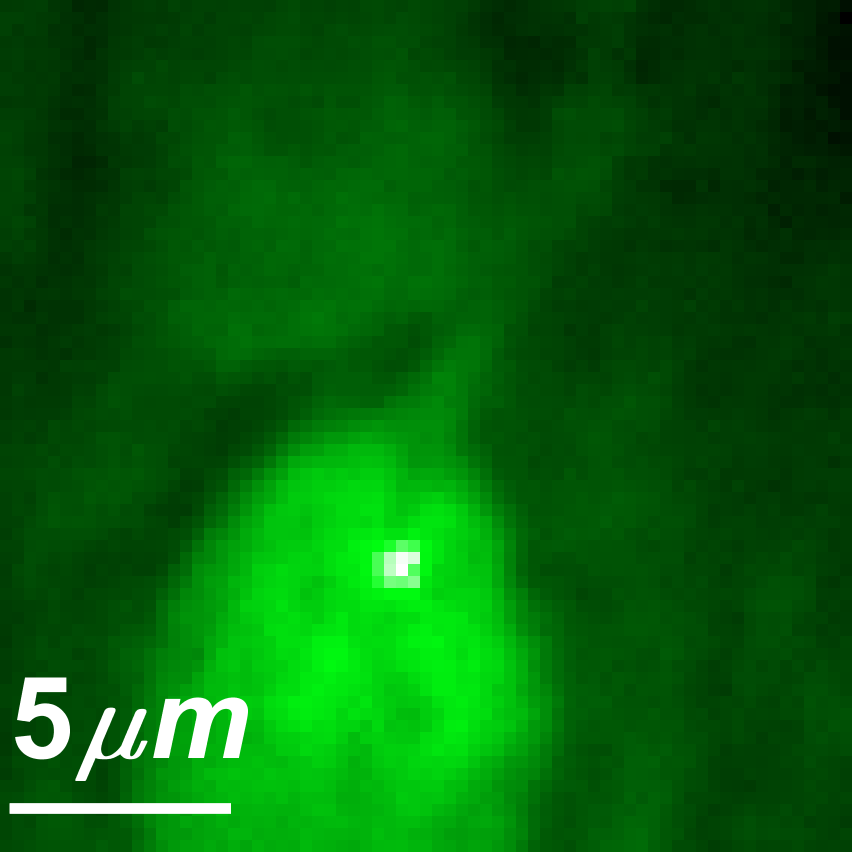}&
			\includegraphics[height= 0.22\textwidth]{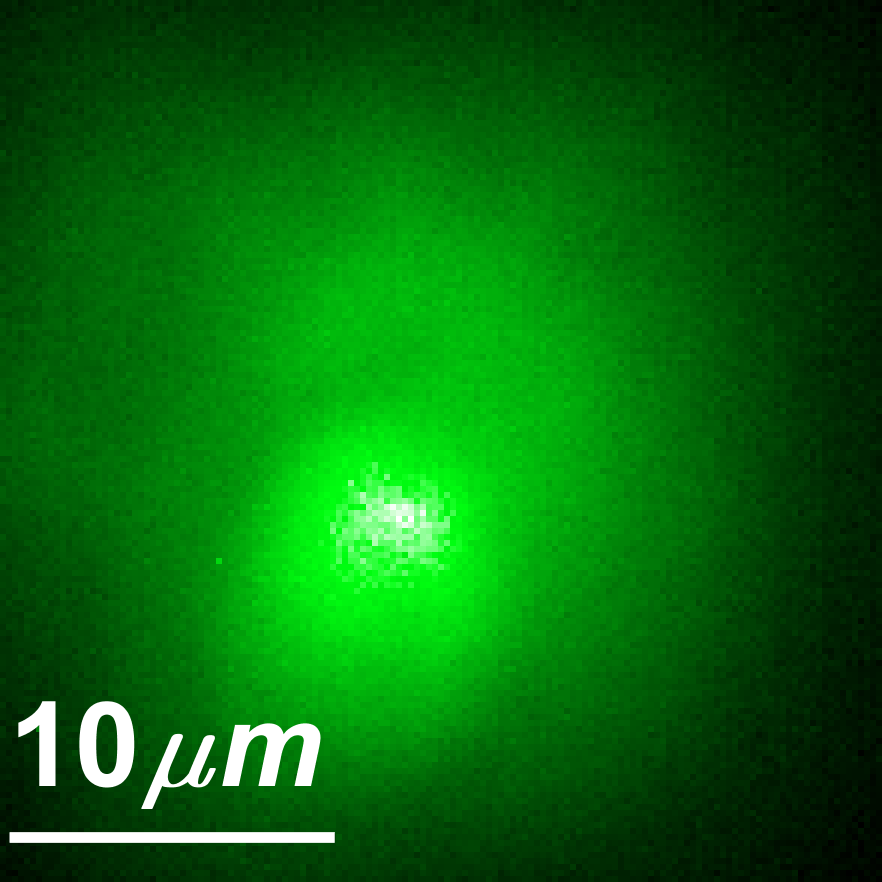}\\
			
			\includegraphics[height= 0.22\textwidth]{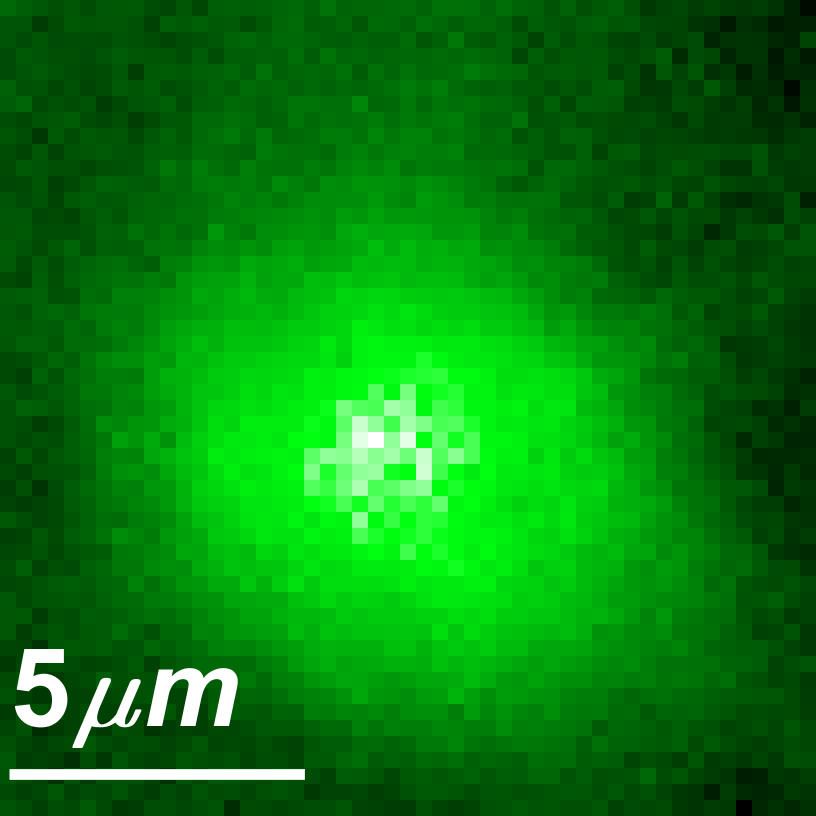}&
			\includegraphics[height= 0.22\textwidth]{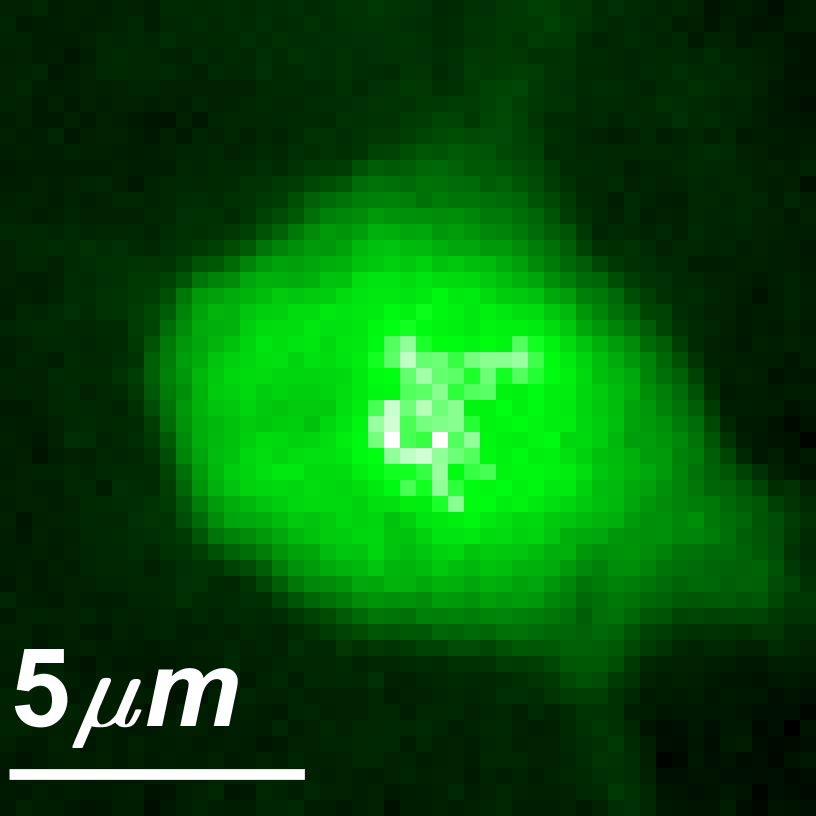}&
			\includegraphics[height= 0.22\textwidth]{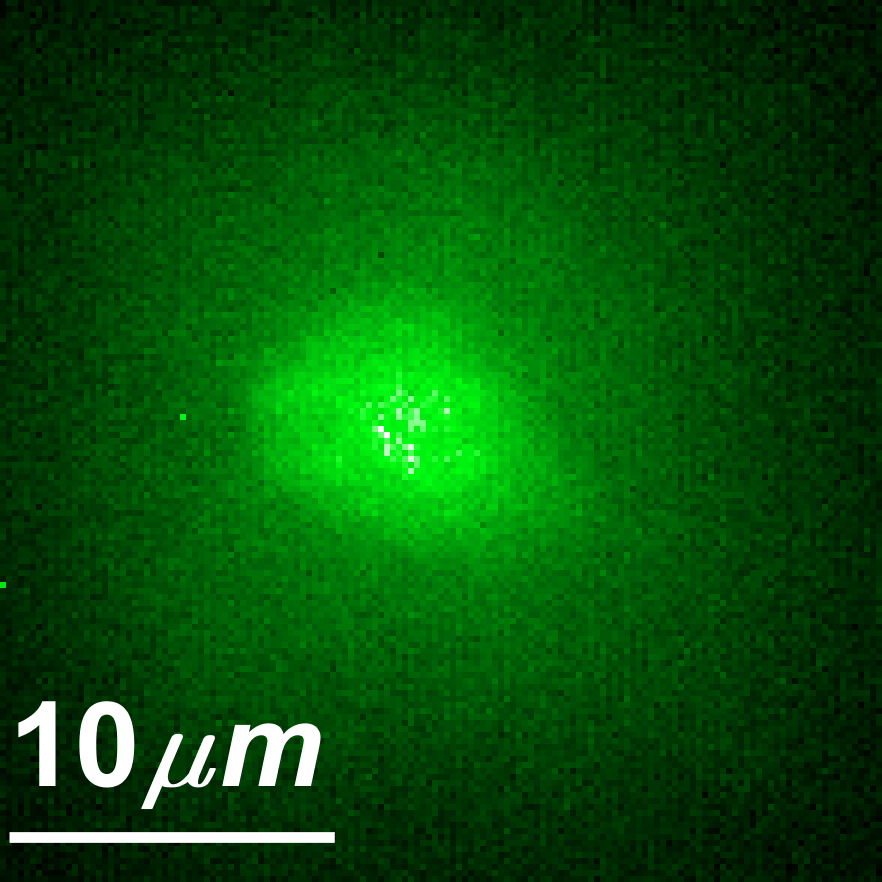}\\
			
			\includegraphics[height= 0.22\textwidth]{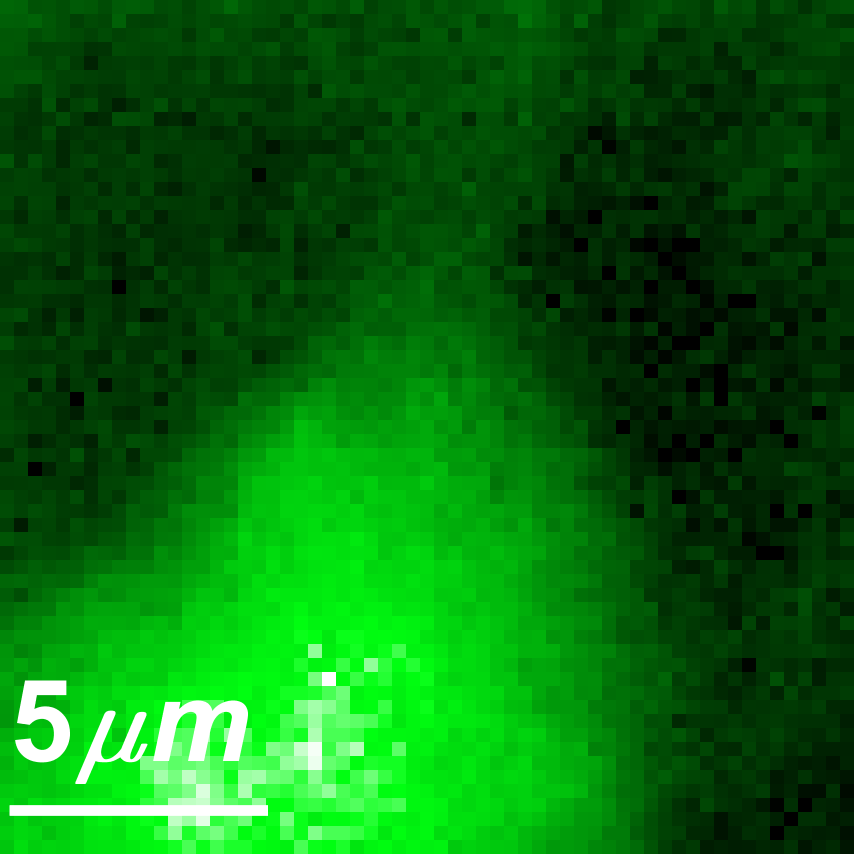}&
			\includegraphics[height= 0.22\textwidth]{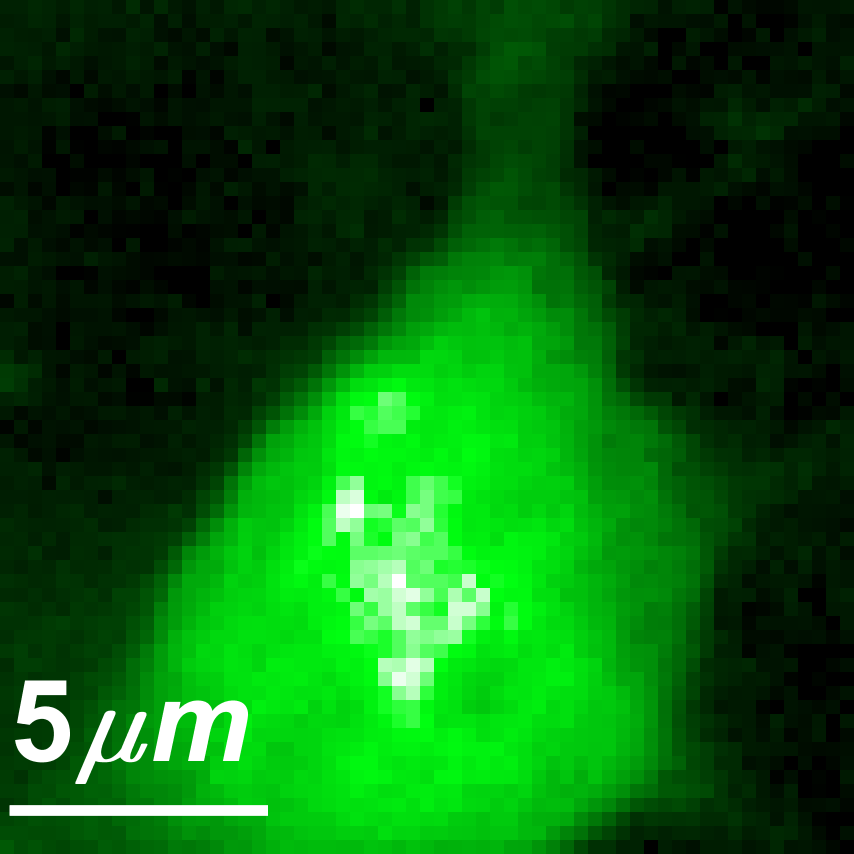}&
			\includegraphics[height= 0.22\textwidth]{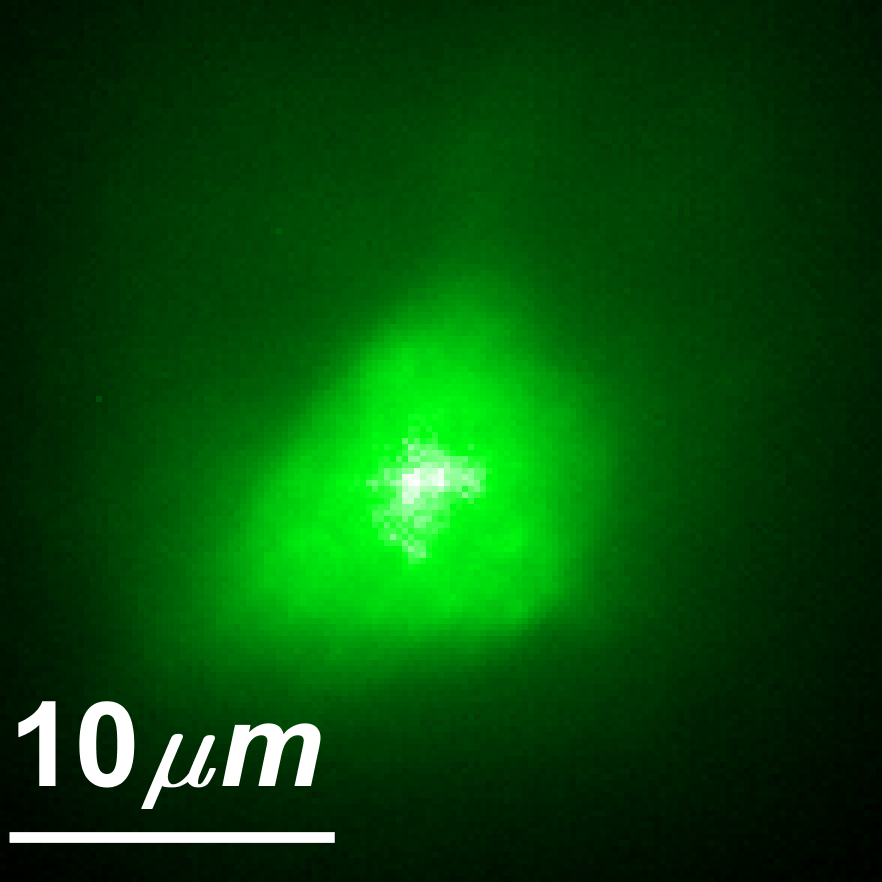}\\
		
			\includegraphics[height= 0.22\textwidth]{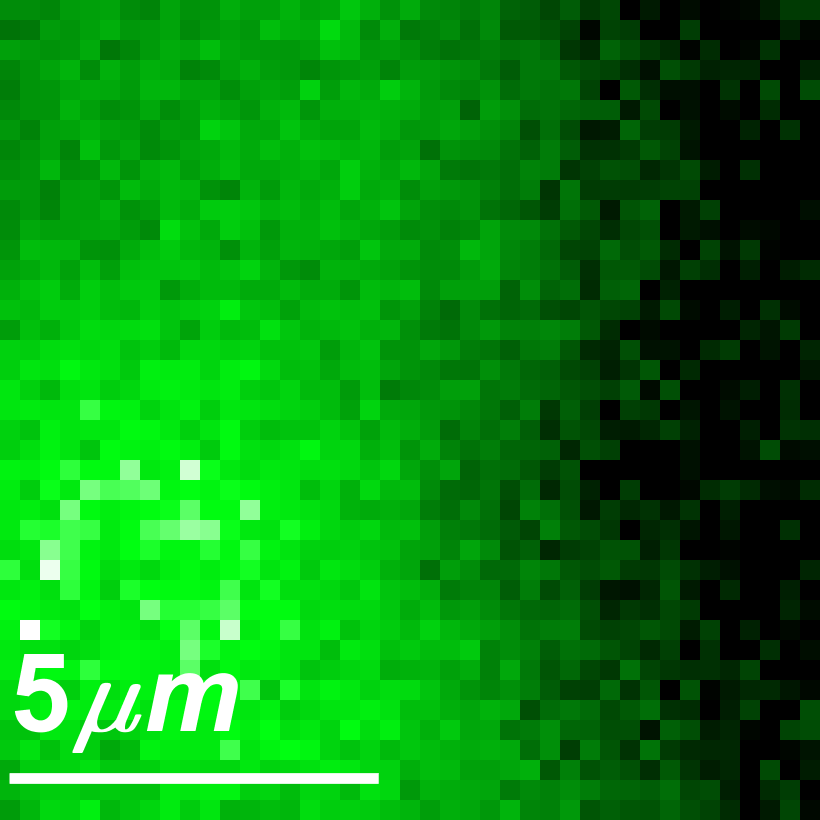}&	
			\includegraphics[height= 0.22\textwidth]{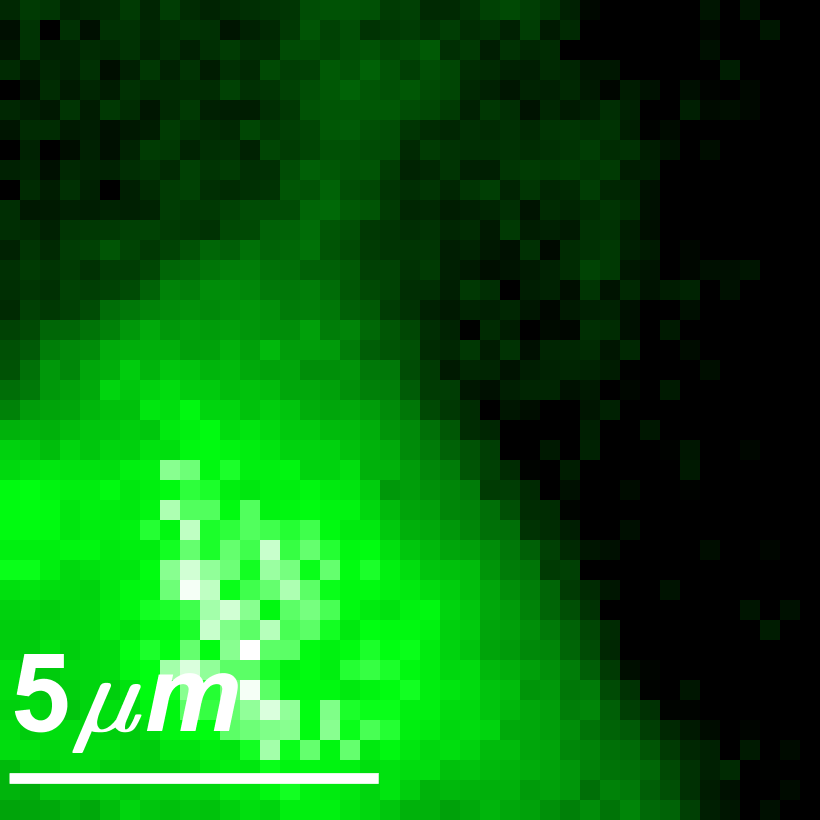}&
			\includegraphics[height= 0.22\textwidth]{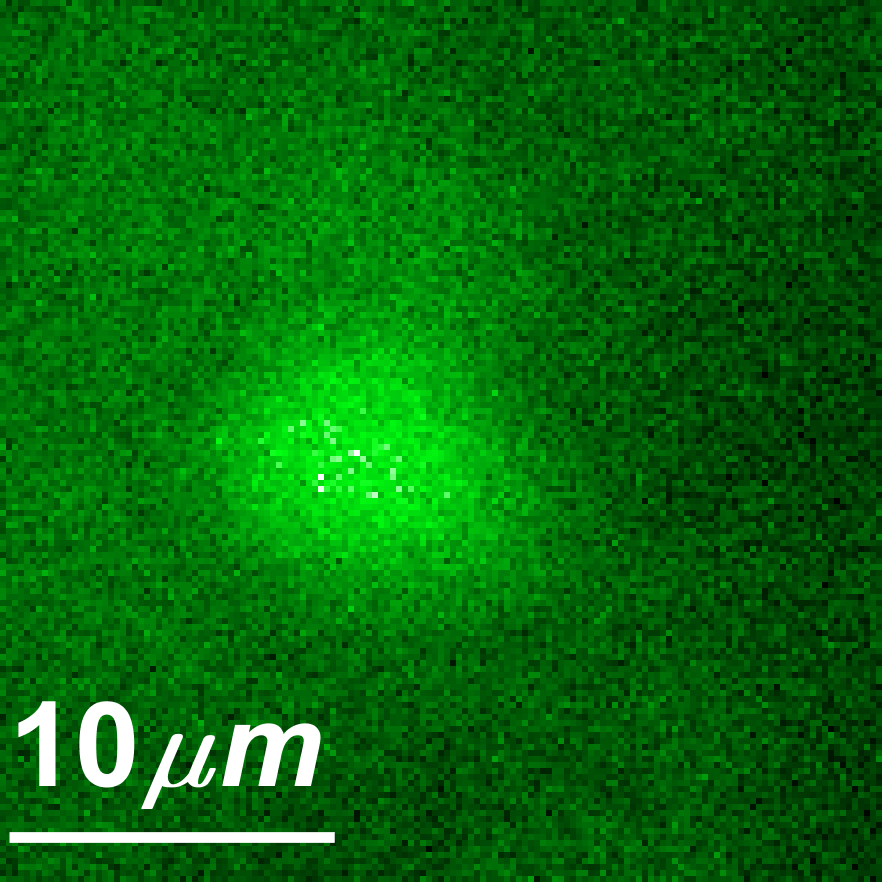}\\

			{\small (a) Uncorrected confocal } &	{\small  (b) Corrected confocal} & 	{\small (c) Reference}\\
			{\small Main camera} & 	{\small  Main camera }&	{\small   Validation camera}\\
		\end{tabular}
		\caption{\blue{Additional results: imaging through a $400\mu m$ thick fluorescent brain slice.   (a) A confocal image of the neuron from the main camera with no correction, strong scattering is present and the neuron structure is lost. (b) A confocal image with our modulation correction, the neuron shape as well as some of the axons are revealed. (c) A  reference  image of the same neuron, from the validation camera. Due to the 3D spreading of the fluorescent components, the validation camera cannot always capture an aberration-free image of the target. }}\label{fig:confocal_images_supp}
	\end{center}
\end{figure*}

\begin{figure*}[t!]
	\begin{center}
		\begin{tabular}{@{}c@{~~~~~~~~~~~~~~~~~~}c}

			\includegraphics[height= 0.3\textwidth]{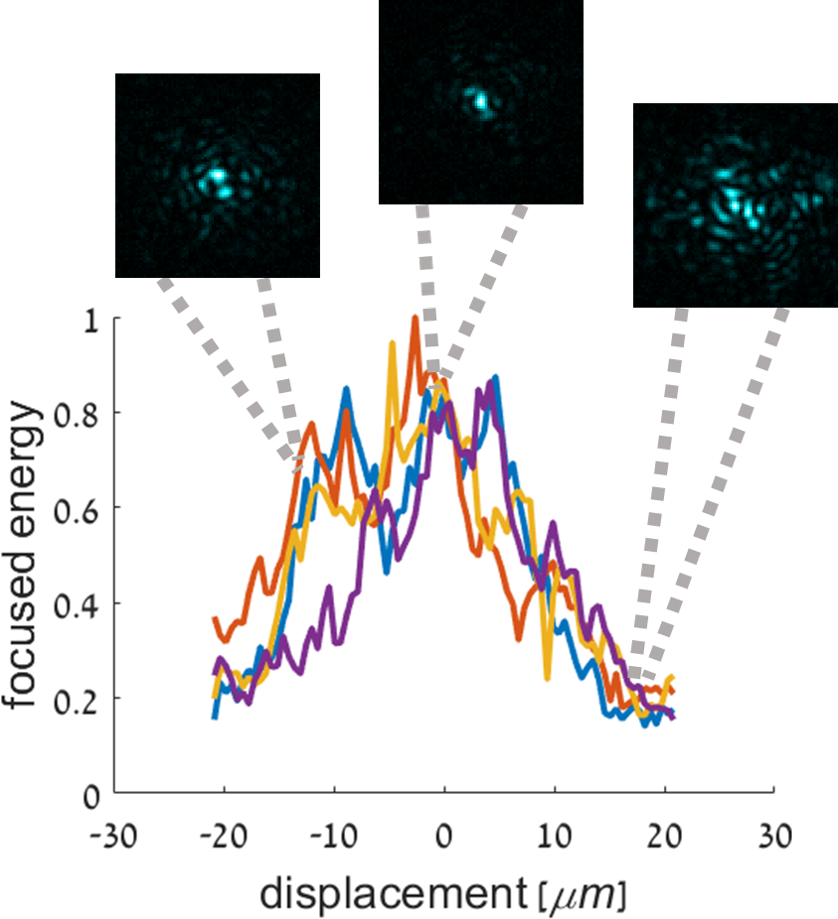}&
			\includegraphics[height= 0.3\textwidth]{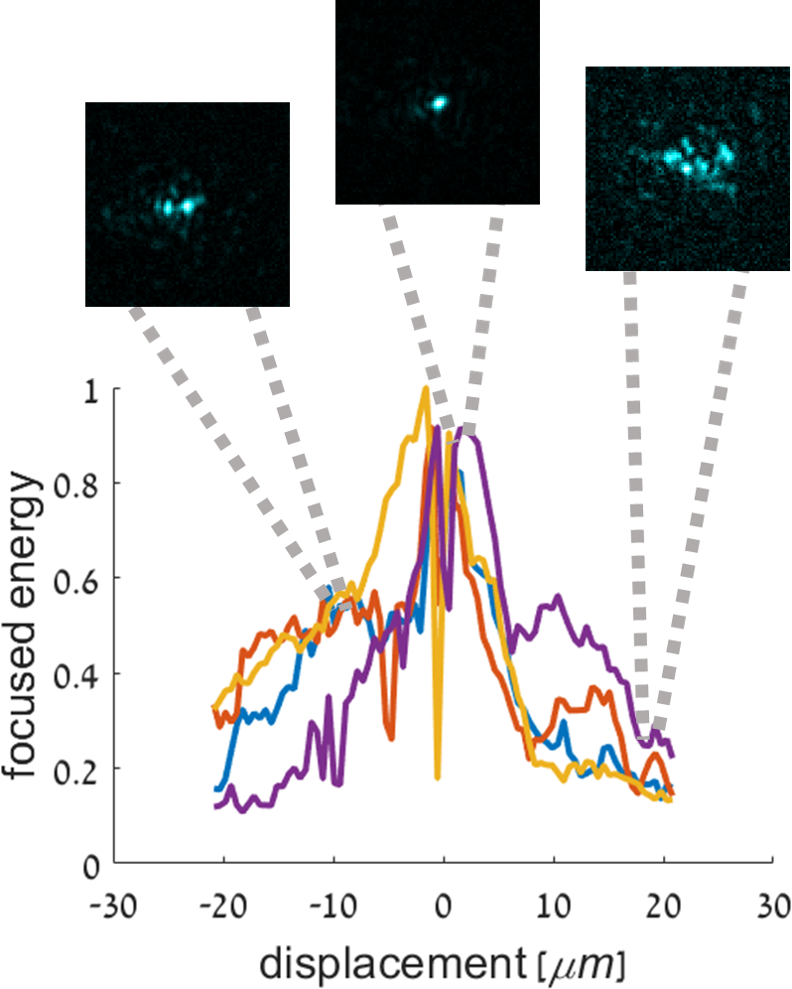}\\
		\end{tabular}

		\caption{\blue{Extent of memory effect correlation: we plot the  decay of  memory effect correlation in our samples, by translating  a recovered modulation to nearby points and measuring the energy of the focused spot behind the tissue, at the validation camera. Insets show sample shapes of the focused spots.  As translation is larger, more speckles arise around the desired focused spot.   Left: focusing through a $400\mu m$ brain slice. Right: focusing through a layer of parafilm. We plot 4 different line scans to demonstrate the variance of such curves.}}\label{fig:ME-decay}

	\end{center}
\end{figure*}

\section{Additional results}

\blue{In \figref{fig:converging2} we show additional images of the focused spot achieved with our modulation. In particular, in the last row we show one failure example where the algorithm has converged on two spots rather than one. }

Additional results imaging a thin brain layer with neurons behind chicken breast or parafilm layers are presented in \figpref{fig:big_area_sup_chicken}{fig:big_area_sup_parafilm}. While the modulation we recover focuses at a single spot, due to the memory effect we can use it to image an area behind the tissue rather than a single spot. In \secref{sec:tilt-shift} below we explain the tilt-shift adjustments required.

\blue{In \figref{fig:confocal_images_supp} we show additional results imaging through a thick $400\mu m$ brain slice. Since the fluorescent target has 3D variation, we use a confocal scanning to isolate a neuron at a single depth. A confocal scanning of our modulation is significantly better than an uncorrected confocal scan.
Also due to the 3D structure it is not always possible to capture a good full-frame reference from the validation camera. }

\blue{In \figref{fig:ME-decay} we test the actual extent of the memory effect for such samples. We run the algorithm until convergence and then start to tilt-shift the modulation so that the focal spot translates along a line at the back of the tissue. We capture the intensity of the translating spot from the validation camera and plot it as a function of distance from the original focal spot. To demonstrate the variance of such curves we plot 4 curves shifting the point at 4 different directions. We include two examples, one focusing through a $400\mu m$ brain slice and one  focusing through a layer of parafilm. We include a few insets visualizing how speckles around the focal spot increase as it is translates.}

\begin{figure*}[t!]
	 \begin{adjustwidth}{-0.7cm}{}
				
		\begin{tabular}{@{~~~~~~}c@{~~}c@{~}c@{~~}c@{~}c@{~~}c@{~}c@{~~}c@{~}c@{~~}c@{~~}c@{}}
	
			\small{(a) Score}& \multicolumn{2}{l}{\!\!\!\!\small{(b)Init main}}	&	\small{(c)2 modul}&&	\small{(d)1 modul}&&\multicolumn{2}{l}{\small{\!\!\!\!(e)1 modul mid}}&\small{(f)Val. init}& \small{(g)Val. fin}\\
			
			\includegraphics[height= 0.10\textwidth]{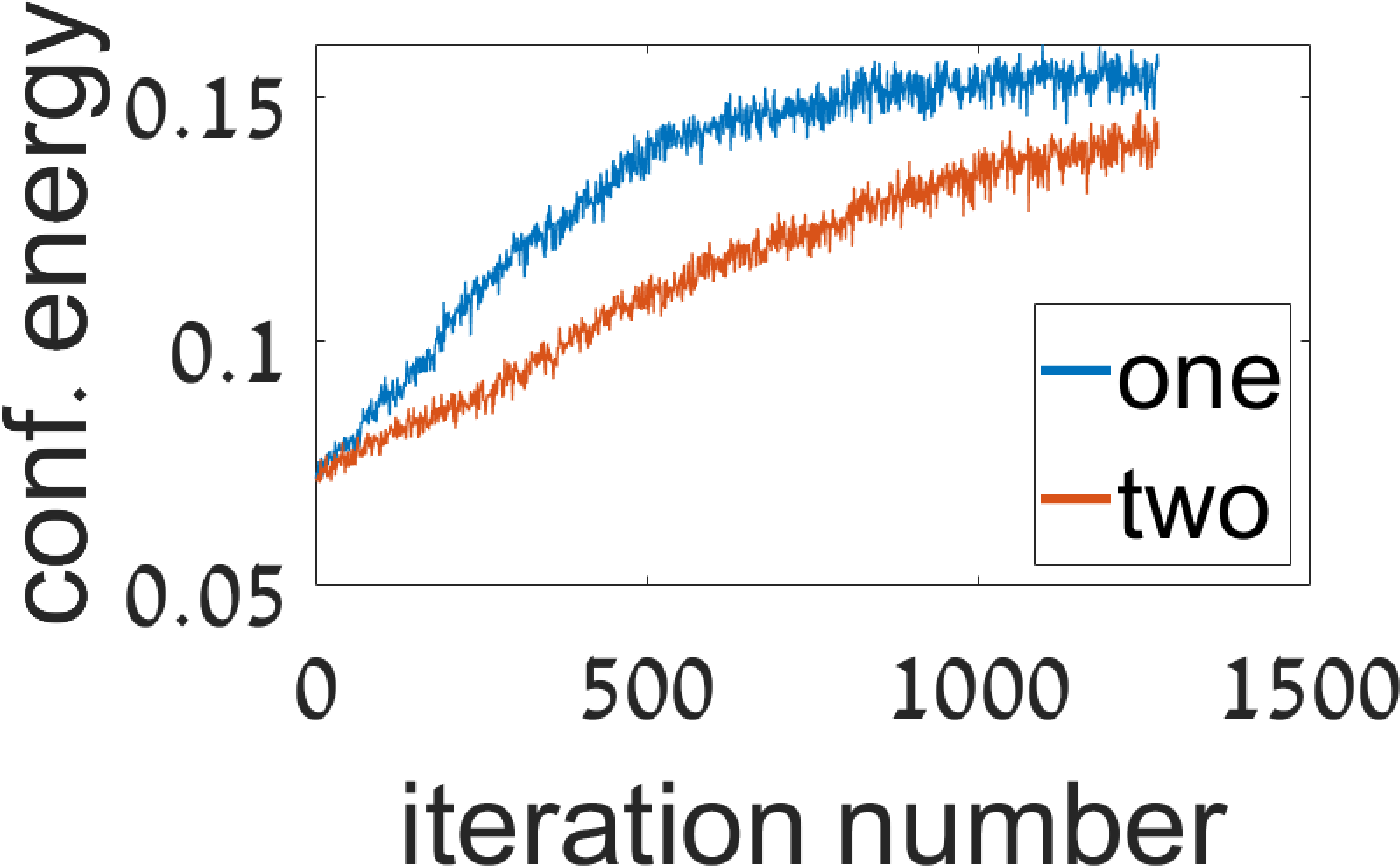}&
			\includegraphics[height= 0.10\textwidth]{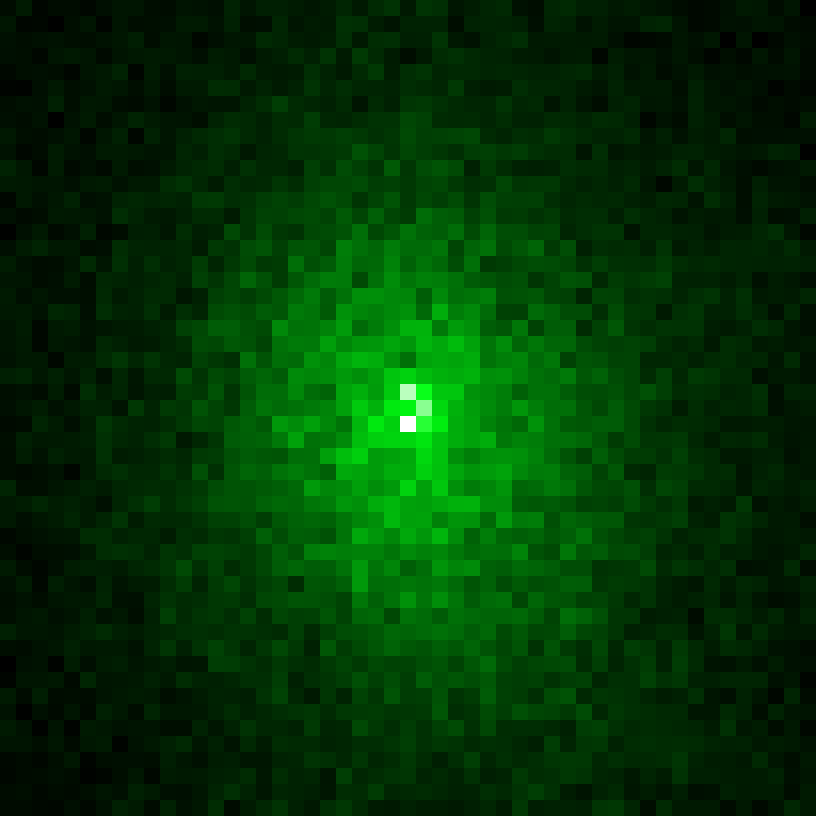}&
			\includegraphics[height= 0.10\textwidth]{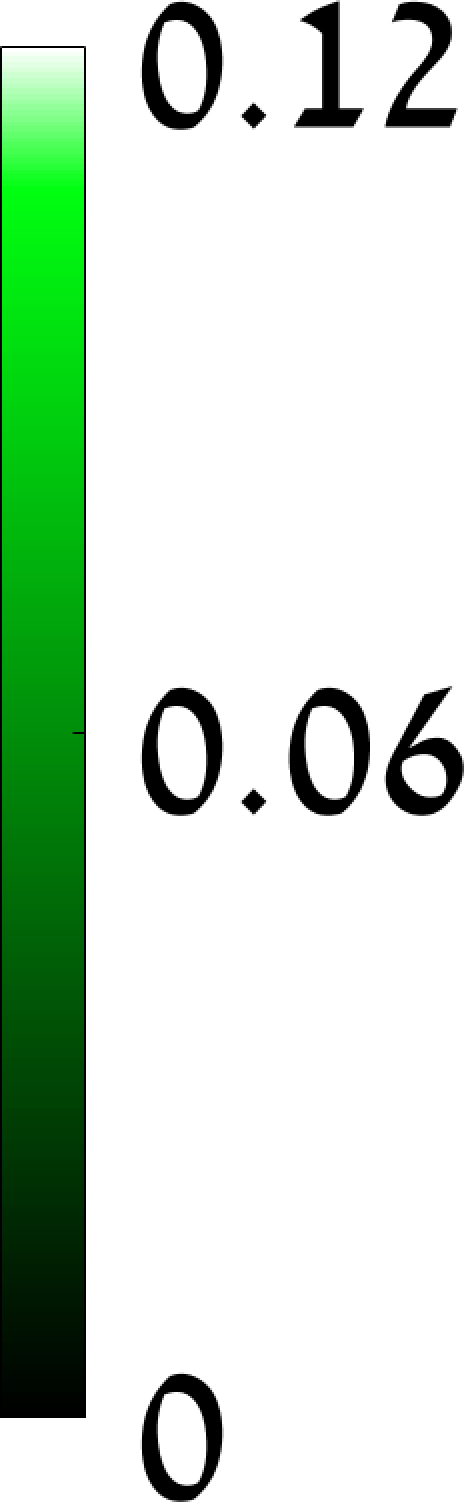}&		
			\includegraphics[height= 0.10\textwidth]{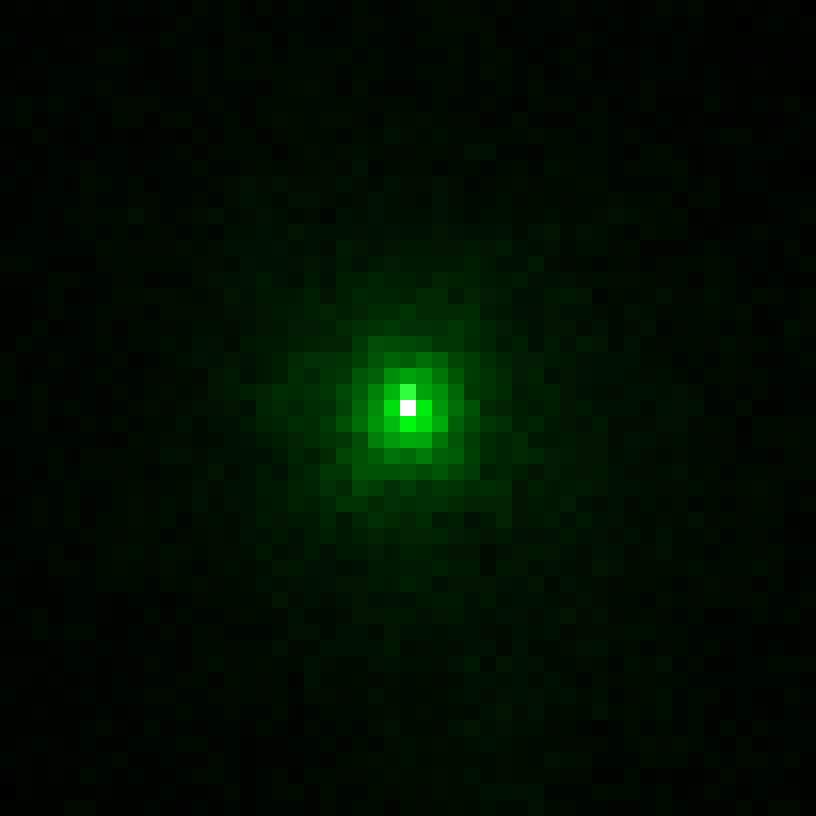}&		
			\includegraphics[height= 0.10\textwidth]{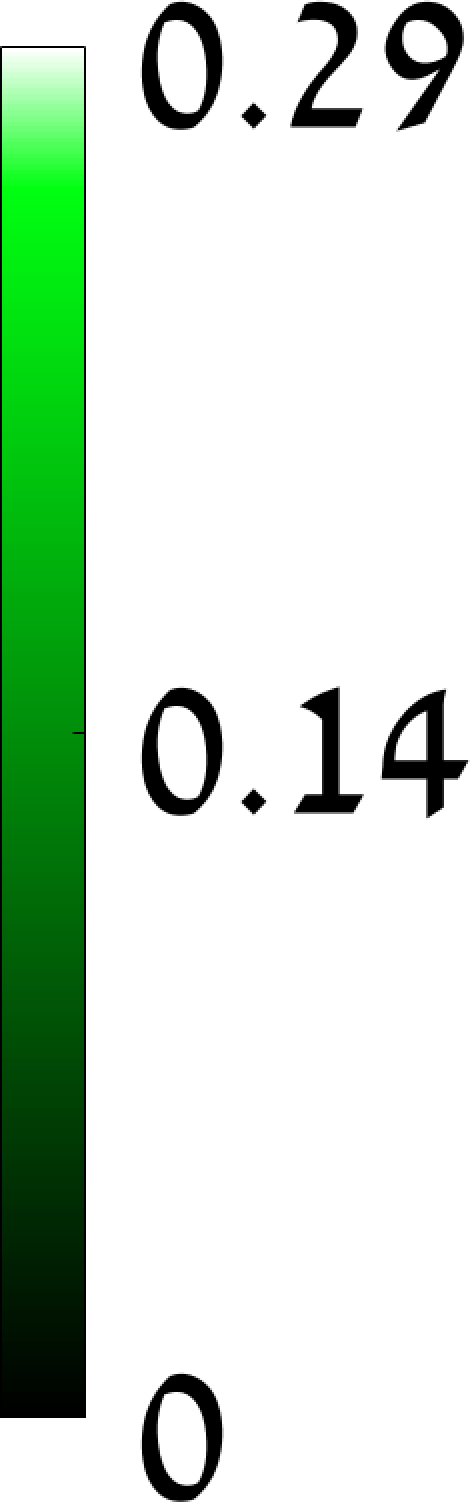}&		
			\includegraphics[height= 0.10\textwidth]{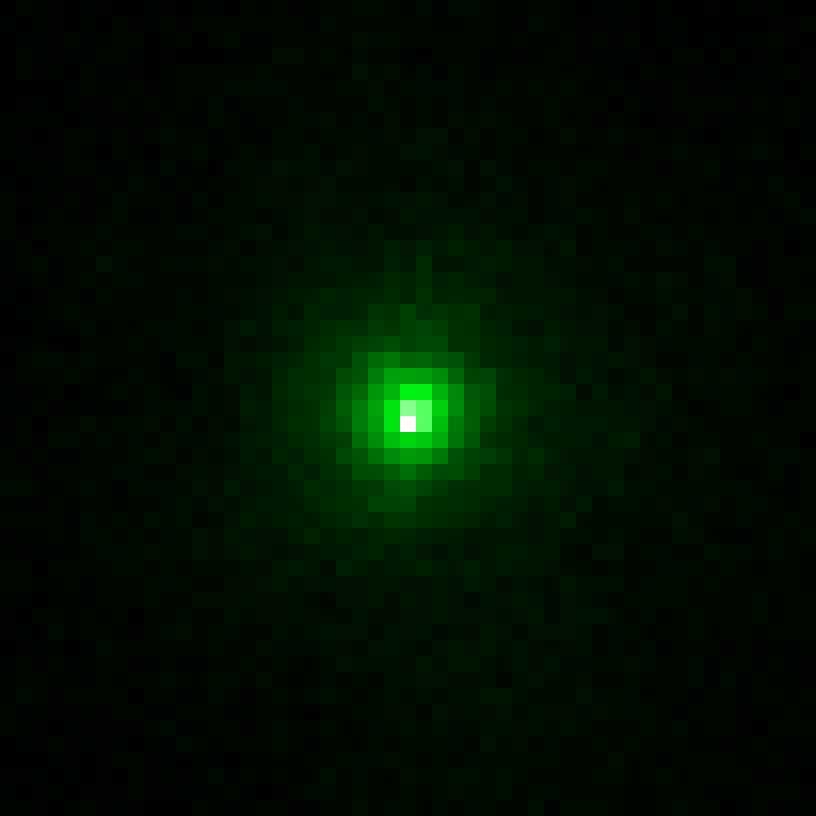}&		
			\includegraphics[height= 0.10\textwidth]{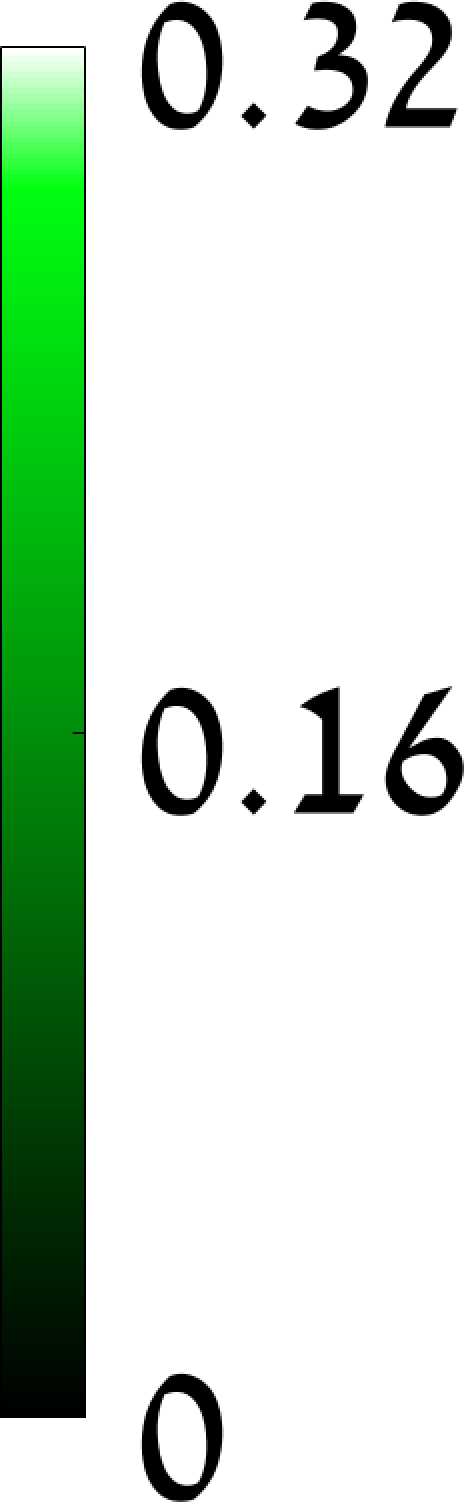}&
			\includegraphics[height= 0.10\textwidth]{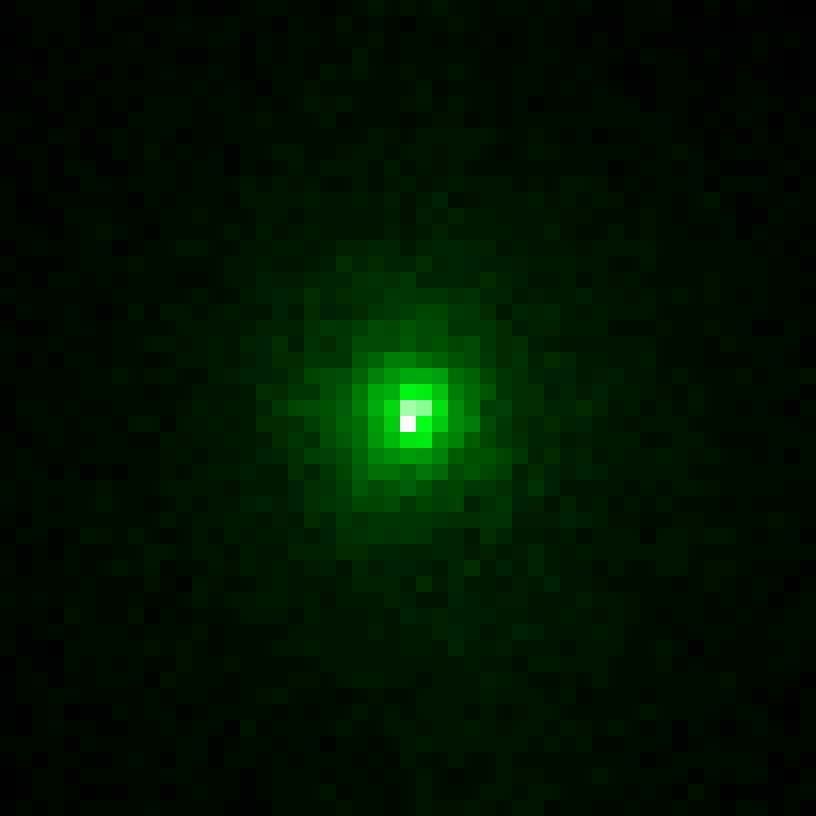}&
			\includegraphics[height= 0.10\textwidth]{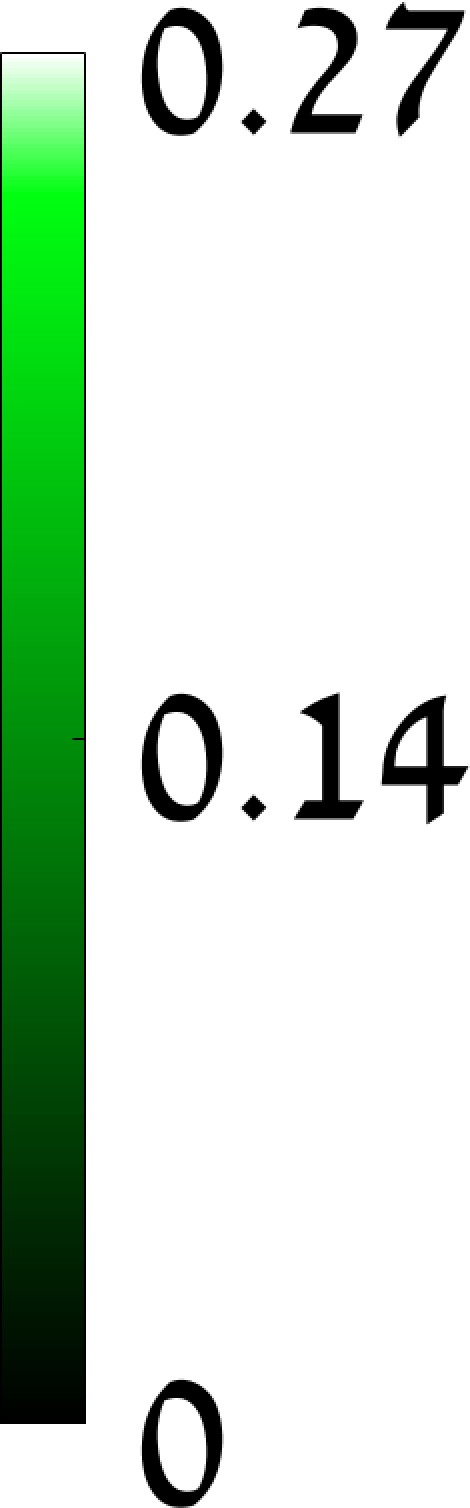}&
			\includegraphics[height= 0.10\textwidth]{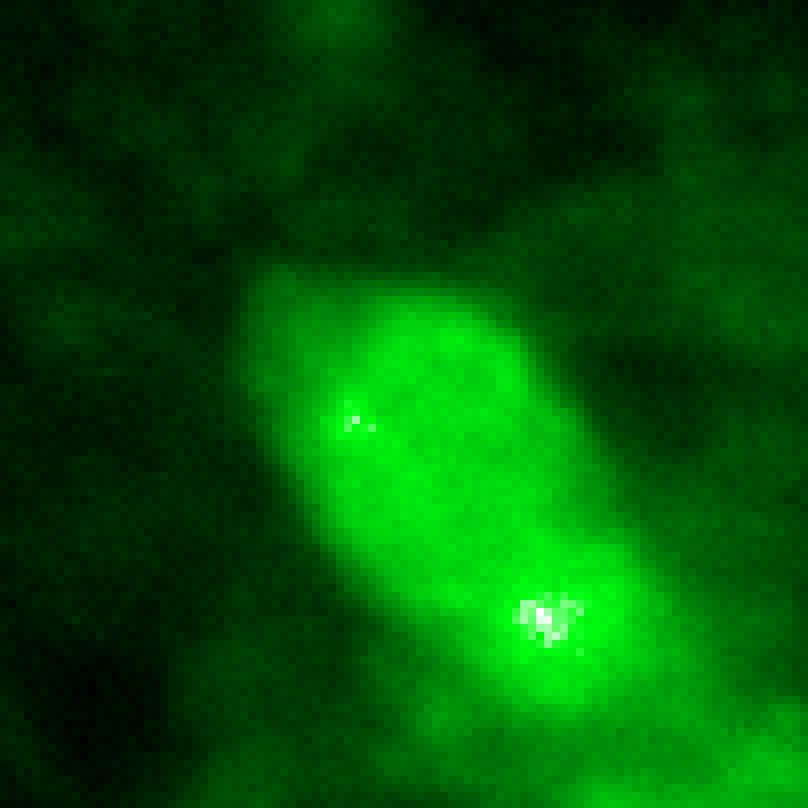}&
			\includegraphics[height= 0.10\textwidth]{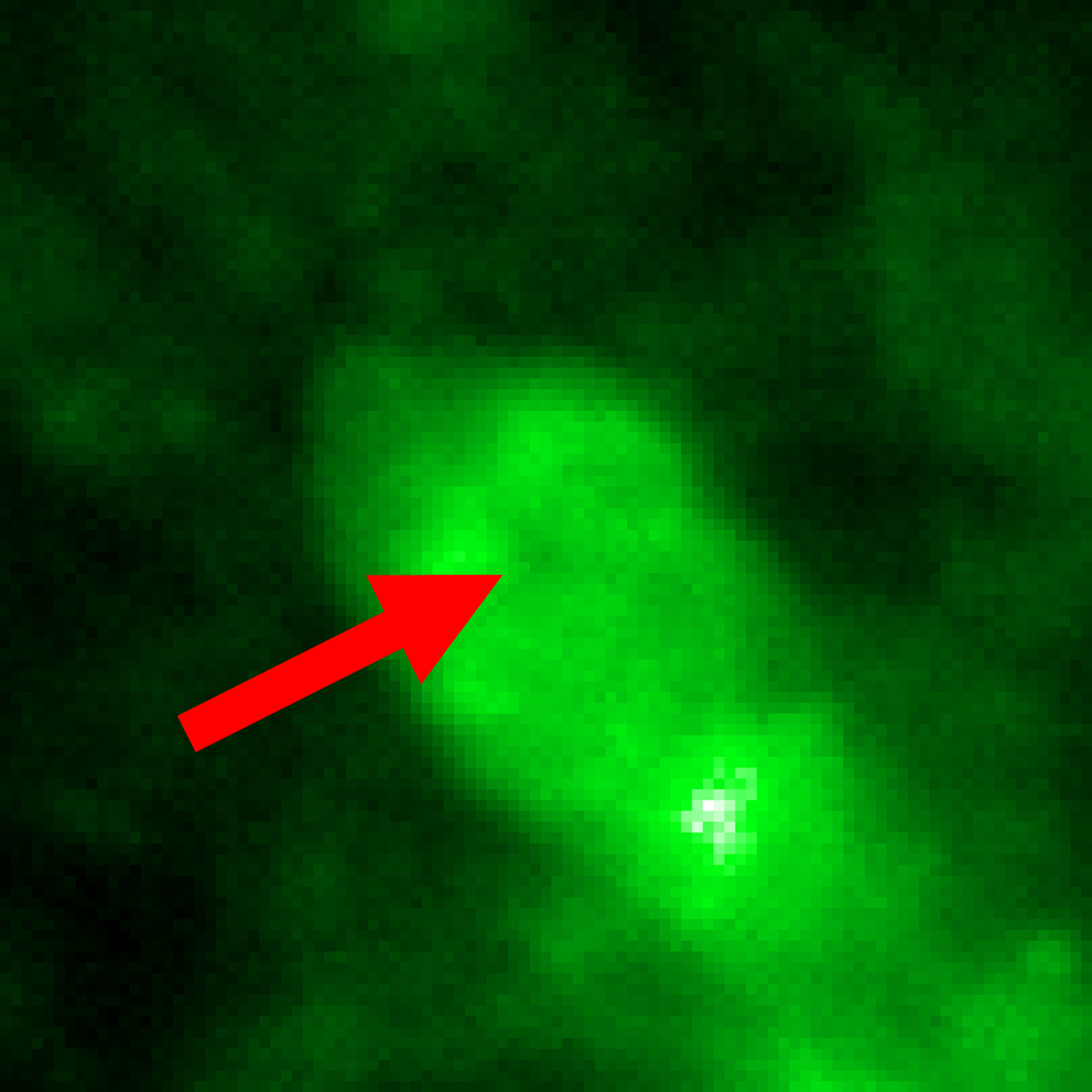}
			\vspace{0.1cm}\\

			\includegraphics[height= 0.10\textwidth]{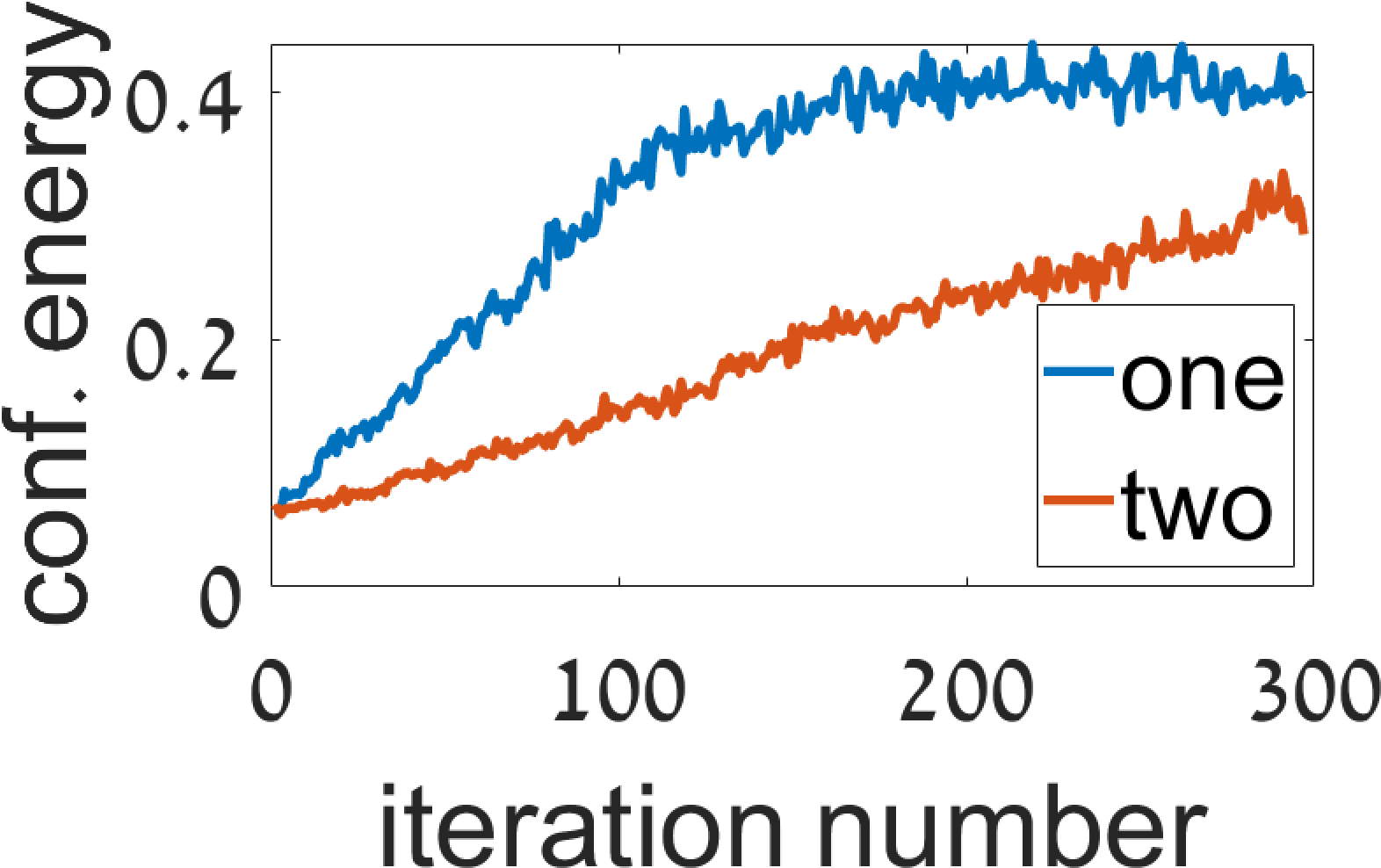}&
			\includegraphics[height= 0.10\textwidth]{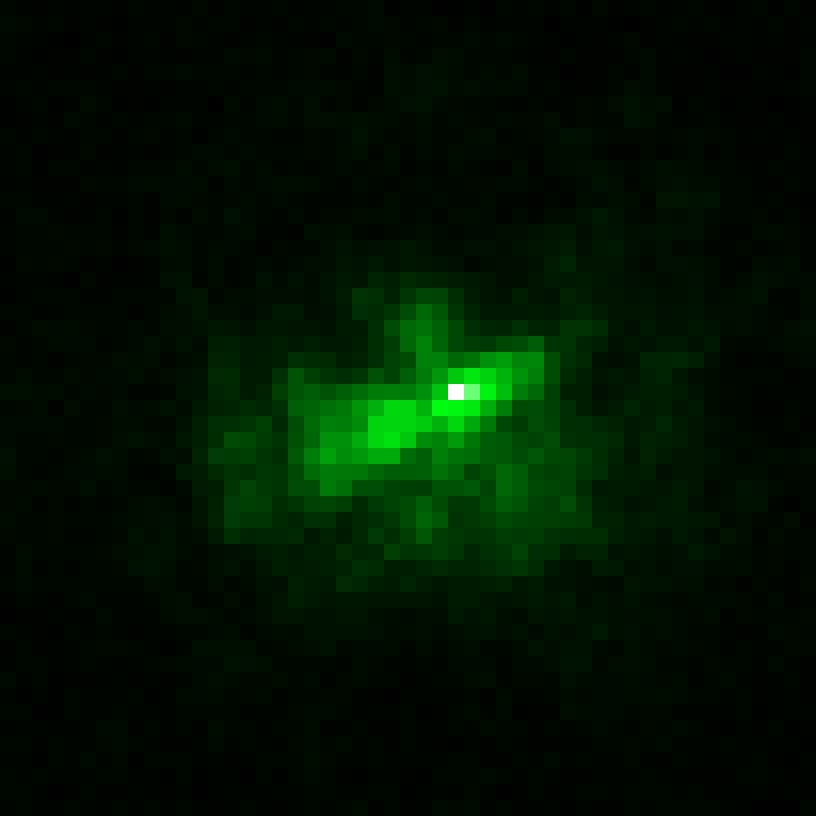}&
			\includegraphics[height= 0.10\textwidth]{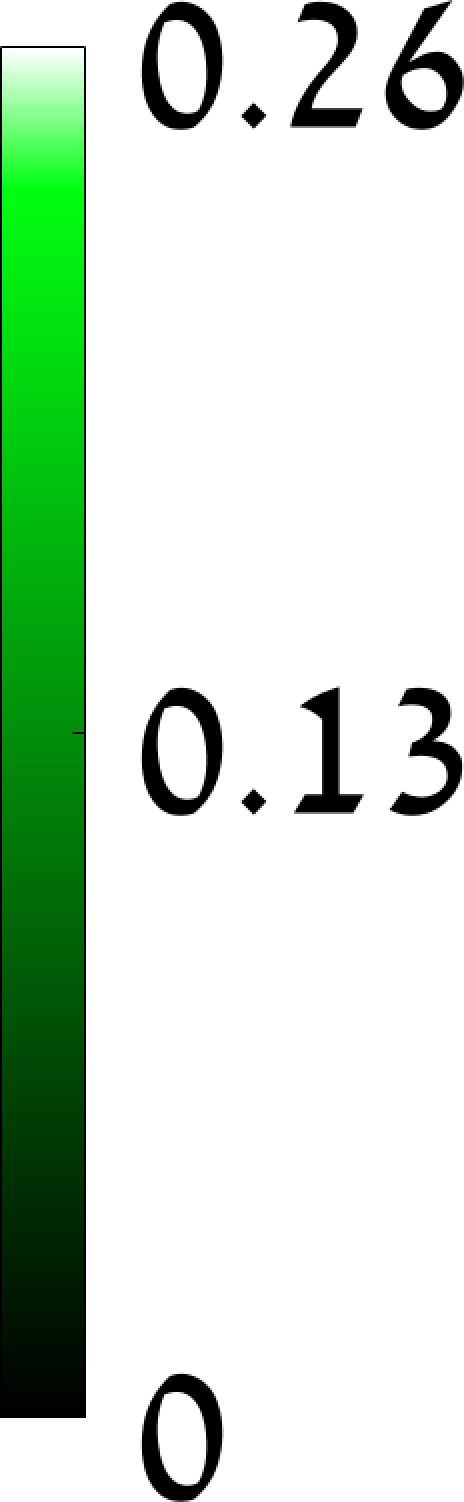}&		
			\includegraphics[height= 0.10\textwidth]{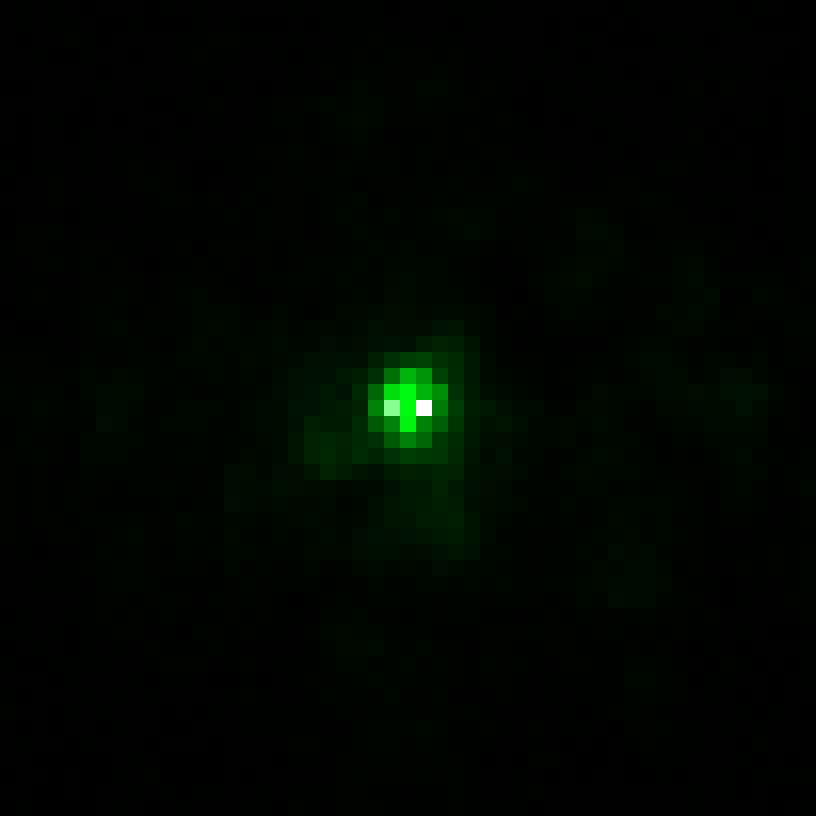}&		
			\includegraphics[height= 0.10\textwidth]{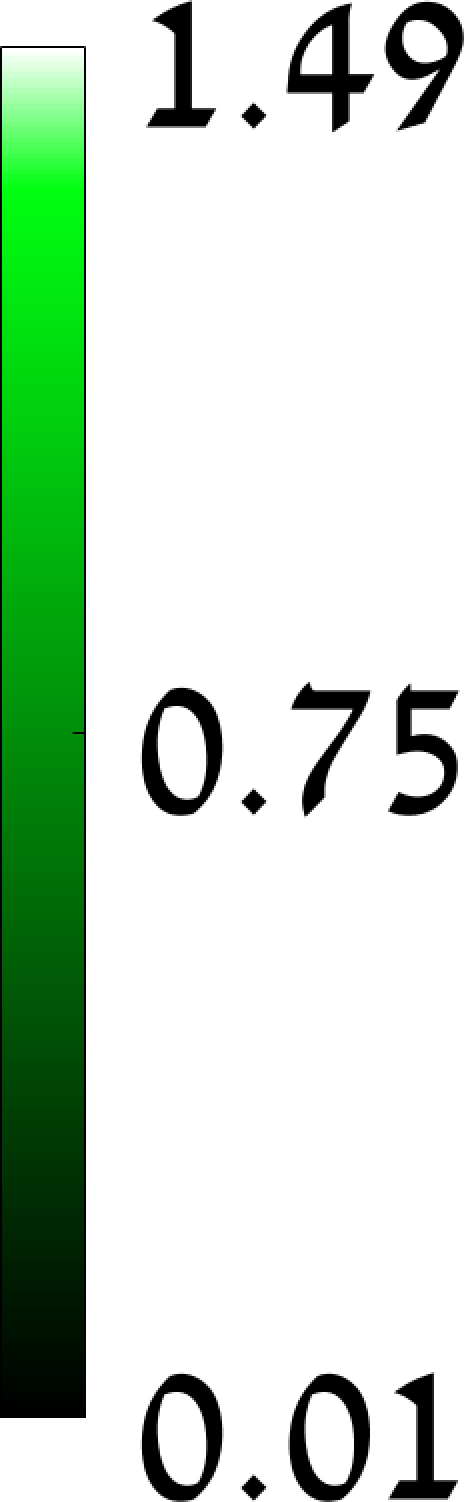}&		
			\includegraphics[height= 0.10\textwidth]{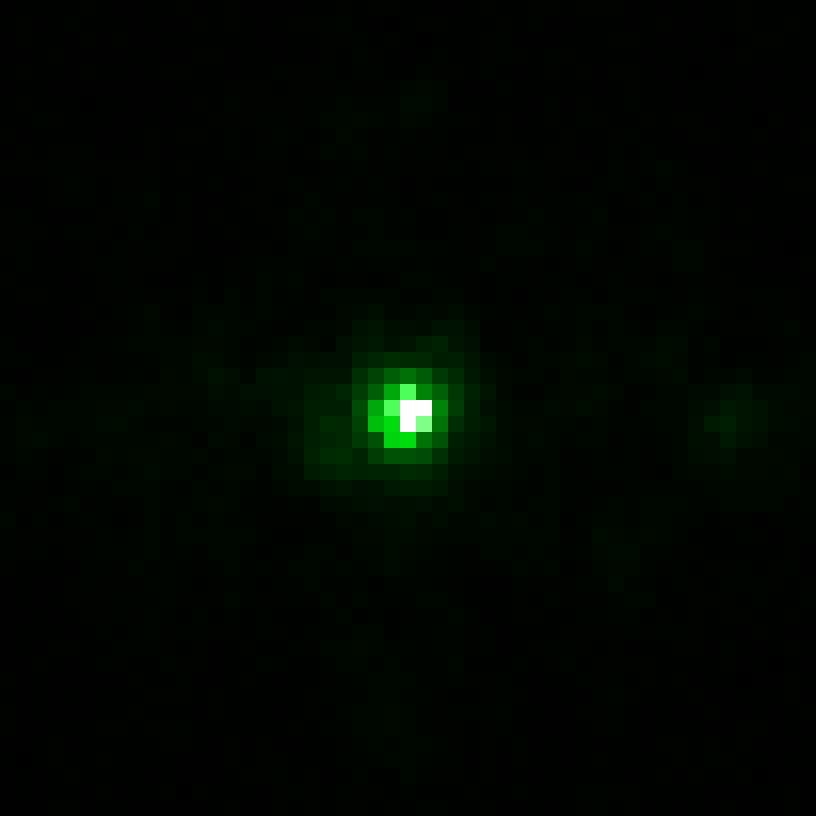}&		
			\includegraphics[height= 0.10\textwidth]{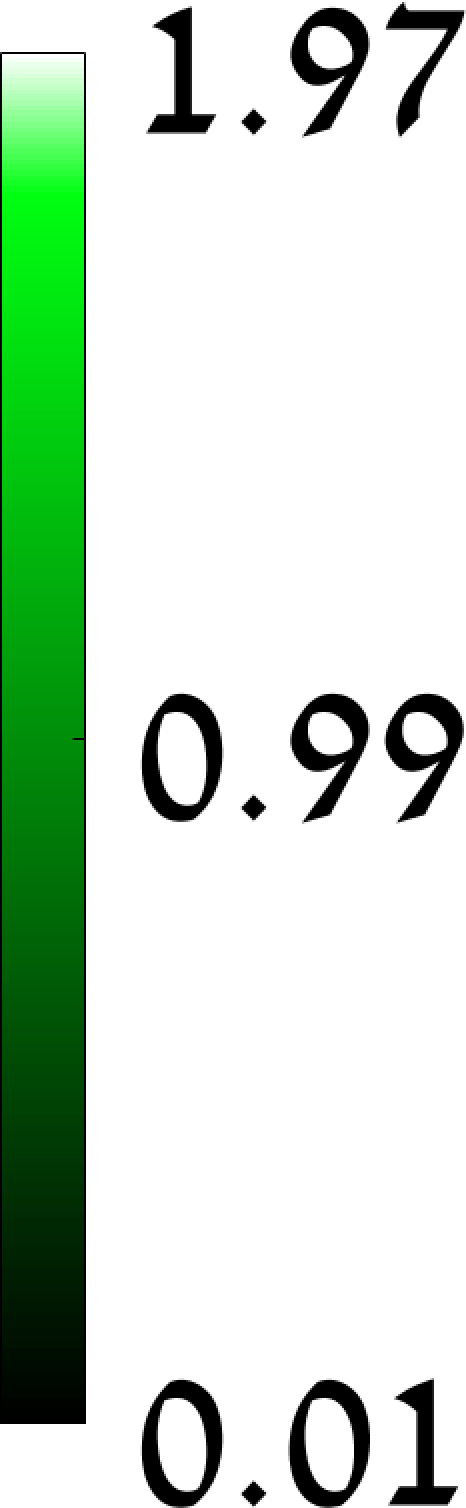}&
			\includegraphics[height= 0.10\textwidth]{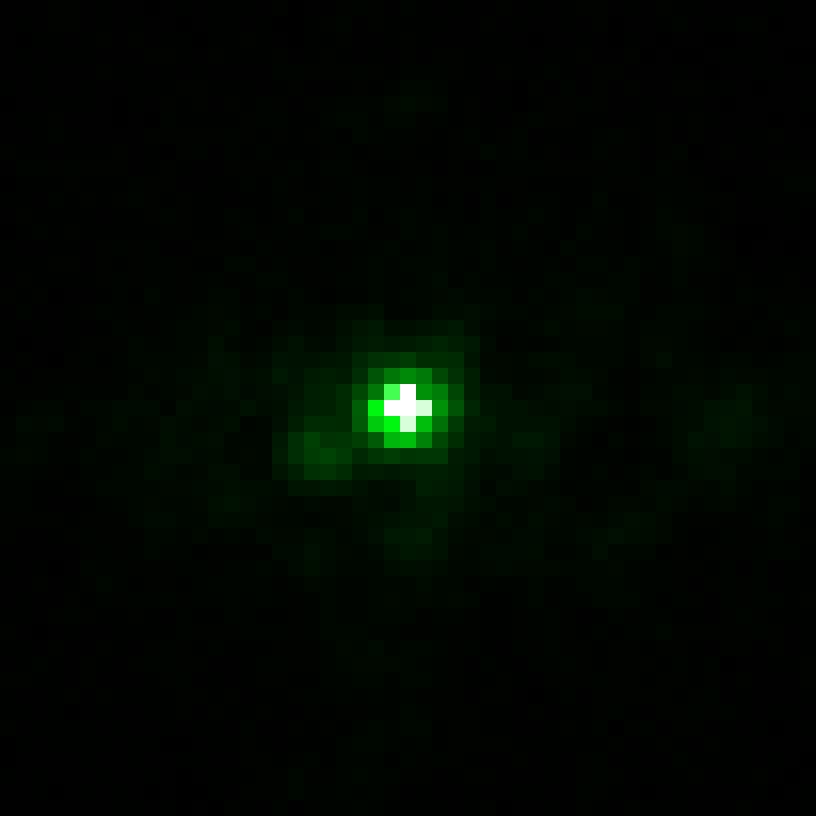}&
			\includegraphics[height= 0.10\textwidth]{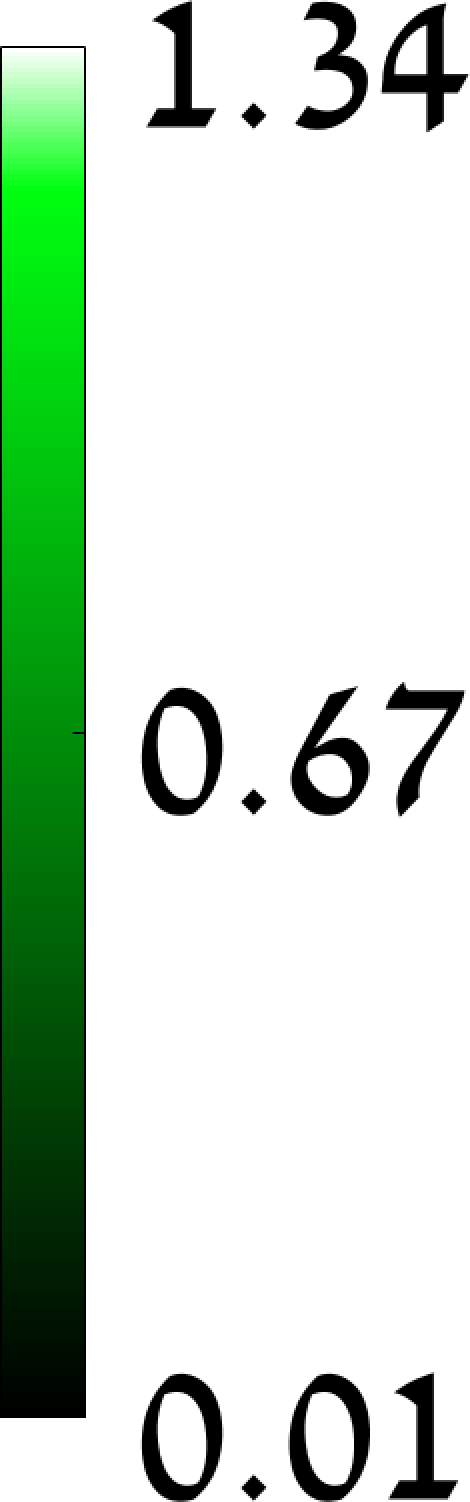}&
			\includegraphics[height= 0.10\textwidth]{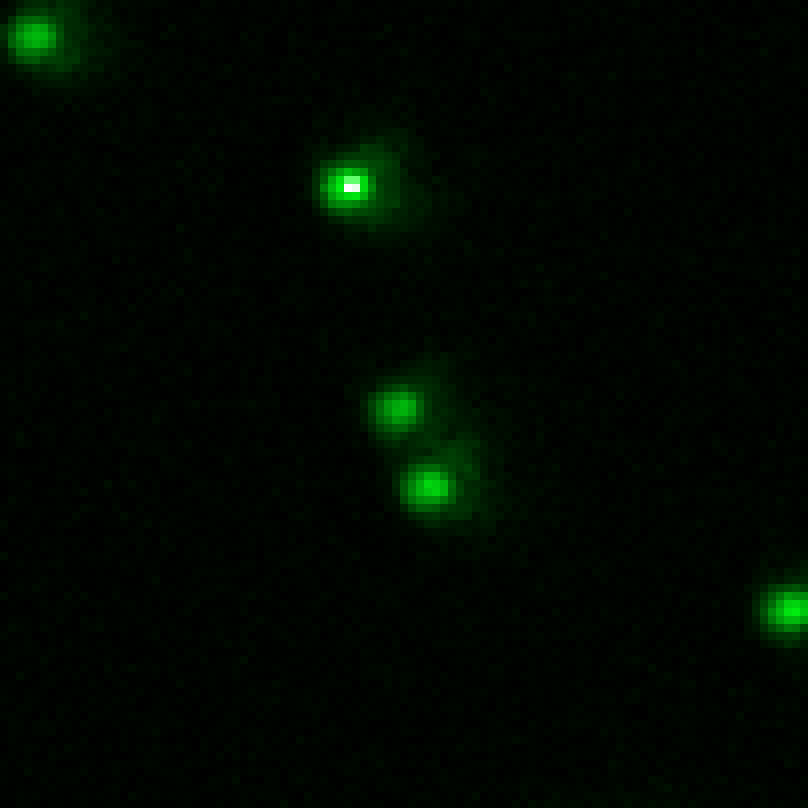}& 
			\includegraphics[height= 0.10\textwidth]{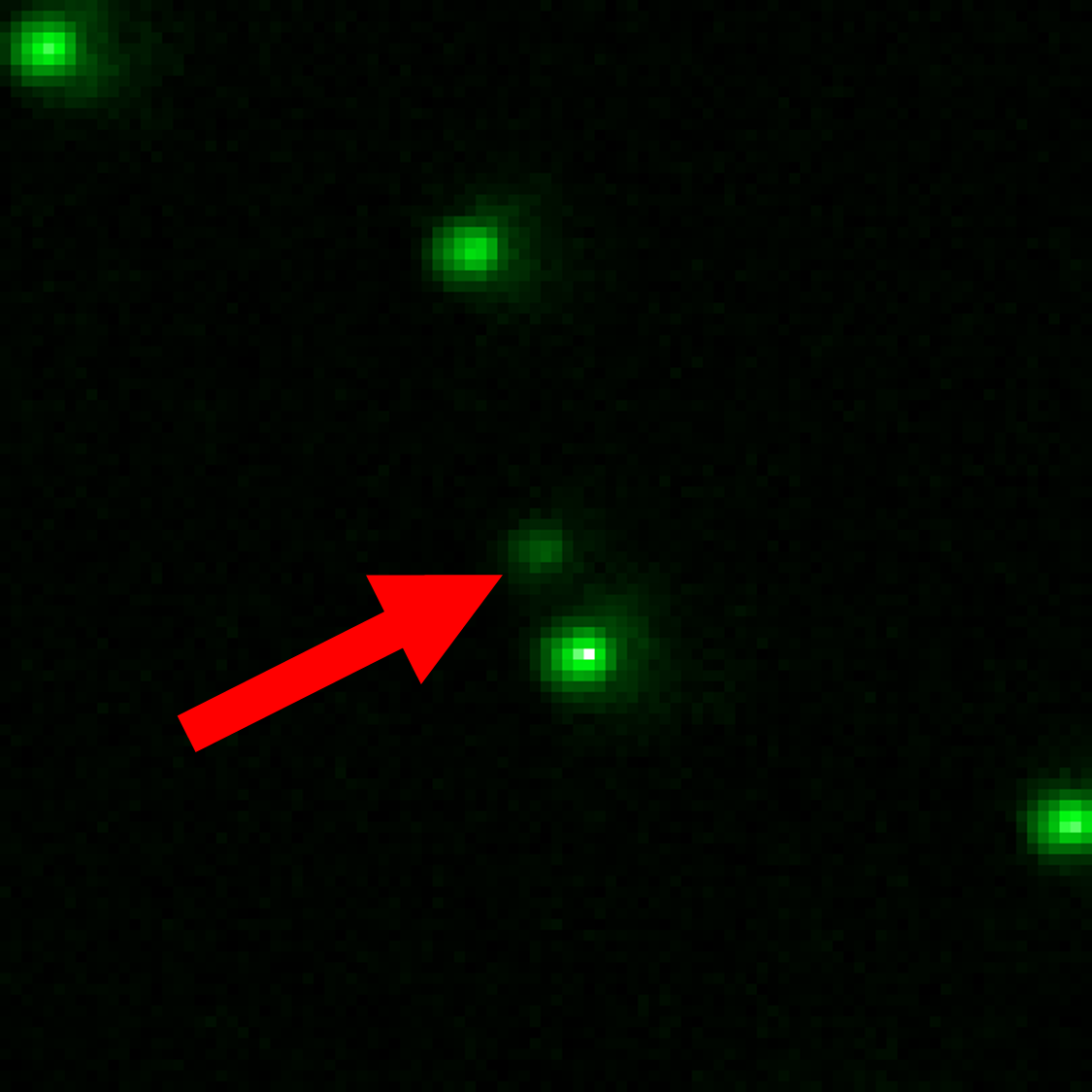}
			\vspace{0.1cm}\\

		\end{tabular}
 \end{adjustwidth}

		\caption{\blue{Comparing the usage of two different modulations on the emission and excitation wavelengths, vs. a common modulation for both. The top row shows results on a brain slice and the lower one uses as the target a set of fluorescent beads behind parafilm. (a) The confocal intensity at the central pixel as a function of the number of iterations. Since the two-mask approach doubles the number of measurements for each dictionary element, its convergence is slower. (b) The initial image at the main camera. (c) The final confocal spot at the main camera using 2 different modulations. (d) The final spot using the same modulation in both channels. Since it scans the dictionary elements more times, the result is better (a higher energy is measured at the central pixel). (e) The result of one mask in the middle of the optimization when the number of dictionary elements is equivalent to what the two-modulation approach has scanned at the end of the optimization. When scanning the dictionary for the same number of times, the one-modulation result is somewhat lower than what was achieved by two modulations.  (f-g) The fluorescent target from the validation camera at the beginning of the optimization and at the end, notice the strong bleaching at the focusing bead, marked with an arrow. We show the final modulations in \figref{fig:2_masks_vs_1_masks}.} }\label{fig:2_masks_vs_1}
\end{figure*}

\begin{figure*}[t!]
	\begin{center}
		\begin{tabular}{@{}c@{~~}c@{~~}c@{~~}c@{~}c@{}}
			
			\small{1 modulation final}&\small{1 modulation mid}&\small{2 modul. excitation}&\small{2 modul. emission}&\\

			\includegraphics[height= 0.16\textwidth]{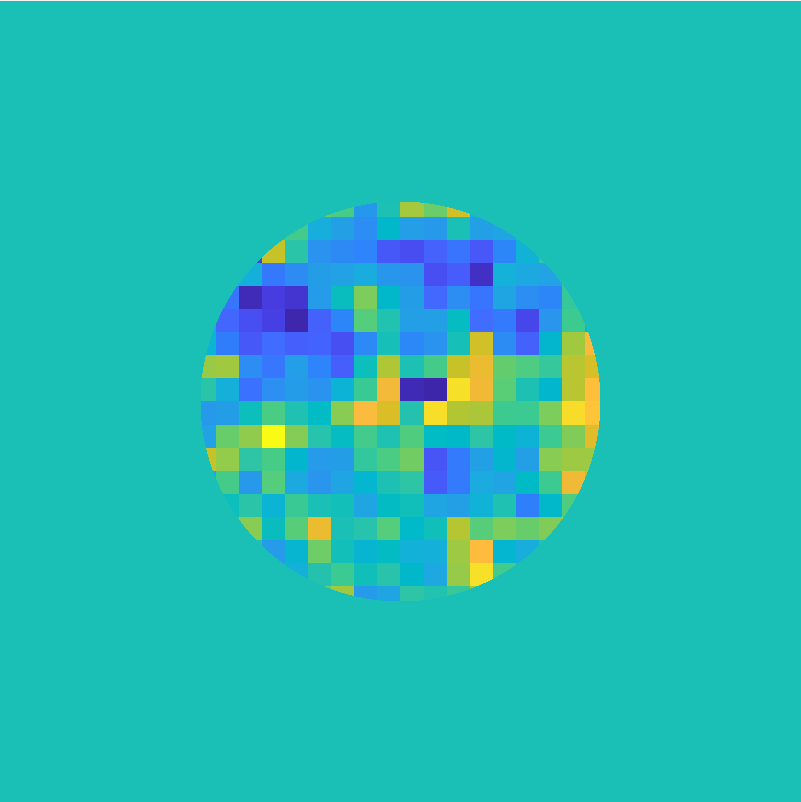}&
			\includegraphics[height= 0.16\textwidth]{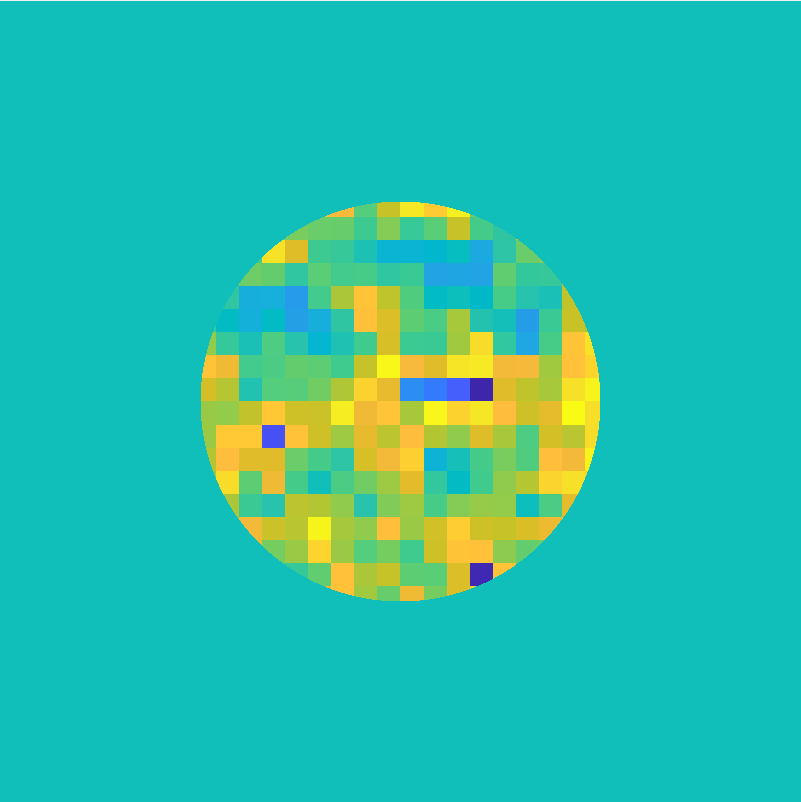}&
			\includegraphics[height= 0.16\textwidth]{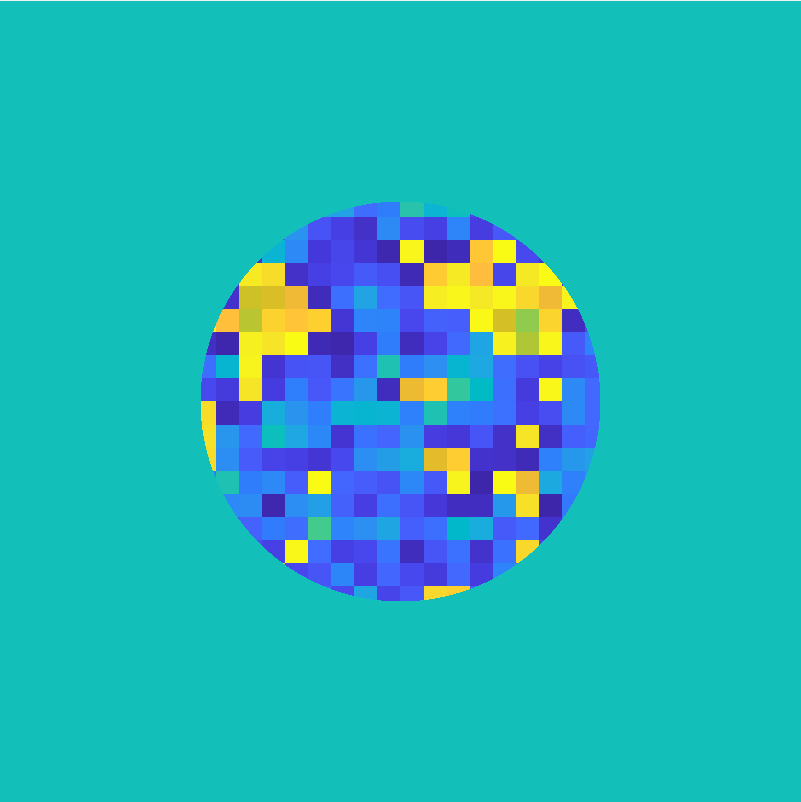}&		
			\includegraphics[height= 0.16\textwidth]{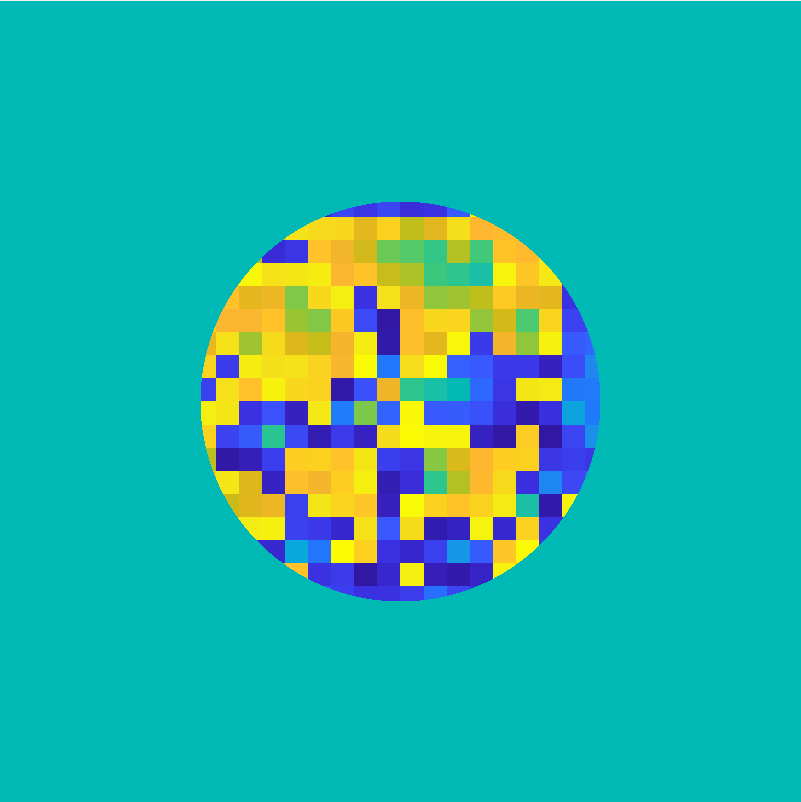}&		
			\includegraphics[height= 0.16\textwidth]{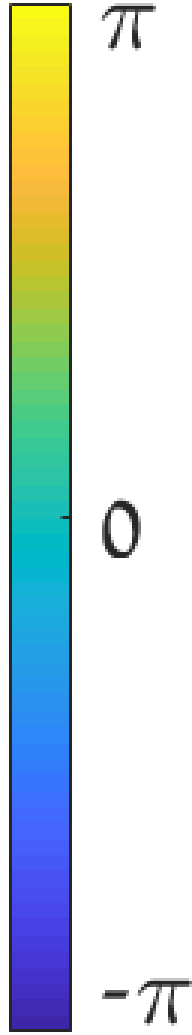}\\

			
			\includegraphics[height= 0.16\textwidth]{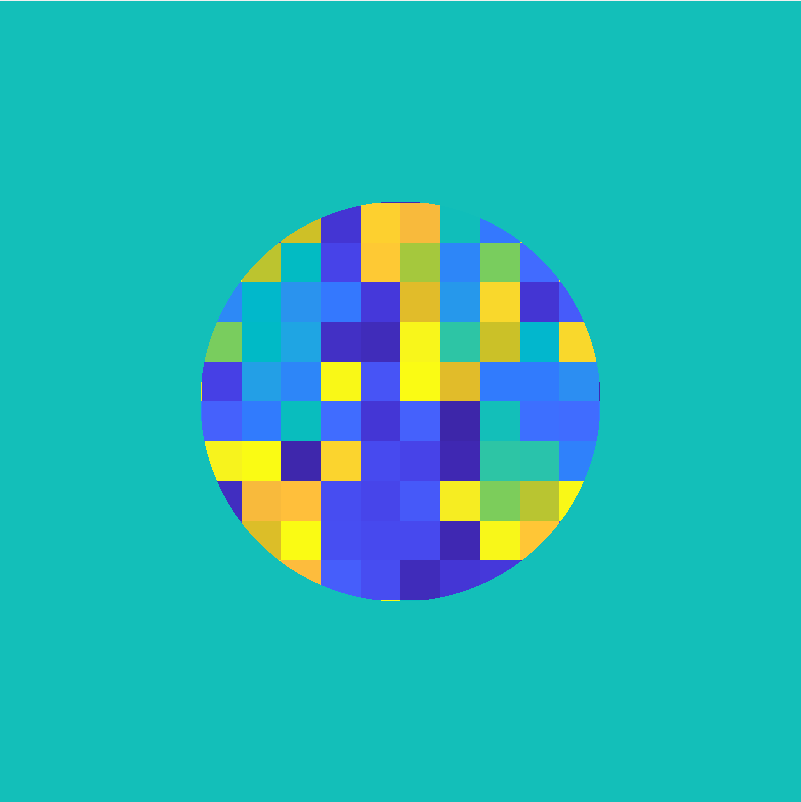}&
			\includegraphics[height= 0.16\textwidth]{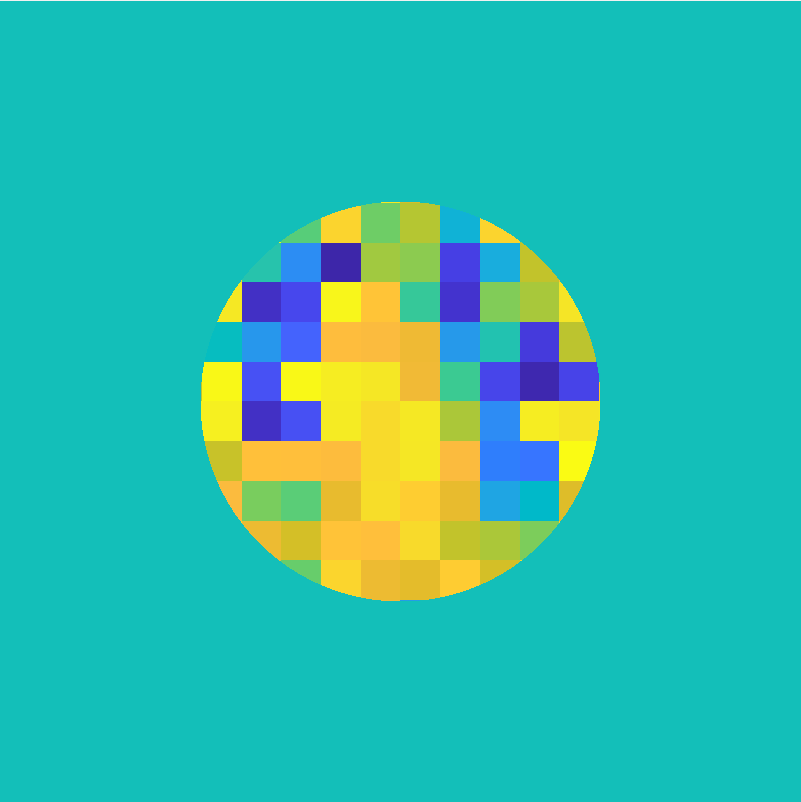}&
			\includegraphics[height= 0.16\textwidth]{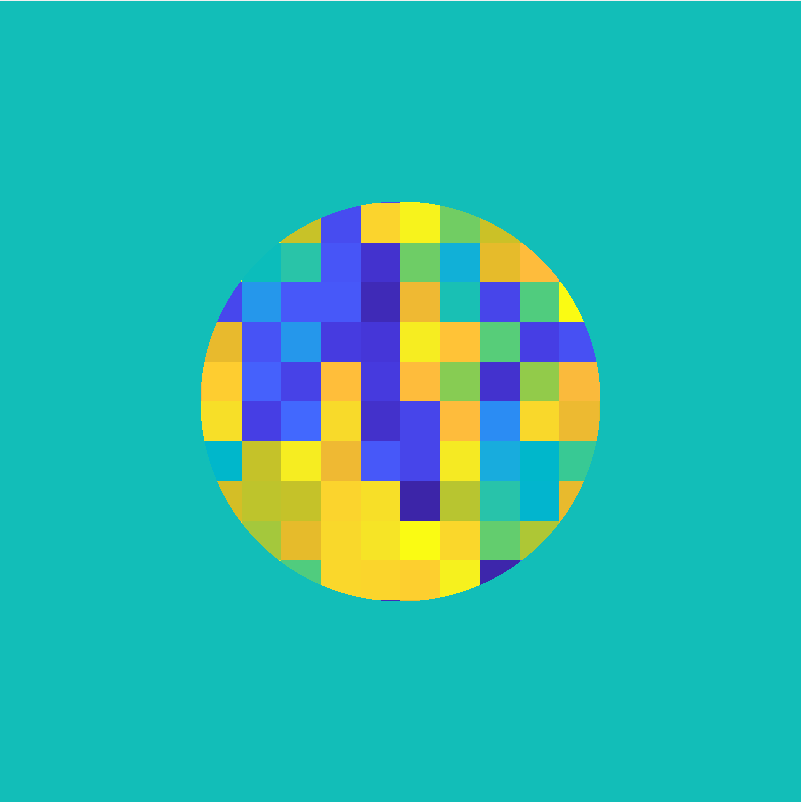}&		
			\includegraphics[height= 0.16\textwidth]{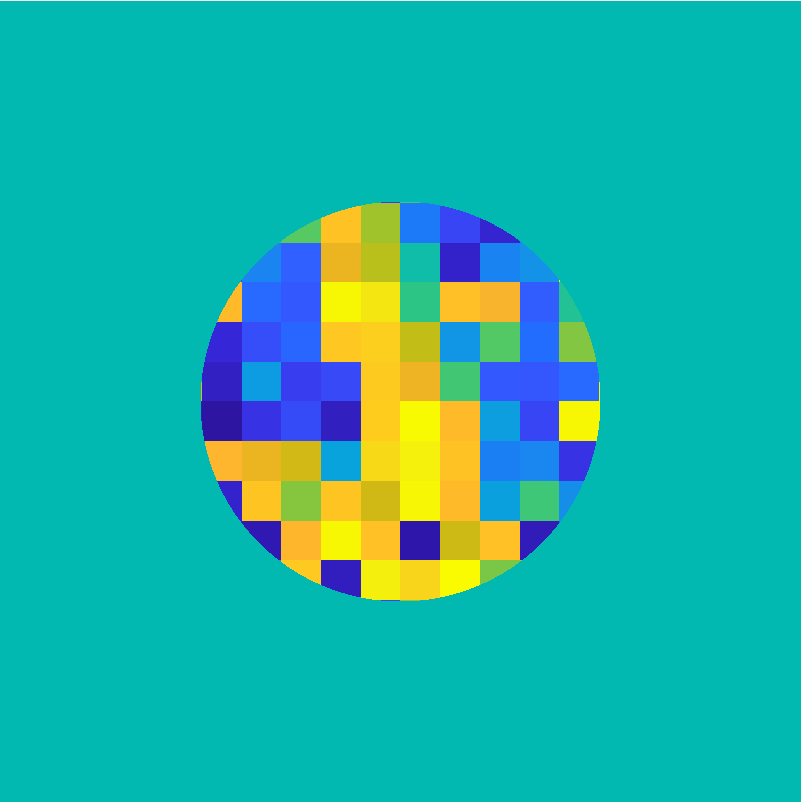}&		
			\includegraphics[height= 0.16\textwidth]{figs/fig_2_masks_vs_one/colorbar.png}

		\end{tabular}
		\caption{\blue{Modulation masks resulting from \figref{fig:2_masks_vs_1}, in the Fourier plane. The correlation between the masks of the two different wavelengths is typically  within the range $[0.6,0.8]$. The correlation between the one-mask approach and the two-mask approach is also in the same range.} }\label{fig:2_masks_vs_1_masks}
	\end{center}
\end{figure*}

\section{One modulation against two modulations}\label{sec:one-vs-two-modul}
\blue{As mentioned above, in our single-photon fluorescence case the emission and excitation wavelengths are similar and we can approximately use the same modulation in both excitation and emission arms.  Alternatively we can solve for a different modulation in each arm, but this doubles the number of exposures. Below we compare these two approaches experimentally and learn that while two different modulations can lead to a better correction, they also require a longer acquisition with more  photo-bleaching. In the presence of photo-bleaching  the faster, one-modulation approach usually leads to better results.}

\blue{Comparing two algorithms on the same sample is challenging, because if we try to run
two algorithms sequentially on the same sample, the second one would have worse results just because more photo-bleaching took place.
To avoid this we run the two algorithms in alternating order.
The $n$'th iteration is composed of 4 steps:
\begin{enumerate}
	\item Test basis element $2n$ for the single modulation case.
	\item Test  basis element $n$ for the excitation arm in the two-modulation case.
	\item Test basis element $2n+1$ for the single modulation case.
	\item Test  basis element $n$ for the emission arm in the two-modulation case.
\end{enumerate} 
The implication is that the single modulation approach scans the elements of the Hadamard basis faster than the two-modulation approach.  
In the results of \figref{fig:2_masks_vs_1} we scan the Hadamard basis 4 times in the single modulation optimization and only twice in the two-modulation optimization. 
We separate the measurement of the single and two-modulation approaches and plot them as two separate curves  in  \figref{fig:2_masks_vs_1}(a). 
The single modulation approach leads to higher energy at the focused spot.
To evaluate the quality of the modulation independently of photo-bleaching, at the end of the optimization we capture 3 images of the focus spot. 1)~With two different modulations, scanning each basis element twice. 
2)~With a single modulation with the same number of exposures, effectively scanning each element 4 times. 
3)~An image of a single modulation in the middle of the optimization, which effectively scans the basis elements only twice as  the  two-modulation optimization. While this image corresponds to the modulation in the middle of the optimization, we capture it at the end, under equal photo-bleaching conditions. 
The single modulation that scanned the basis elements 4 times gives the best results. However if we compare the two modulations to a single modulation that has scanned the elements for the same number of times, the results are better.}
 
\blue{In \figref{fig:2_masks_vs_1} we demonstrate this comparison twice, once when the fluorescent target is a neuron behind a layer of parafilm, and once when we spread a set of fluorescent beads behind the parafilm.  In \figref{fig:2_masks_vs_1}(f-g) we show images of the target from the validation camera in the beginning of the optimization and at the end.  For the beads example, one can see that the bead  at which the algorithm has converged (marked by an arrow) is significantly dimmer at the end of the optimization, showing the strong bleaching. }

\blue{In \figref{fig:2_masks_vs_1_masks} we show the phase of the modulation masks we found with the different approaches described above. The correlation between the modulations at the two different wavelengths is typically within the range $[0.6,0.8]$. The correlation between modulations found at two different wavelengths to the same modulation for both wavelengths is also in the same range. It is unclear if this is really a measure of the chromatic memory effect correlation, or if this difference is a result of the noisy  optimization. }

	\section{Comparison with alternative wavefront-shaping scores}

 \blue{We compare our confocal score with the variance maximization approach of~\cite{Boniface:19}, showing that our approach can converge using a significantly smaller number of photons. We also compare against  one of the non-local approaches in~\cite{YeminyKatz2021}. This approach assumes that a single modulation can correct a wide image region rather than a single spot. Our evaluation shows that when memory-effect exists over a wide extent this algorithm can indeed recover good modulations, but the quality of the results degrades for a short ME, where the size of the iso-planatic patches that can be corrected with a single modulation is small.}
 
 \blue{We note that  it is very hard to run two algorithms on the same data under equal noise conditions, as due to photo-bleaching the 2nd algorithm would run on a  significantly weaker fluorescent source.
 To support a controlled evaluation we used simulated transmission matrices synthesized using a multi-plane propagation model. We assume the tissue has a thickness of $L=250\mu m$ and the aberration of light propagating through this volume is formed by  equally spaced, planar pseudo-random phase masks. We use different number of aberration layers to test different memory-effect extents as described below.
 Given a transmission matrix we can simulate the images formed under any modulation of choice and add a different amount of noise. Hence we can evaluate the results of different algorithms while varying any parameter of interest.}

	\begin{figure*}[t!]
	\begin{center}
		\begin{tabular}{@{}c@{~~}c@{~~}c@{~~}c@{~~}c@{~}c@{~~}c@{~}c@{~}r@{}}
			\small{(a)Ground truth}&\small{(b)Speckle sup.}&\small{(c)Input main}&\small{(d)Laser power}&\small{(e)Final main}&&\small{(f)Final val.}&
			&\\
			\includegraphics[width= 0.13\textwidth]{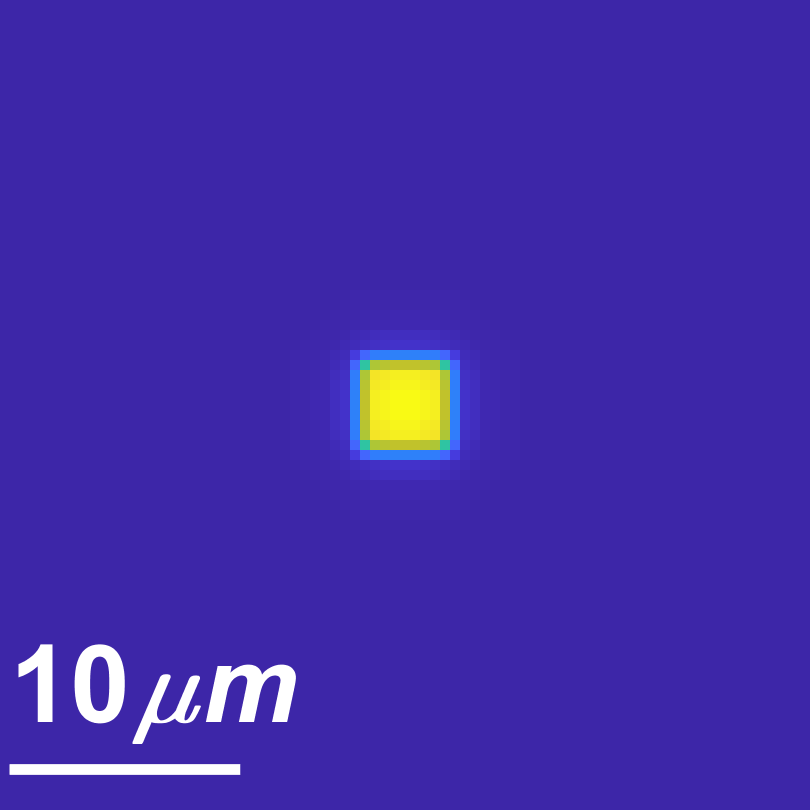}&
			\includegraphics[width= 0.13\textwidth]{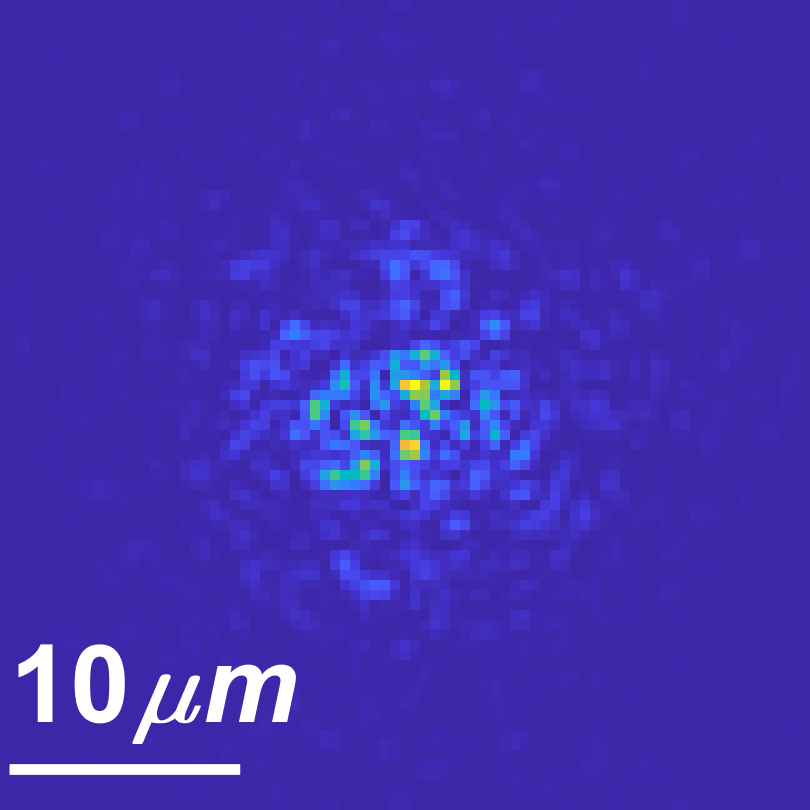}&
			\includegraphics[width= 0.13\textwidth]{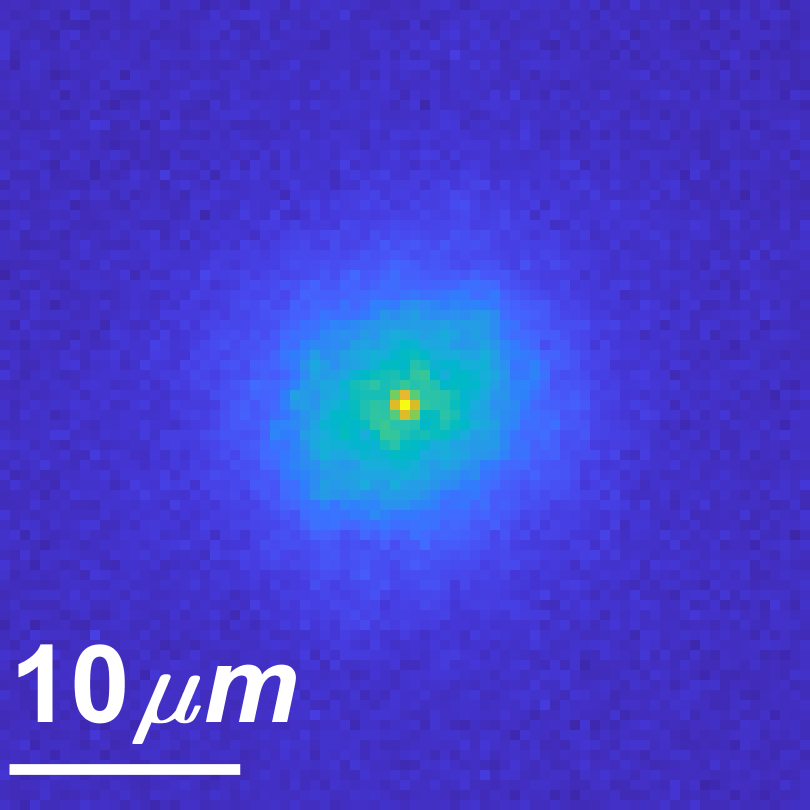}&		
			\includegraphics[height= 0.13\textwidth]{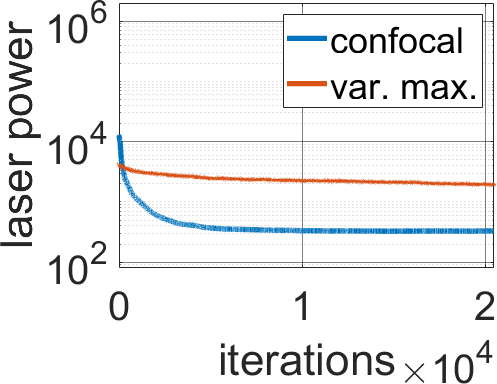}&		
			\includegraphics[width= 0.13\textwidth]{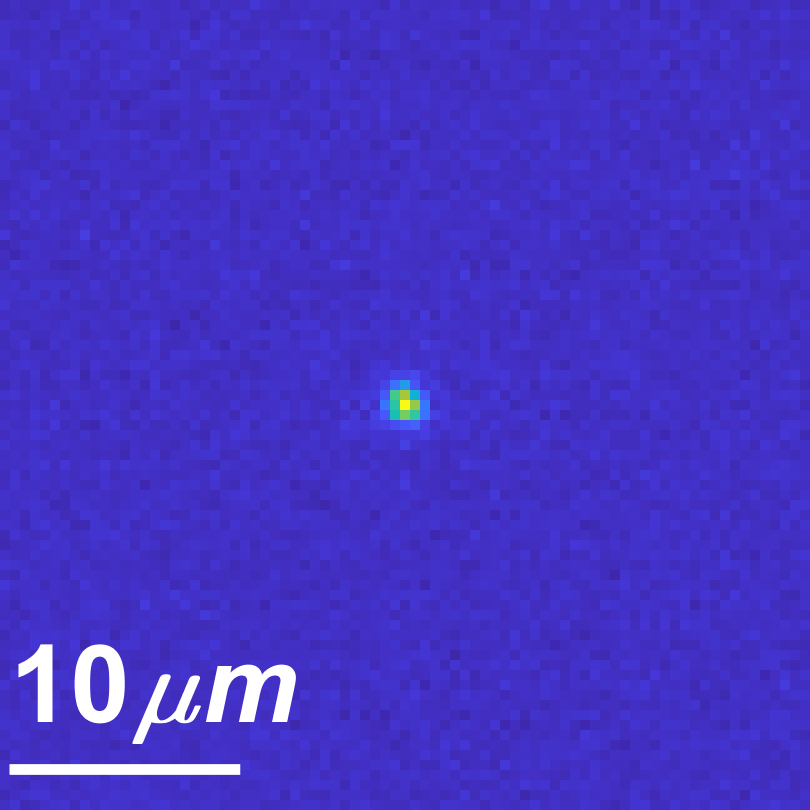}&		
			\includegraphics[height= 0.13\textwidth]{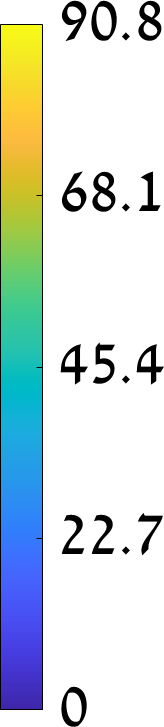}&		
			\includegraphics[width= 0.13\textwidth]{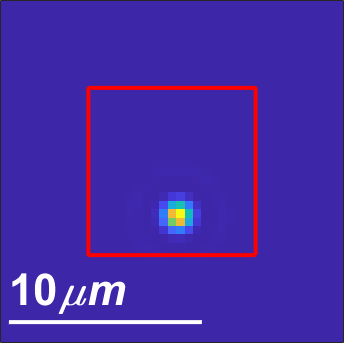}&
			\includegraphics[height= 0.13\textwidth]{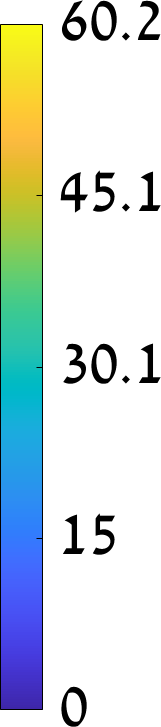}&\multirow[c]{1}{*}[1.6 cm]{\rotatebox[origin=c]{90}{Confocal}}
			\\
			&&&&
			\includegraphics[width= 0.13\textwidth]{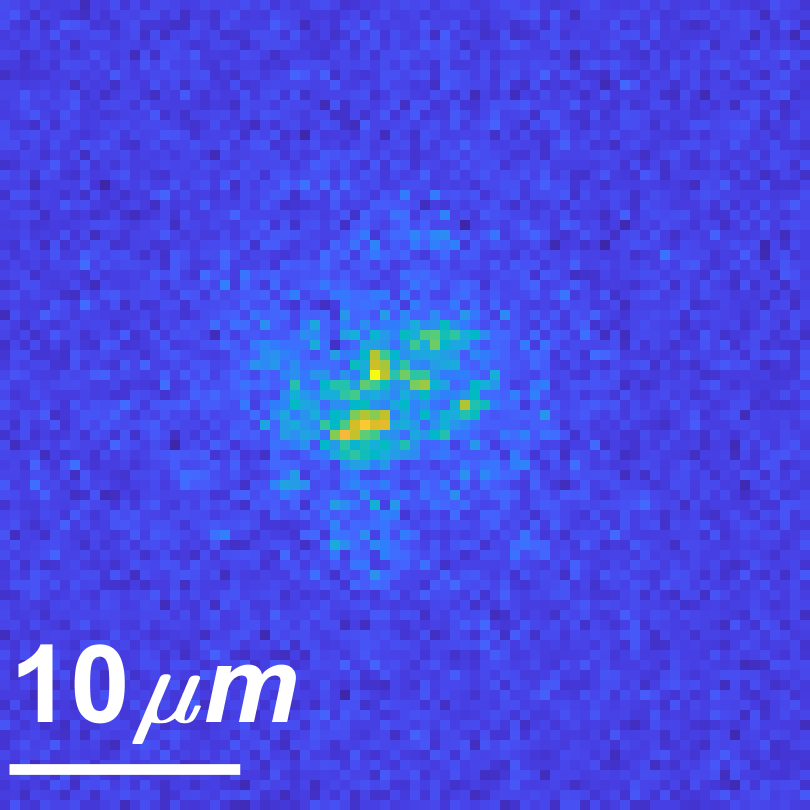}&		
			\includegraphics[height= 0.13\textwidth]{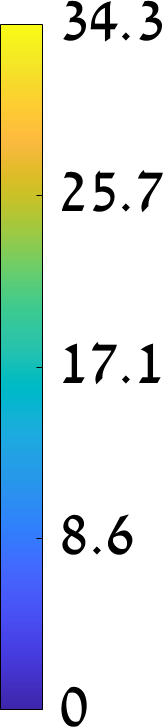}&		
			\includegraphics[width= 0.13\textwidth]{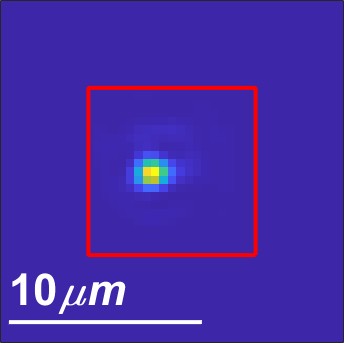}&
			\includegraphics[height= 0.13\textwidth]{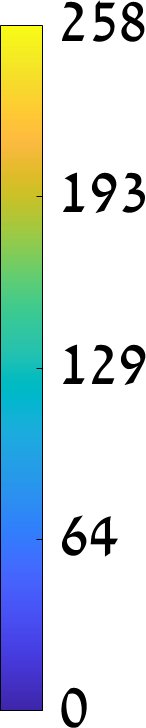}&
			\multirow[c]{1}{*}[1.7 cm]{\rotatebox[origin=c]{90}{Var. max.}}
			\\
			
			&
			\includegraphics[width= 0.13\textwidth]{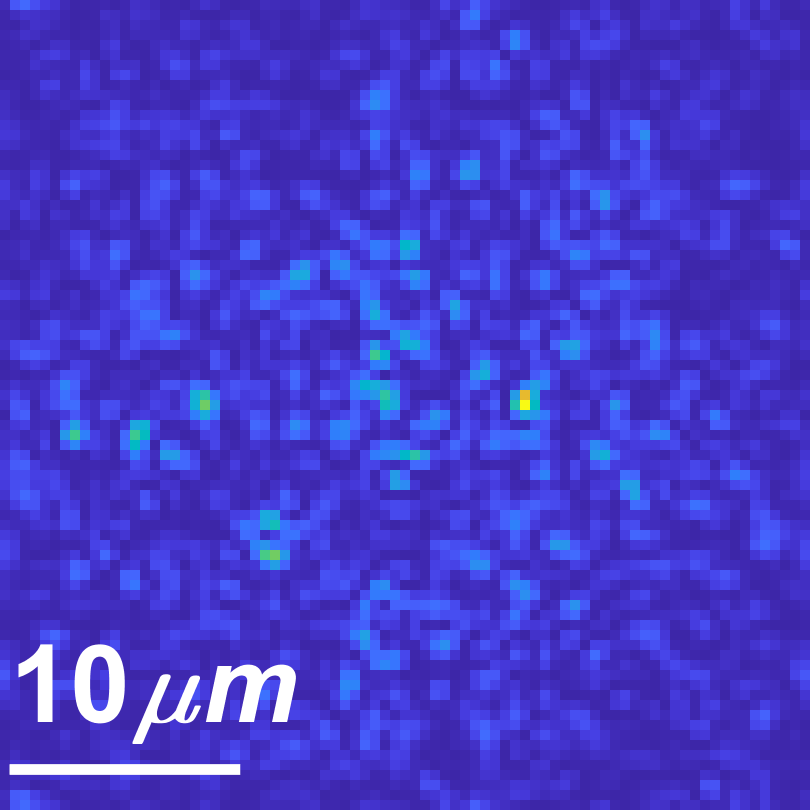}&
			\includegraphics[width= 0.13\textwidth]{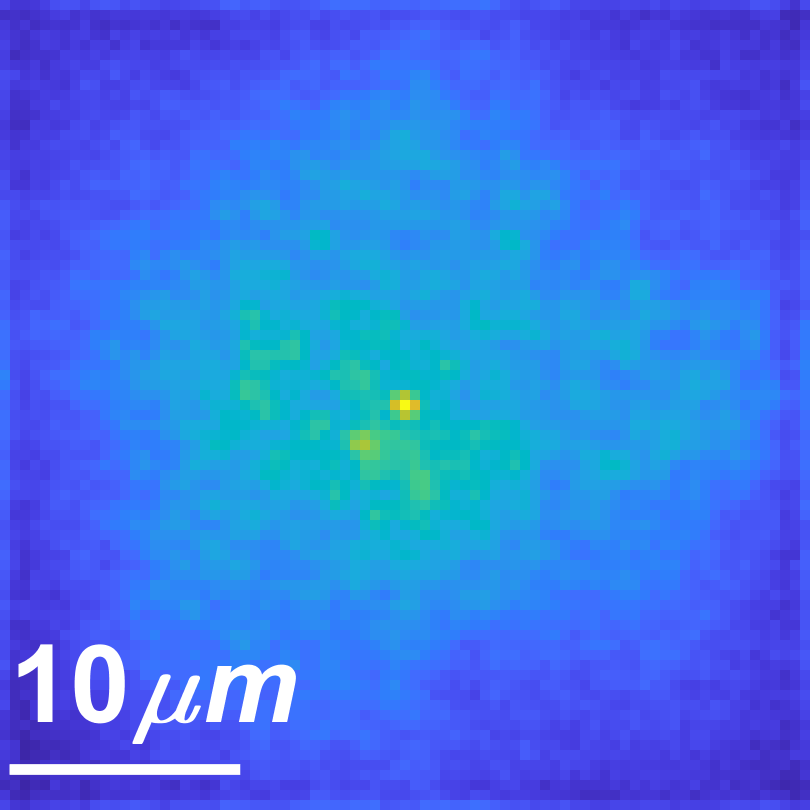}&		
			\includegraphics[height= 0.13\textwidth]{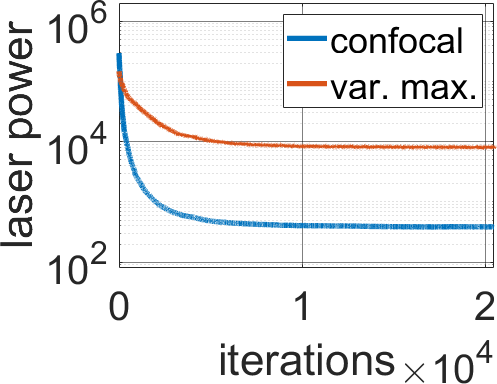}&		
			\includegraphics[width= 0.13\textwidth]{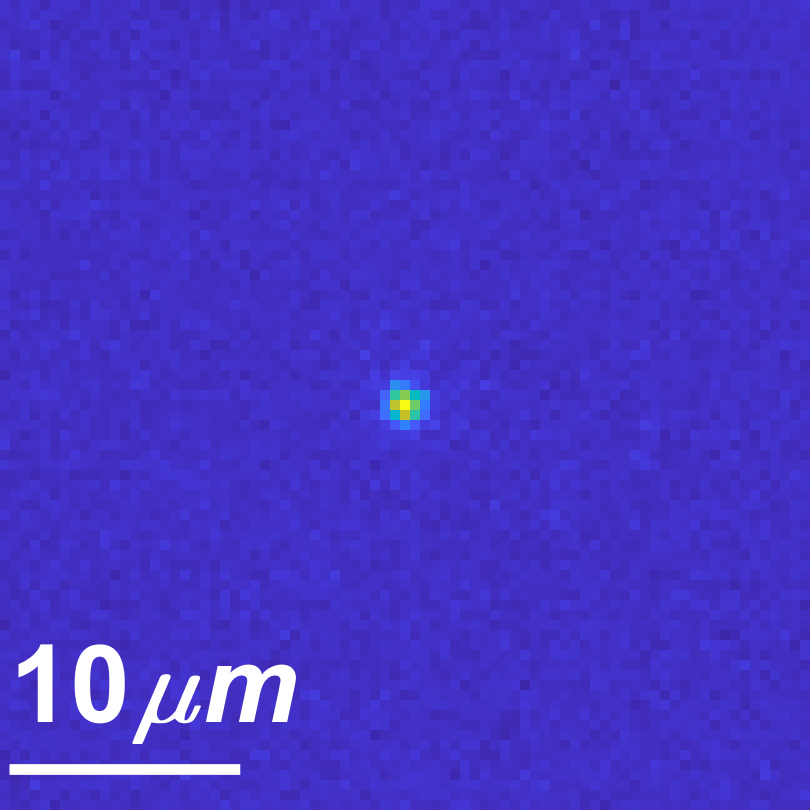}&		
			\includegraphics[height= 0.13\textwidth]{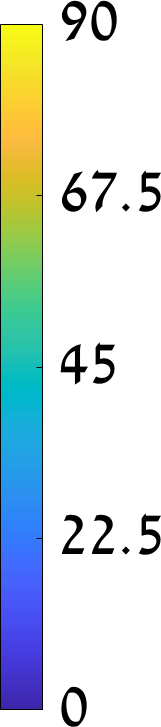}&		
			\includegraphics[width= 0.13\textwidth]{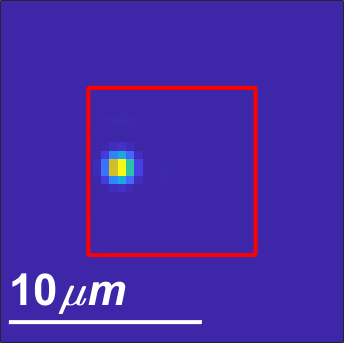}&
			\includegraphics[height= 0.13\textwidth]{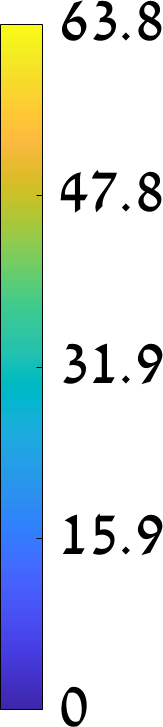}&\multirow[c]{1}{*}[1.6 cm]{\rotatebox[origin=c]{90}{Confocal}}
			\\
			&&&&
			\includegraphics[width= 0.13\textwidth]{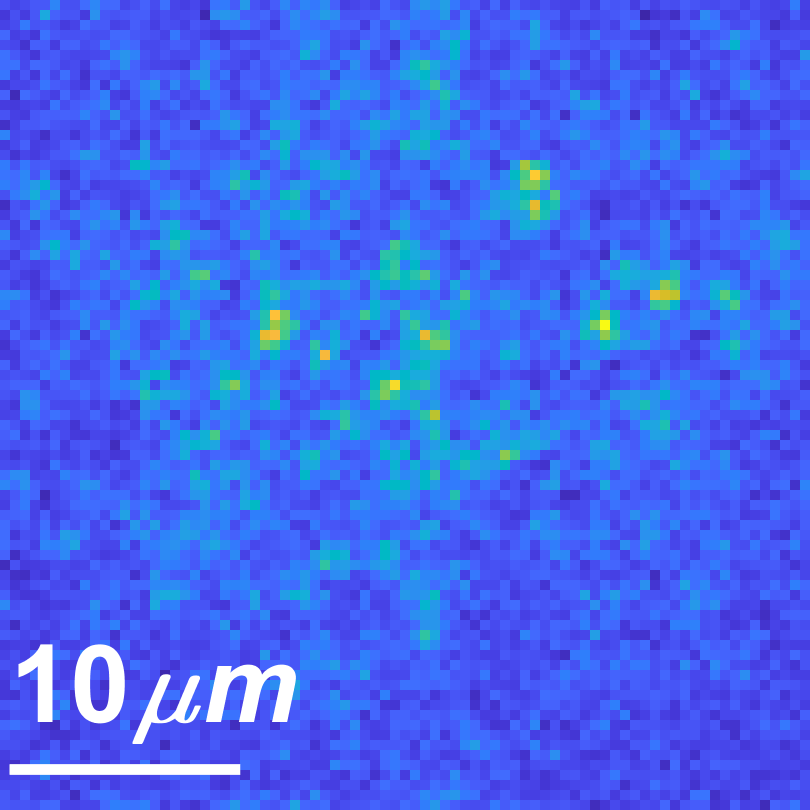}&		
			\includegraphics[height= 0.13\textwidth]{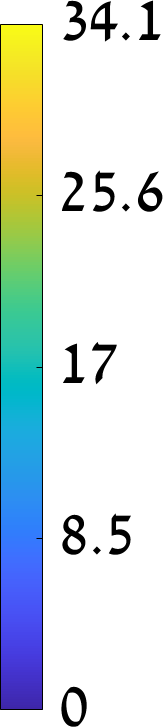}&		
			\includegraphics[width= 0.13\textwidth]{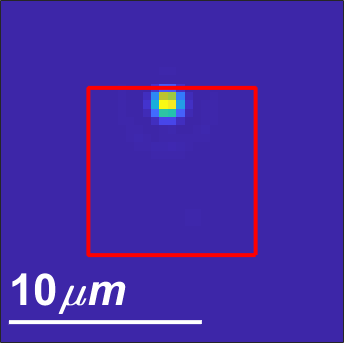}&
			\includegraphics[height= 0.13\textwidth]{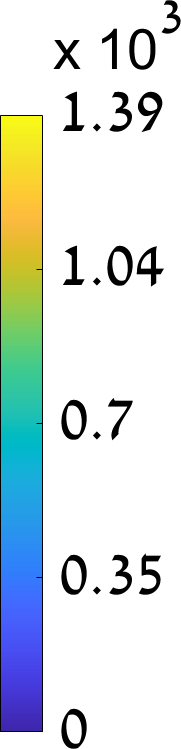}&
			\multirow[c]{1}{*}[1.7 cm]{\rotatebox[origin=c]{90}{Var. max.}}
			\\
			&
			\includegraphics[width= 0.13\textwidth]{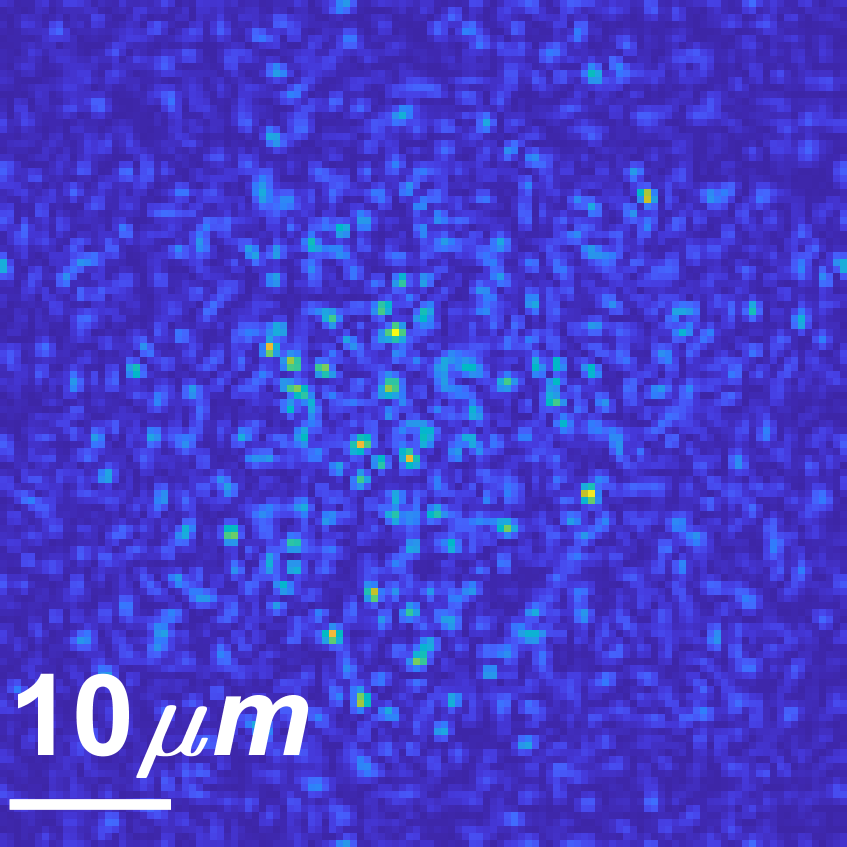}&
			\includegraphics[width= 0.13\textwidth]{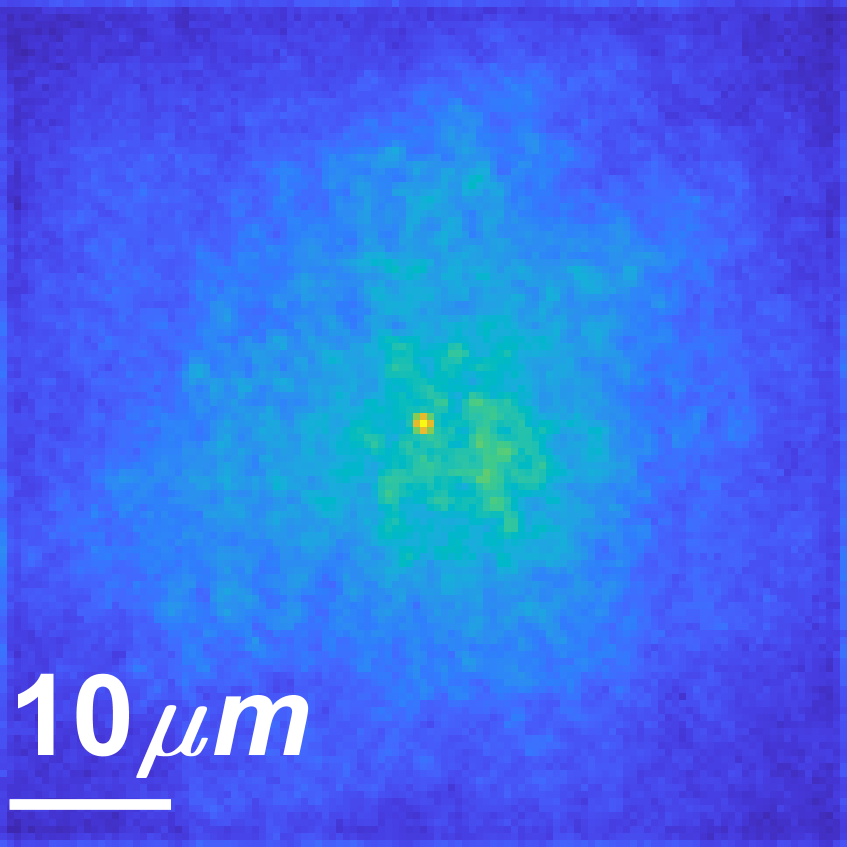}&		
			\includegraphics[height= 0.13\textwidth]{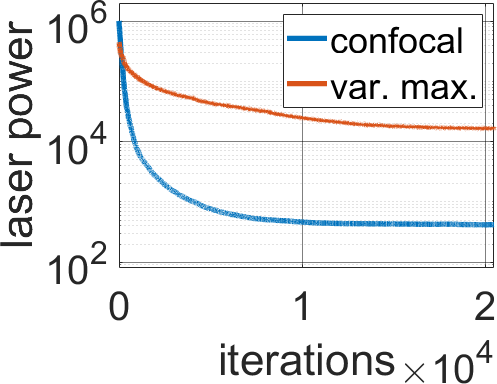}&		
			\includegraphics[width= 0.13\textwidth]{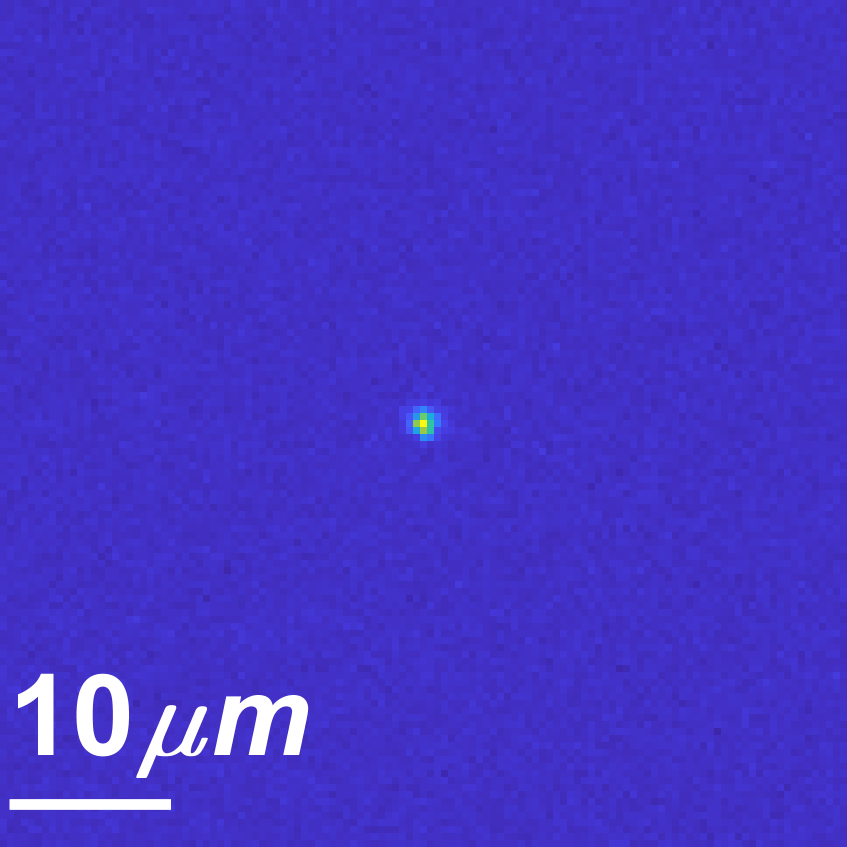}&		
			\includegraphics[height= 0.13\textwidth]{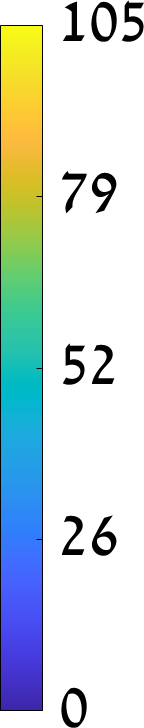}&		
			\includegraphics[width= 0.13\textwidth]{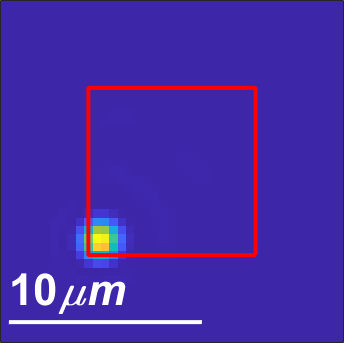}&
			\includegraphics[height= 0.13\textwidth]{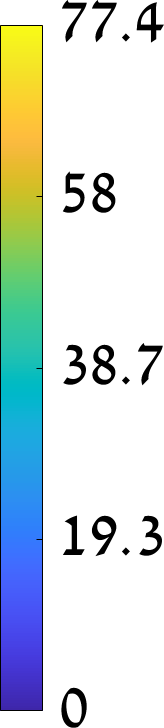}&\multirow[c]{1}{*}[1.6 cm]{\rotatebox[origin=c]{90}{Confocal}}\\
			&&&&
			\includegraphics[width= 0.13\textwidth]{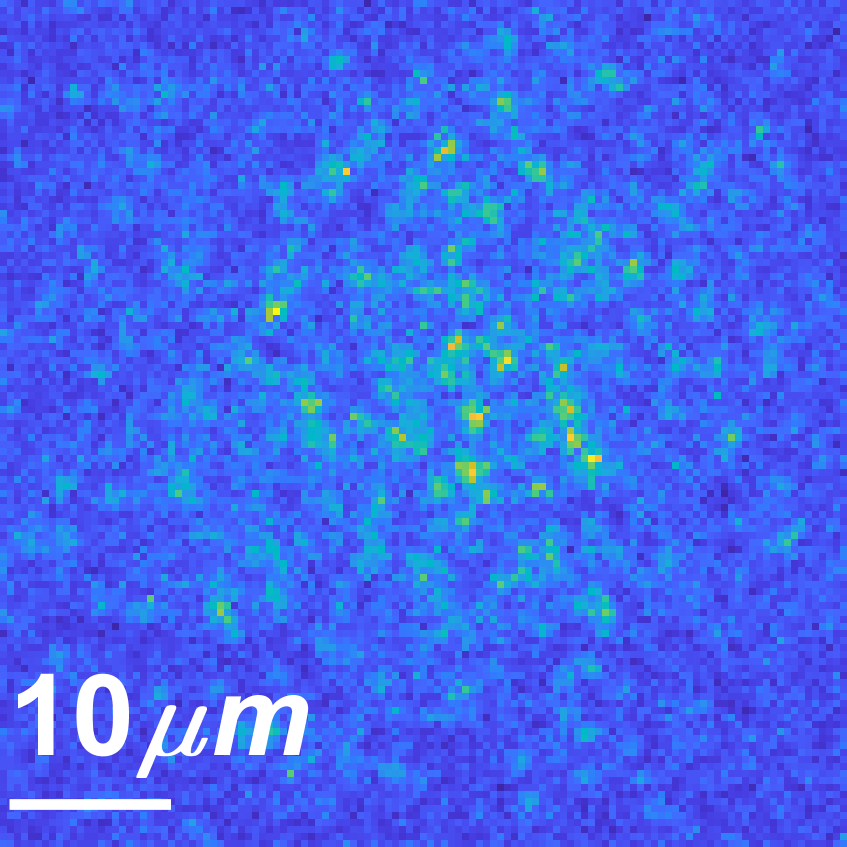}&		
			\includegraphics[height= 0.13\textwidth]{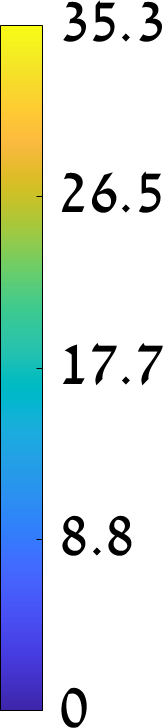}&		
			\includegraphics[width= 0.13\textwidth]{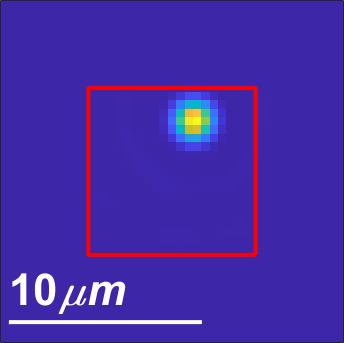}&
			\includegraphics[height= 0.13\textwidth]{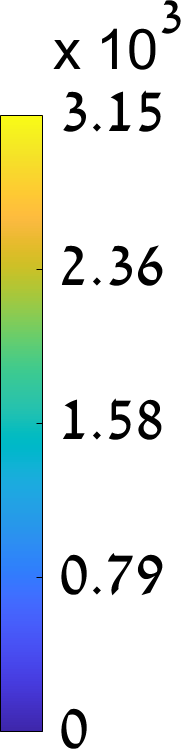}&\multirow[c]{1}{*}[1.7 cm]{\rotatebox[origin=c]{90}{Var. max.}}
			\\
		
		\end{tabular}
		\caption{\blue{Comparing the variance maximization score against the confocal score. During optimization we reduce the laser power so that the SNR of the captured images remains fixed. Using the confocal score, as the estimated modulation improves, more photons are brought into a single pixel and hence we can capture a high SNR image with a weaker laser power, reducing photo-bleaching. 
			(a) Fluorescent target from the validation camera. (b) Scattering by a single spot (one row of the transmission matrix), illustrating the speckle spread. (c) Input image from the main camera. (d) Laser power at each  iteration in both algorithms. (e) Final image from the main camera. The confocal algorithm results in a spot, and the variance maximization algorithm results in a speckle pattern. (f) Final image from the validation camera where both algorithms excite a single spot. The original area of the fluorescent target is marked with a square. The position of the focused spot inside the fluorescent area can vary.
			The different rows simulate 3 different speckle supports. When scattering is wider the advantage of the confocal score is more dominant, whereas for the smallest support we require $ 5.8\times$ fewer photons and for the wider one $43\times$. }		}\label{fig:compare-spk-var-conf}
	\end{center}
\end{figure*}

\begin{figure*}[t!]
	\begin{center}
		\begin{tabular}{@{}c@{~~~~}c@{~}c@{~~}c@{~}c@{~~}c@{~}c@{~~}c@{~}c@{~~}c@{}}
			& \multicolumn{3}{c}{Short range ME}&&\multicolumn{3}{c}{Wide range ME}&&\\
			
			(a)Ground truth&(b)Non-local&&(c)Confocal&&(d)Non-local&&(e)Confocal&&\\

			\includegraphics[width= 0.15\textwidth]{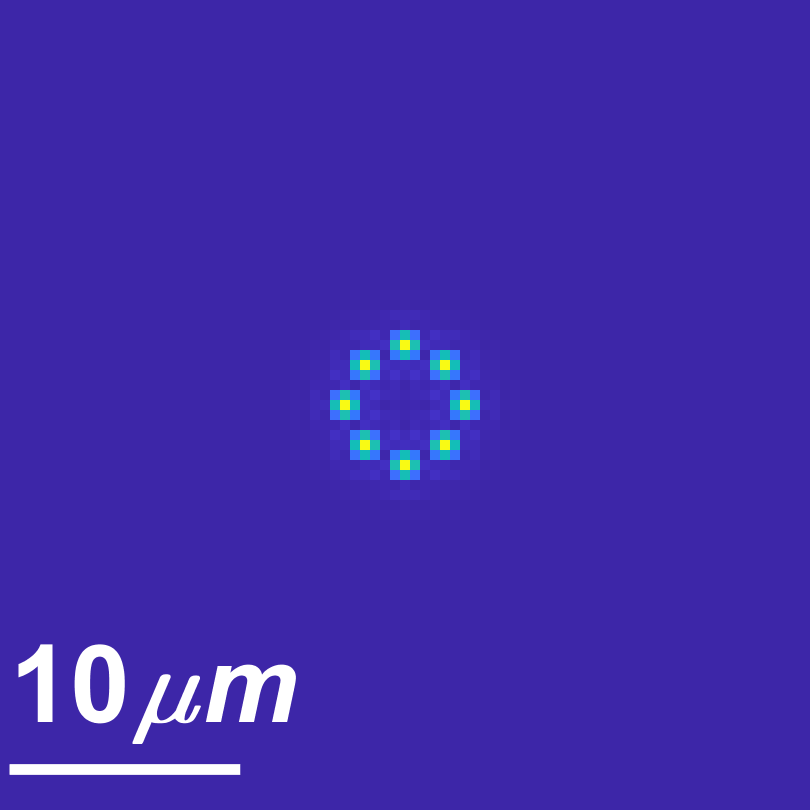}&
			\includegraphics[width= 0.15\textwidth]{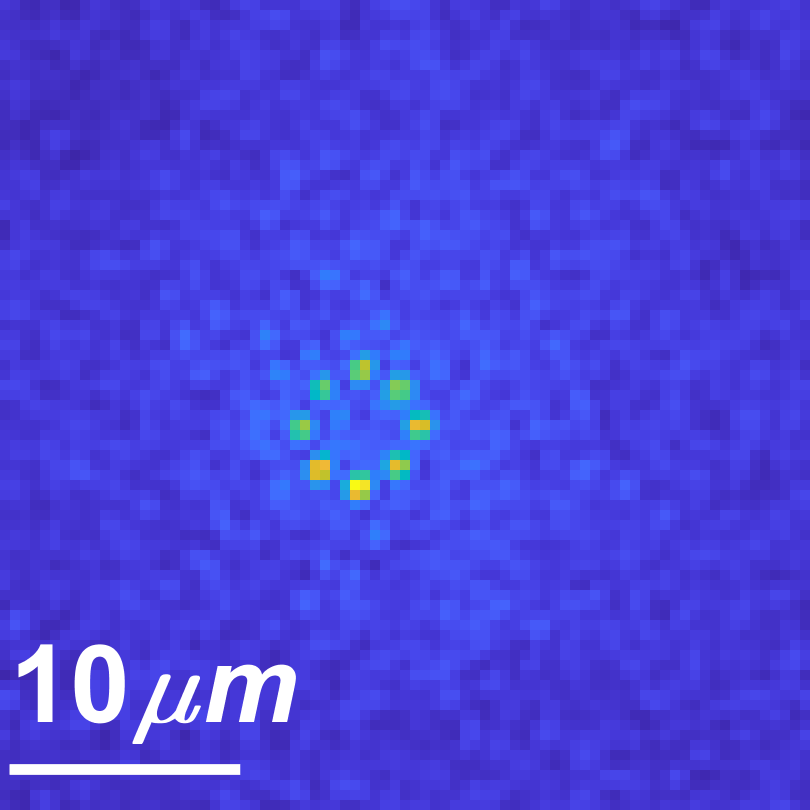}&
			\includegraphics[height= 0.15\textwidth]{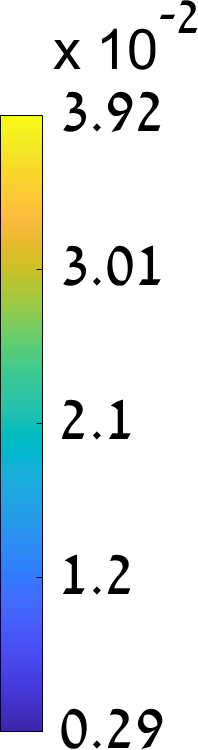}&	
			\includegraphics[width= 0.15\textwidth]{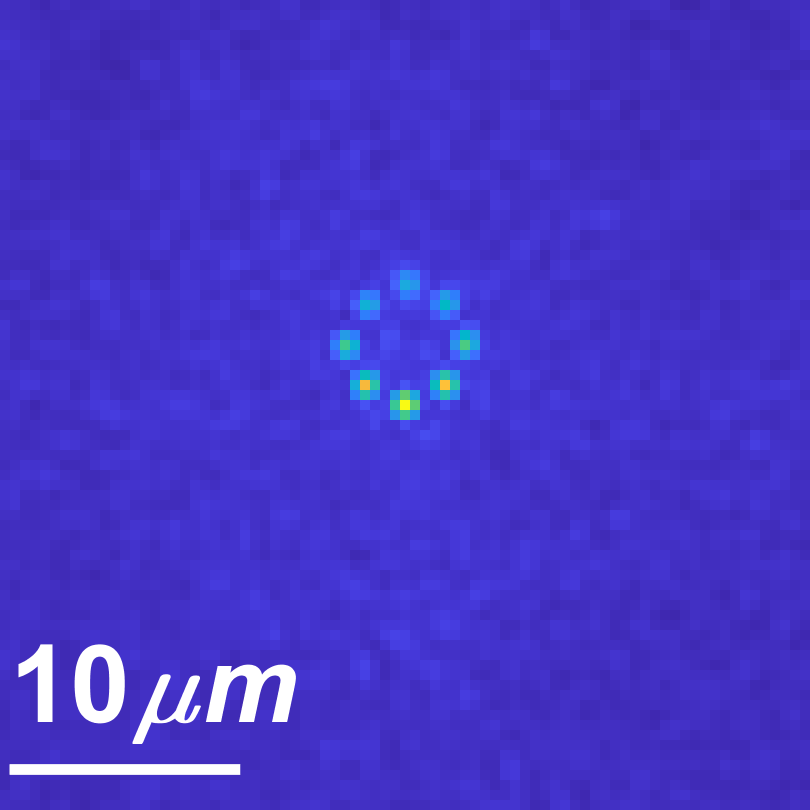}&
			\includegraphics[height= 0.15\textwidth]{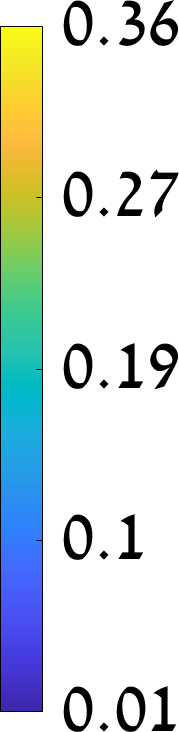}&			
			\includegraphics[width= 0.15\textwidth]{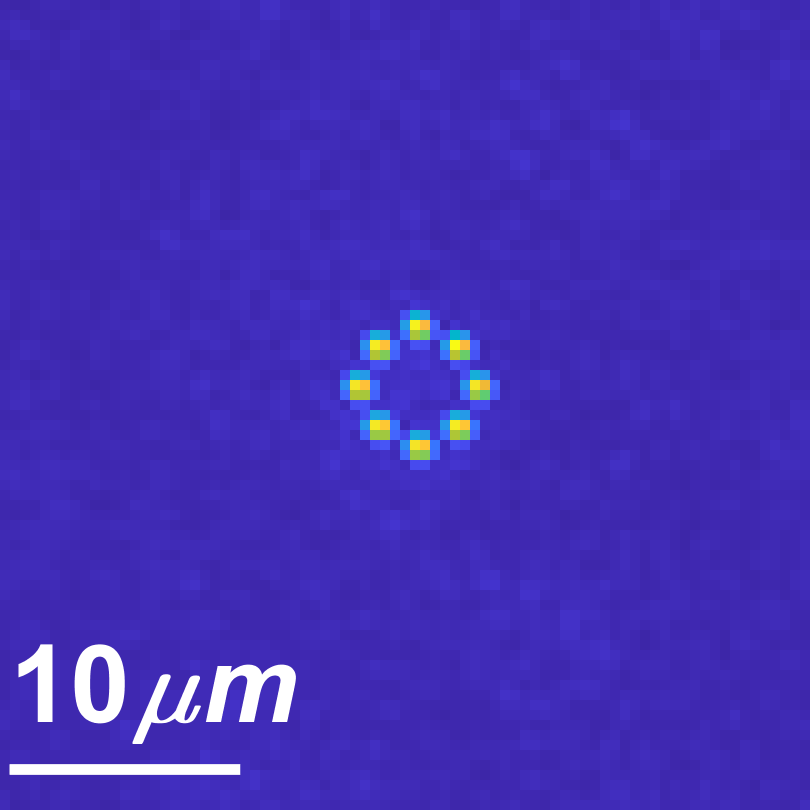}&
			\includegraphics[height= 0.15\textwidth]{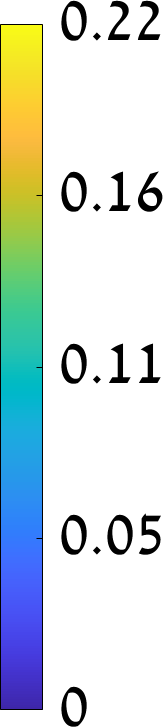}&		
			\includegraphics[width= 0.15\textwidth]{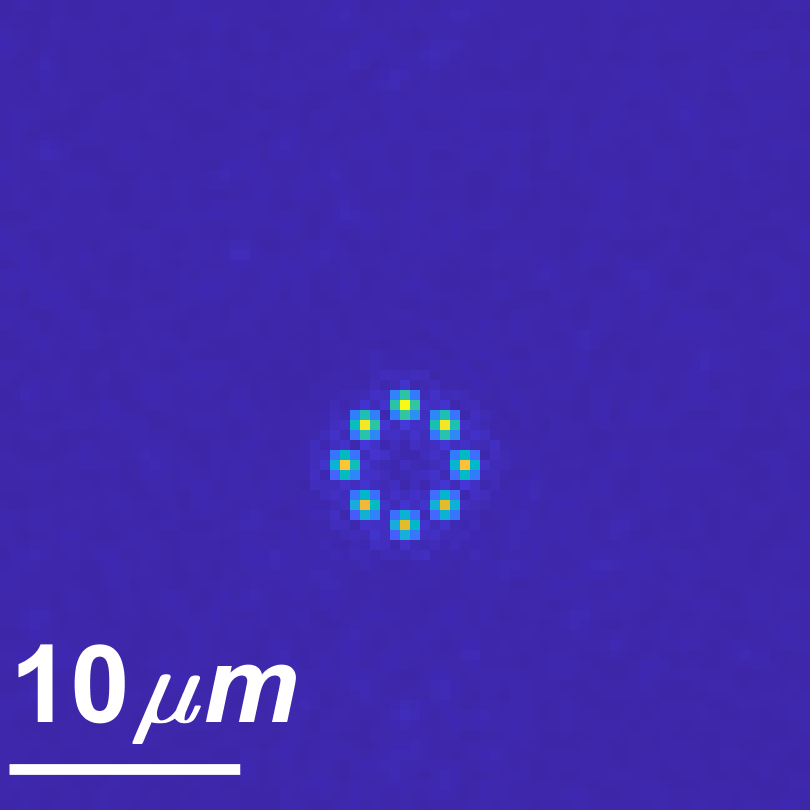}&
			\includegraphics[height= 0.15\textwidth]{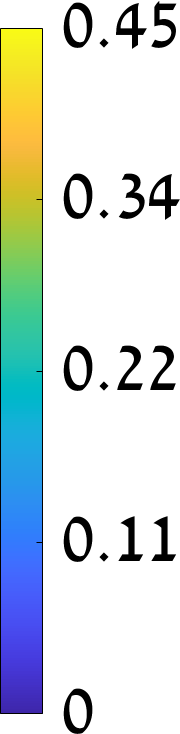}&	
			\multirow[c]{1}{*}[2.1 cm]{\rotatebox[origin=c]{90}{Main camera}}
			\vspace{0.1cm}
			\\

			&\includegraphics[width= 0.15\textwidth]{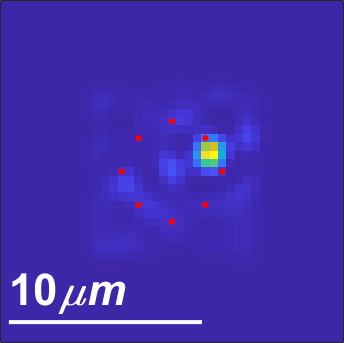}&
			\includegraphics[height= 0.15\textwidth]{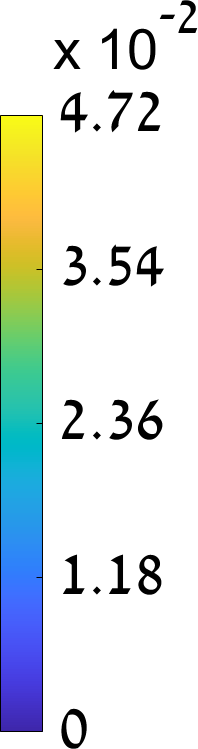}&	
			\includegraphics[width= 0.15\textwidth]{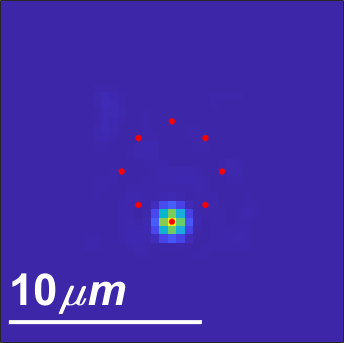}&
			\includegraphics[height= 0.15\textwidth]{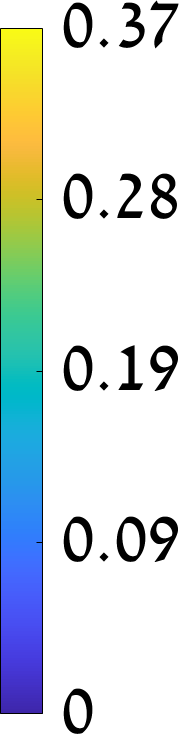}&	
			\includegraphics[width= 0.15\textwidth]{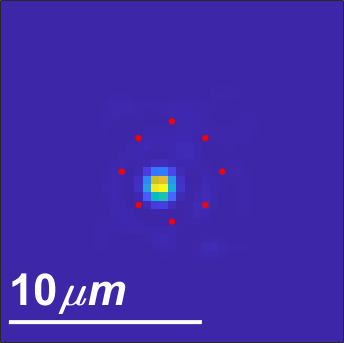}&
			\includegraphics[height= 0.15\textwidth]{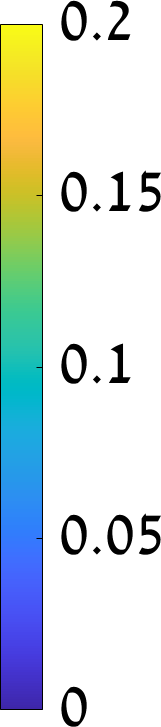}&	
			\includegraphics[width= 0.15\textwidth]{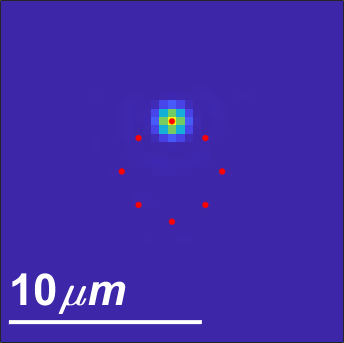}&
			\includegraphics[height= 0.15\textwidth]{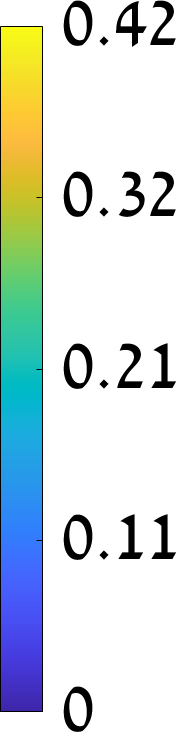}&
			\multirow[c]{1}{*}[2.1 cm]{\rotatebox[origin=c]{90}{Valid. camera}}\vspace{0.1cm}\\

			\includegraphics[width= 0.15\textwidth]{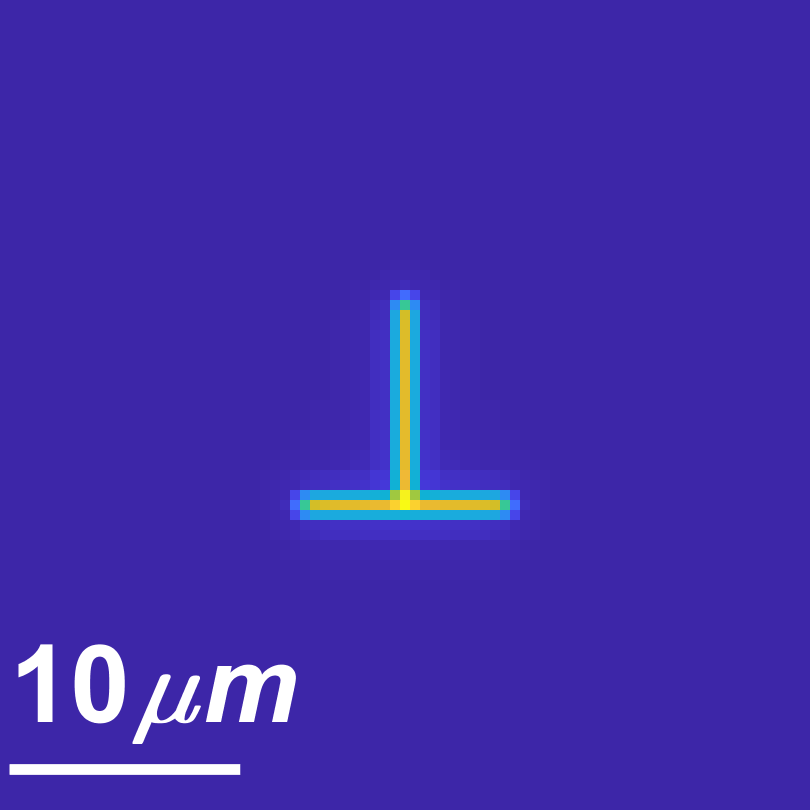}&
			\includegraphics[width= 0.15\textwidth]{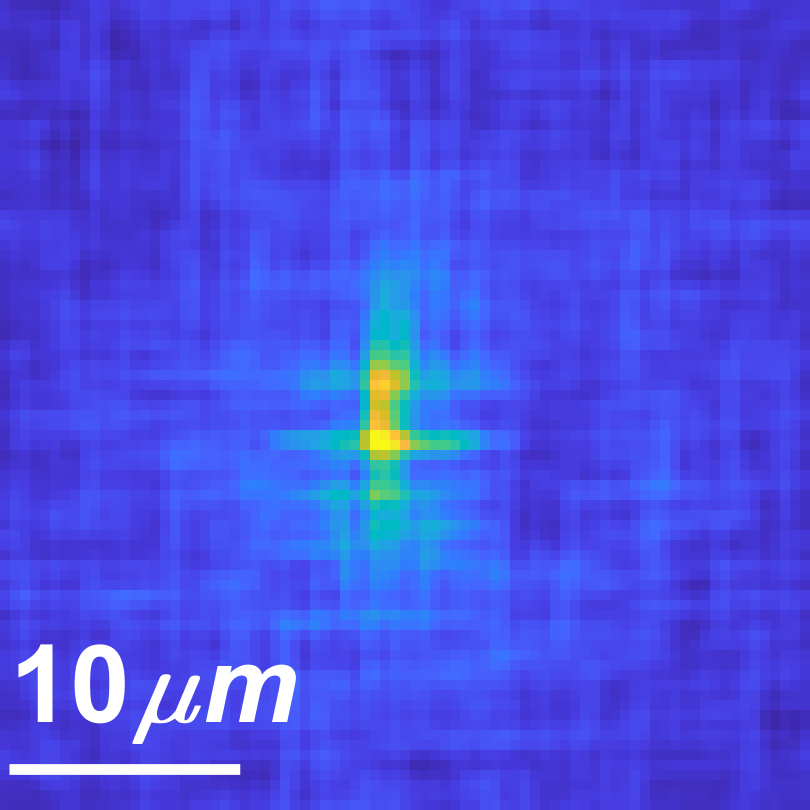}&
			\includegraphics[height= 0.15\textwidth]{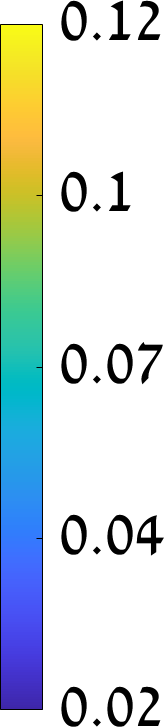}&	
			\includegraphics[width= 0.15\textwidth]{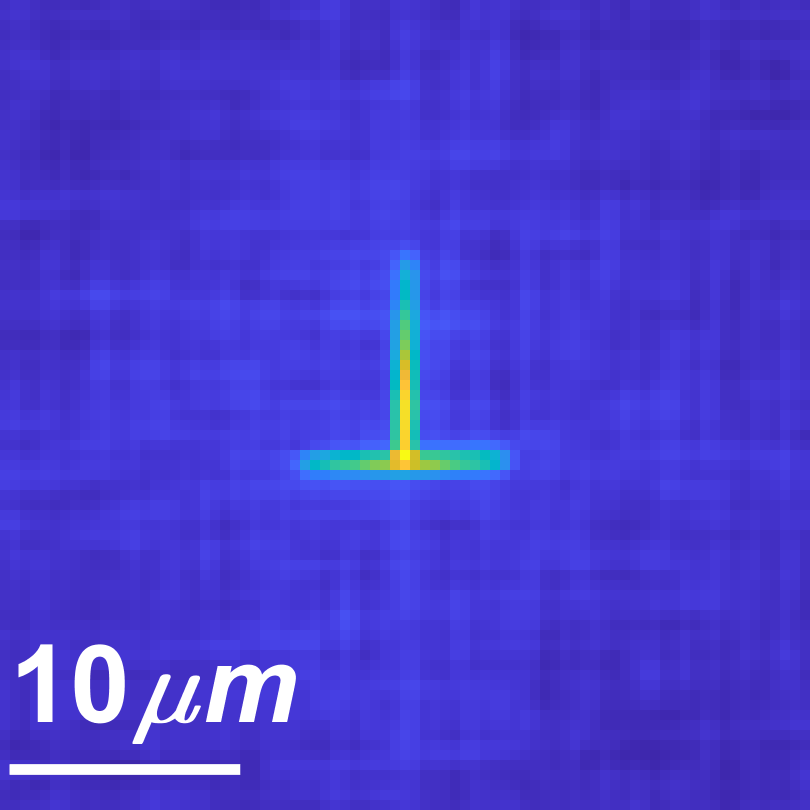}&
			\includegraphics[height= 0.15\textwidth]{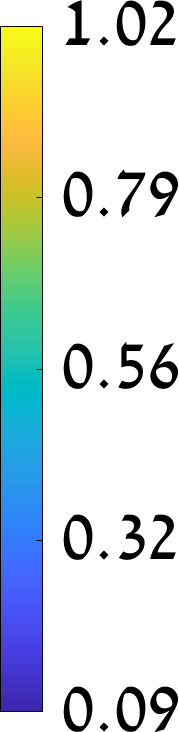}&			
			\includegraphics[width= 0.15\textwidth]{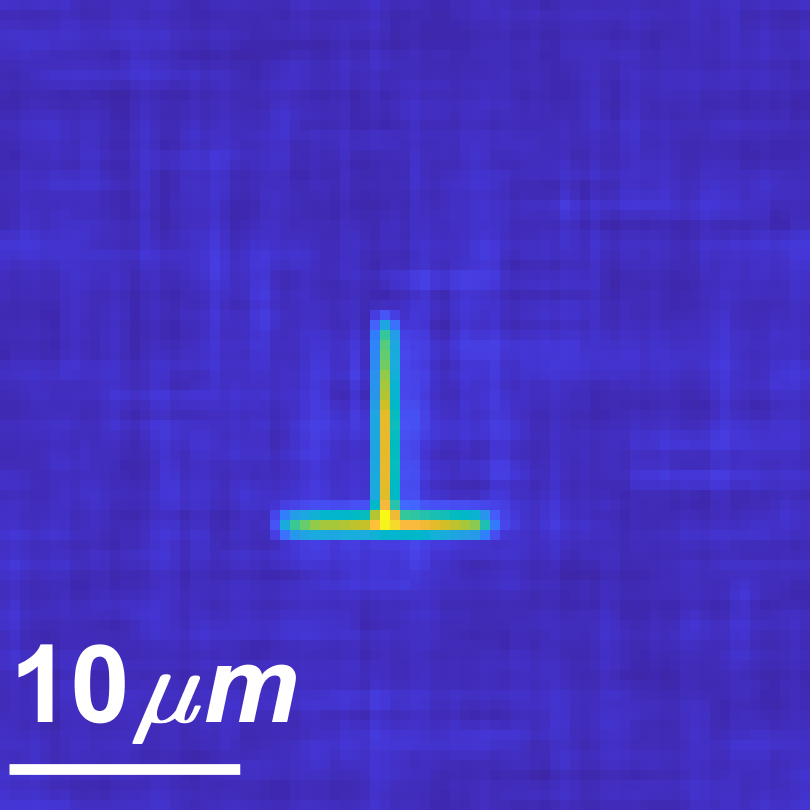}&
			\includegraphics[height= 0.15\textwidth]{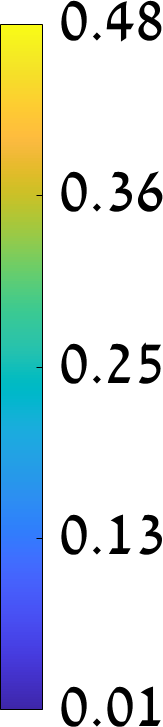}&		
			\includegraphics[width= 0.15\textwidth]{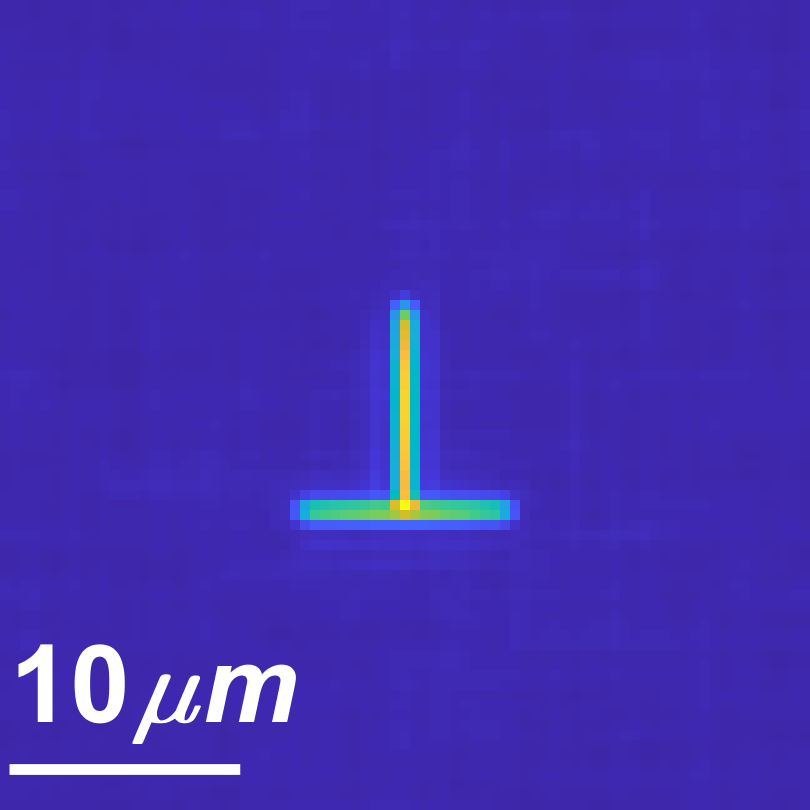}&
			\includegraphics[height= 0.15\textwidth]{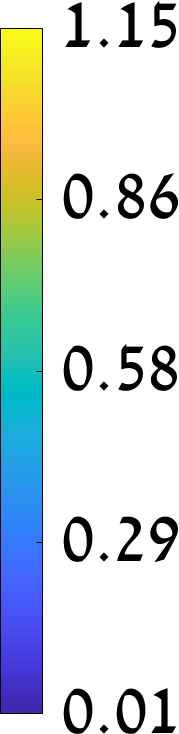}&
			\multirow[c]{1}{*}[2.1 cm]{\rotatebox[origin=c]{90}{Main camera}}\vspace{0.1cm}\\

			&\includegraphics[width= 0.15\textwidth]{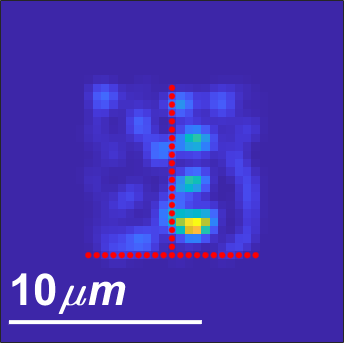}&
			\includegraphics[height= 0.15\textwidth]{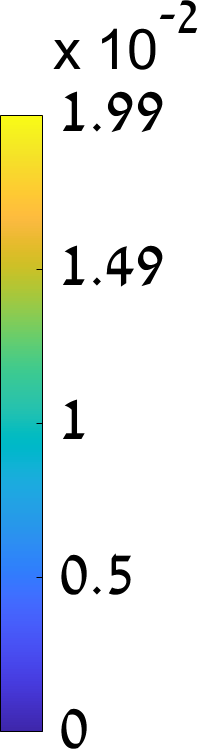}&	
			\includegraphics[width= 0.15\textwidth]{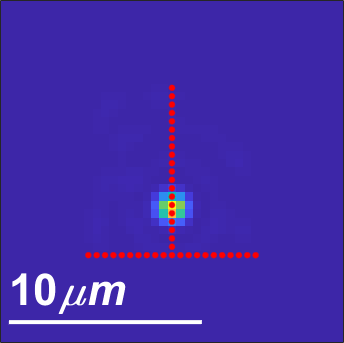}&
			\includegraphics[height= 0.15\textwidth]{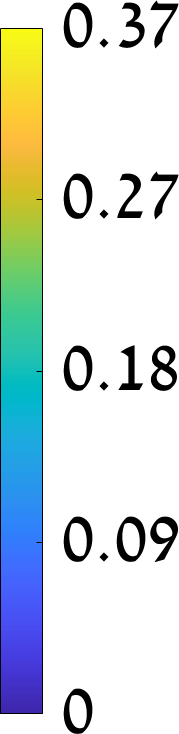}&	
			\includegraphics[width= 0.15\textwidth]{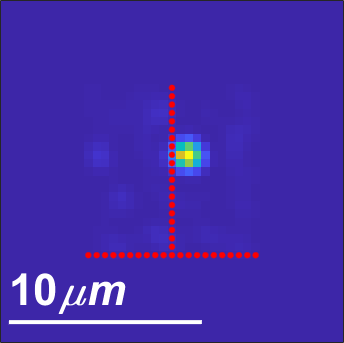}&
			\includegraphics[height= 0.15\textwidth]{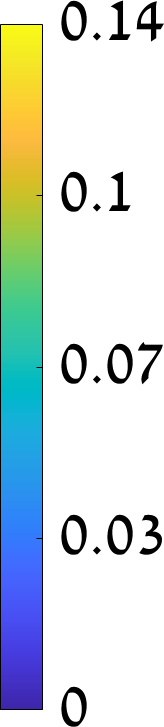}&	
			\includegraphics[width= 0.15\textwidth]{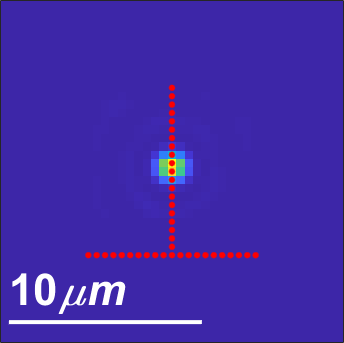}&
			\includegraphics[height= 0.15\textwidth]{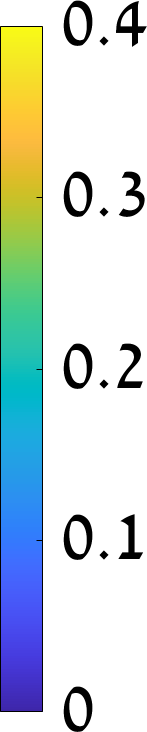}&
			\multirow[c]{1}{*}[2.1 cm]{\rotatebox[origin=c]{90}{Valid. camera}}	\vspace{0.1cm}\\\\
		
			&\multicolumn{3}{c}{\includegraphics[height= 0.15\textwidth]{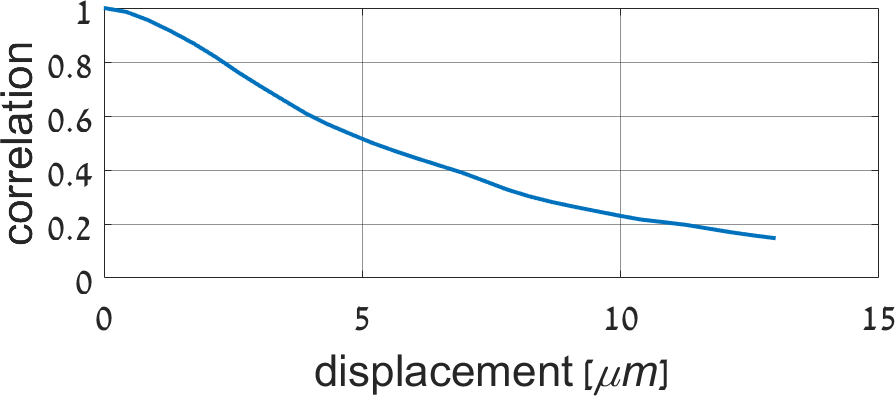}}&
			&\multicolumn{3}{c}{\includegraphics[height= 0.15\textwidth]{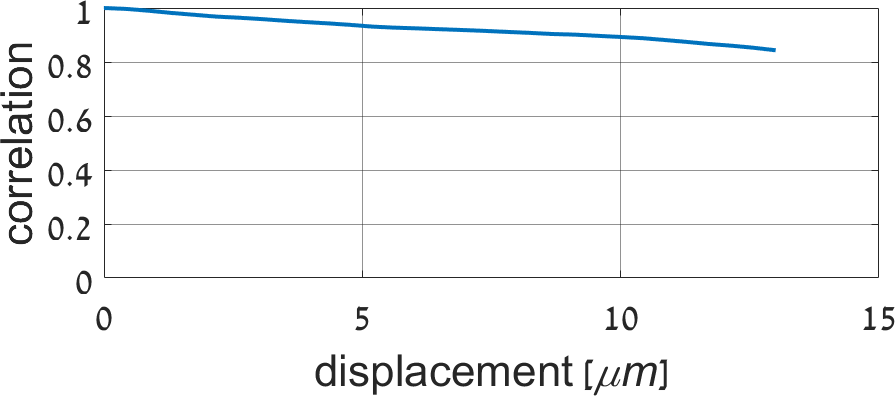}}&
			&\multirow[c]{1}{*}[2.2 cm]{\rotatebox[origin=c]{90}{Mem. effect}}\\

		\end{tabular}
		\caption{\blue{Comparison against the non-local score of~\cite{YeminyKatz2021}. (a) The ground truth fluorescent target. (b-c) Results using a transmission matrix with short range memory effect. In this case a single correction cannot explain the full image and the non-local approach leads to degraded results. (e-f) Results in the presence of long-range memory effect, where the non-local approach is successful. We simulate two fluorescent targets, the top one is sparser and easier to handle and the lower one is denser. In each example the top row shows images from the main camera under a wide illumination (for the confocal algorithm, we run it until convergence and then use a wide illumination while only correcting the imaging arm). In the lower row we use the recovered modulation in the imaging arm and show a view from the validation camera. If the modulation is good it should bring all light into a single sharp spot. The overlaid red dots illustrate the actual position of the fluorescent target. The lowest row shows a plot of the decay of memory effect correlation in each of the transmission matrices.} 	}\label{fig:non-local-score-compare}
	\end{center}
\end{figure*}

 \subsection{Speckle-variance maximization}
 \blue{As mentioned in the main paper both our confocal score and the variance maximization score seek to optimize the same non-linear function of fluorescent intensity, yet they are not equivalent in terms of SNR. 
 To demonstrate this we run both algorithms, and in each iteration, as we update the modulation we also vary the laser power such that the SNR of the score we measure in each algorithm will be kept fixed. Since our algorithm attempts to bring all photons to one sensor pixel, the images we measure are less noisy then speckle-variance score (which does not use a correction on the emitted speckles).  Alternatively, we can achieve the same SNR as the speckle variance score with a weaker excitation power. }
 
 \blue{\figref{fig:compare-spk-var-conf}(d) demonstrates the curve of laser power as a function of iteration number in each algorithm. One can see that our algorithm can work with a significantly lower number of photons, meaning that the algorithm has a much better chance to converge without much photo-bleaching. 
We have repeated the evaluation with three different transmission matrices, where we changed the statistics of the aberration layers to be less forward scattering, hence the speckle support is wider. We can see that for a wider aberration the gain of the confocal score is higher (compare the different rows of \figref{fig:compare-spk-var-conf}, for the small support we require $ 5.8\times$  fewer photons, for the wide one we requiere $43\times$ fewer photons). This is because for a wider speckle pattern the speckles are spread between more pixels, hence the images are noisier and estimating the speckle variance without modulating the emission path is harder.}

\subsection{Non-local scores}

\blue{Next, we compare against a recent non-local approach by~\cite{YeminyKatz2021}. This  assumes that a single modulation can correct a wide image region rather than a single spot. It uses a wide illumination and only correct the imaging arm, where a good modulation should lead to a sparse image,  measured using a   combination of a modified entropy and a variance maximization scores}. 

\blue{In \figref{fig:non-local-score-compare} we evaluate this approach using two different transmission matrices. In the first case we used a transmission matrix generated by a single aberration layer. In this case the memory effect correlation holds over a wide extent, and a single modulation can correct a wide iso-planatic region. In the second case we simulated a transmission matrix with $3$ different aberration layers equally spaced between depth $0$ to $250\mu m$. This is a more realistic approximation of a scattering tissue exhibiting volumetric aberration, but the extent of the memory effect is much shorter. In this case there is no single modulation which can fully  correct the entire image. 
\figref{fig:non-local-score-compare} shows a comparison of this algorithm against our confocal score with two structures of hidden neurons. 
When memory-effect exists over a wide extent  the non-local score of ~\cite{YeminyKatz2021} can indeed recover good modulations, but the quality of the results degrades when ME range is short, and the size of the iso-planatic patches that can be corrected with a single modulation is small.}

\section{Tilt-shift correction}\label{sec:tilt-shift}
\begin{figure}[t!]
	\begin{center}
		\begin{tabular}{c}
			\begin{tabular}{ccc}
				\includegraphics[width= 0.13\textwidth]{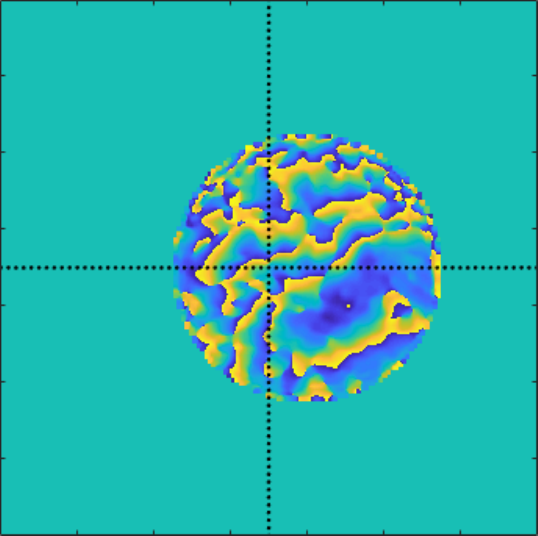} &
				\includegraphics[width= 0.13\textwidth]{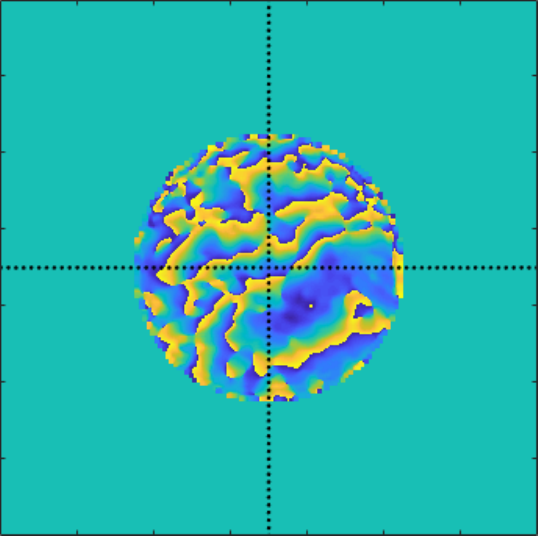} &
				\includegraphics[width= 0.13\textwidth]{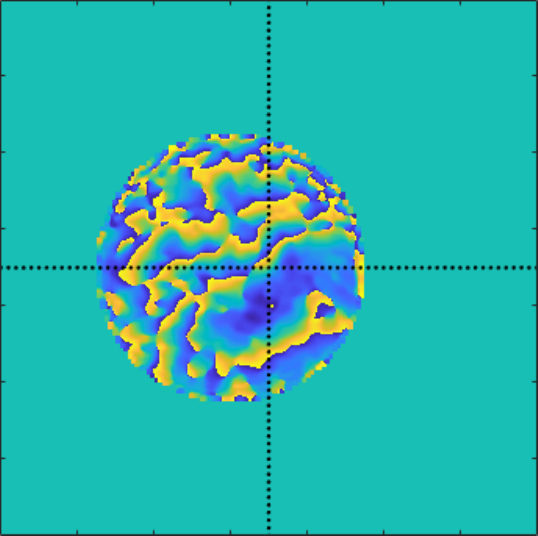}\\
				\includegraphics[width= 0.13\textwidth]{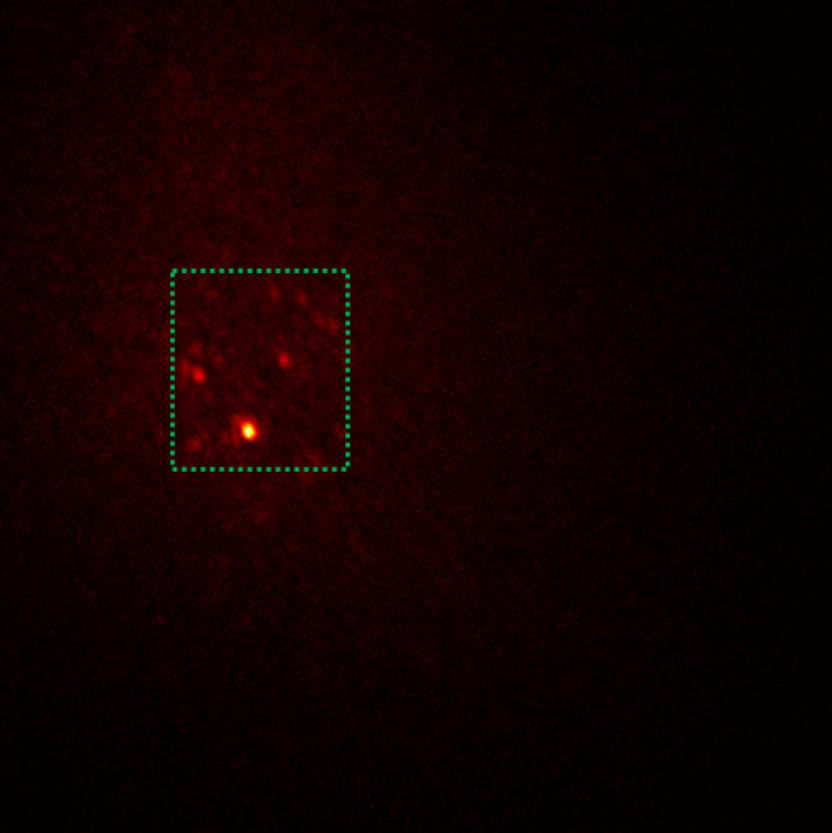} &
				\includegraphics[width= 0.13\textwidth]{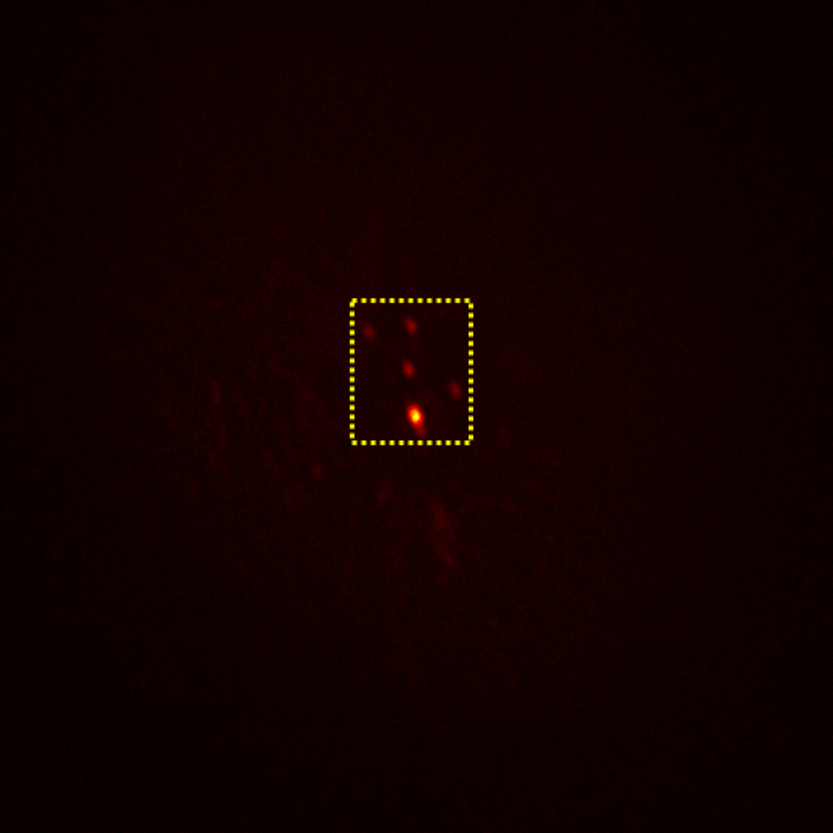} &
				\includegraphics[width= 0.13\textwidth]{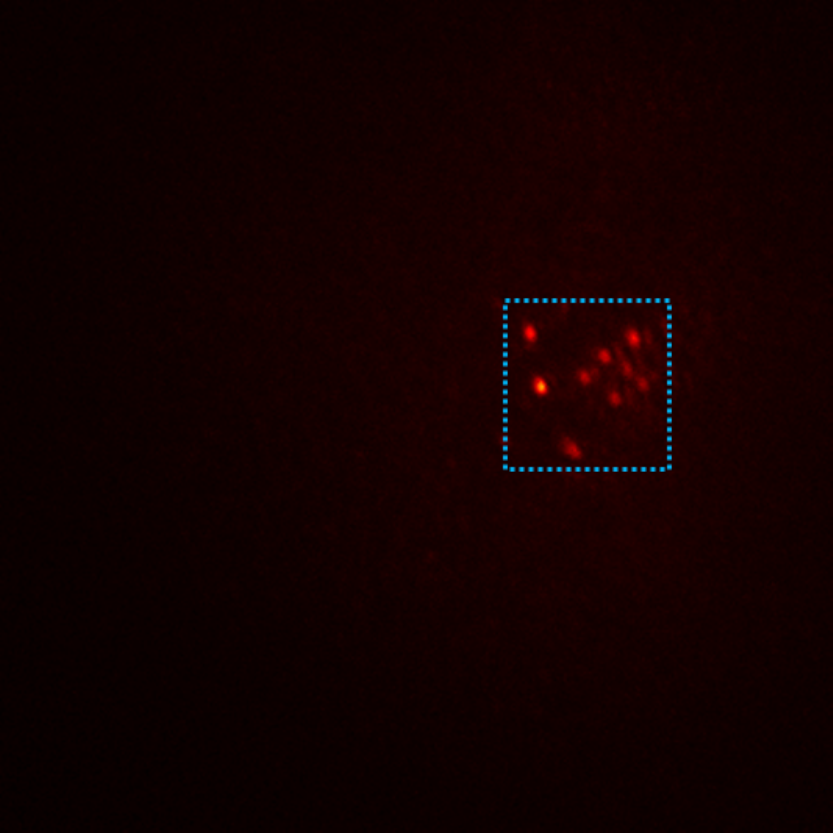}\\
				\multicolumn{3}{c}{\footnotesize{Three shifts of  correction pattern and the resulting images.}}
			\end{tabular}
			\\\\
			\begin{tabular}{cc}
				\includegraphics[width= 0.13\textwidth]{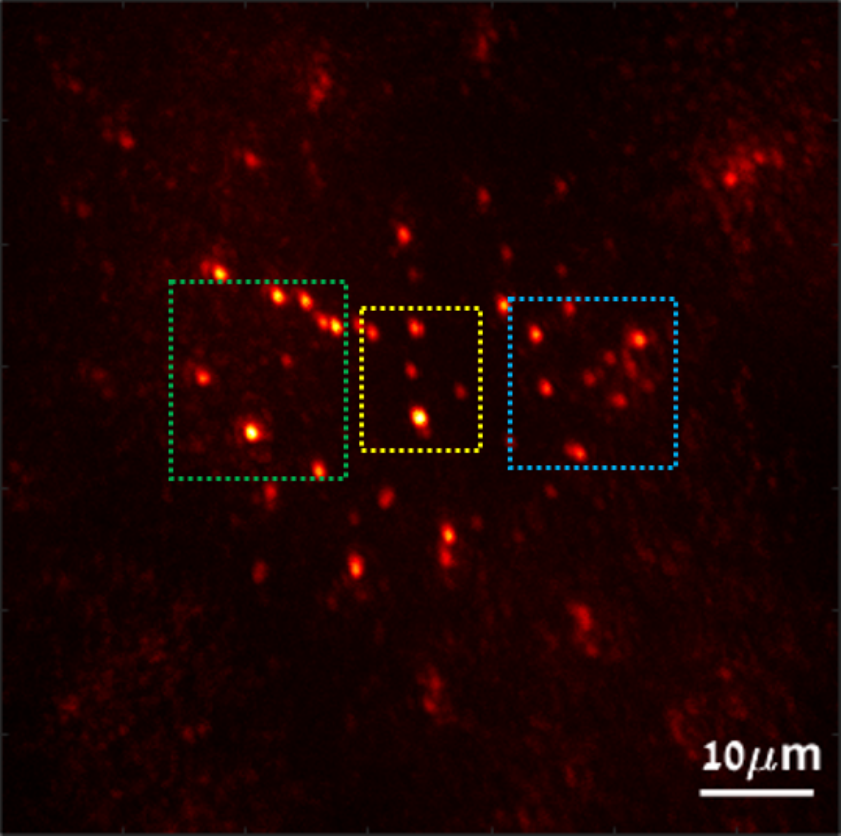} &
				\includegraphics[width= 0.13\textwidth]{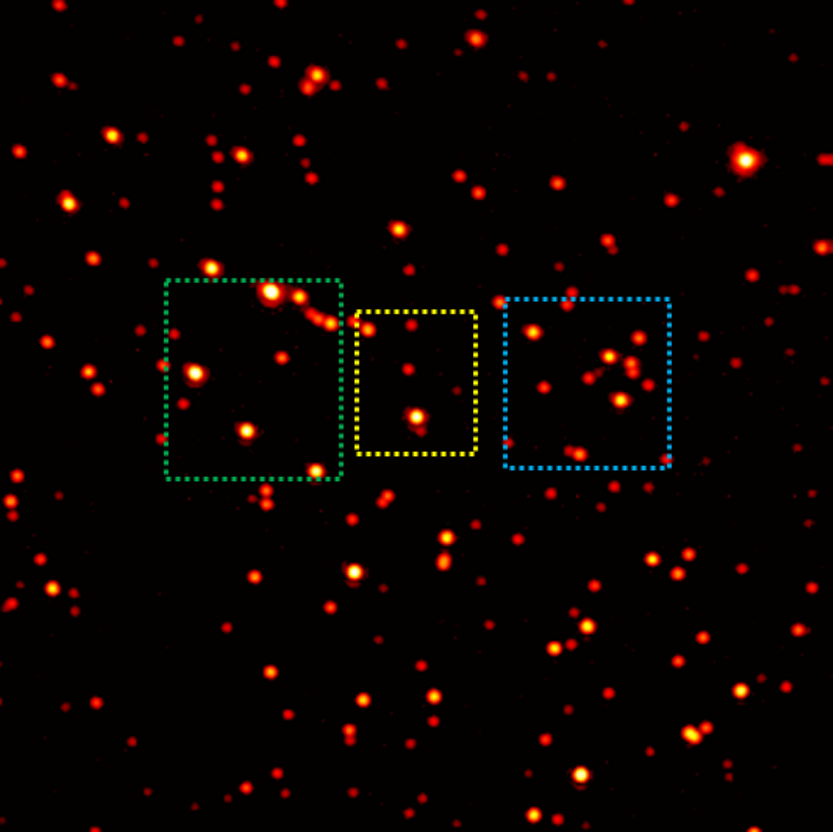}\\
				\footnotesize{Combined reconstruction} & \footnotesize{Reference}
			\end{tabular}
		\end{tabular}
					\captionof{figure}
			{Using the tilt-shift memory effect to see a wide area behind the tissue. The top row demonstrates three different shifts of the recovered correction pattern. The second row demonstrates the image we capture by placing this shifted mask on the SLM of the imaging arm. Each shift allows us to see a different sub-region of fluorescent sources. By merging $21\times 21$ such shifts we get the wider image in the lowest row.  We  compare this reconstruction against the reference from the validation camera.
		}\label{fig:tilt_shift}
	\end{center}
\end{figure}

Below we explain the acquisition of wide area images of \figpref{fig:big_area_sup_chicken}{fig:big_area_sup_parafilm}. 
Given a wavefront shaping modulation that applies to one fluorescent particle inside the tissue sample, we correct nearby ones using the tilt-shift memory effect. 
For that, we denote by $u^{\ptd_1}_x,u^{\ptd_2}_x$ two speckle fields obtained on the sensor plane of our main camera (where $x$ denotes spatial position on this plane), generated by fluorescent particles at $\ptd_1,\ptd_2$. We focus the objective such that the sensor plane is conjugate to the plane containing the fluorescent sources. The tilt-shift memory effect~\cite{osnabrugge2017generalized,single-sct-iccp-21} implies that, for small displacements, $u^{\ptd_1}$ is correlated with a tilted and shifted version of $u^{\ptd_2}$ and thus can be approximated as:
\BE\label{eq:tilt-shift-adj}
u^{\ptd_1}_x\approx u^{\ptd_2}_{x+\Dl}e^{ik \alpha <\Dl, x>}
\EE
with $\Dl=\ptd_2-\ptd_1$ the displacement between the sources. 
If there was no tilt, and the speckle at the image plane could be explained by pure shift, placing in the Fourier plane the Fourier transform of $u^{\ptd_1}_x$ would correct the emission from $\ptd_1$ and the emission from nearby points $\ptd_2$.  Given the tilt, the Fourier correction for $\ptd_2$ should be a shifted version of the Fourier correction of $\ptd_1$. 
To account for this, we place in the Fourier plane of our imaging arm shifted versions of our recovered mask. \figref{fig:tilt_shift} illustrates that using a fluorescent beads target. Each shift allows us to see the fluorescent particles in a different local region. By scanning multiple shifts of the modulation mask, we construct a wider image of the fluorescent particles inside the tissue sample, as shown in the last row of \figref{fig:tilt_shift}.

\section{Calibration and alignment}\label{sec:calibration}
Below we elaborate on various calibration and alignment details.

First, to correctly modulate the Fourier transform of the wave, the  illumination SLM needs to be at the focal plane of the lens right after it ($L3$ in the system figure), and the imaging SLM at the focal plane of the lens before it ($L6$). We do this alignment  using another camera focused at infinity. We use this camera to view the SLM through the relevant lens, forming a relay system. We adjust the distance between the SLM and the lens  until the calibration camera can see a sharp image of the SLM plane. We also ensure that the distance between the sensor of the main/validation cameras  and the lenses $L7/L8$ attached to them is set such that the cameras focus at infinity.

A second step of the alignment is to focus the excitation laser and the system camera on the same target plane. 
	In our setup the sample and the objective of the validation camera are mounted on two motorized z-axis (axial) translation stages.
	We  use fluorescent beads with no aberrating tissue and adjust the axial distance between the beads and the objective of the main camera (Obj1 in the setup figure) such that the main camera sees a sharply focused image of the bead. Then we adjust the distance  of the
  validation objective (Obj2 in the setup figure) from the beads so that we see a sharp image of the same beads in the validation camera. We then  want the laser to generate its sharpest spot on the same plane. Assuming the validation and main camera are focused at the same plane,  we adjust the position of the  lens $L4$ until the validation camera sees a sharp laser spot. 

	After the system has been aligned, we need to determine two mappings. The first one is between frequencies to pixels on the SLM. A second, more challenging one  is the registration between the two SLMs,  so that we can map a pixel on the imaging SLM to a pixel on the illumination SLM controlling the same frequency. We start with a mapping between frequencies to the SLM on the imaging arm. We first put a calibration camera that can image the camera SLM plane directly when it receives laser light. Since the SLM is conjugate to the aperture of the objective we see an illuminated circle on the SLM plane, corresponding to the numerical aperture of the imaging system. The center of this circle gives us a first estimate of the zero (central) frequency of the Fourier transform. Assuming we know the focal length of $L6$, the SLM pitch and the wavelength of the emitted light, we can map frequencies to SLM pixels using simple geometry.  Alternatively we can display on the SLM sinusoidals  of various frequencies. This shifts the image on the sensor plane. By measuring the shift resulting from each sinusoidal we can calibrate the mapping between frequencies and SLM pixels. 
To align between the two SLMs we place a fluorescent bead behind a scattering tissue. We capture a few images of the speckles resulting from this bead (at emission wavelength) and use a phase diversity approach \cite{MUGNIER20061} to estimate the complex wavefront that emerges from this bead. The Helmholtz reciprocity (phase conjugation) principle  states that if the conjugate of this wavefront is placed on the illumination SLM, it will focus into a point behind the tissue. 
However, we need to determine how to position this modulation on the  SLMs keeping in mind that tilt and shift on these planes may impact the results.
For the imaging SLM this is less of an issue because we have already marked the zero frequency and because  a
 tilt of the imaging SLM only shifts the position of the spot on the sensor.  
 However, if the illumination SLM is not registered correctly, we may  see a sharp spot behind the tissue, but it will be shifted from the bead of interest and will not excite it.
 Thus, we  tilt and shift the modulation on the illumination SLM   until
  the intensity we measure on the main camera (when the modulation correction is on) is maximized. After this is achieved we can fine-tune the shift  on the imaging SLM, which is equivalent to the  position of the zero (central) frequency that we have previously marked by looking at the illuminated circle.

To use a recovered modulation pattern to image a larger region of the fluorescent target, we leverage the tilt shift memory effect. To apply the scan we need to recover the parameter $\alpha$ of \equref{eq:tilt-shift-adj}, determining the ratio between the tilt and shift. For that, after we recover the modulation pattern, we place it on the illumination SLM and use the validation camera to view the focused spot. 
	We then adjust the ratio between tilt and shift of the modulation pattern so that we can move the focused spot in the validation camera, while preserving maximal intensity.

	%
	
	\bibliographystyle{unsrt}	
	\bibliography{biblio_anat,proposal}

\begin{thebibliography}{10}

\bibitem{Booth2014}
Martin Booth.
\newblock Adaptive optical microscopy: The ongoing quest for a perfect image.
\newblock {\em Light: Science and Applications}, 3:e165, 04 2014.

\bibitem{Ji2017review}
Na~Ji.
\newblock Adaptive optical fluorescence microscopy.
\newblock {\em Nature Methods}, 14:374--380, 03 2017.

\bibitem{HampsonBooth21review}
Karen Hampson, Raphael Turcotte, Donald Miller, Kazuhiro Kurokawa, Jared Males,
  Na~Ji, and Martin Booth.
\newblock {\em Nature Reviews Methods Primers}, 1:68, 10 2021.

\bibitem{Horstmeyer15}
Roarke Horstmeyer, Haowen Ruan, and Changhuei Yang.
\newblock Guidestar-assisted wavefront-shaping methods for focusing light into
  biological tissue.
\newblock {\em Nature Photonics}, 2015.

\bibitem{YU2015632}
Hyeonseung Yu, Jongchan Park, KyeoReh Lee, Jonghee Yoon, KyungDuk Kim, Shinwha
  Lee, and YongKeun Park.
\newblock Recent advances in wavefront shaping techniques for biomedical
  applications.
\newblock {\em Current Applied Physics}, 15(5):632--641, 2015.

\bibitem{Gigan22}
Sylvain Gigan, Ori Katz, et~al.
\newblock Roadmap on wavefront shaping and deep imaging in complex media.
\newblock {\em arXiv preprint arXiv:2111.14908}, 2021.

\bibitem{Vellekoop:07}
I.~M. Vellekoop and A.~P. Mosk.
\newblock Focusing coherent light through opaque strongly scattering media.
\newblock {\em Opt. Lett.}, 2007.

\bibitem{Yaqoob2008}
Zahid Yaqoob, Demetri Psaltis, Michael Feld, and Changhuei Yang.
\newblock Optical phase conjugation for turbidity suppression in biological
  samples.
\newblock {\em Nature photonics}, 2008.

\bibitem{Vellekoop2010}
Ivo~M. Vellekoop, Aart Lagendijk, and Allard~P. Mosk.
\newblock Exploiting disorder for perfect focusing.
\newblock {\em Nature Photonics}, 2010.

\bibitem{Vellekoop2012}
Ivo~M. Vellekoop, Meng Cui, and Changhuei Yang.
\newblock Digital optical phase conjugation of fluorescence in turbid tissue.
\newblock {\em Applied Physics Letters}, 2012.

\bibitem{Conkey:12}
Donald~B. Conkey, Albert~N. Brown, Antonio~M. Caravaca-Aguirre, and Rafael
  Piestun.
\newblock Genetic algorithm optimization for focusing through turbid media in
  noisy environments.
\newblock {\em Opt. Express}, 2012.

\bibitem{PopoffPhysRevLett2010}
S.~M. Popoff, G.~Lerosey, R.~Carminati, M.~Fink, A.~C. Boccara, and S.~Gigan.
\newblock Measuring the transmission matrix in optics: An approach to the study
  and control of light propagation in disordered media.
\newblock {\em Phys. Rev. Lett.}, 2010.

\bibitem{chen20203PointTM}
Yujun Chen, Manoj~Kumar Sharma, Ashutosh Sabharwal, Ashok Veeraraghavan, and
  Aswin~C. Sankaranarayanan.
\newblock {3PointTM: F}aster measurement of high-dimensional transmission
  matrices.
\newblock In {\em Euro. Conf. Computer Vision (ECCV)}, 2020.

\bibitem{Tang2012}
Jianyong Tang, Ronald~N. Germain, and Meng Cui.
\newblock Superpenetration optical microscopy by iterative multiphoton adaptive
  compensation technique.
\newblock {\em Proceedings of the National Academy of Sciences},
  109(22):8434--8439, 2012.

\bibitem{Katz:14}
Ori Katz, Eran Small, Yefeng Guan, and Yaron Silberberg.
\newblock Noninvasive nonlinear focusing and imaging through strongly
  scattering turbid layers.
\newblock {\em Optica}, 1(3):170--174, Sep 2014.

\bibitem{Wang20142PAdaptive}
Chen Wang, Rui Liu, Daniel Milkie, Wenzhi Sun, Zhongchao Tan, Aaron Kerlin,
  Tsai-Wen Chen, Douglas Kim, and Na~Ji.
\newblock Multiplexed aberration measurement for deep tissue imaging in vivo.
\newblock {\em Nature methods}, 11, 08 2014.

\bibitem{Liu2018}
Tsung-Li Liu, Srigokul Upadhyayula, Daniel~E. Milkie, Ved Singh, Kai Wang,
  Ian~A. Swinburne, Kishore~R. Mosaliganti, Zach~M. Collins, Tom~W. Hiscock,
  Jamien Shea, Abraham~Q. Kohrman, Taylor~N. Medwig, Daphne Dambournet, Ryan
  Forster, Brian Cunniff, Yuan Ruan, Hanako Yashiro, Steffen Scholpp, Elliot~M.
  Meyerowitz, Dirk Hockemeyer, David~G. Drubin, Benjamin~L. Martin, David~Q.
  Matus, Minoru Koyama, Sean~G. Megason, Tom Kirchhausen, and Eric Betzig.
\newblock Observing the cell in its native state: Imaging subcellular dynamics
  in multicellular organisms.
\newblock {\em Science}, 360(6386):eaaq1392, 2018.

\bibitem{Fiolka:12}
Reto Fiolka, Ke~Si, and Meng Cui.
\newblock Complex wavefront corrections for deep tissue focusing using low
  coherence backscattered light.
\newblock {\em Opt. Express}, 20(15):16532--16543, Jul 2012.

\bibitem{Jang:13}
Jaeduck Jang, Jaeguyn Lim, Hyeonseung Yu, Hyun Choi, Jinyong Ha, Jung-Hoon
  Park, Wang-Yuhl Oh, Wooyoung Jang, SeongDeok Lee, and YongKeun Park.
\newblock Complex wavefront shaping for optimal depth-selective focusing in
  optical coherence tomography.
\newblock {\em Opt. Express}, 21(3):2890--2902, Feb 2013.

\bibitem{Xu11}
X.~Xu, H.~Liu, and L.V. Wang.
\newblock Time-reversed ultrasonically encoded optical focusing into scattering
  media.
\newblock {\em Nature Photonics}, 5:154--157, 2011.

\bibitem{Wang2012}
Ying~Min Wang, Benjamin Judkewitz, Charles~A. DiMarzio, and Changhuei Yang.
\newblock Deep-tissue focal fluorescence imaging with digitally time-reversed
  ultrasound-encoded light.
\newblock {\em Nature Communications}, 3, 2012.

\bibitem{Kong:11}
Fanting Kong, Ronald~H. Silverman, Liping Liu, Parag~V. Chitnis, Kotik~K. Lee,
  and Y.~C. Chen.
\newblock Photoacoustic-guided convergence of light through optically diffusive
  media.
\newblock {\em Opt. Lett.}, 36(11):2053--2055, Jun 2011.

\bibitem{Boniface:19}
Antoine Boniface, Baptiste Blochet, Jonathan Dong, and Sylvain Gigan.
\newblock Noninvasive light focusing in scattering media using speckle variance
  optimization.
\newblock {\em Optica}, 2019.

\bibitem{Dror22}
Dror Aizik, Ioannis Gkioulekas, and Anat Levin.
\newblock Fluorescent wavefront shaping using incoherent iterative phase
  conjugation.
\newblock {\em Optica}, 9(7):746--754, Jul 2022.

\bibitem{Booth2002}
Martin~J. Booth, Mark A.~A. Neil, Rimas Ju{\v s}kaitis, and Tony Wilson.
\newblock Adaptive aberration correction in a confocal microscope.
\newblock {\em Proceedings of the National Academy of Sciences},
  99(9):5788--5792, 2002.

\bibitem{Choi2015}
Sungsam Kang, Seungwon Jeong, Wonjun Choi, Hakseok Ko, {Taeseok D.} Yang, {Jang
  Ho} Joo, Jae-Seung Lee, {Yong Sik} Lim, {Q Han} Park, and Wonshik Choi.
\newblock Imaging deep within a scattering medium using collective accumulation
  of single-scattered waves.
\newblock {\em Nature Photonics}, 2015.

\bibitem{Bonora:13}
S.~Bonora and R.~J. Zawadzki.
\newblock Wavefront sensorless modal deformable mirror correction in adaptive
  optics: optical coherence tomography.
\newblock {\em Opt. Lett.}, 38(22):4801--4804, Nov 2013.

\bibitem{Antonello:20}
Jacopo Antonello, Aur\'{e}lien Barbotin, Ee~Zhuan Chong, Jens Rittscher, and
  Martin~J. Booth.
\newblock Multi-scale sensorless adaptive optics: application to stimulated
  emission depletion microscopy.
\newblock {\em Opt. Express}, 28(11):16749--16763, May 2020.

\bibitem{YeminyKatz2021}
Tomer {Yeminy} and Ori {Katz}.
\newblock {Guidestar-free image-guided wavefront-shaping}.
\newblock {\em Science Advances}, 7(21):eabf5364, 2021.

\bibitem{Stern:19}
Galya Stern and Ori Katz.
\newblock Noninvasive focusing through scattering layers using speckle
  correlations.
\newblock {\em Opt. Lett.}, 2019.

\bibitem{Daniel:19}
Anat Daniel, Dan Oron, and Yaron Silberberg.
\newblock Light focusing through scattering media via linear fluorescence
  variance maximization, and its application for fluorescence imaging.
\newblock {\em Opt. Express}, 2019.

\bibitem{Metzler23NeuWS}
Brandon~Y. Feng, Haiyun Guo, Mingyang Xie, Vivek Boominathan, Manoj~K. Sharma,
  Ashok Veeraraghavan, and Christopher~A. Metzler.
\newblock Neuws: Neural wavefront shaping for guidestar-free imaging through
  static and dynamic scattering media.
\newblock {\em Science Advances}, 9(26):eadg4671, 2023.

\bibitem{kang2023coordinatebased}
Iksung Kang, Qinrong Zhang, Stella~X. Yu, and Na~Ji.
\newblock Coordinate-based neural representations for computational adaptive
  optics in widefield microscopy, 2023.

\bibitem{PanJi2023}
Daisong Pan, Xinxin Ge, Lydia Liu, Leah Ferger, Ehud Isacoff, and Na~Ji.
\newblock {Frequency-multiplexed aberration measurement and correction for
  confocal microscopy (Conference Presentation)}.
\newblock In Thomas~G. Bifano, Na~Ji, and Lei Tian, editors, {\em Adaptive
  Optics and Wavefront Control for Biological Systems IX}, volume PC12388, page
  PC1238806. International Society for Optics and Photonics, SPIE, 2023.

\bibitem{Judkewitz14}
Benjamin Judkewitz, Roarke Horstmeyer, Ivo Vellekoop, and Changhuei Yang.
\newblock Translation correlations in anisotropically scattering media.
\newblock {\em Nature Physics}, 2014.

\bibitem{osnabrugge2017generalized}
Gerwin Osnabrugge, Roarke Horstmeyer, Ioannis~N. Papadopoulos, Benjamin
  Judkewitz, and Ivo~M. Vellekoop.
\newblock Generalized optical memory effect.
\newblock {\em Optica}, 2017.

\bibitem{Aizik_code:23}
Dror Aizik.
\newblock Non-invasive and noise-robust light focusing using confocal wavefront
  shaping code, 2023.

\bibitem{DrorArxiv}
Dror Aizik and Anat Levin.
\newblock Non-invasive and noise-robust light focusing using confocal wavefront
  shaping.
\newblock {\em preprint at arXiv:2301.11421}, 2023.

\bibitem{RevModPhys.89.015005}
Stefan Rotter and Sylvain Gigan.
\newblock Light fields in complex media: Mesoscopic scattering meets wave
  control.
\newblock {\em Rev. Mod. Phys.}, 2017.

\bibitem{single-sct-iccp-21}
C.~Bar, M.~Alterman, I.~Gkioulekas, and A.~Levin.
\newblock Single scattering modeling of speckle correlation.
\newblock In {\em ICCP}, 2021.

\bibitem{MUGNIER20061}
Laurent~M. Mugnier, Amandine Blanc, and Jérôme Idier.
\newblock Phase diversity: A technique for wave-front sensing and for
  diffraction-limited imaging.
\newblock volume 141 of {\em Advances in Imaging and Electron Physics}, pages
  1--76. Elsevier, 2006.

\end{thebibliography}
\end{document}